\documentclass[12pt]{article}

\usepackage[usenames,dvipsnames]{color}
\usepackage[abs]{overpic}
\usepackage[table]{xcolor}

\usepackage{mathptm,graphicx,rotate,color,rotating,multirow,caption,array,multicol,
  fullpage,amsmath,cite,amssymb,abstract,wrapfig,longtable,forloop,colortbl,tabu,tabularx,xcolor,comment}
 \definecolor{myRed}{rgb}{0.9019, 0.0274,  0.1647058}
 \definecolor{myGreen}{rgb}{0.4274,   0.7529,   0.28235}
 \definecolor{myBlue}{rgb}{0.2588,   0.3098,   0.643137}
 \definecolor{myCyan}{rgb}{0.4901,   0.80392,   0.8627}
 \definecolor{myMagenta}{rgb}{0.70588,   0.29019,   0.61960}
 \definecolor{myYellow}{rgb}{1,1,.1}
 \definecolor{myOrange}{rgb}{ 1,.756,.028}
 \definecolor{myBlack}{rgb}{0,0,0}
\usepackage{times,comment,graphicx,amsmath,amssymb,multicol,overpic,color,colortbl,tabularx,xcolor,cite,authblk}

\definecolor{myRed}{rgb}{1,0.8,0.8}%{0.9019, 0.0274,  0.1647058}
\definecolor{myGreen}{rgb}{0.4274,   0.7529,   0.28235}
\definecolor{myBlue}{rgb}{0.2588,   0.3098,   0.643137}
\definecolor{myCyan}{rgb}{0.88218 ,  1.447056,  1.55286}%{0.4901,   0.80392,   0.8627}
\definecolor{myMagenta}{rgb}{0.70588,   0.29019,   0.61960}
\definecolor{myYellow}{rgb}{1,1,.1}

\newcommand{\onet}{O$^*$NET }

\title{Small cities face greater impact from automation} 

\author[1]{\scriptsize Morgan R. Frank}
\author[1]{Lijun Sun} 
\author[1,2]{Manuel Cebrian}
\author[3]{Hyejin Youn}
\author[1,4,*]{Iyad Rahwan}

\affil[1]{\scriptsize Media Laboratory, Massachusetts Institute of Technology, Cambridge, Massachusetts, USA}
\affil[2]{Data61 Unit, Commonwealth Scientific and Industrial Research Organization, Melbourne, Victoria, AUS}
\affil[3]{Kellogg School of Management, Northwestern University, Evanston, Illinois, USA}
\affil[4]{Institute for Data, Systems, \& Society, Massachusetts Institute of Technology, Cambridge, Massachusetts, USA}
\affil[*]{corresponding author: irahwan@mit.edu}

\date{}

\begin{document} 

%\baselineskip24pt


\maketitle 

\begin{abstract}
    The city has proven to be the most successful form of human agglomeration and provides wide employment opportunities for its dwellers. 
    As advances in robotics and artificial intelligence revive concerns about the impact of automation on jobs, a question looms: How will automation affect employment in cities?
    Here, we provide a comparative picture of the impact of automation across U.S. urban areas.
    Small cities will undertake greater adjustments, such as worker displacement and job content substitutions.
    We demonstrate that large cities exhibit increased occupational and skill specialization due to increased abundance of managerial and technical professions. 
    These occupations are not easily automatable, and, thus, reduce the potential impact of automation in large cities.
    Our results pass several robustness checks including potential errors in the estimation of occupational automation and sub-sampling of occupations.
    Our study provides the first empirical law connecting two societal forces: urban agglomeration and automation's impact on employment.
\end{abstract}

Cities, which accommodate over half of the world's population~\cite{migrationReport}, are modern society's hubs for economic productivity~\cite{kraas2013megacities,ash2008reimagining,montgomery2008urban} and innovation~\cite{bettencourt1,bettencourt2,hyejin}.
Since job migration is the leading factor in urbanization~\cite{migrationReport,rozenblat2007firm}, policy makers are increasingly concerned about the impact of Artificial Intelligence and automation on employment in cities~\cite{carlsson2012technological,olsen2014rise,acemoglu2017robots}.
While researchers have investigated automation in national economies and individual employment, it remains unclear \emph{a priori} how cities naturally respond to this threat. In a world struggling between localism and globalism, a question emerges: \emph{How will different cities cope with automation?} Answering this question has implications on everything from urban migration to investment, and from social welfare policy to educational initiatives.

%, or what factors can maximize efficiency benefits while minimizing technological unemployment in cities.
%In particular, several urban indicators exhibit pervasive relationships to city size~\cite{hyejin,mace2008reproducing,batty2008size}, which motivates an investigation across U.S. cities into the mechanisms that determine the impact of automation.

%%% how to estimate automation's impact? Why do skills matter?
%The differential impact of automation across cities remains unknown, and 

To construct a comparative picture of automation in cities, our first challenge is to get reliable estimates of how automation impacts workers.
Existing estimates are wide ranging. 
Frey and Osborne~\cite{freythe2013} estimate that 47\% of U.S. employment is at ``high risk of computerization" in the foreseeable future, while an alternative OECD study concludes a more modest 9\% of employment is at risk~\cite{arntz2016risk}.
Note that these results do not tell us about the impact of automation in cities as they are presented at a national level.
Differences in these predictions arise from discrepancies over two main skill dynamics: the substitution of routine skills, and complementarity of non-routine and communication skills~\cite{david2001skill,autor2015there,BrynjolfssonMcAfee}. 
Additionally, technology-driven efficiency may redefine the skill requirements of occupations and actually increase employment in low-skilled jobs~\cite{bessen2015computer,bessen2015learning}.
%We can proceed using these estimates to analyze cities knowing that the robustness of our results will need to be validated under the potential for error in these estimates.

Nevertheless, even if we take current estimates of the \emph{absolute} risk of computerization of jobs with skepticism, these estimates can provide useful guidance about \emph{relative} risk to different cities that is robust to errors in the estimates provided by \cite{freythe2013} and \cite{arntz2016risk}. 
We can interpret the `risk of computerization' estimates as an educated guess about which occupations will experience greater adjustment due to machine substitution of a large portion of their content.
These adjustments represent a significant cost to an urban system from both technological unemployment and expensive worker retraining programs.

%%% Both labor dynamics may be apparent in cities, and they both come with a cost
%Combined, these factors detail a complicated picture.
% via Manuel...
%Will cities generate non-automatable employment through the skill complementarity of their workforce, thus sustaining current urbanization? Or, to the contrary, will automation substitute for human labor, thus diminishing job availability and ultimately diminishing urbanization? Either way, both technological unemployment and costly worker retraining represent a shock to the \emph{status quo} in cities, and must be accounted for when quantifying automation's impact on the sustainability of urban labor systems.

%%% Are economically successful cities diverse or specialized?
A priori, it is not obvious whether large cities will experience more or less impact from automation. 
On one hand, an influx of occupational diversity explains the wealth-creation, innovation, and success of cities~\cite{glaeser2011triumph,quigley1998urban,henderson1991urban,pan2013urban}.
%, and is evidenced by the increased number of industries, and occupation types in large cities~\cite{hyejin, bettencourt2014professional} (see Supplementary Materials, Fig. S1B).
On the other hand, cities connect people with greater efficiency~\cite{pan2013urban,sim2015great}.
This enables a greater division of labor that increases overall productivity~\cite{smith1976inquiry,bettencourt2014professional,sveikauskas1975productivity} through occupational specialization.
%As evidence, R\&D employment has been shown to scale superlinearly with city size~\cite{bettencourt2}.
However, the division of labor may facilitate automation as it identifies routine tasks and encourages worker modularity. 
If these modular jobs are at greater risk of computerization, then more workers may be impacted by automation in large cities. 
These observations pose a puzzle: \emph{are the forces of diversity, specialization, and the division of labor shaping a city's ability to accommodate automation?}

%It is not clear \emph{a priori} how cities naturally respond to automation.
%Will cities generate non-automatable employment through the skill complementarity of their workforce, thus sustaining current urbanization?
%Or, to the contrary, will automation substitute for human labor, thus diminishing job availability and ultimately diminishing urbanization?
Here, we undertake a comparative examination of cities while measuring the relative impact of automation on employment. 
We also contextualize these measurements through a detailed analysis of the skill composition of different cities.
Note that \emph{impact} includes unemployment, but may also manifest itself through the changing skill demands of occupations as automation diminishes the need for individual types of skills~\cite{bessen2015computer,bessen2015learning}. 
In light of imminent automation technology, we highlight a complicated relationship between labor diversity and specialization in cities, and discover that small cities are susceptible to the negative impact of automation.

%%%%%%%%%%%%%%%%%%%%%%%%%%%%%%%%%%%%%%%%%%%%
\section*{Results}
\subsection*{The Expected Job Impact of Automation in Cities}
\begin{figure*}[!p]
	\centering
	\begin{overpic}[width=.47\textwidth]{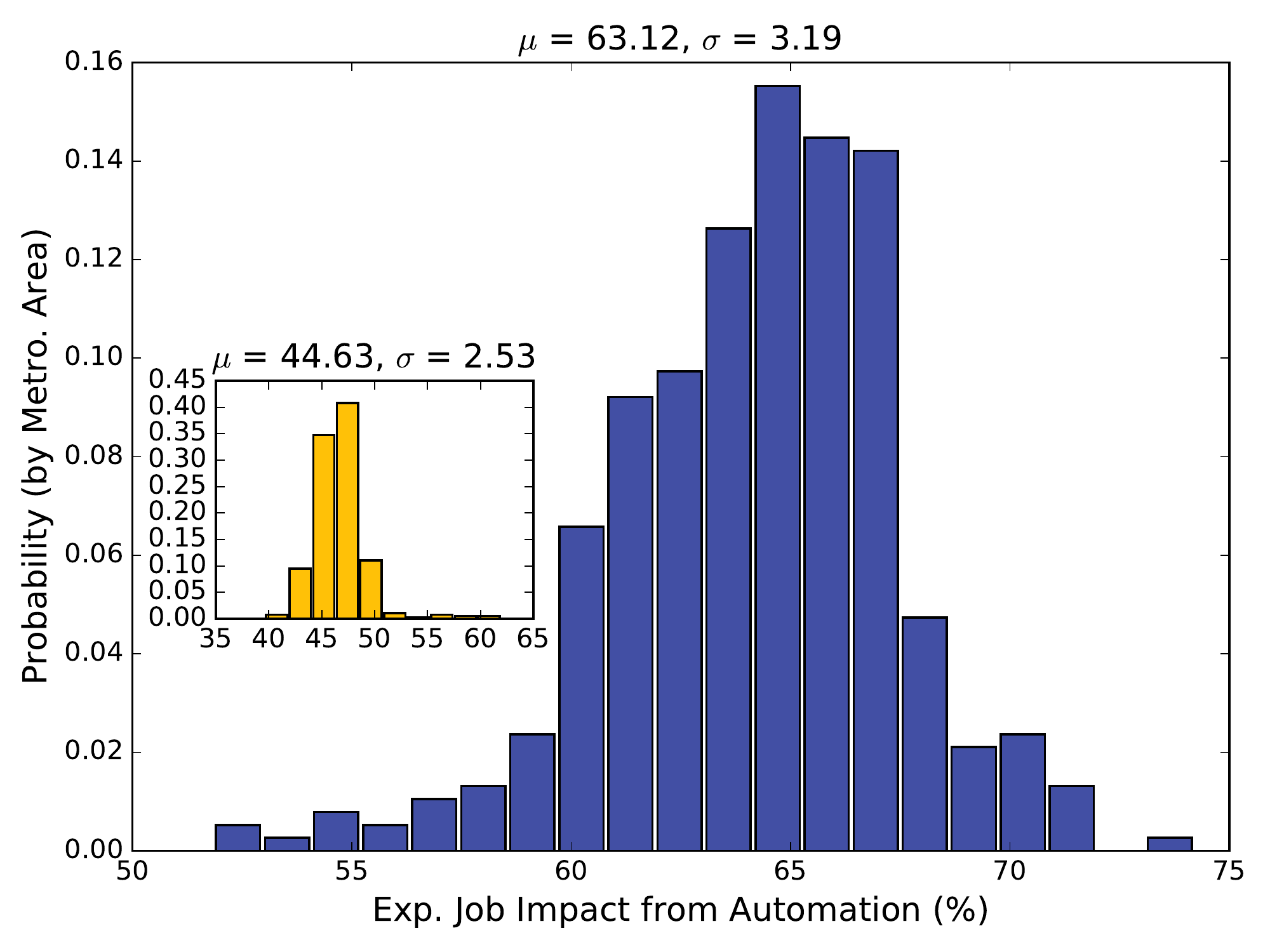}
		\put(30,140){\fbox{\small A}}
	\end{overpic}
	\begin{overpic}[width=.47\textwidth]{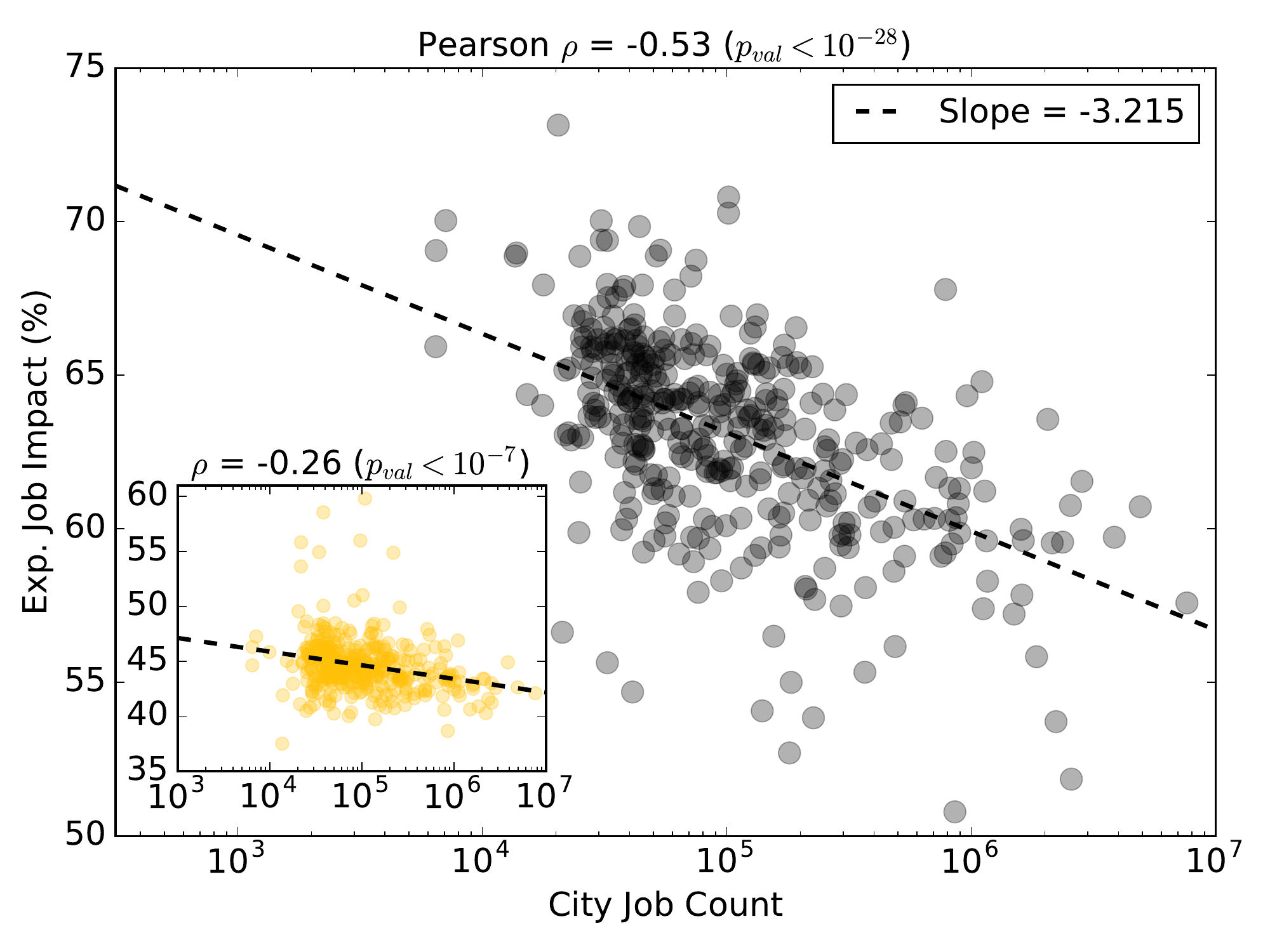}
		\put(30,140){\fbox{\small B}}
	\end{overpic}\\
	\begin{overpic}[width=.95\textwidth,trim=0cm .5cm 0cm 0cm,clip]{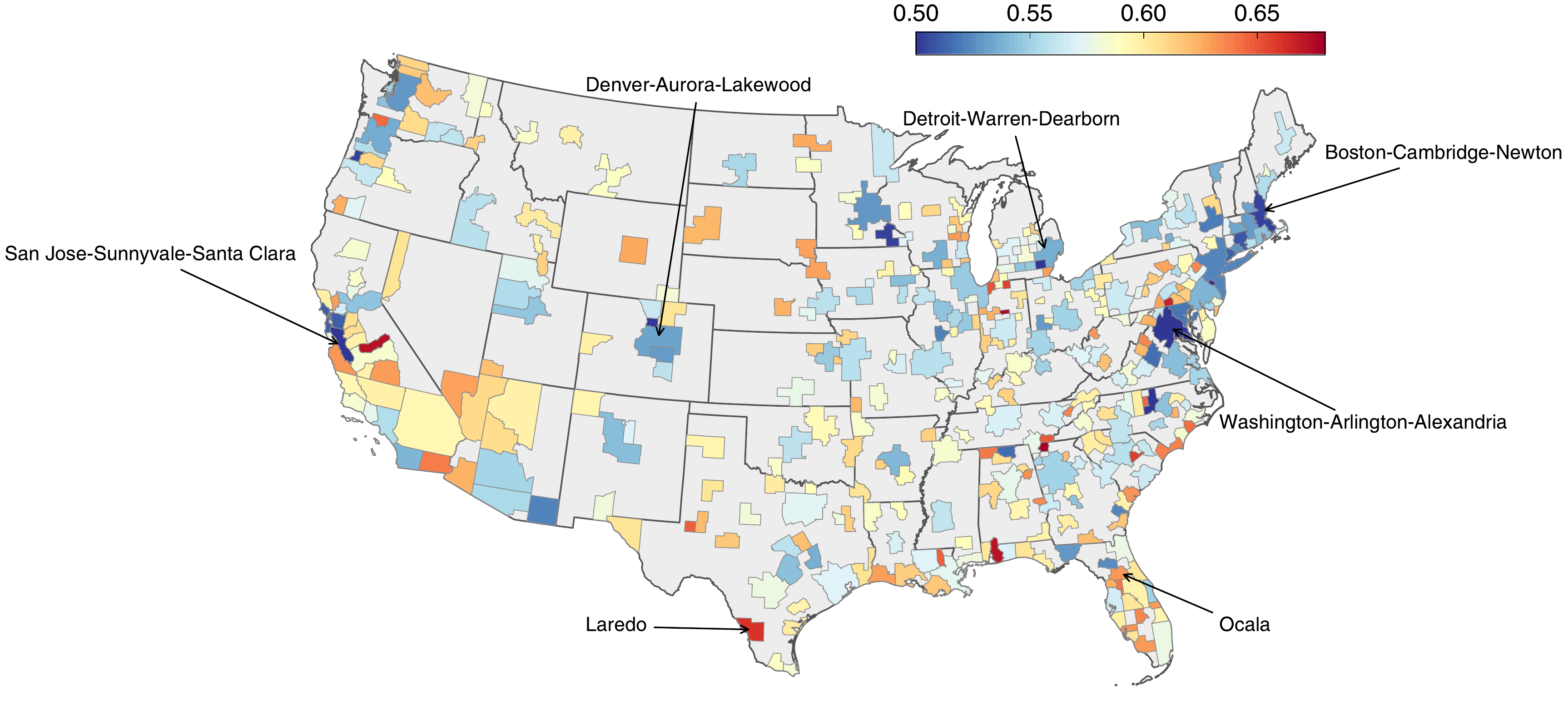}
	    \put(27,140){\fbox{\small C}}
    \end{overpic}
	\caption{
		The impact of automation in U.S. cities.
		{\bf (A)} The distribution of expected job impact ($E_m$) from automation across U.S. cities using estimates from \cite{freythe2013}. 
	    (Inset) The distribution using alternative estimates~\cite{arntz2016risk}.
		{\bf (B)} Expected job impact decreases logarithmically with city size using estimates from \cite{freythe2013}.
		We provide the line of best fit ($\text{Slope}=-3.215$) with Pearson correlation to demonstrate significance.
		(Inset) Decreased expected job impact with increased city size is again observed using alternative estimates~\cite{arntz2016risk} (best fit line has slope $-1.24$, Pearson $\rho=-0.26$, $p_{val}<10^{-7}$).
		{\bf (C)} A map of U.S. metropolitan statistical areas colored according to expected job impact from automation.
	}
	\label{jobDispFig}
\end{figure*}

We estimate automation's expected impact on jobs in cities according to 
\begin{equation}
    E_m =\displaystyle\sum_{j\in Jobs}p_{auto}(j)\cdot share_m(j), 
    %\frac{\displaystyle\sum_{j\in Jobs_m}p_{auto}(j)\cdot f_m(j)}{\displaystyle\sum_{j\in Jobs}f_m(j)},
\end{equation}
where $Jobs$ denotes the set of occupations, $share_m(j)$ denotes the employment share in city $m$ with occupation $j$ according to the U.S. Bureau of Labor Statistics (BLS), and $p_{auto}(j)$ denotes the probability of computerization for occupation $j$ as estimated by \cite{freythe2013} (see Supplementary Materials S3 for more details).
%using the probabilities of computerization produced in \cite{freythe2013} and data from the U.S. Bureau of Labor Statistics (BLS) in Figure~\ref{jobDispFig} (see S3 for calculation).
Each city should expect between one-half and three-quarters of their current employment to be affected in the foreseeable future due to improvements in automation (see Fig.~\ref{jobDispFig}A. Also note that this estimate differs from \cite{freythe2013} which focused on national statistics).
While this calculation omits potential job creation or job redefinition which typically accompany innovation~\cite{mazzucato2013financing,archibugi2013economic}, it highlights the differential impact of automation across cities and smooths potential noise in the predicted automation of individual jobs.
Expected job impact may represent employment loss or changes in the type of work performed by those workers (e.g. see \cite{bessen2015computer,bessen2015learning,acemoglu2017robots}).

What differentiates cities' resilience to automation?
Figure~\ref{jobDispFig}B demonstrates that expected job impact decreases according to $E_m\propto-3.2\times\log_{10}(\text{city size}),$ which suggests that larger cities are more resilient to the negative effects of automation.
This relationship is significant with a Pearson correlation $\rho=-0.53$ ($p_{val}<10^{-28}$), and shows that laborers in smaller cities are susceptible to the impact of automated methods ($R^2=0.28$).
We confirm our finding using separate conservative skill-based estimates of the automatability of jobs~\cite{arntz2016risk} ($\text{Pearson }\rho=-0.26\text{ }(p_{val}<10^{-7})$ and $E_m\propto-1.24\times\log_{10}(\text{city size})$. See Fig.\ref{jobDispFig}B inset and Supplementary Materials S3.2).
Despite the conservative nature of these alternative probabilities, we again observe increased resilience with city size.
Furthermore, we demonstrate in the Supplementary Materials S3.1 that the observed negative trend relating city size to expected job impact from automation is robust to errors in the probabilities of computerization (i.e. $p_{auto}$) produced by \cite{freythe2013} and robust to random removal of occupations from the analysis.
%This prompts us to quantify urban factors which explain a city's resilience further.
%
%Existing work identifies three strategies available to individual workers facing automation~\cite{maccroy}. 
%First, laborers can compete with automation by performing tasks better than automated methods.
%Second, they can complement automation by procuring skills required to use and develop automation technology.
%Finally, they can seek industries where automation has no foreseeable impact.
%Having observed that cities grow to be resilient, we now examine which strategy cities employ.

\subsection*{Labor Specialization in Large Cities}
\begin{figure*}[!p]
	\centering
	\begin{overpic}[width=.45\textwidth]{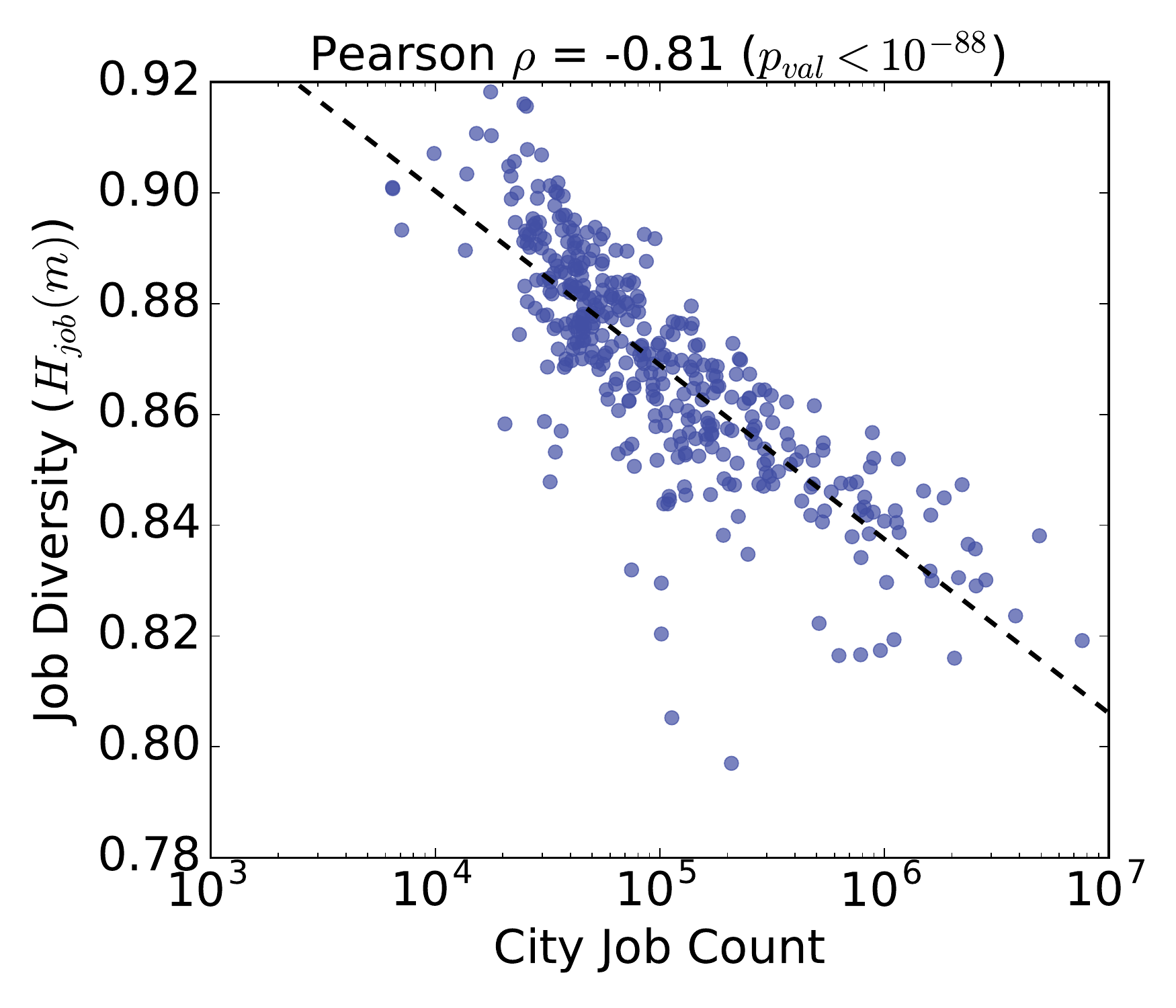}
		\put(40,145){\fbox{\small A}}
	\end{overpic}
	\begin{overpic}[width=.45\textwidth]{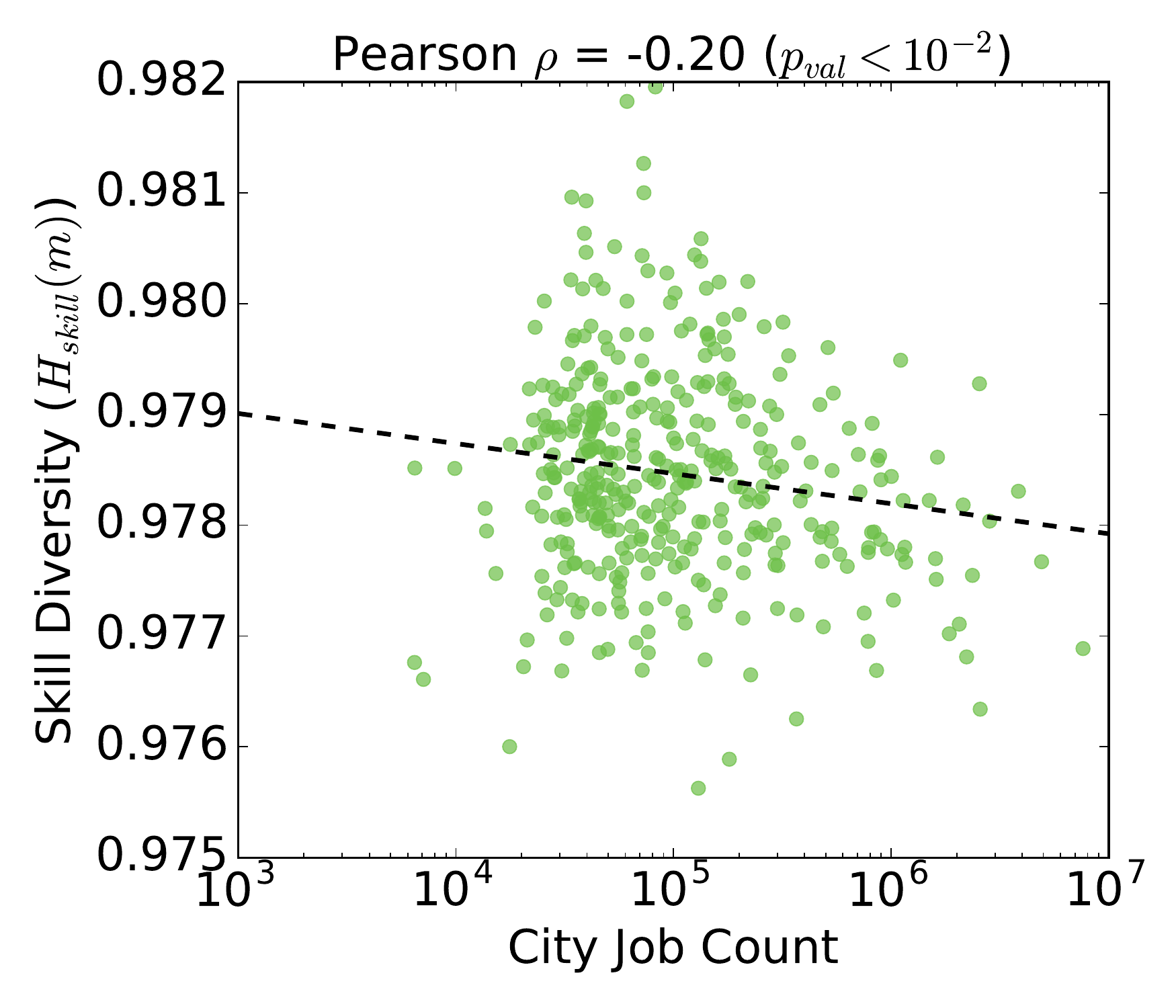}
		\put(45,145){\fbox{\small B}}
	\end{overpic}
	\begin{overpic}[width=.45\textwidth]{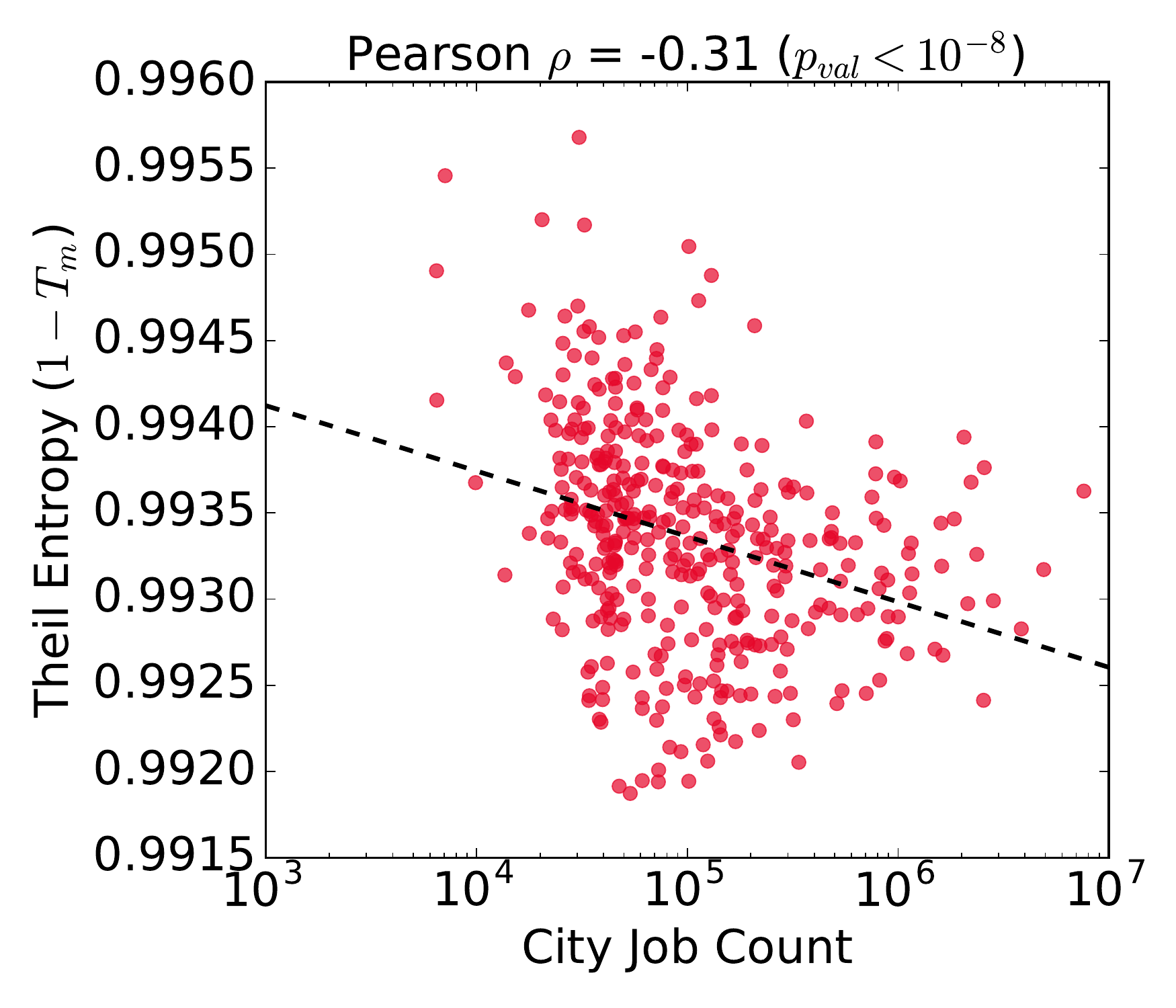}
		\put(50,145){\fbox{\small C}}
	\end{overpic}
%    \begin{overpic}[height=5cm]{figures/MLS_all.pdf}
%        \put(17,60){\fbox{\small D}}
%    \end{overpic}
%    \begin{overpic}[height=4.75cm]{figures/residualPlot.pdf}
%        \put(20,50){\fbox{\small E}}    
%    \end{overpic}
	\caption{
		Large cities reveal increased occupational specialization through both job and skill distributions.
		{\bf (A)} Shannon entropy of job distributions, $H_{job}(m)$, decreases with city size.
		{\bf (B)} Shannon entropy of the \onet skill distributions, $H_{skill}(m)$, decreases with city size.
		{\bf (C)} Theil entropy, $T_m$, reveals the proportion of specialized jobs increases with city size.
		For plots (A), (B), \& (C), we provide the line of best fit for reference.
%		{\bf (D)} A multiple linear regression model utilizing measures for labor specialization accounts for 66\% of the variance in expected job impact from automation across U.S. cities (see SI Appendix, Section 4 for details). 
%        {\bf (E)} The distribution of residuals between the actual and predicted values of the model, and the rank of some example cities.
	}
	\label{specializationEvidence}
\end{figure*}

We explore the mechanisms underpinning resilience to automation by examining the most distinctive characteristics of urban economies: diversification and specialization.
Since automation typically targets workplace skills~\cite{arntz2016risk}, we consider the \onet skill dataset, which relates occupations to their constituent workplace tasks and skills, in addition to employment data.
For large cities, specialization (i.e. decreased Shannon entropy) appears in the employment distributions across occupations (Fig.~\ref{specializationEvidence}A) and, separately, in the aggregate distributions of skills  (Fig.~\ref{specializationEvidence}B).
Additionally, we use Theil entropy to measure the proportion of specialized jobs (in terms of skills) in comparison to the skill specialization of the city on whole.
Figure~\ref{specializationEvidence}C demonstrates an increasing proportion of specialized jobs in large cities (i.e. $1-T_m$ decreases).
See Materials and Methods for calculations of entropy measures.

\begin{figure*}[!p]
    \centering
    \tiny
    \begin{tabular}{ccccccccc}
        \hline
        Model & 
            (1) & 
                (2) & 
                    (3) & 
                        (4) & 
                            (5) & 
                                (6) & 
                                    (7) & 
                                        (8) \\ \hline
        Variable & \multicolumn{8}{c}{Coefficient (Standard Error)} \\ \hline
        $size_m$ & 
            $0.009$ &
                &
                    &
                        & 
                            -0.016 & 
                                0.011 & 
                                    0.012 & 
                                        -0.013 \\
                 & 
                 ($<10^{-4}$)&
                 & 
                    &
                        & ($<10^{-4}$) & 
                            ($<10^{-4}$)& 
                                ($<10^{-4}$)& 
                                    ($<10^{-4}$) \\
        $income_m$ & 
            $-0.016$ &
                &
                    &
                        & 
                            $-0.013$ & 
                                $-0.013$ & 
                                    $-0.015$ & 
                                        $-0.012$ \\
                 & 
                    ($<10^{-5}$) &
                        &
                            &
                                & 
                                    ($<10^{-5}$) & 
                                        ($<10^{-5}$) & 
                                            ($<10^{-5}$) & 
                                                ($<10^{-5}$) \\
        $bachelor_m$  & 
            $-0.005$ &
                &
                    &
                        & 
                            -0.005 & 
                                -0.002 & 
                                    -0.004 & 
                                        -0.002 \\
                 & 
                    ($10^{-5}$) &
                        &
                            &
                                & 
                                    ($<10^{-5}$) & 
                                        ($<10^{-5}$) & 
                                            ($<10^{-5}$) & 
                                                ($<10^{-5}$) \\
        $GDP_m$ & 
            $0.001$ &
                &
                    &
                        & 
                            0.003 & 
                                -0.001 & 
                                    -0.001 & 
                                        0.001 \\
                & 
                    ($<10^{-5}$) &
                        &
                            &
                                & 
                                    ($<10^{-5}$) & 
                                        ($<10^{-5}$)& 
                                            ($<10^{-5}$) & 
                                                ($<10^{-5}$) \\
        $jobs_m$ & 
            $-0.015$ &
                &
                    &
                        & 
                            -0.002 & 
                                -0.017 & 
                                    -0.020 & 
                                        0.001 \\
                 & 
                    ($<10^{-4}$) &
                        &
                            &
                                & 
                                    ($<10^{-4}$)& 
                                        ($<10^{-4}$) & 
                                            ($<10^{-4}$) & 
                                                ($<10^{-4}$) \\
        $H_{job}(m)$ &
            & 
                $0.006$ &
                    &
                        & 
                            -0.012 &
                                &
                                    & 
                                        -0.010 \\
        &
            & 
                ($<10^{-5}$)&
                    &
                        & 
                            ($<10^{-4}$) &
                                &
                                    & 
                                        ($<10^{-5}$)\\
        $H_{skill}(m)$ &
            &
                & 
                    $0.014$ &
                        &
                            & 
                                0.009 &
                                    & 
                                        0.019 \\
        &
            &
                & 
                    ($<10^{-5}$)&
                        &
                            & 
                                ($<10^{-5}$) &
                                    & 
                                        ($<10^{-5}$) \\
        $(1-T_m)$ &
            &
                &
                    & 
                        $-8\times10^{-5}$ &
                            &
                                & 
                                    -0.005 & 
                                        0.012 \\ 
        &
            &
                &
                    & 
                        ($<10^{-5}$)&
                            &
                                & 
                                    ($<10^{-4}$) & 
                                        ($<10^{-5}$) \\
        %Intercept &&&&&&&& \\ 
        \hline
        Sample Size & 302 & 302 & 302 & 302 & 302 & 302 & 302 & 302 \\
        p-value &$<10^{-10}$ & 0.0003 & $<10^{-10}$& 0.96 & $<10^{-10}$ & $<10^{-10}$ & $<10^{-10}$ & $<10^{-10}$ \\
        $R^2$ & 0.534 & 0.0429 & 0.197 & $<10^{-5}$& 0.570 & 0.60 & 0.557 & 0.660 \\
        Adjusted $R^2$ & 0.534 & 0.0429 & 0.197 & $<10^{-5}$ & 0.570 & 0.60 & 0.557& 0.660 \\ \hline
        %\multicolumn{9}{c}{$^{*}$ p-value $< .1$, $^{**}$ p-value $< .01$. $^{***}$ p-value $< .001$} \\ \hline
    \end{tabular}
    \begin{overpic}[width=.43\textwidth,trim=0cm 0cm -1cm 0cm]{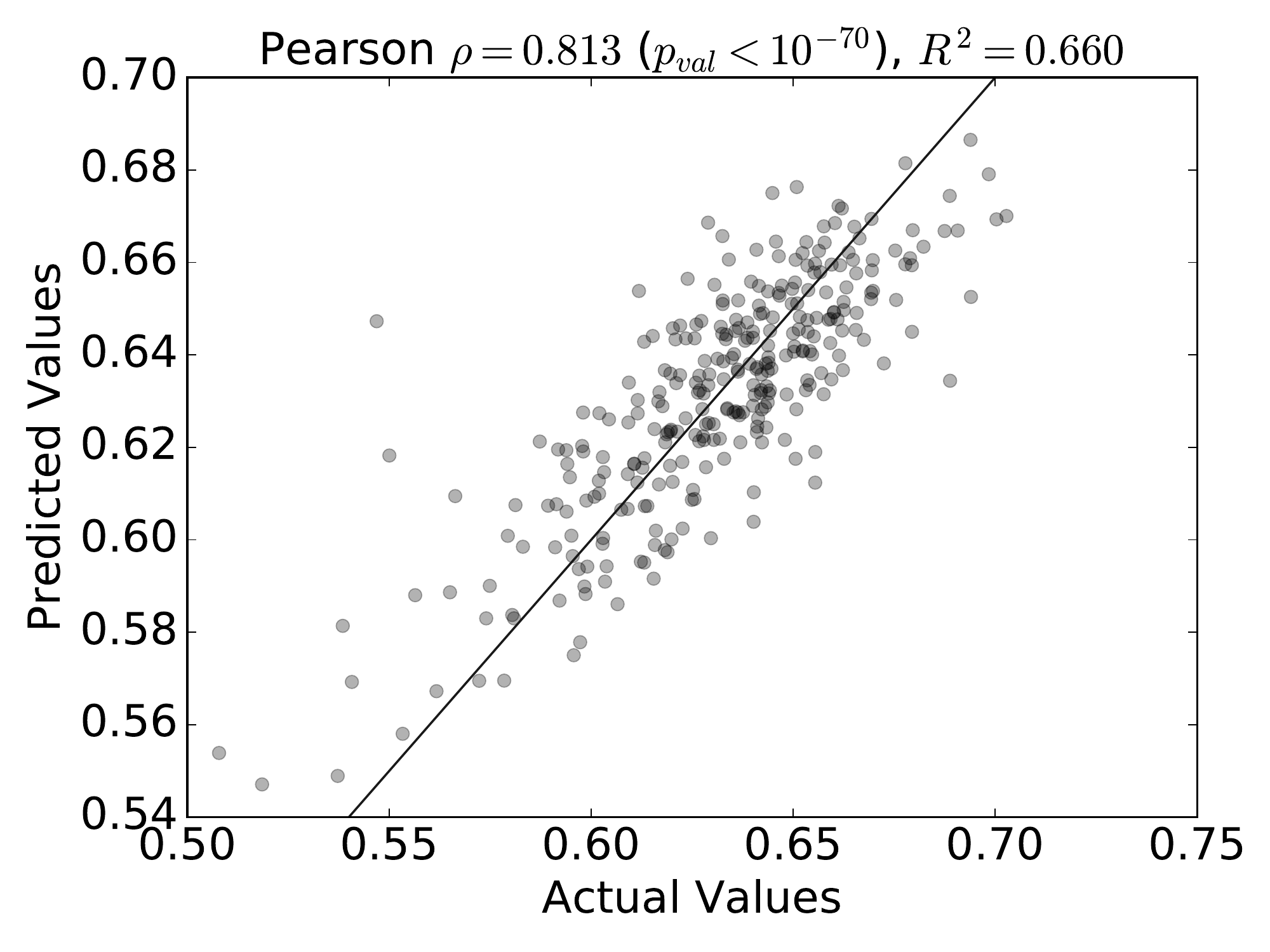}
        \put(35,110){\fbox{\small B}}
        \put(25,270){\fbox{\small A}}
    \end{overpic}
    \begin{overpic}[width=.47\textwidth]{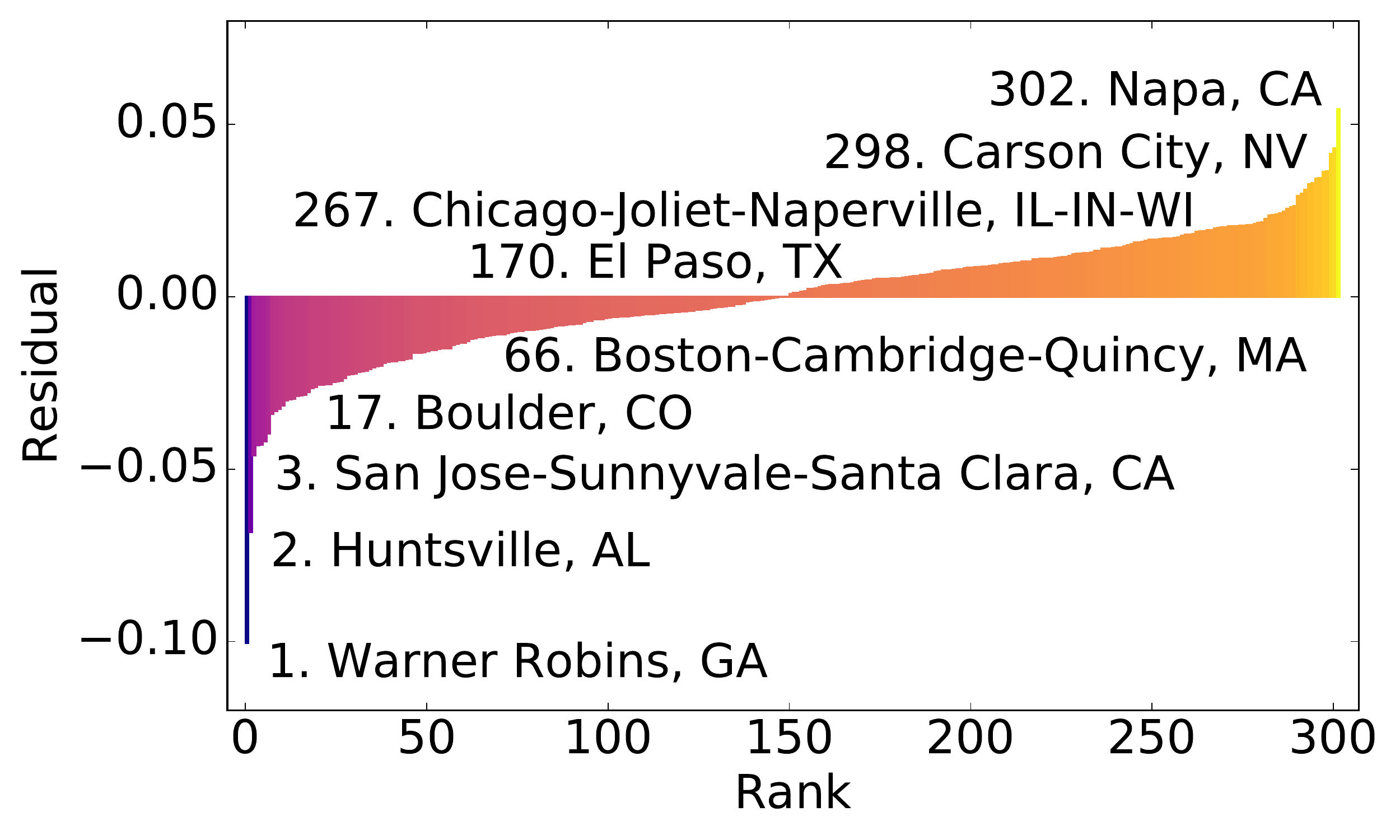}
        \put(40,110){\fbox{\small C}}    
    \end{overpic}
    \caption{
        %Examining the predictive power of specialization measures in multiple linear regression models of expected job impact in cities ($E_m$).
        Labor specialization can model expected job impact ($E_m$) in cities.
        {\bf (A)} A multiple linear regression analysis for predicting $E_m$ that considers generic urban indicators including $\log_{10}$ city total employment ($size_m$), median annual household income ($income_m$), percentage of population with a bachelor's degree ($bachelor_m$), $\log_{10}$ GDP per capita ($GDP_m$), and the number of unique occupations ($jobs_m$).
        All variables have been standardized.
        %All reported coefficients are significant with $p_{val}<10^{-3}$.
        {\bf (B)} The actual $E_m$ values for each city plotted against the predicted values using Model 8 from (A), which captures 66\% of the variance in expected job impact from automation across U.S. cities (see Supplementary Materials, S4 for additional analysis). 
        {\bf (C)} The distribution of residuals between the actual and predicted values from Model 8, and the rank of some example cities.
    }
    \label{MlsVarImportance}
\end{figure*}

Using these specialization measures in three separate linear regression models reveals that skill specialization is predictive of expected job impact in cities ($R^2 = 0.20$).
%see SI Appendix, Section 4).
Multiple linear regression produces the most predictive model accounting for 66\% of the variance across cities (see Fig.~\ref{MlsVarImportance}A\&B).
%~\ref{specializationEvidence}D).
This model relies most strongly on skill specialization while controlling for several generic urban factors, such as per capita GDP, city size, and education levels, and improves on the base model using only these generic urban factors without specialization measures ($R^2=0.53$).
We confirm the stability of our regression results by alternatively training the regression model on half of the cities and measuring the performance of the regression on the remaining cities as validation (see Supplementary Materials S4).

The residuals between the actual and modelled values highlight notably resilient cities (given the model), such as Boulder, C.O. and Warner Robins, G.A., and notably susceptible cities, such as Napa, C.A. and Carson City, N.V. (see Fig.~\ref{MlsVarImportance}C).
%~\ref{specializationEvidence}E).
Examining these cities more closely may allow urban policy experts with a nuanced understanding of the policies in these cities to more easily identify causal mechanisms.
The predictive power of this model and its reliance on workplace skills justifies our inclusion of skills data in addition to occupation data, and motivates us to characterize urban resilience to automation from the skills in cities.

\subsection*{How Occupations and Workplace Skills Change with City Size}

How do different types of occupations change with city size~\cite{pumain2004scaling}, and how do these changes contribute to the differential impact of automation across cities?
%Given the differential impact of automation and levels of specialization across cities, we want to know how employment in different types of occupations changes with city size~\cite{pumain2004scaling}.
While it is tempting to look only for the largest changes in employment share, more subtle differences for very automatable, or very not automatable, occupations can also produce big changes in expected job impact.
We capture this confounding effect by decomposing the difference in expected job impact of cities $m$ and $n$ according to
\begin{align}
    \begin{split}
        E_m-E_n &= \displaystyle\sum_{j\in Jobs}p_{auto}(j)\cdot\big(share_m(j)-share_n(j)\big)\\
            &= \sum_{j\in Jobs}\big(p_{auto}(j)-E_n\big)\cdot\big(share_m(j)-share_n(j)\big),
    \end{split}
    \label{occShift}
\end{align}
where we have profited from $\sum E_n\cdot(share_m(j)-share_n(j))=0$.
%, and introduced the notation $\delta_{m,n}(j)$ to denote the contribution of occupation $j$ to the difference in expected job impact.
We consider the percentage of the difference explained by occupation $j$ according to
\begin{equation}
    \delta_{m,n}(j) = 100\cdot\frac{\big(p_{auto}(j)-E_n\big)\cdot\big(share_m(j)-share_n(j)\big)}{E_m-E_n}.
\end{equation}

Occupation $j$ can increase or decrease the overall difference in expected job impact depending on the sign of the corresponding term in equation (\ref{occShift}), or, equivalently, the sign of $\delta_{m,n}(j)$.
In turn, this sign depends on the relative automatability of the occupation and the relative employment share.
More details for this calculation and an example analysis comparing individual cities are provided in the Supplementary Materials S3.4.

In Figure~\ref{citySizeShift}, we employ an ``occupation shift" to visualize the contributions of each occupation to the difference in expected job impact in large and small cities.
After adding the employment distributions for the 50 largest cities and 50 smallest cities together, respectively, we calculate $\delta(j)$ for each occupation.
Each occupation is assigned a quadrant and color based on the sign of $\delta(j)$ and the relative automatability of occupation $j$.
This visualization identifies both occupations that increase the differential impact (i.e. occupations on the right) and occupations that decrease the differential impact (i.e. occupations on the left).
For example, increased employment for Cashiers, which is relatively susceptible to automation, in small cities contributes the most to the overall difference in expected job impact.
Likewise, differences in employment for Software Developers, a relatively resilient occupation, also increases the overall difference.
On the other hand, increased employment for Elementary School Teachers, which is another relatively resilient occupation, in small cities decreases the difference.
On aggregate, differences in employment for occupations that are relatively resilient to automation contribute the most to the differential impact of automation in large and small cities (see Fig.~\ref{citySizeShift} inset). 

To explore the role of resilient occupations further, we focus on how employment for different occupation types changes with city size.
We use K-means clustering algorithm (i.e. occupations are instances and raw \onet skill importance are features) to identify five clusters of jobs according to skill similarity (see Fig.~\ref{citySizeShift} occupation labels and Fig.~\ref{jobScaling}A. The complete list of occupations is provided in Supplementary Materials S6.3) and examine the scaling relationship between job clusters and city size according to $(\text{number of workers})\propto (\text{city size})^\beta$ in Figure~\ref{jobScaling}B.
Note that the exponent, $\beta$, entirely describes the growth rate of these job clusters relative to city size.
The job cluster comprised of highly specialized jobs, such as Mathematician and Chemist, exhibits a notably superlinear scaling relationship with city size ($\beta=1.39$).
This scaling exponent is similar to the scaling relationship observed for \emph{Private R\&D employment} ($\beta=1.34$) found in \cite{bettencourt2} and is in good agreement with similar studies on job growth~\cite{bessen2015computer}.
Furthermore, our finding of one job cluster exhibiting notably larger scaling than the other job clusters is stable to sub-sampling occupations at various rates (see Supplementary Material S6.3.2).
Managerial jobs also grow superlinearly, but to a weaker extent ($\beta=1.08$).
The job cluster exhibiting the slowest growth ($\beta=0.94$) is comprised of entertainment and service jobs.
We check the robustness of these scaling relationships using methods from \cite{leitao2016scaling} (see Supplementary Materials S6.3.3).

In Figure~\ref{jobScaling}C, we quantify each job cluster's contribution to the differential impact of automation across large and small cities according to
\begin{equation}
    %\frac{\displaystyle\sum_{j\in \text{Job Cluster}}\big(p_{auto}(j)-E_{\text{Large Cities}}\big)\cdot\big(share_{\text{Small Cities}}(j)-share_{\text{Large Cities}}(j)\big)}{E_{\text{Small Cities}}-E_{\text{Large Cities}}},
    \Delta_{\text{Small Cities},\text{Large Cities}}(\text{\scriptsize Job Cluster}) =\sum_{j\in\text{Job Cluster}}\delta_{\text{Small Cities},\text{Large Cities}}(j).
    \label{occTypeShift}
\end{equation}
%where have added the employment distributions of the 50 largest cities and the 50 smallest cities together, respectively.
The low automatability and high difference in employment of highly specialized job cluster (represented by purple) in large and small cities indeed explains a significant amount of the difference in expected job impact.
However, we also find that the more susceptible occupations represented by the blue job cluster in Figure~\ref{jobScaling} accounts for a similar proportion of the difference.
Interestingly, the differences in occupations from the yellow job cluster serve to decrease the differential impact of automation between large and small cities.
These conclusions are supported by the analysis on individual occupations presented in Figure~\ref{citySizeShift}.

We confirm that the fastest growing job cluster is indeed comprised of ``technical'' jobs from their constituent workplace skills.
We employ K-means clustering (i.e. \onet skills are instances and the correlation of raw \onet importance of skills across occupations are features) to simplify the complete space of \onet skills to ten skill types based on the co-occurrence of skills across jobs (see Supplementary Materials S6.5 for complete description of skill clusters). 
These simplified skill types allow us to intuitively explore which skills indicate specialization or indicate resilience in cities.
Computational/Analytical skills and Management skills are more likely in faster growing (i.e. superlinear) jobs, while physical skills, such as Physical Coordination and Control/Perceptual skills, indicate notably slower job growth with city size (Fig.~\ref{jobScaling}D).
We confirm our findings using alternative definitions for aggregate workplace tasks and skills (see Supplementary Materials S5).

%To answer this question, we employ K-means clustering to simplify the space of skills from 230 \onet skills to ten skill types based on the co-occurrence of skills across jobs (further details in S5, and the complete set of skills comprising each skill type is presented in the S6.6). 
%These simplified skill types allow us to intuitively explore which skills indicate specialization (Fig.~\ref{specializationEvidence}D\&E) or indicate resilience (Fig.~\ref{specializationEvidence}F\&G) in cities.
The skills which are relied on by fast-growing technical jobs suggest mechanisms for resilience and growth in cities.
%Furthermore, examining workplace skills links technical job growth to resilience to automation in cities.
Existing work~\cite{maccroy} identifies that individual workers can gain skills to compete with automation, gain skills to complement automation, or seek industries removed from the impacts of automation.
Similar to individual workers, the division of labor in large cities allows them to specialize in skills removed from the threat of automation.
Computational/Analytical, Managerial, Organization, and Relational skills are more likely to be present in specialized and resilient cities (Fig.~\ref{skillTable}A\&C), while Physical Coordination and Control/Perceptual skills indicate both decreased specialization and decreased resilience in cities (Fig.~\ref{skillTable}B\&D).
We confirm our results using alternative groups of workplace tasks~\cite{kok2014cities} provided by \onet (see Supplementary Materials S5.1) and again by examining the routineness of workplace tasks~\cite{david2001skill} (see Supplementary Material S5.2).
Figure~\ref{skillTable}E reflects the same conclusion by comparing the relationship of each skill type to city size (right column) and expected job impact (middle column) (see Supplementary Materials S6.4 for comparison with raw \onet skills).
Effectively, large cities employ workers whose skills better prepare them to interface with automation technology, while small cities rely more prominently on physical workers, who are more susceptible to automation.
%These trends support the division of labor theory by highlighting the increased abundance of technical and managerial skills with city size.
%through the positive effects indicated by managerial and organizational skills, along with specialized technical skills.

\section*{Discussion}
Cities are modern society's hubs for economic productivity and innovation.
However, the impact of automation on employment in cities threatens to alter urbanization, which is largely driven by employment opportunity.
Fortunately, urbanization itself appears to contain a mitigating solution.
% for the detrimental impact of automation on employment. 
It is difficult to concretely identify causal mechanisms at the scale of this investigation, but, despite this difficulty, we highlight evidence for the division of labor in large cities and show its importance as a piece of the automation and urbanization puzzle.

%Our analysis explores the role of labor diversity and specialization across cities of different sizes.
In particular, large cities have more unique occupations and industries~\cite{hyejin}, but distribute employment less uniformly across those occupations.
This juxtaposition of both diversity and specialization in large cities is reconcilable through the division of labor theory~\cite{smith1976inquiry}.
Under the division of labor argument, large firms have better ability to support specialized workers along with the management required to coordinate them~\cite{coase1937nature}.
To this end, we find that average number of workers per firm increases logarithmically with city size (see Supplementary Materials Fig. S1A).
At the same time, workers possessing specialized skills seek specific employment opportunities which maximize their financial return~\cite{bloom2008urbanization,schich2014network}.
The demand for specific specialized jobs increases occupational specialization while also increasing the number of unique job types and industries in a city~\cite{rozenblat2007firm}.

What do large cities specialize in and why? 
The division of labor encourages worker modularity, which has the potential to impact whole groups of workers who are competing with automation technology.
Therefore, specialization alone is not enough to explain the resilience to automation impact that we observe across cities.
For example, Detroit, which is famous for its specialization in automotive manufacturing, has experienced economic down turn~\cite{klier2009tail}, while the San Francisco Bay area, epicenter of the information technology industry, continues to flourish despite the dot-com bubble (perhaps due to its support of a ``creative class" of workers~\cite{florida2004rise}).
%and did not suffer during the dot-com bubble~\cite{florida2004rise}.
Our analysis highlights specific occupations, such as Mathematician and Chemist, as well as specific types of skills, such as Computational/Analytical skill, that explain the increased resilience of large cities.
These occupations and skills may inform policy makers in small cities as they identify new industries and design worker retraining programs to mitigate the negative effects of automation on employment.

%It is important to put our results in context. 
By quantifying relative \emph{impact}, we provide an upper bound for \emph{technological unemployment} in cities.
Changing labor demands produce systemic effects, which make it difficult to precisely predict employment loss~\cite{autor2015there}.
%For example, automation's historical effects on total employment have been driven by systemic effects, in addition to job-level effects~\cite{autor2015there}. 
For example, the introduction of Automated Teller Machines (ATMs) suggested a likely decrease in human bank teller employment. 
However, contrary to this prediction, ATM technology cut the cost to banks for opening and operating new branches, and, as a result, national bank teller employment \emph{increased}~\cite{bessen2015computer,bessen2015learning}.
However, these bank tellers performed different tasks, such as relationship management and investment advice, which required very different skills.
%However, contrary to this prediction, ATM technology cut the cost to banks for opening and operating new branches. 
%This shifted the supply curve outwards, leading banks to open more branches at market equilibrium. 
%The result was an \emph{increase} in bank teller employment~\cite{bessen2015computer,bessen2015learning}, but those tellers performed different tasks, such as relationship management and investment advice, which required very different skills.
Hence, by \emph{impact}, we refer to the \emph{magnitude} of the skill substitution shocks that cities must respond to.

The actual technological unemployment in a city will be shaped both by free market dynamics (e.g. shifts in supply and demand curves) and by economic and educational policy (e.g. worker re-training, or skilled migration). 
Nevertheless, we observe a strong trend relating city size to automation impact that is robust to errors in the automatability of individual occupations and occupational sub-sampling.
For example, the estimates of occupational automation, which we employ in our analysis, would need to be severely flawed (errors over 50\%) for the negative dependency on city size to disappear.
Recognizing that small cities will experience larger adjustments to automation calls on policy-makers to pay special attention to the pronounced risks we have identified.

%Despite being seemingly unrelated societal forces, we uncover a positive interplay between urbanization and automation.
%Larger cities not only tend to be more innovative~\cite{bettencourt1,bettencourt2}, but also attract the workers who are prepared to both use and improve cutting-edge technology.
%In turn, these workers are more specialized in their workplace skills and less likely to be replaced by automated methods in the foreseeable future.
%This study highlights an important piece of the puzzle of how automated processes impact urbanization and informs new investigations into humanity's progression towards cities in the face of a dynamic occupation market.
Despite being seemingly unrelated societal forces, we uncover a positive interplay between urbanization and automation.
Larger cities not only tend to be more innovative~\cite{bettencourt1,bettencourt2}, but also harbor the workers who are prepared to both use and improve cutting-edge technology.
In turn, these workers are more specialized in their workplace skills and less likely to be replaced by automated methods in the foreseeable future.
%It is difficult to concretely identify causal mechanisms at the scale of this investigation, but, despite this difficulty, we highlight evidence for the division of labor in large cities and show its importance as a piece of the automation and urbanization puzzle.
These findings open the door for more controlled investigations with input from policy makers.

\section*{Materials and Methods}
% The materials and methods section should provide sufficient information to allow replication of the results. Begin with a section titled Experimental Design describing the objectives and design of the study as well as pre-specified components.
% 
% In addition, include a section titled Statistical Analysis at the end that fully describes the statistical methods with enough detail to enable a knowledgeable reader with access to the original data to verify the results. The values for N, P, and the specific statistical test performed for each experiment should be included in the appropriate figure legend or main text.
The U.S. Bureau of Labor Statistics (BLS) data identifies the employment distribution of about 700 different occupations across each of 380 U.S. metropolitan statistical areas (also referred to as ``cities") in 2014. 
From these employment distributions, we calculate the probability of a worker in city $m$ having job $j$ according to
\begin{equation}
	\label{pj}
	p_m(j) = \frac{f_m(j)}{\displaystyle\sum_{j\in Jobs_m}f_m(j)},
\end{equation}
where $Jobs_m$ denotes the set of job types in city $m$ according to BLS data, and $f_m(j)$ denotes the number of workers in city $m$ with job $j$.

We assess the specialization or diversity of the employment distribution in city $m$ by calculating the normalized Shannon entropy.
% (denoted $H_{job}(m)$).
Shannon entropy, an information theoretic measure for the expected information in a distribution, can be normalized according to
\begin{equation}
	\label{entropy}
	H_{job}(m) = -\displaystyle\sum_{j\in Jobs_m}p_m(j)\cdot\frac{\log(p_m(j))}{\log\big(|Jobs_m|\big)}.
\end{equation}
%Here, Shannon entropy is a measure for the predictability of an employment distribution relative to the number of unique jobs in a city.
This quantity measures the predictability of an employment distribution given the set of unique occupations in a city.
The measure is maximized when the distribution is least predictable (i.e. the distribution is uniform).
Therefore, the denominator of $\log(|\text{Jobs}_m|)$ normalizes the entropy score so that we can compare the distributions of jobs in cities with different sets of job categories (see Supplementary Materials S2.1 for further discussion).
The values for normalized Shannon entropy lie between 0 (specialization) and 1 (diversity).

For each occupation, the BLS \onet dataset details the importance of 230 different workplace skills, such as Manual Dexterity, Finger Dexterity, Complex Problem Solving, Time Management, and Negotiation.
BLS obtains this information through several separate surveys which group the raw \onet skills into the following categories: Abilities, Education/Training/Experience, Interests Knowledge, Skills, Work Activities, and Work Context.
We normalize the raw survey responses to obtain a value between 0 (irrelevant to the occupation) and 1 (essential to the occupation) indicating the absolute importance of that skill to that occupation. 
We refer to these values of skill importance as \emph{raw skill values}.

For a given occupation, we normalize each raw skill value by the sum of the values to obtain the relative importance of each skill to that occupation (denoted $p_j(s)$).
Similar to above, we measure the normalized Shannon entropy of the relative skill distribution of job $j$ according to
\begin{equation}
	\label{jobEntropy}
	H_j = -\sum_{s\in Skills_j}p_j(s)\cdot\frac{\log(p_j(s))}{\log\big(|Skills_j|\big)},
\end{equation}
where $Skills_j$ denotes the set of \onet skills with non-zero importance to job $j$.
We employ normalized Shannon entropy here to facilitate a fair comparison of relative skill distributions between jobs which may have received the same raw \onet value for a given skill, but have different numbers of non-zero raw \onet skills.

We obtain a distribution of relative skill importance for a city according to
\begin{equation}
	\label{citySkillDist}
	p_m(s) = \sum_{j\in Jobs_m}p_j(s)\cdot p_m(j),
\end{equation}
where $p_m(s)$ is the relative importance of skill $s$ in city $m$.
Again, we use normalized Shannon entropy to assess the skill specialization in a city according to 
\begin{equation}
	\label{citySkillEntropy}
	H_{skill}(m) = -\sum_{s\in Skills_m} p_m(s)\cdot \frac{\log(p_m(s))}{\log\big(|Skills_m|\big)},
\end{equation}
where $Skills_m$ represents the set of \onet skills with non-zero importance in city $m$.

These aggregate skill distributions for a city may obfuscate the specialization of skills through the relative abundance of jobs in that city.
For example, the city-level aggregation of skills may appear diverse, while the jobs within the city are actually specialized.
The Theil entropy~\cite{entropy} of a city is a multi-level information theoretic measure defined by
\begin{equation}
	\label{theil}
	T_m = \sum_{j\in Jobs_m} p_m(j)\cdot\frac{H_{skill}(m)-H_j}{H_{skill}(m)}.
\end{equation}
$T(m)=1$ indicates that each job specializes in exactly one skill, and $T(m)=0$ indicates that the specialization of skills among jobs is equal to the specialization of skills on the city-level aggregation.
We do not observe any jobs relying on exactly one skill, and so we expect the Theil entropy of any given city to be well below 1.
%This metric is traditionally used to account for neighborhood-level segregation within cities.
We present $1-T_m$ throughout the study for easy comparison to Shannon entropy.

\bibliographystyle{naturemag}
\bibliography{citimation_arxiv}

\section*{Acknowledgements:} 
This work is supported, in part, by a gift from the Siegel Family Endowment.
The authors would like to thank David Autor and Lorenzo Coviello for their feedback during the undertaking of this study.

\section*{Author Contributions} 
M.R.F., M.C., H.Y., \& I.R. conceived the study.
M.R.F., \& L.S. performed calculations. 
M.R.F. \& L.S. produced figures. 
All authors wrote the manuscript.

\section*{Competing Interests} 
The authors declare that they have no competing financial interests.\\
%\noindent \textbf{Data and materials availability:} Additional data and materials are available online.

% For your review copy (i.e., the file you initially send in for
% evaluation), you can use the {figure} environment and the
% \includegraphics command to stream your figures into the text, placing
% all figures at the end.  For the final, revised manuscript for
% acceptance and production, however, PostScript or other graphics
% should not be streamed into your compliled file.  Instead, set
% captions as simple paragraphs (with a \noindent tag), setting them
% off from the rest of the text with a \clearpage as shown  below, and
% submit figures as separate files according to the Art Department's
% instructions.

\clearpage
\clearpage

\begin{figure}[!p]
    \centering
    \includegraphics[width=\textwidth,trim=1.25cm .25cm .25cm .25cm,clip]{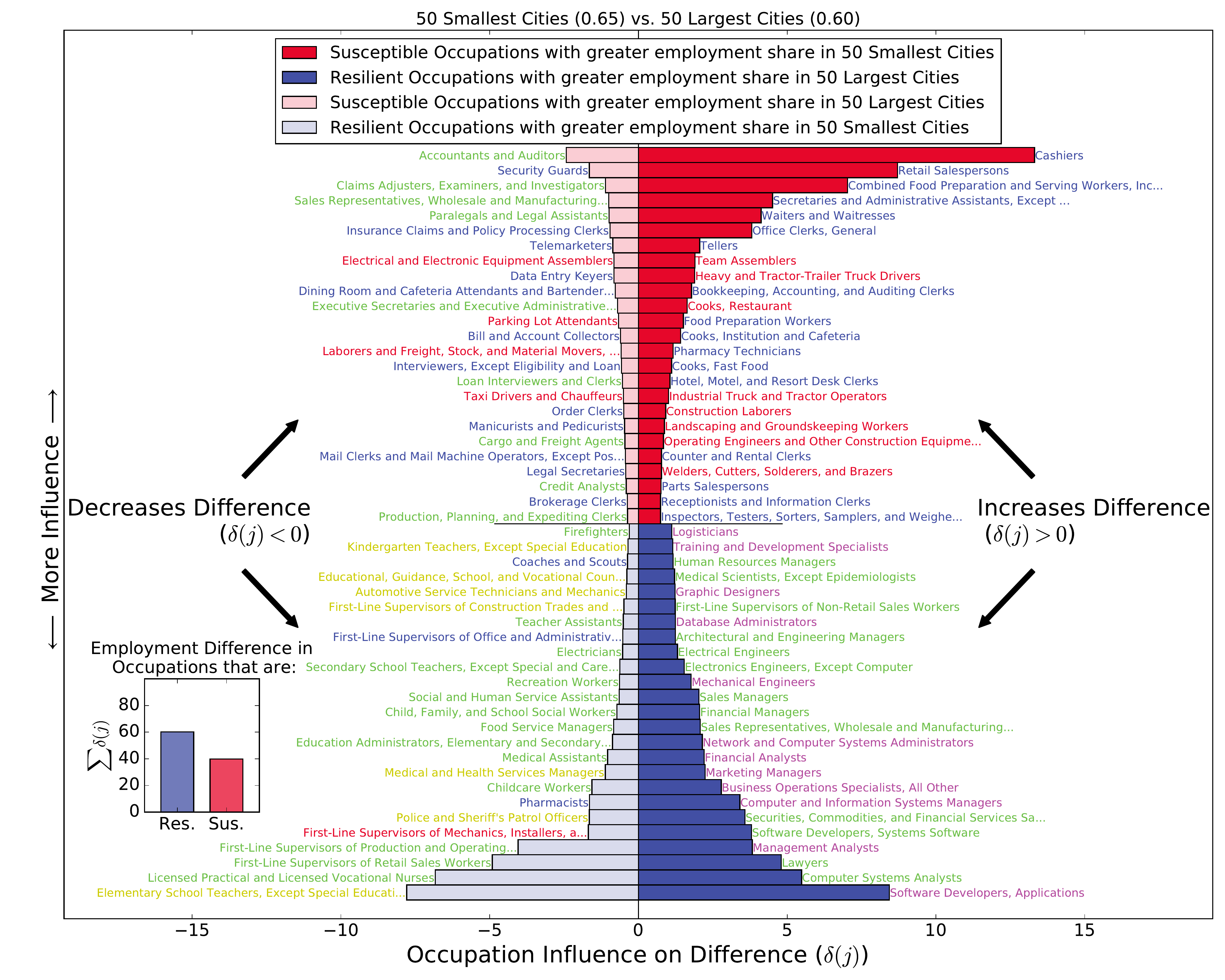}
    \caption{
        An occupation shift explaining the difference in expected job impact for the 50 largest cities (impact: $0.60$) compared to the 50 smallest cities (impact: $0.65$) using equation (\ref{occShift}).
        Each horizontal bar represents $\delta_{(\text{Small Cities},\text{Large Cities})}(j)$.
        The occupation title is provided next to the corresponding bar and colored according to its job cluster.
        Red bars represent occupations with higher risk of computerization compared to the expected job impact in large cities.
        Blue bars represent occupations with lower risk of computerization compared to the expected job impact in large cities.
        Dark colors represent occupations that increase the difference, while pale colors represent occupations that decrease the difference in expect job impact.
        Bars in each of the quadrants are vertically ordered according to $|\delta_{(\text{Small Cities},\text{Large Cities})}(j)|$.
        The inset in the bottom left of the plot summarizes the overall influence of resilient occupations compared to occupations that are at risk of computerization.
    }
    \label{citySizeShift}
\end{figure}

\begin{figure*}[!p]
	\centering
    \begin{overpic}[width=.49\textwidth,trim=0cm 2cm 3cm 1.5cm,clip]{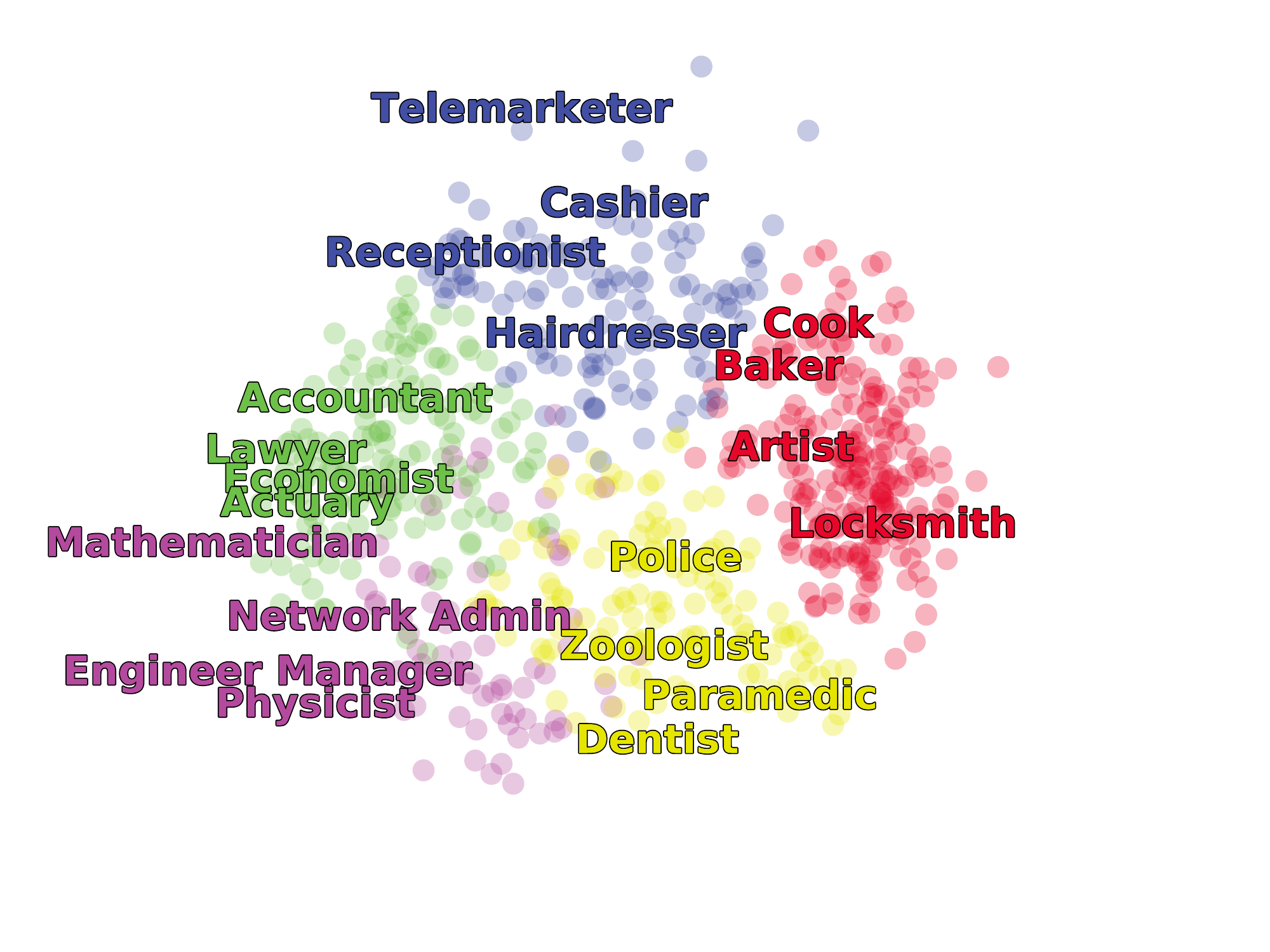}
        \put(20,145){\fbox{\small A}}
    \end{overpic}   
    \begin{overpic}[width=.49\textwidth,trim=0cm 0cm 0cm 0cm]{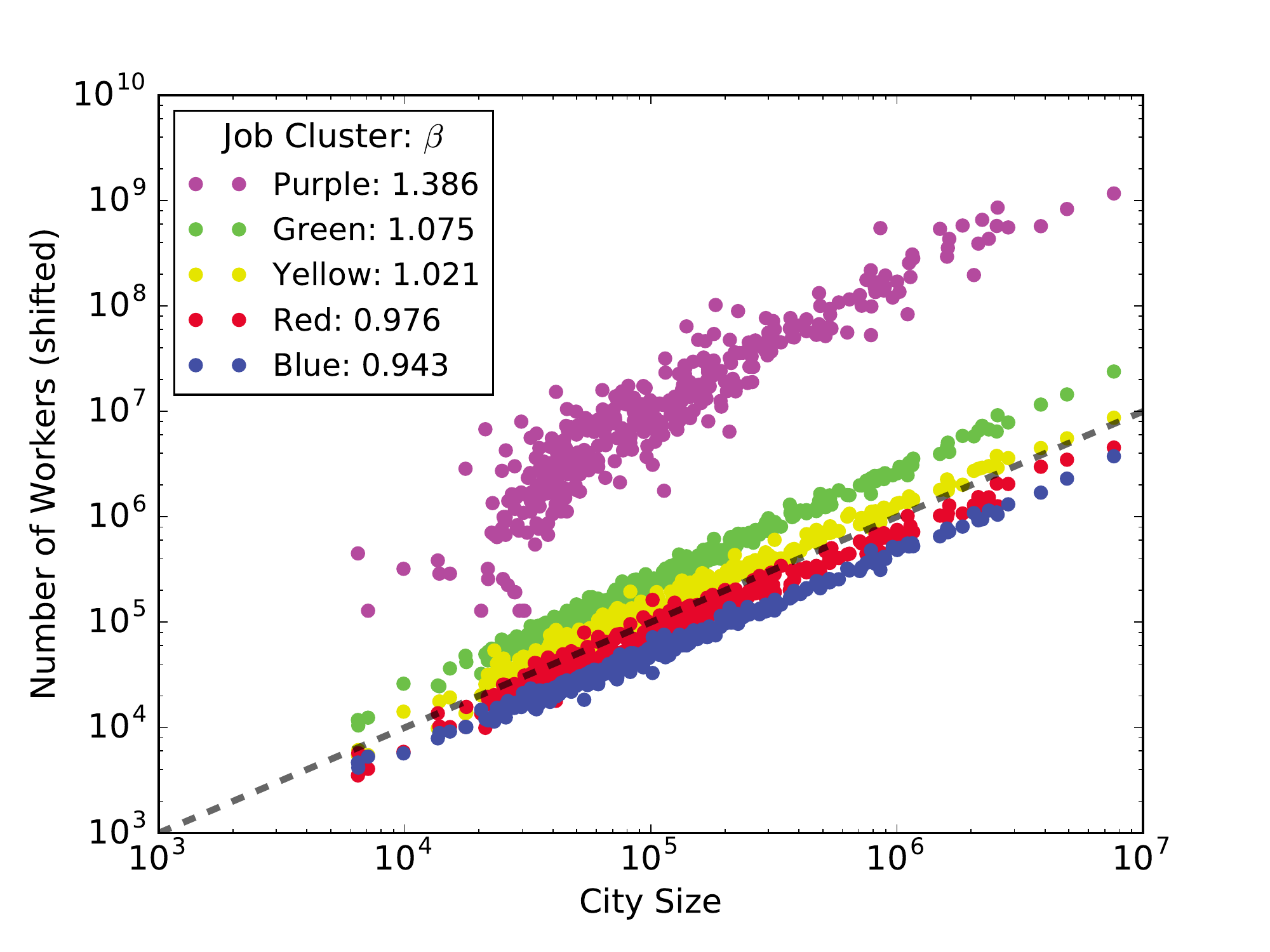}	
    	\put(-10,145){\fbox{\small B}}
    \end{overpic}
    \begin{overpic}[width=.49\textwidth,trim=0cm 0cm 0cm 0cm]{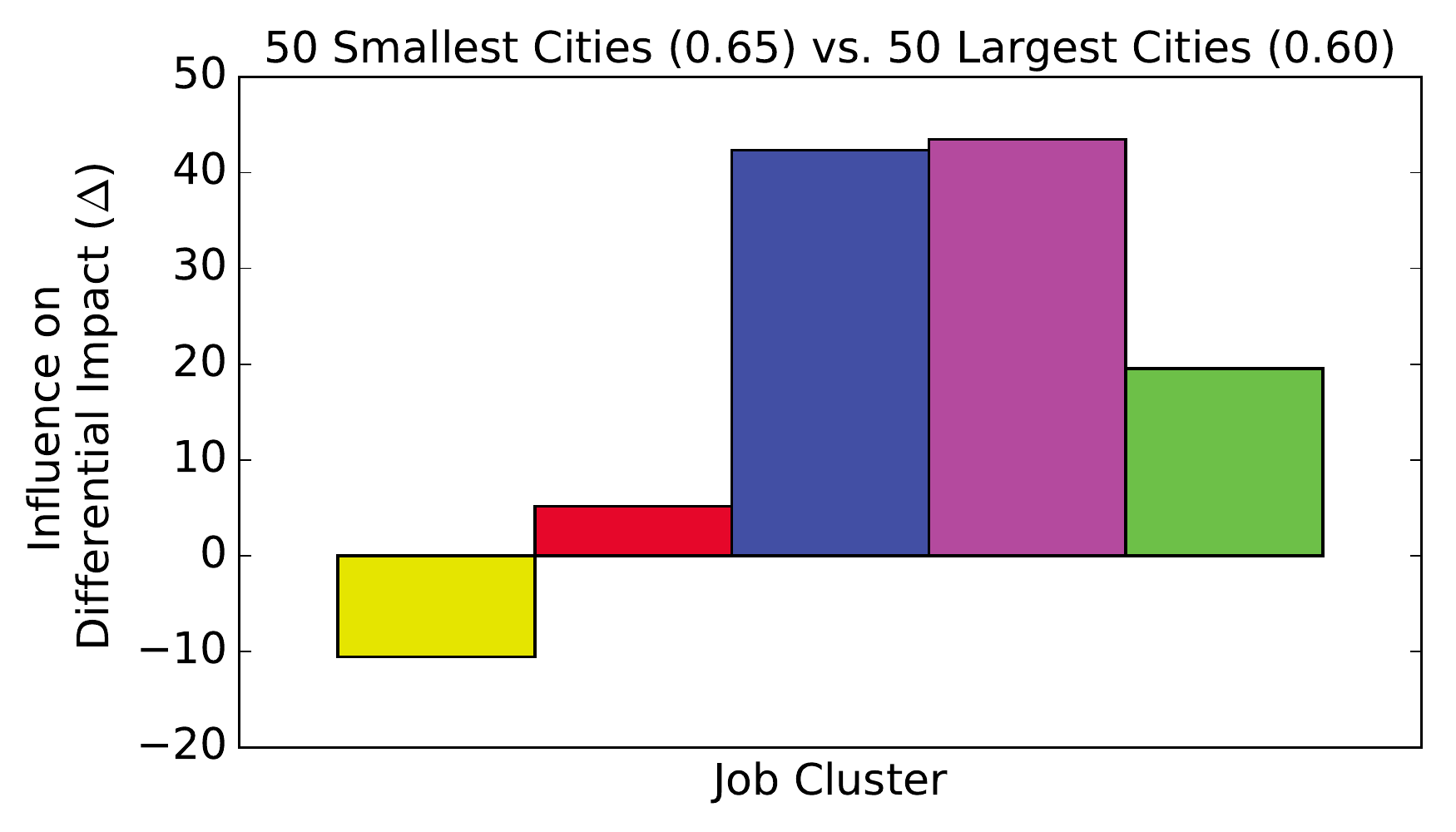}
        \put(40,90){\fbox{\small C}}
    \end{overpic}
	\begin{overpic}[width=.49\textwidth,trim=0cm 0cm 0cm 0cm]{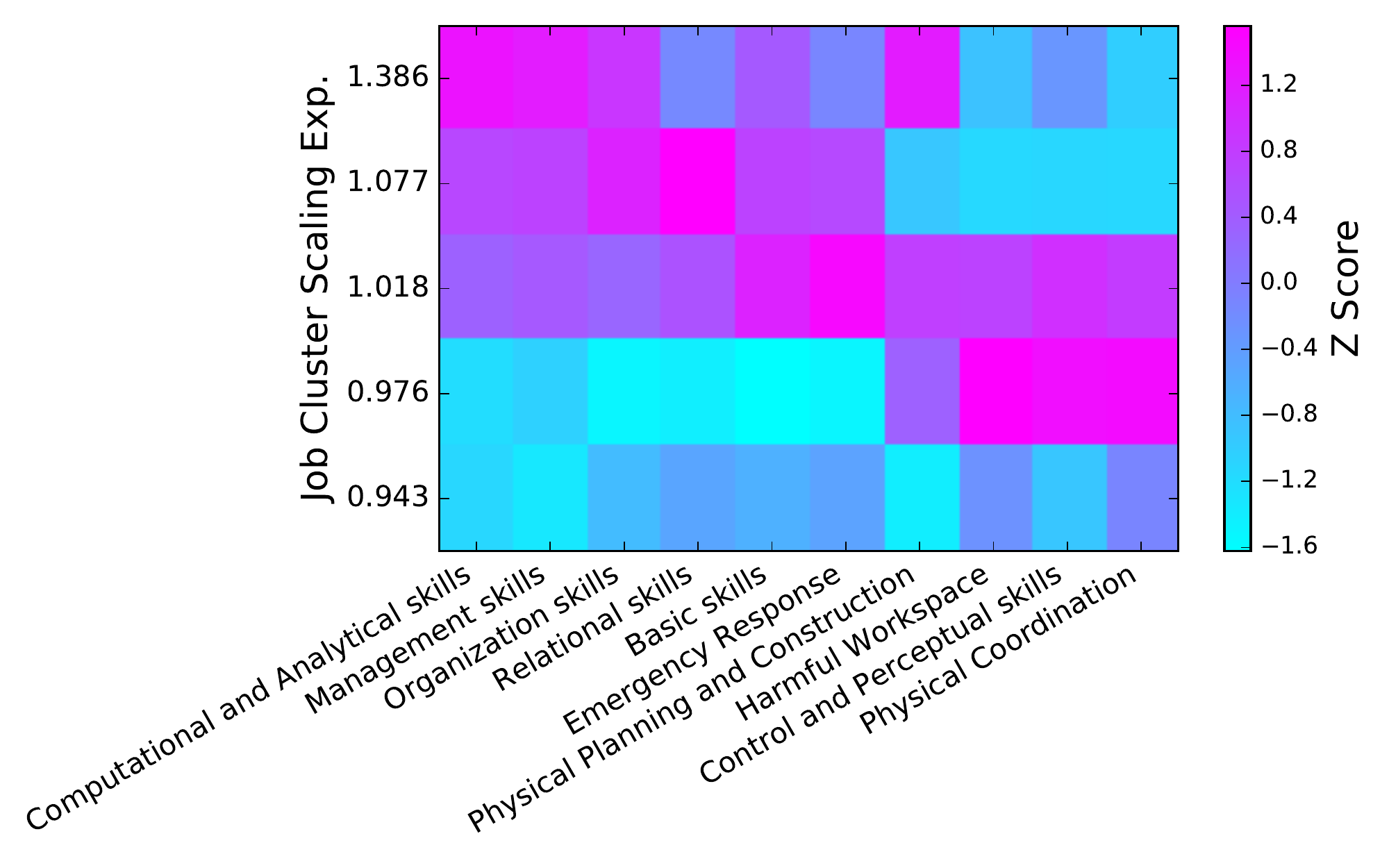}
        \put(15,90){\fbox{\small D}}
    \end{overpic}
    \caption{
    Technical occupations grow superlinearly with city size.
    {\bf (A)} We project jobs onto a 2-D plane using principal component analysis.
	A few representative jobs from each cluster are highlighted (color). 
    %See Supplementary Material S6.3 for the complete list of jobs comprising each job type.
    {\bf (B)} We plot the employment (y-axis) in a given job cluster (color) versus the total employment in a city (x-axis), and vertically shift points according to the linear fit in log scale.
    The black dashed line has a slope of 1 for reference.
	{\bf (C)} The influence of each job cluster on the difference in expected job impact of the 50 largest cities ($E_{\text{Large Cities}}=0.60$) compared to the 50 smallest cities ($E_{\text{Small Cities}}=0.65$) according to equation (\ref{occTypeShift}).
	{\bf (D)} 
	%By summing the importance of each skill type to each job cluster, we assess how strongly those skills indicate a scaling relationship according to its z score.
	%For a given skill type, 
	After summing the importance of each skill type to each job cluster, we calculate z scores for a skill type according to the distribution of importance across job clusters. 
	}
	\label{jobScaling}
\end{figure*}

\begin{figure*}[!p]
	\centering
	\begin{overpic}[width=.45\textwidth,trim=0cm .5cm 0cm 1cm,clip]{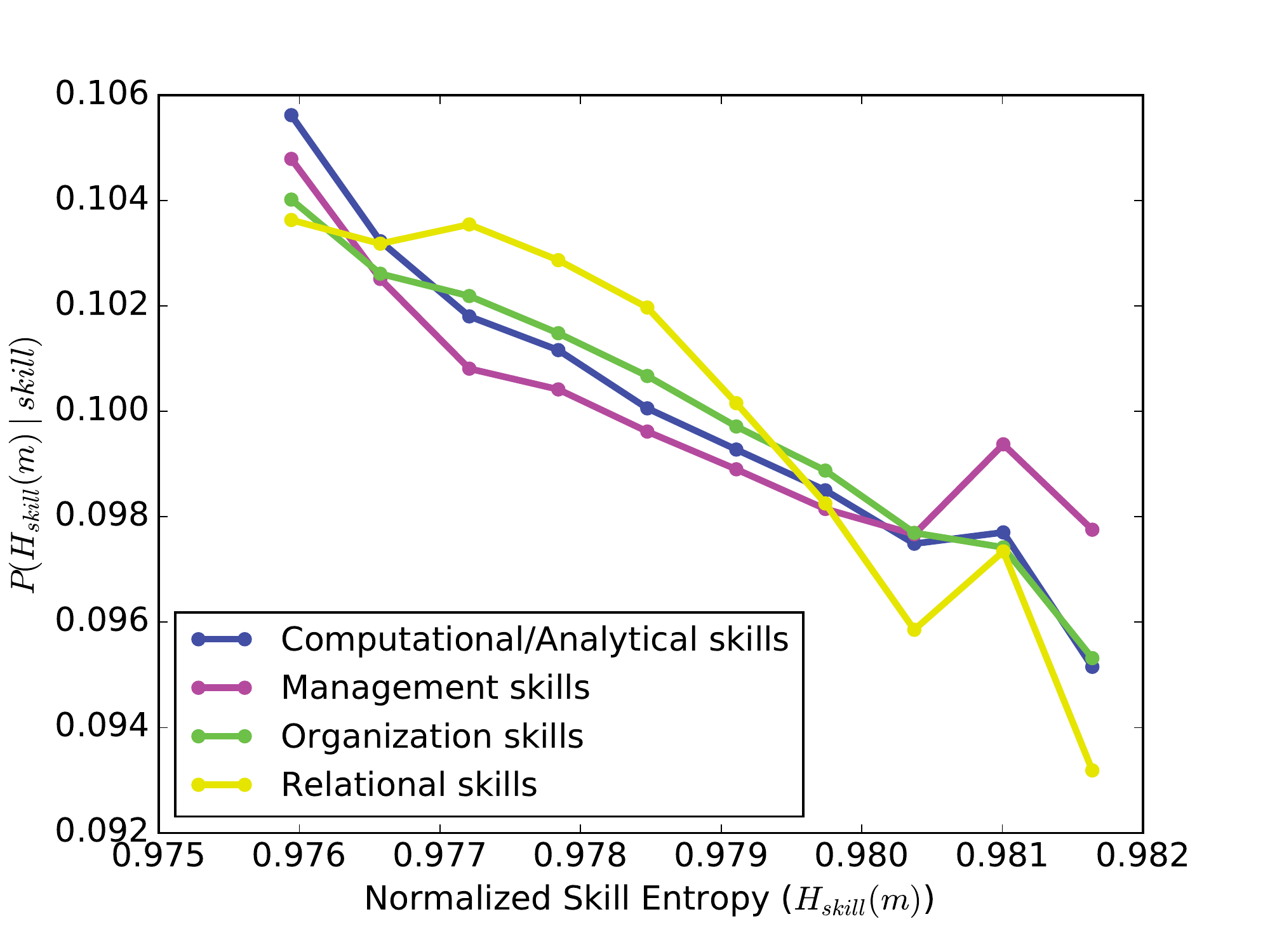}
		\put(160,125){\fbox{\small A}}
	\end{overpic}
	\begin{overpic}[width=.45\textwidth,trim=0cm .5cm 0cm 1cm,clip]{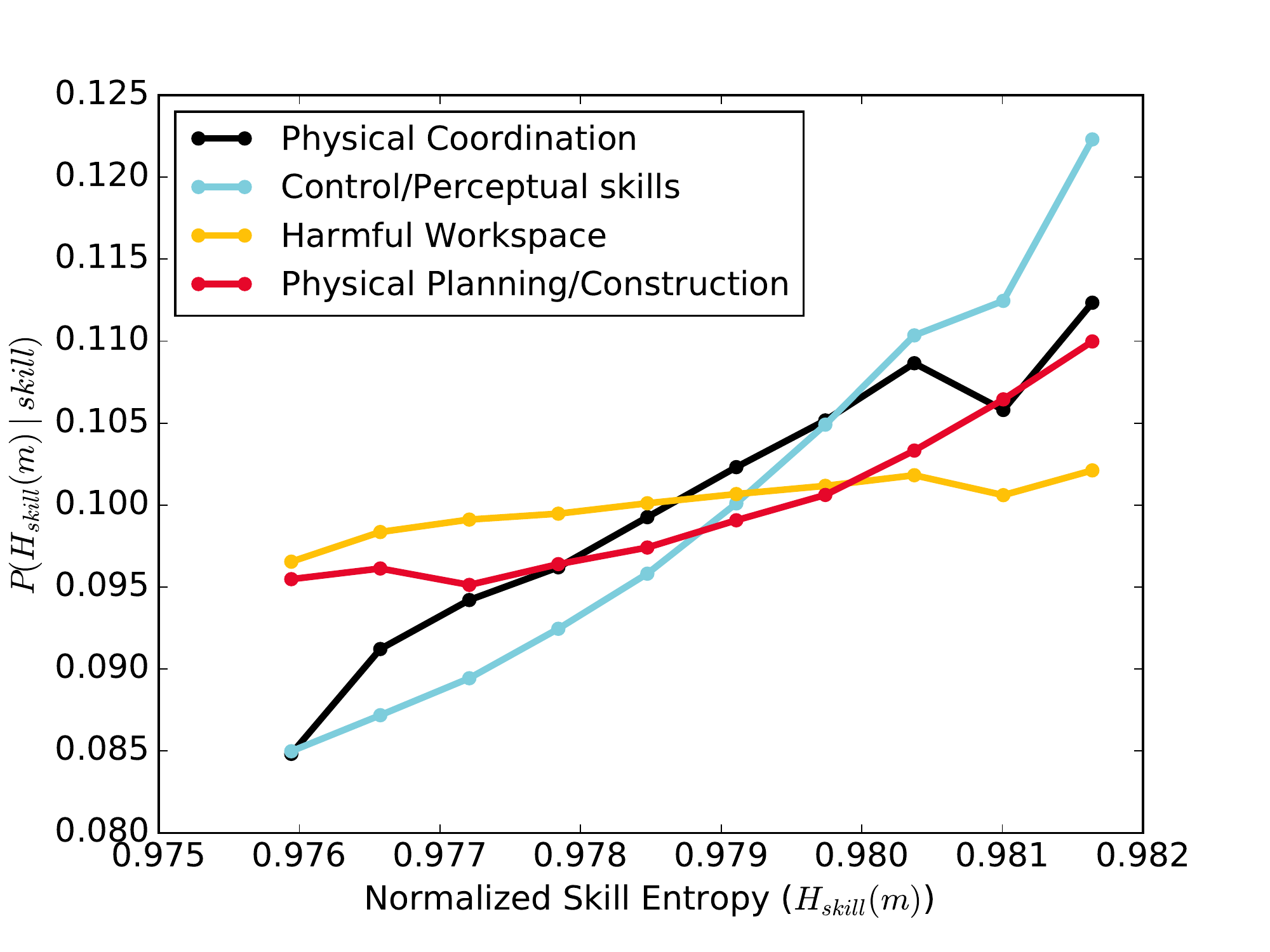}
		\put(160,35){\fbox{\small B}}
	\end{overpic} 
	\begin{overpic}[width=.45\textwidth,trim=0cm .5cm 0cm 1cm,clip]{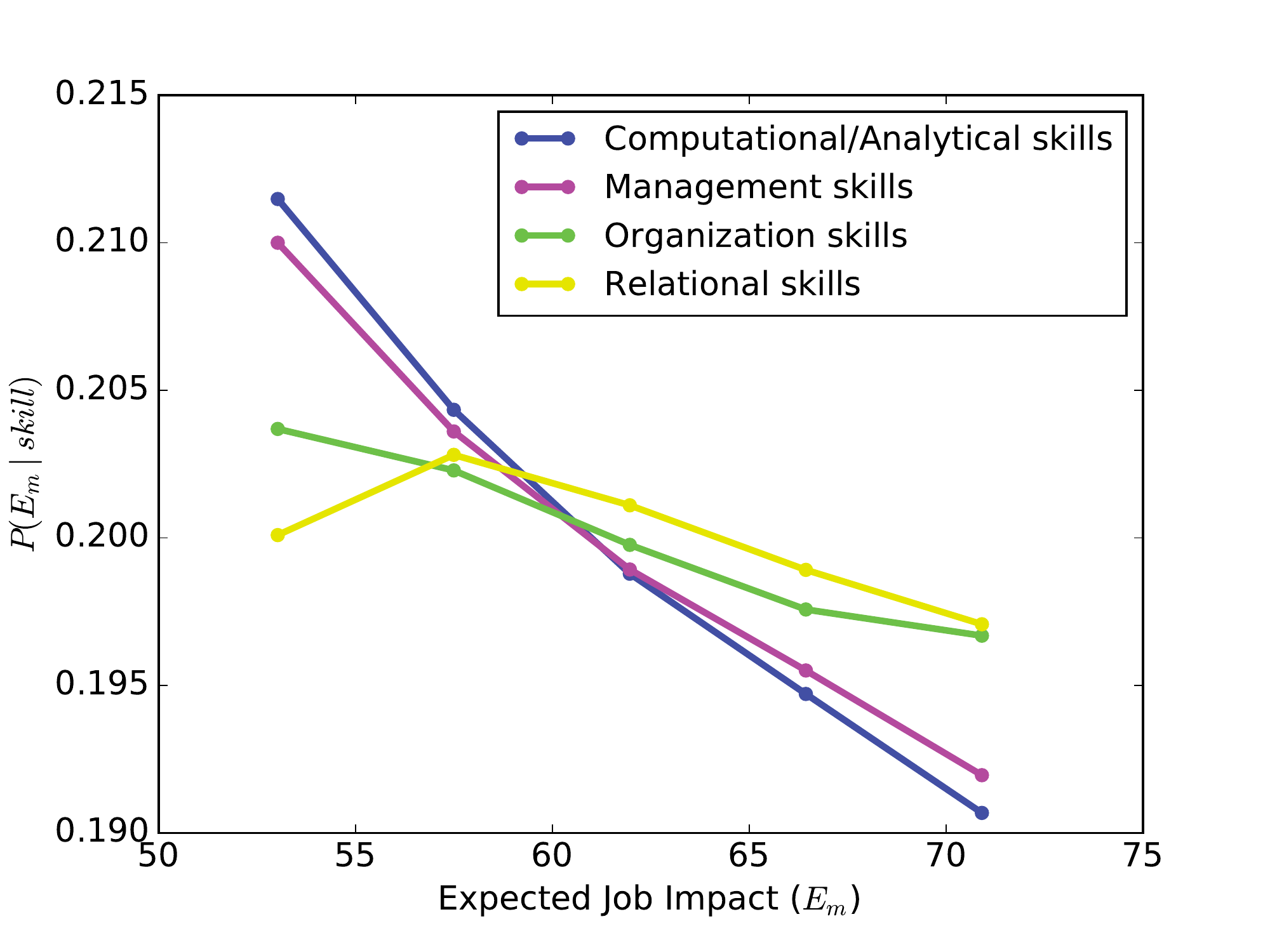}
		\put(40,35){\fbox{\small C}}
	\end{overpic}
	\begin{overpic}[width=.45\textwidth,trim=0cm .5cm 0cm 1cm,clip]{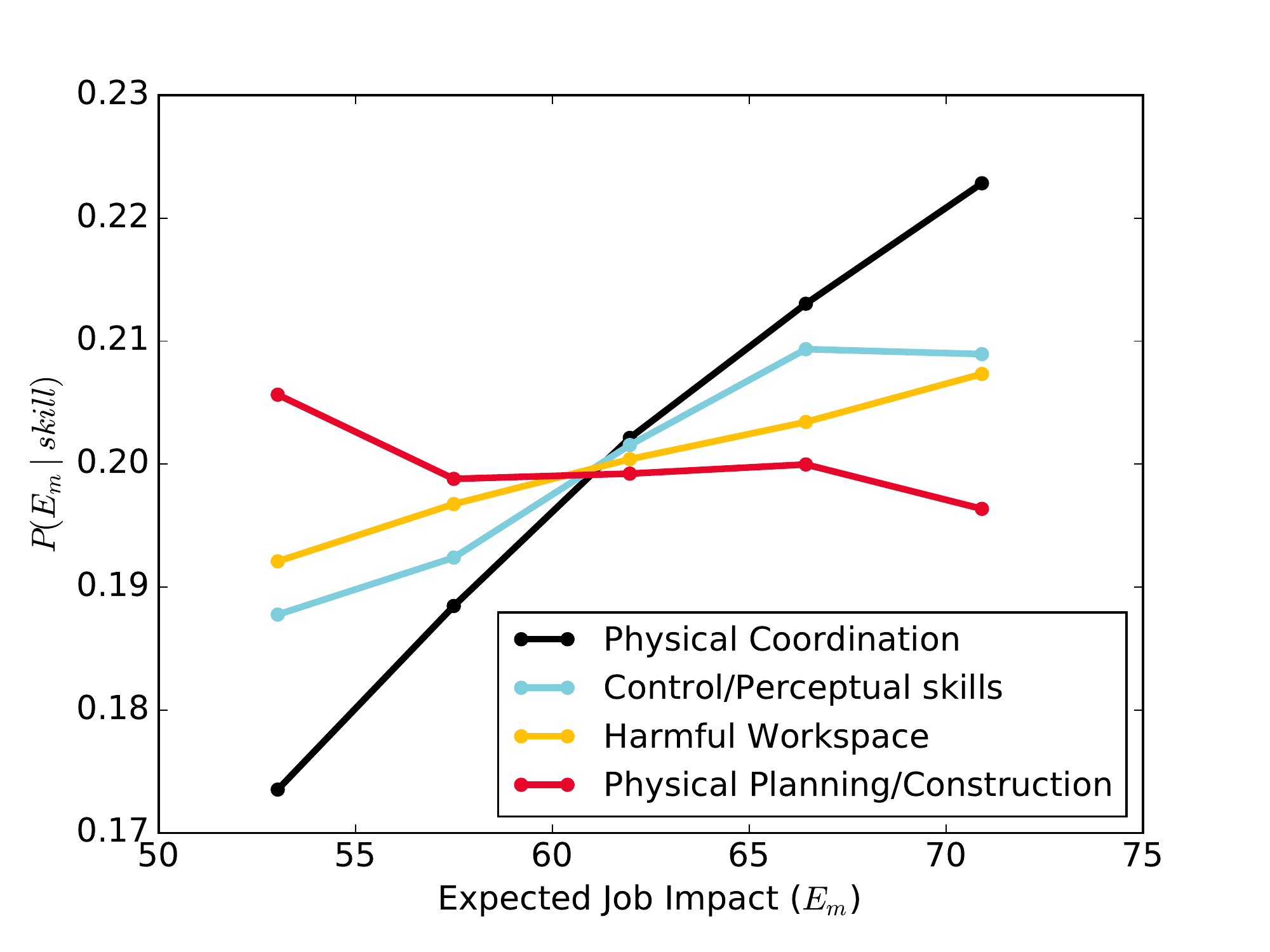}
		\put(40,125){\fbox{\small D}}
		\put(-230,5){\fbox{\small E}}
	\end{overpic}
	%\scriptsize
	\small
	\begin{tabularx}{.97\textwidth}{|l|X|X|}
		\hline
		\bf Skill Type & \bf Impact Correlation &\bf Log$_{10}$ City Size Correlation\\ \hline                                                                                                                
		\rowcolor{myCyan}Computational/ Analytical & -0.88 ($<10^{-124}$) & 0.58 ($<10^{-34}$) \\ \hline
		\rowcolor{myCyan}Management & -0.87 ($<10^{-120}$) & 0.52 ($<10^{-27}$) \\ \hline
		\rowcolor{myCyan}Organization & -0.62 ($<10^{-41}$) & 0.35 ($<10^{-11}$) \\ \hline
		Relationship & -0.26 ($<10^{-6}$) & -0.06 (0.3) \\ \hline
		Physical Planning & -0.07 (0.21) & 0.18 (0.0006) \\ \hline
		Basic/General & 0.13 (0.01) & -0.24 ($<10^{-5}$) \\ \hline
		Control \& Perceptual & 0.45 ($<10^{-19}$) & -0.14 (0.005) \\ \hline
		\rowcolor{myRed}Emergency Response & 0.46 ($<10^{-20}$) & -0.32 ($<10^{-9}$) \\ \hline
		\rowcolor{myRed}Physical Coordination & 0.83 ($<10^{-99}$) & -0.52 ($<10^{-26}$) \\ \hline
		\rowcolor{myRed}Harmful Workspace & 0.90 ($<10^{-140}$) & -0.54 ($<10^{-28}$) \\ \hline
	\end{tabularx}
	\caption{
    Workplace skills explain occupational specialization and job impact in cities.
    {\bf (A) \& (B)} 
    %Skill type importance as a function of skill specialization ($H_{skill}(m)$). 
    Skill types in (A) indicate specialized cities, while skill types in (B) indicate occupational diversity.
    {\bf (C) \& (D)} 
    %Revealed skill importance as a function of expected job impact ($E_m$). 
    Skill types in (C) indicate resilient cities, while skill types in (D) indicate increased job impact from automation.
    {\bf (E)} 
	The Pearson correlation of skill type abundance to the expected job impact and to $\log_{10}$ city size with p-values in parentheses. 
    See Supplementary Materials S6.4 for a similar table for raw \onet skills.
	}
	\label{skillTable}
\end{figure*}

\clearpage
\section*{Supplementary Materials}
\begin{figure}[t]
	\centering
	\begin{overpic}[width=.45\textwidth]{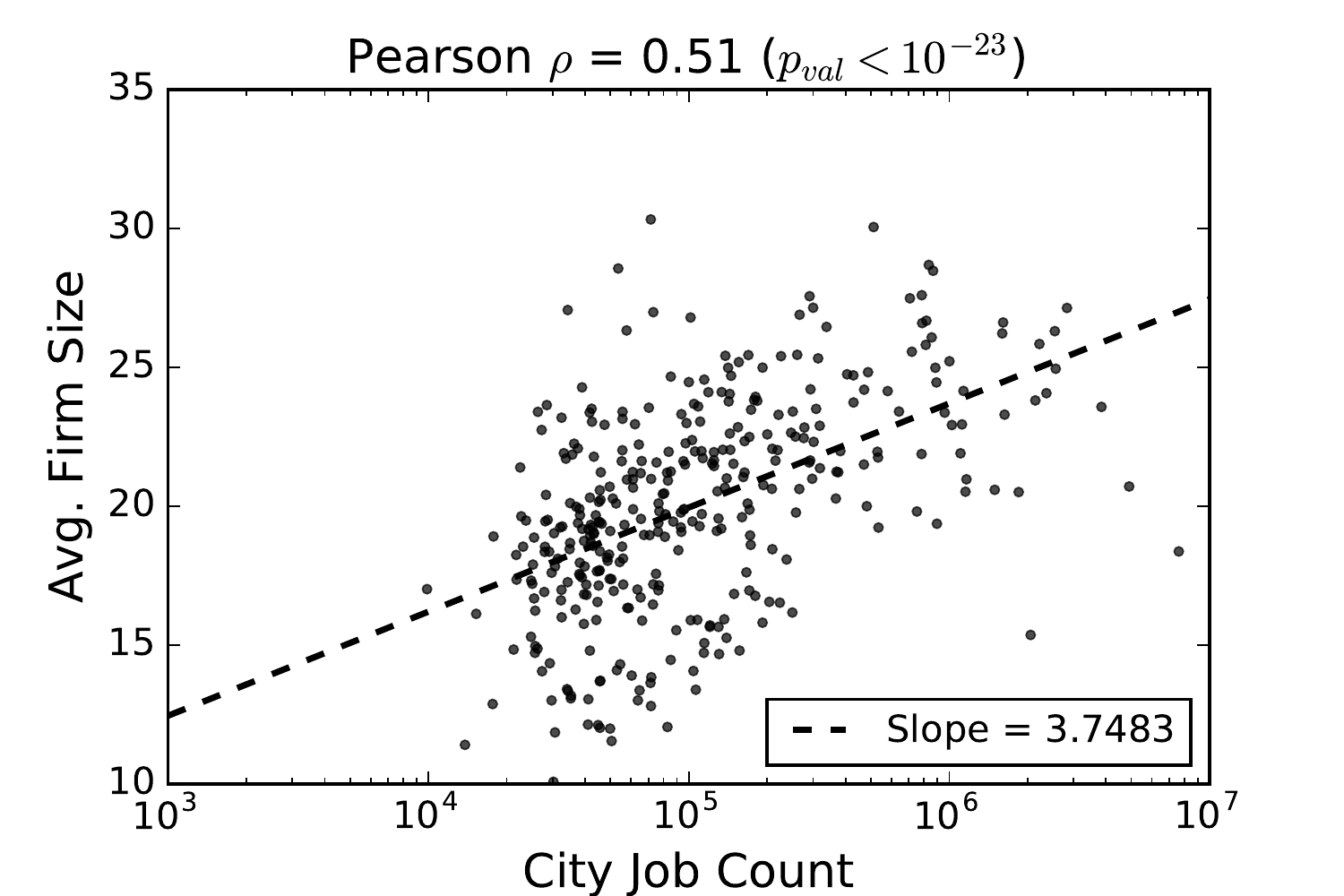}
		\put(30,100){\fbox{\small A}}
	\end{overpic}
	\begin{overpic}[width=.45\textwidth]{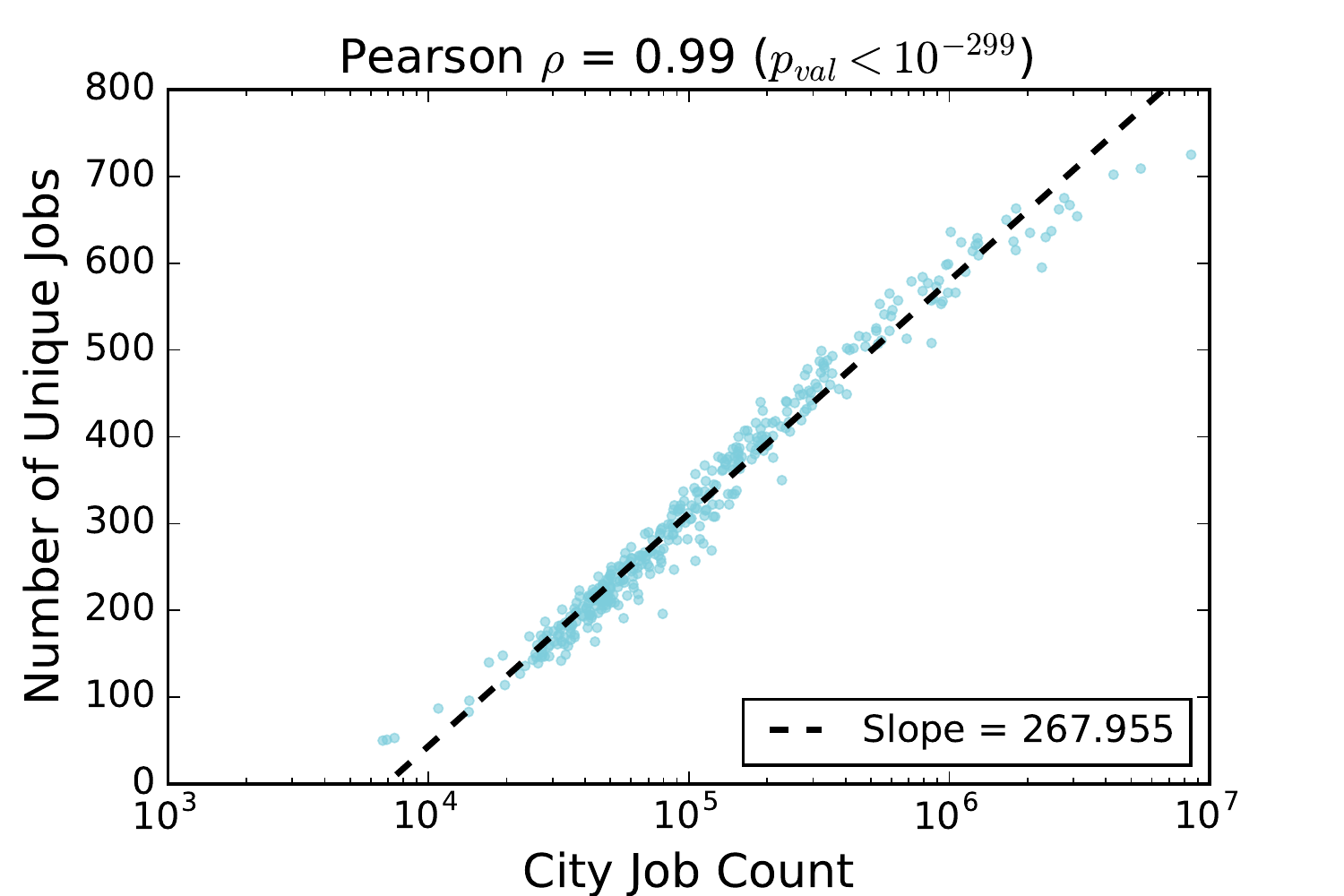}
		\put(30,100){\fbox{\small B}}
	\end{overpic}
	\caption{
	{\bf (A)} The average number of workers per firm grows logarithmically with city size.
	{\bf (B)} Consistent with \cite{hyejin}, we find the number of unique jobs grows logarithmically with city size.
	}
	\label{firmSize}
\end{figure}
%\vspace{.5cm}
%Several figures presented in the main text and in the Supplementary Material are available online as interactive plots at \url{http://web.media.mit.edu/~mrfrank/citimation.html}.

\section{Firm Size Increases with City Size}
	\label{firms}
\indent The U.S. Bureau of Labor Statistics (BLS) uses the annual tax filings of companies to produce a yearly census of those companies.
Unfortunately, the data available to the public doesn't include the specific distribution of BLS jobs comprising each firm.
Previous work has shown that firm sizes nation wide follow a Zipf distribution~\cite{axtell2001zipf} indicating that a majority of firms are small, but surprisingly large ones also exist infrequently.
Figure~\ref{firmSize} shows that the average number of workers per firm increases logarithmically with city size.
Larger firms have more capital with which to hire specialized workers along with organization/managerial staff to coordinate those workers.
According to our theory, there exists a positive feedback loop where large firms provide demand for specialized workers and cities provide a richer market of skilled workers to meet that demand.

\section{Measuring Labor Specialization}
	\label{specialization}
\begin{comment}
\indent We rely on 2014 U.S. Bureau of Labor Statistics (BLS) data for identifying the distribution of 700 different jobs across each of 380 U.S. metropolitan statistical areas. 
We use $p_m(j)$ to denote the probability of a worker in city $m$ having job $j$ according to
\begin{equation}
	\label{pj}
	p_m(j) = \frac{f_m(j)}{\displaystyle\sum_{j\in Jobs_m}f_m(j)},
\end{equation}
where $Jobs_m$ denotes the set of job types in city $m$ according to BLS data, and $f_m(j)$ denotes the number of workers in city $m$ with job $j$.\\
\indent In particular, we can assess the specialization or diversity of the distribution of jobs in a given city ($m$) by calculating the normalized Shannon entropy of the distribution (denoted $H_{job}(m)$).
Shannon entropy is an information theoretic measure for the expected information in a distribution, and normalized Shannon entropy is calculated according to 
\begin{equation}
	\label{entropy}
	H_{job}(m) = -\displaystyle\sum_{j\in Jobs_m}p_m(j)\cdot\frac{\log(p_m(j))}{\log\big(|Jobs_m|\big)}.
\end{equation}
Shannon entropy is a measure for the predictability of a distribution relative to the number of jobs in a city (in this case).
The measure is maximized when the distribution is least predictable (i.e. the distribution is uniform).
Therefore, the denominator of $\log(|\text{Jobs}_m|)$ normalizes the entropy score so that we can compare the distributions of jobs in cities with different sets of job categories.
The values for normalized Shannon entropy lie between 0 (specialization) and 1 (diversity).
\end{comment}

\subsection{Normalized Shannon Entropy}
We employ normalized Shannon entropy, as opposed to the standard Shannon entropy definition, to control for size effects on the distributions in cities. 
For example, it has been shown that the number of different occupations grows with city size (see \cite{hyejin}, and SM Fig. \ref{firmSize}B), but this result may be due to randomness as more people (e.g. sampled from a long-tailed distribution) are added to a city, and does not account for how workers are distributed amongst these occupations. 
Large cities may have only a few workers of otherwise absent occupations in small cities, but, perhaps, this distinction does not mean much qualitatively. 
This motivates us to consider the distribution of workers amongst different occupations, rather than only considering the number of occupations, and to apply relevant information theoretic methods to measure the diversity/specialization of these distributions. 
We normalize the standard Shannon entropy calculation to control for the number of different occupations in a city, or, equivalently, we normalize Shannon entropy by the maximum possible Shannon entropy given the number of different occupations in the city (i.e. given a number of occupations, Shannon entropy is maximized for the uniform distribution). 

This normalization is a standard practice for comparing the diversity or information of systems of different sizes. 
For a summary of normalized entropy, see \cite{kumar1986normalized}.
In particular, normalized Shannon entropy has been used in a variety of fields, including virology~\cite{cabot2000nucleotide}, climatology~\cite{wijesekera1997shannon}, and city science~\cite{eagle2010network}.
To understand this normalization, consider that a sufficient number of roles of a fair 6-sided dice and, separately, of a fair 20-sided dice should each produce uniform distributions with maximized Shannon entropy.
However, the Shannon entropy of the distribution for the 6-sided dice is $-\sum \frac{1}{6}\cdot\log(\frac{1}{6}) = 1.79$ and the entropy of the distribution for the 20-sided dice is $-\sum\frac{1}{20}\cdot\log(\frac{1}{20})=3.00$; specifically, they are not equivalent despite both being discrete uniform distributions because the distributions have a different number of bins.
This is analogous to cities having a different number of unique occupations due, potentially, to randomness that occurs with increased city size.
We control for this effect by normalizing Shannon entropy according to the maximum possible Shannon entropy given the number of bins in the discrete distribution.
Specifically, given a discrete system with $N$ bins (i.e. $N$-sided dice, or a city with $N$ unique occupations), Shannon entropy is maximized when the distribution is uniform, and the maximum value is given by
\begin{equation}
    -\sum_{i=1}^N \frac{1}{N}\cdot\log(\frac{1}{N}) = -N\cdot\frac{1}{N}\cdot\log(1/N) = -\log(1/N)=\log(N).
\end{equation}
Therefore, to normalize Shannon entropy according to the maximum possible Shannon entropy, we divide the standard Shannon entropy calculation by $\log(N)$ to obtain $$-\sum_{i=1}^N p_i\cdot\frac{\log(p_i)}{\log(N)},$$
where $p_i$ is the probability of bin $i$.
This normalized Shannon entropy produces a value of 1 for discrete uniform distributions regardless of the number of bins (i.e. regardless of $N$).
In particular, this normalization allows us to control for the number of unique occupations across cities of different sizes to determine the uniformity of job and skill distributions in cities.

\subsection{The Labor Specialization of Individual Jobs}
	\label{jobTableSection}
\begin{figure}[h]
	\centering
	\begin{overpic}[width=.45\textwidth]{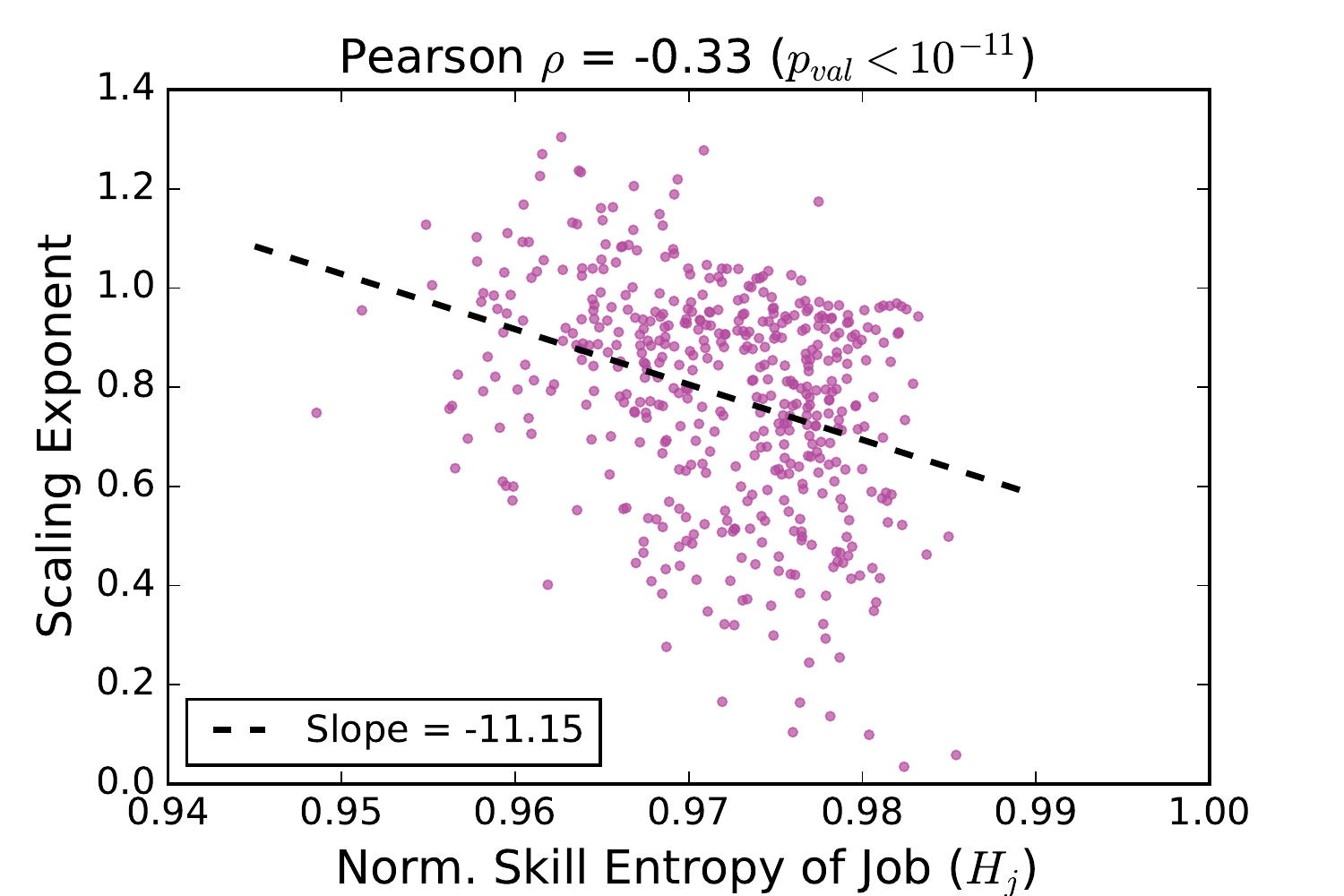}
		\put(160,110){\fbox{\small A}}
	\end{overpic}
	\begin{overpic}[width=.45\textwidth]{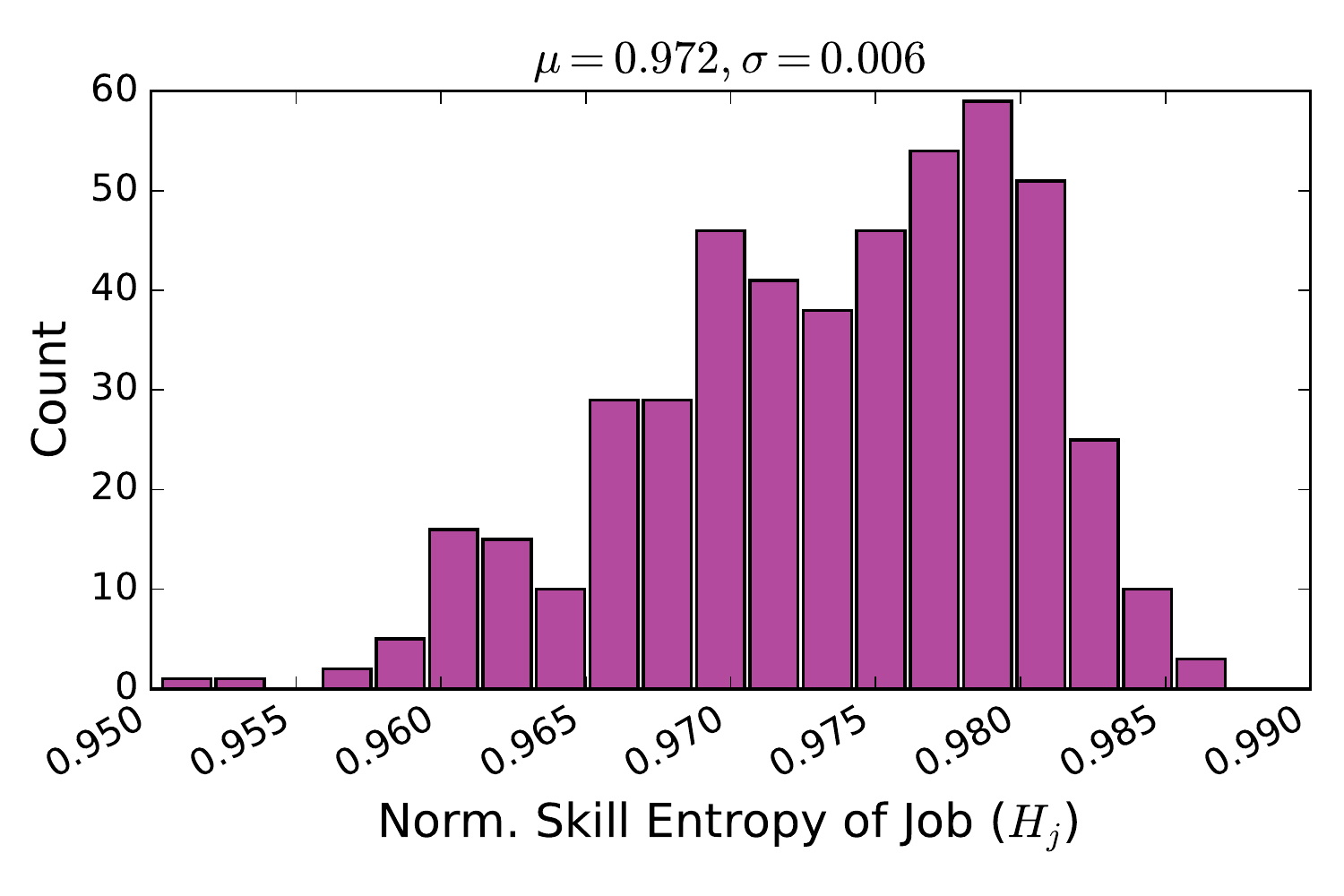}
		\put(180,110){\fbox{\small B}}
	\end{overpic}
	\caption{
	Characterizing the skill specialization of individual jobs.
	{\bf (A)} Skill specialization indicates larger scaling exponents with city size for individual jobs.
	{\bf (B)} The distribution of skill specialization across BLS jobs.
	}
	\label{jobEntropyPlots}
\end{figure}

We present BLS jobs ordered by decreasing skill specialization in Table~\ref{jobTableSection}.
We also provide the scaling exponent of each BLS job, along with the Pearson correlation of the relative abundance of each job to the expected job impact from automation (discussed below) across cities.
Figure~\ref{jobEntropyPlots}A shows that specialized jobs tend to have larger scaling exponents.
Figure~\ref{jobEntropyPlots}B shows the distribution of job specialization.% according to Eq.~\ref{jobEntropy}.
%%%%%%%%%%%%
\subsection{Characterizing Specialization through \onet Skills}
	\label{skillsSection}
\begin{figure}[h]
	\centering
	\begin{overpic}[width=.32\textwidth]{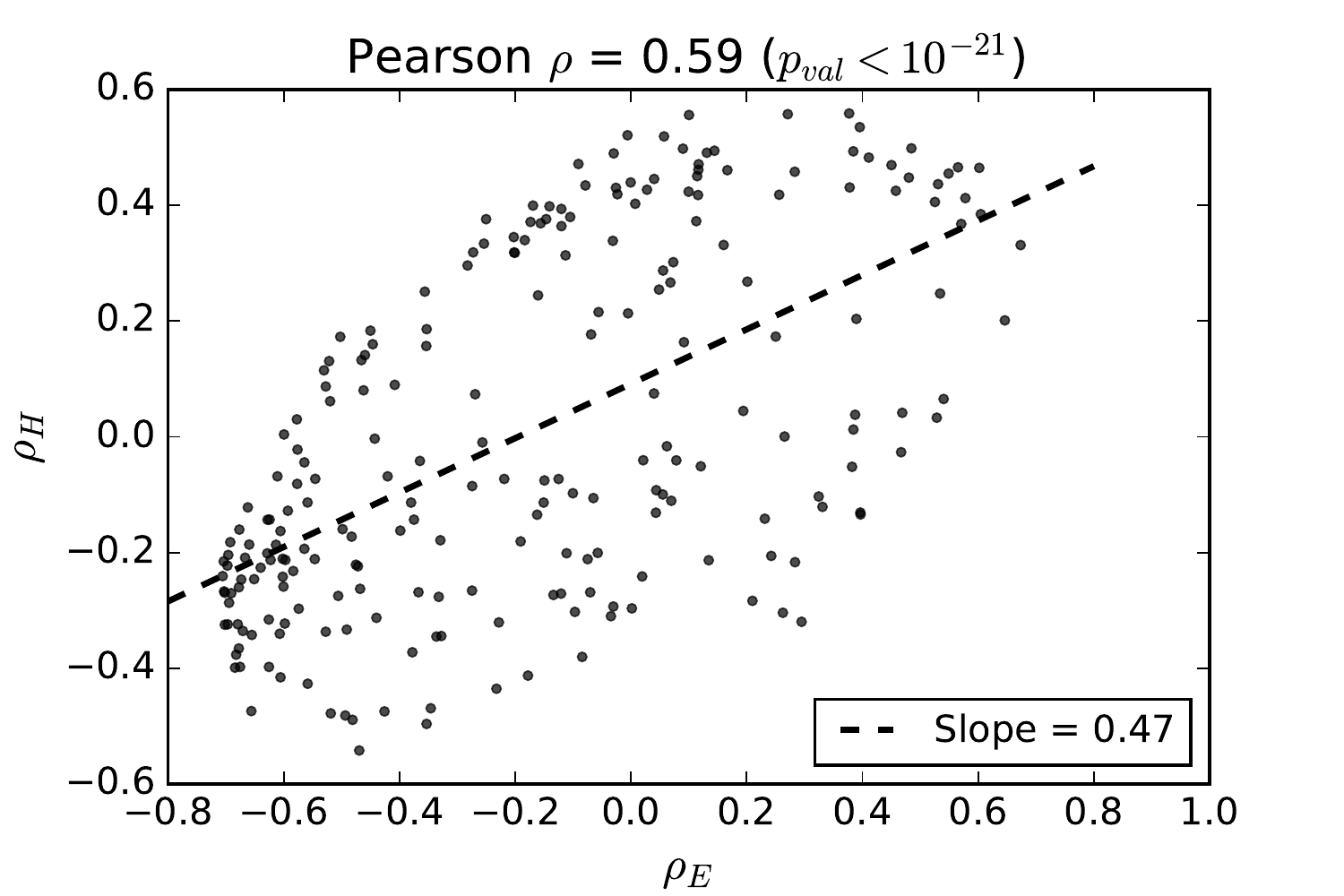}
		\put(20,75){\fbox{\small A}}
	\end{overpic}
	\begin{overpic}[width=.32\textwidth]{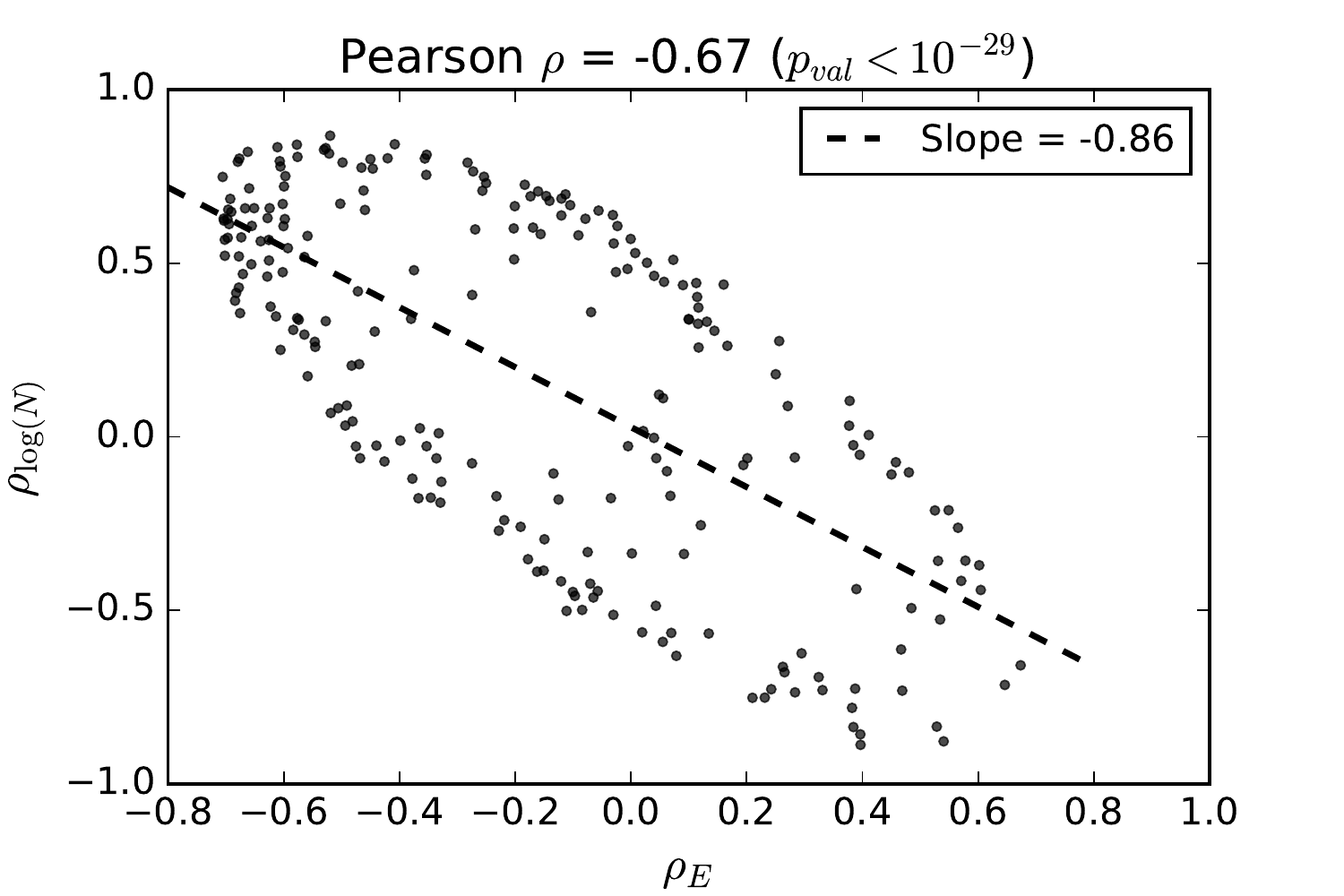}
		\put(20,30){\fbox{\small B}}
	\end{overpic}
	\begin{overpic}[width=.32\textwidth]{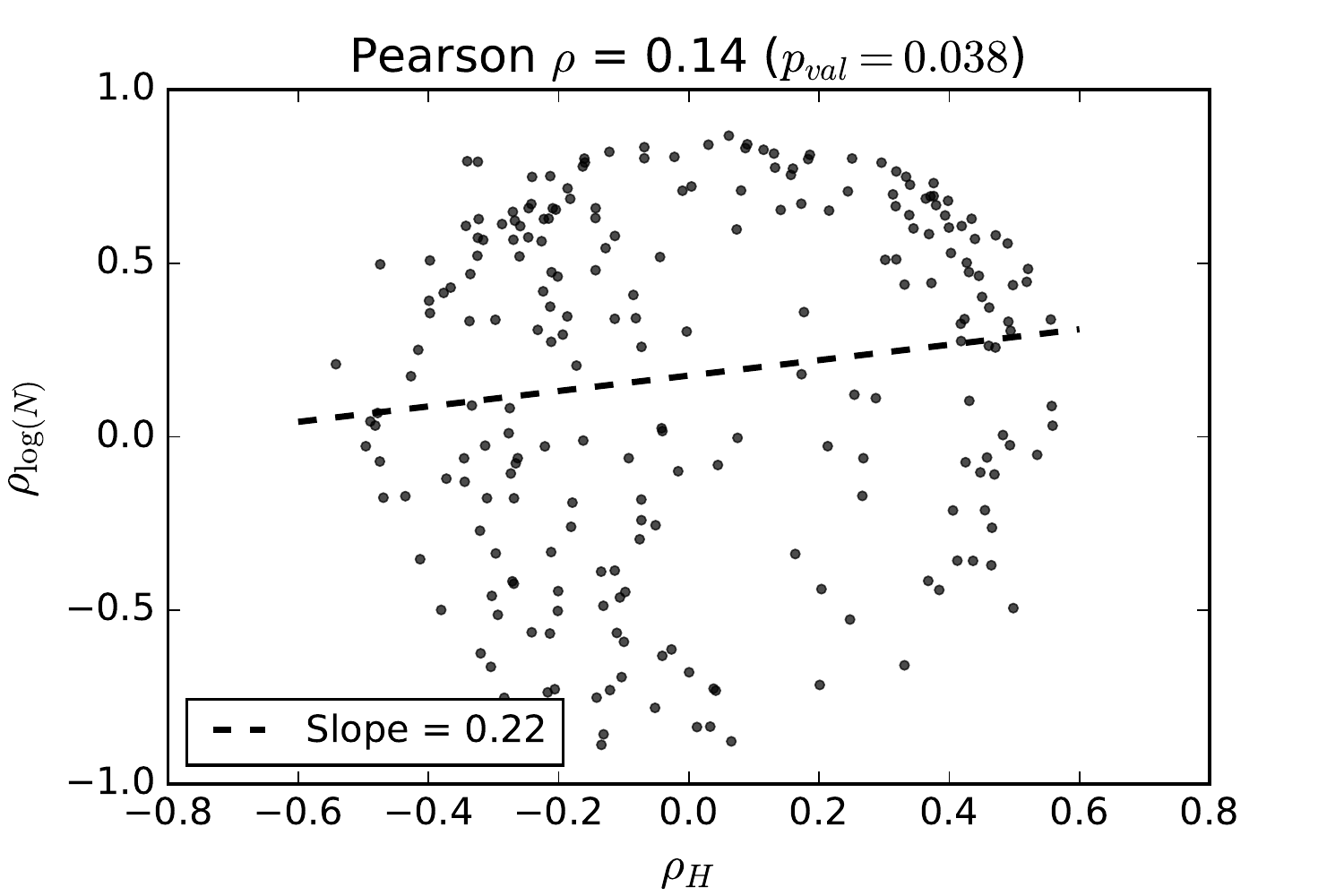}
		\put(20,30){\fbox{\small C}}
	\end{overpic}
	\caption{
	Comparing the relationships of \onet skills to city size, expected job impact, and labor specialization.
	{\bf (A)} We plot Pearson correlation of raw skill importance to expected job impact ($\rho_E$) on the x-axis versus the Pearson correlation of raw skill importance to city skill entropy ($\rho_H$) on the y-axis. 
	We see that which indicate job impact from automation also indicate decreased specialization in cities.
	{\bf (B)} We plot Pearson correlation of raw skill importance to expected job impact ($\rho_E$) on the x-axis versus the Pearson correlation of raw skill importance to city size ($\rho_{\log(N)}$) on the y-axis. 
	We see that which indicate job impact from automation also indicate smaller city sizes.
	{\bf (C)} We plot Pearson correlation of raw skill importance to city skill entropy ($\rho_H$)  on the x-axis versus the Pearson correlation of raw skill importance to city size ($\rho_{\log(N)}$) on the y-axis. 
	The correlation between these two variables is not significant
	}
	\label{skillsPlot}
\end{figure}
\indent We want to understand how each \onet skill contributes to the relationships we observe.
We present our findings in Table~\ref{skillsRelate}.
First, we compare the raw importance of a skill in each city by summing the raw importance of the skill across each job.
We then measure the Pearson correlation of the sum of a given skill compared to the expected job impact of each city (denoted $\rho_E$, second column of table), the skill entropy each city (denoted $\rho_{H}$, third column of table), and the size of each city (denoted $\rho_{\log(N)}$, right-most column of table).
The skills in the Table~\ref{skillsRelate} are ordered according to their correlation with expected job impact in cities.
For each column, the p-value for the correlation is presented in parentheses.\\
\indent Figure~\ref{skillsPlot} allows us to understand how related each correlation is by taking the Pearson correlation of each $\rho$ we described above.
Figure~\ref{skillsPlot}A demonstrates that skills which indicate lower expected job impact in cities also indicate greater skills specializations in cities.
Figure~\ref{skillsPlot}B demonstrates that skills which indicate lower expected job impact in cities also indicate larger cities.
Interestingly, Figure~\ref{skillsPlot}C demonstrates that skills which indicate skill specialization in cities are not significantly related to the skills which indicate city size.
This finding is surprising given the other panels of the figure, and motivates us to consider the relationship between occupational specialization and city size through the jobs in each city (see main text).
%\clearpage
%%%%%%%%%%%%%
\section{Estimating the Affects of Automation}
	\label{automation}
\indent Automation and its impact on labor are increasingly important topics to researchers~\cite{BrynjolfssonMcAfee,carlsson2012technological,olsen2014rise}.
Examples throughout history, such as the industrial revolution and the advent of computers, demonstrate how technological advancement can lead to both job loss and job creation~\cite{mazzucato2013financing,archibugi2013economic}. 
However, it is extremely difficult to predict how quickly a seemingly imminent technology will reach maturity and what the impact of that technology will be.
For example, it's currently topical to discuss self-driving cars, but, while autonomous-capable cars are available for purchase, no self-driving cars are currently operated on the mass market.
On the other hand, early leaders in computer hardware famously offered pessimistic predictions on the impact of computing:
\begin{itemize}
	\item \emph{``There is no reason anyone would want a computer in their home''} - Ken Olsen, founder of Digital Equipment Corporation (1977)
	\item\emph{``I think there is a world market for maybe five computers"} - Thomas Watson, former president of IBM (1943)
\end{itemize}

\subsection{Estimating Automation Impact using Frey/Osborne Data}
Frey and Osborne~\cite{freythe2013} produced probabilities of computerization for each BLS job.
They convened a workshop of leaders in automation to identify which of the BLS jobs were certainly automatable and certainly not-automatable.
They used the \onet skills dataset to identify the raw importance of nine workplace skills to each BLS job: Manual Dexterity, Finger Dexterity, Cramped Workspace/Awkward Positions, Originality, Fine Arts, Social Perceptiveness, Negotiation, Persuasion, and Assisting \& Caring for Others.
These \onet skills represent ``known bottlenecks to computerization.''
Using the importance of these skills to the jobs whose automatability was clear, they used a Gaussian process classifier to produce a probability of computerization for each BLS job.
\\
\begin{figure}[t]
	\centering
%	\begin{overpic}[width=.49\textwidth]{../journalFigures/jobLossDist.pdf}
%		\put(170,120){\fbox{\small A}}
%	\end{overpic}
	\begin{overpic}[width=.49\textwidth]{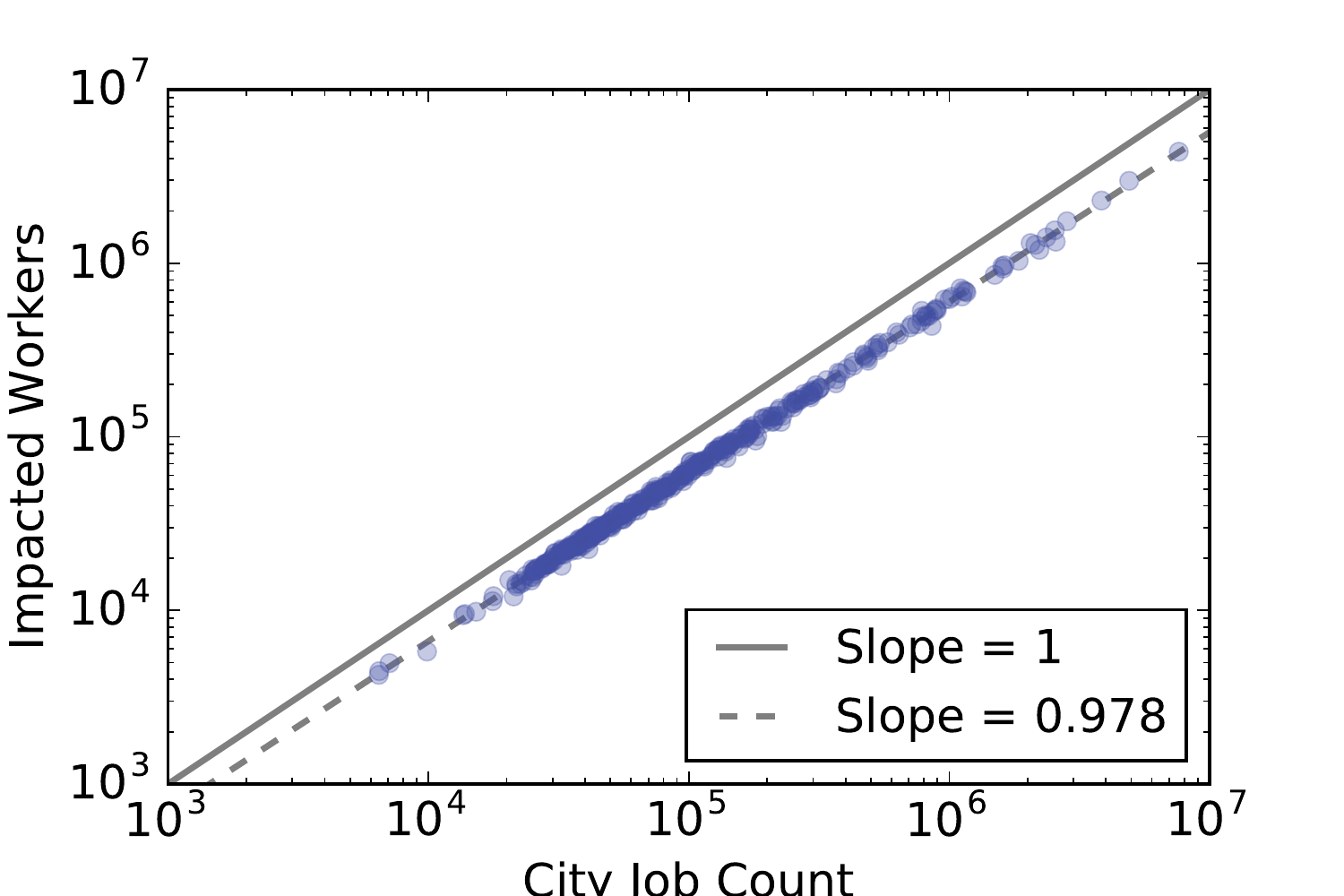}
		\put(40,120){\fbox{\small A}}
	\end{overpic}
	\begin{overpic}[width=.49\textwidth]{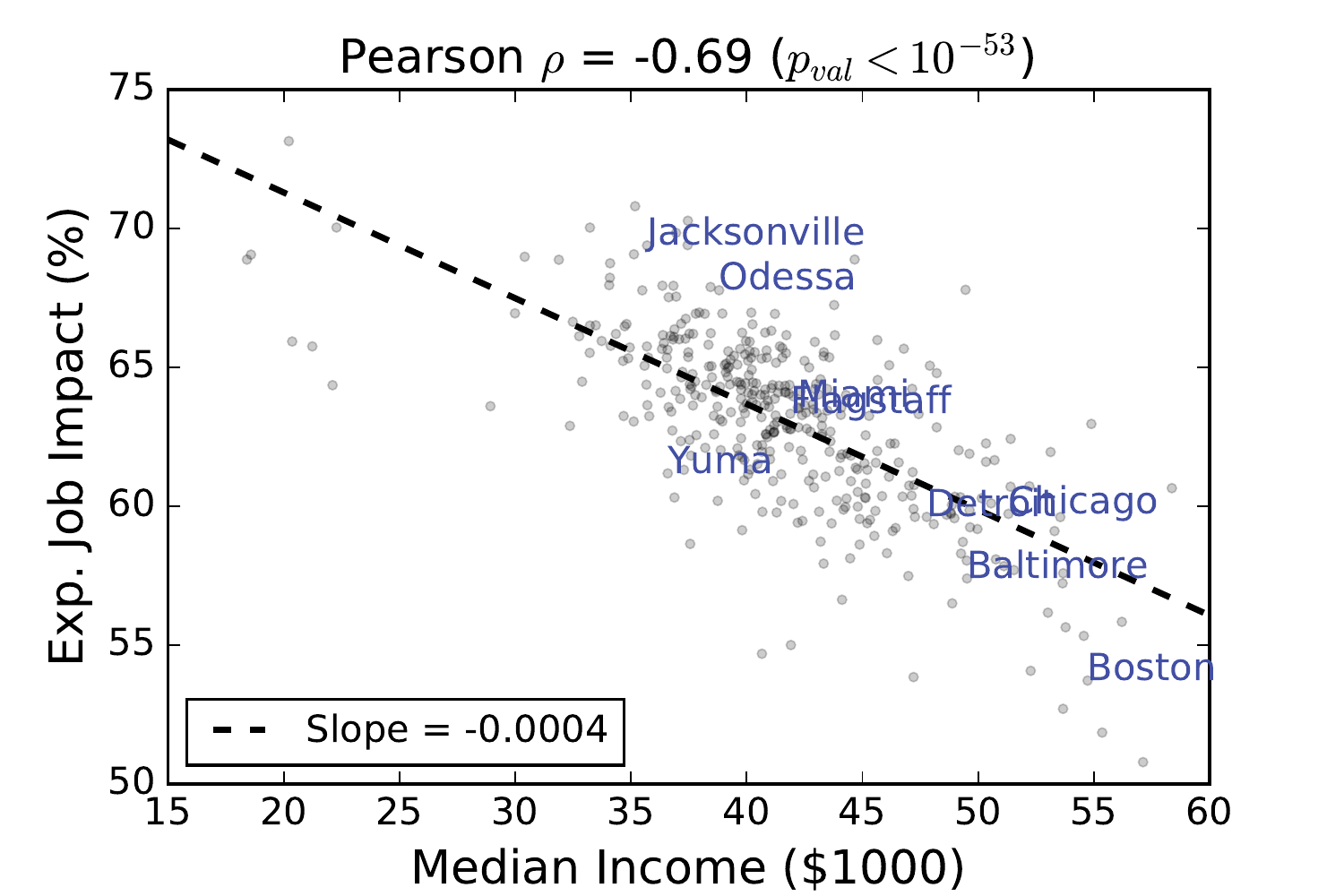}
		\put(170,120){\fbox{\small B}}
	\end{overpic}
	\caption{
		%{\bf (A)} The distribution of expected job displacement from automation across U.S. cities. 
		{\bf (A)} The expected number of displaced workers grows slightly sublinearly ($\beta=0.978$) with city size.
		{\bf (B)} Expected job impact is anti-correlated with median income of cities according to U.S. Census.
	}
	\label{expJobDisp}
\end{figure}
\indent Frey and Osborne used these probabilities to conclude that 47\% of the current U.S. jobs are at ``high risk" of computerization. 
Several studies~\cite{autor2015there,benessia2015sustainability,pajarinencomputerization2015} utilize these same probabilities to investigate the impacts of automation, which highlights the utility of the probabilities despite the difficulty of the prediction undertaken in \cite{freythe2013}.
We use the same probabilities in combination with the distribution of BLS jobs across U.S. cities to add spatial resolution to their findings. 
For a city, $m$, the expected job impact from automation is calculated according to
\begin{equation}
	\label{cityJobDisp}
	E_m = \sum_{j\in Jobs_m} p_{auto}(j)\cdot p_m(j),
\end{equation}
where $p_{auto}(j)$ is the probability of computerization according to \cite{freythe2013} and $p_m(j)$ is the proportion of workers in city $m$ with job $j$.
%Figure~\ref{expJobDisp} shows that each U.S. city can expect between 1/2 and 3/4 of their current workers to be displaced as imminent automation technologies reach maturity.
Table~\ref{orderedCities} demonstrates the ordered list of cities according to expected job impact.
\\
\indent As mentioned above, it's difficult to validate automation predictions.
Nonetheless, our calculations for expected job impact represent an aggregate signal for the types of jobs in a city in relation to imminent automation technology.
In the Table~\ref{orderedCities}, we present the U.S. cities ordered by their expected job impact from automation.
The list produces an ordering that appears to make sense; cities with technology companies and research institutes, such as Boston, M.A., and Boulder, C.O., have the lowest expected job impact, while cities relying on the tourist industry and agriculture, such as Myrtle Beach, S.C., and Napa, C.A., have the highest expected job impact.
While the absolute proportions can only be validated with time, we believe the overall trend embodied in expected job impact in cities represents an underlying true signal. 

\begin{figure}[t]
    \centering
    \begin{overpic}[width=.45\textwidth]{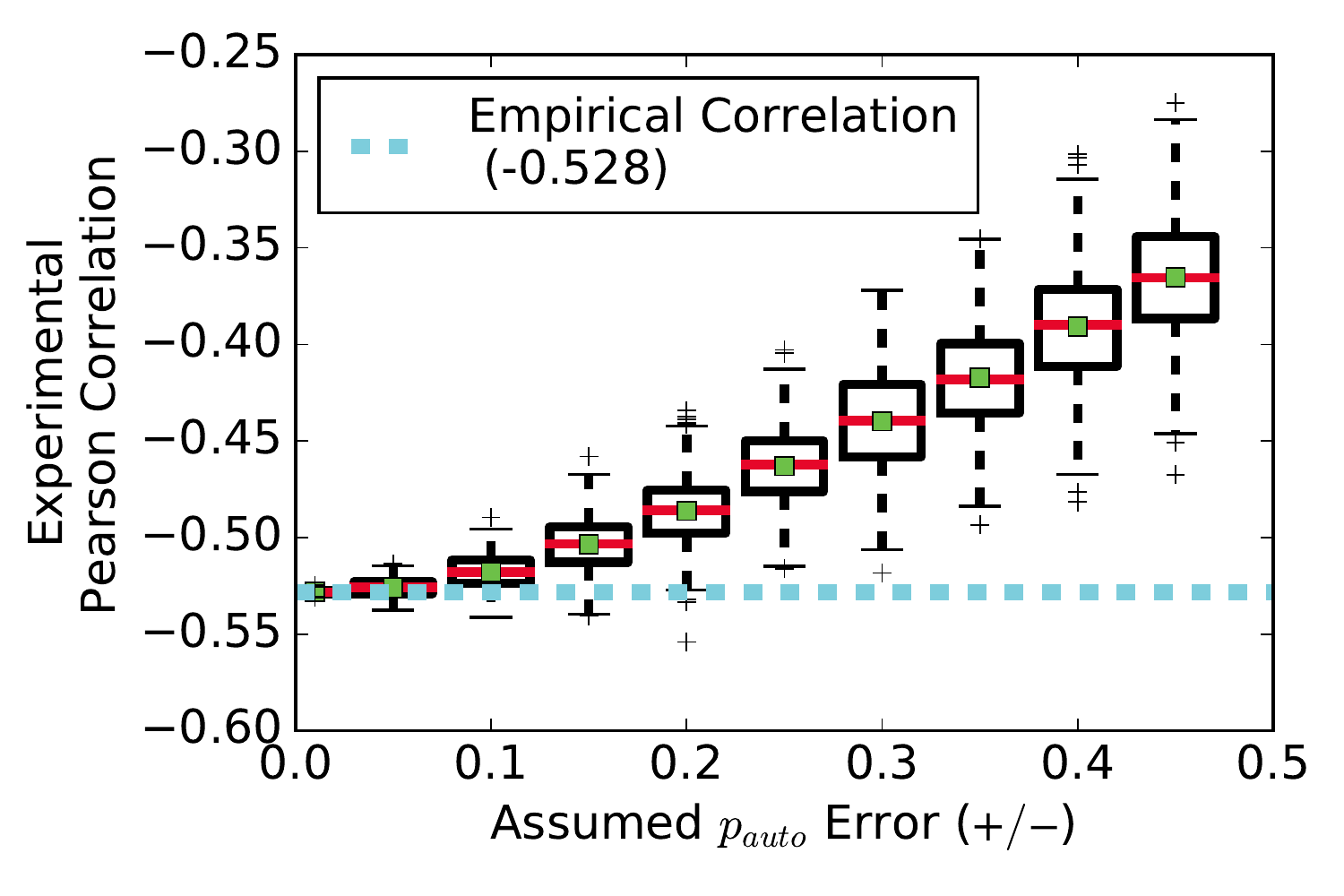}
        \put(70,90){\fbox{\small A}}
    \end{overpic}
    \begin{overpic}[width=.45\textwidth]{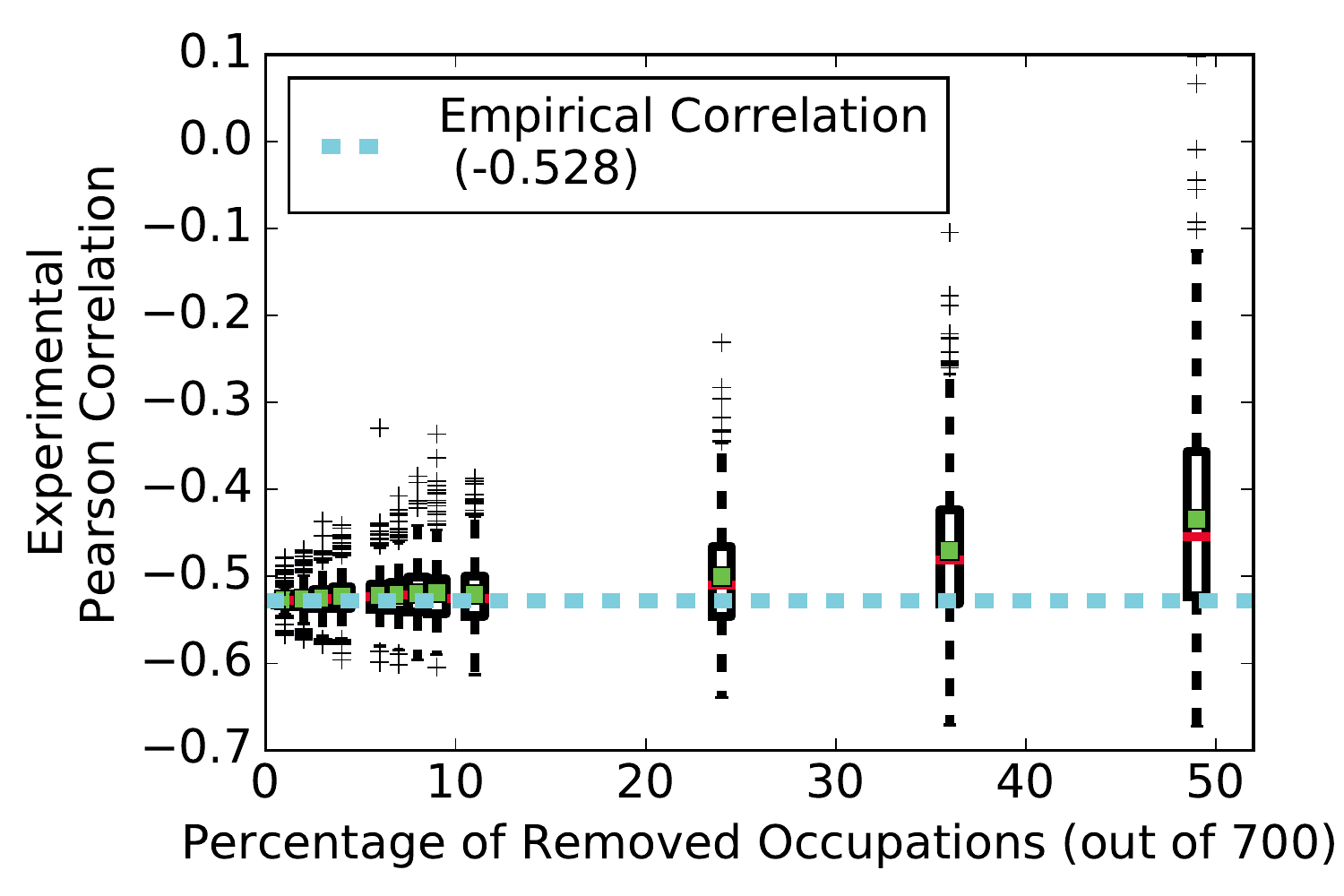}
        \put(70,90){\fbox{\small B}}
    \end{overpic}
    \caption{
        The relationship between city size and expected job impact from automation is robust.
        {\bf (A)} For choice of assumed error in the predictions from \cite{freythe2013}, we perform 500 trials measuring the resulting Pearson correlation between $\log_{10}$ city size and expected job impact from automation after the error has been added to each occupation's probability of computerization (y-axis). 
        {\bf (B)} After selecting a proportion of occupations (x-axis), we perform 500 trials of randomly selecting that many occupations to remove while measuring the resulting Pearson correlation between $\log_{10}$ city size and the expected job impact from automation in cities (y-axis).
    }
    \label{FOrobust}
\end{figure}

To demonstrate the robustness of our results further, we perform two robustness checks to verify the negative trend between city size and expected job impact from automation (see Fig. 1B from the main text).
The probability of computerization (i.e. $p_{auto}(j)$) from \cite{freythe2013} are produced through a machine learning process applied to predictions of the automatability of jobs from experts.
Therefore, we expect some errors in the predictions of these experts, and our task is to demonstrate that the error in the resulting $p_{auto}(j)$ would need to be substantial to invalidate our finding.
We perform this analysis by artificially adding random noise to each $p_{auto}(j)$ according to 
\begin{equation}
    p^*_{auto}(j) = p_{auto}(j)+e_j,
\end{equation}
where $e_j$ is chosen uniformly at random from the interval $[-error,+error]$ for each occupation.
For each choice of $error$, we perform 500 trials calculating a new $p^*_{auto}(j)$ for each occupation and recalculating the expected job impact from automation in each city according to 
\begin{equation}
    E^*_m = \sum_{j\in Jobs_m} p^*_{auto}(j)\cdot p_m(j)
\end{equation}
similar to equation~\ref{cityJobDisp}.
Finally, we measure the Pearson correlation between $\log_{10}$ the total employment in each city and $E^*_m$ so that we can compare to the empirical relationship we observe in Figure 1B of the main text (Pearson $\rho=-0.53$, $p_{val}<10^{-28}$).
Figure~\ref{FOrobust}A demonstrates the results of this exercise.
We find that substantial error ($error\approx0.15$) needs to be added to the empirical probabilities of computerization for each occupation before our result from the main text no longer represents the observed trend.
Even if we make the extremely strong assumption of $error=.5$, we would still observe a strong negative trend, and we would still conclude that small cities face greater impact from automation.

In the second robustness check, we test the robustness of our observed relationship between city size and expected job impact if a randomly selected subset of occupations are removed from the analysis.
For each proportion of occupations to be removed, we perform 500 trials of randomly selecting occupations to be ignored and recalculate $E_m$ using the $p_{auto}(j)$ presenting in \cite{freythe2013}.
We then measure the resulting Pearson correlation between these new $E_m$ and $\log_{10}$ the total employment in each city.
Figure~\ref{FOrobust}B demonstrates that our empirical observation from Figure 1B in the main text holds even if very large proportions of occupations are ignored. 
In fact, only when we ignored half of all occupations did we observe any trials demonstrating a trend contrary to the one presented in the main text.
Therefore, we conclude that small cities face greater impact from automation.

\subsection{Estimating Automation Impact using OECD Data}

\begin{figure}[!ht]
	\centering
	\begin{overpic}[width=.49\textwidth]{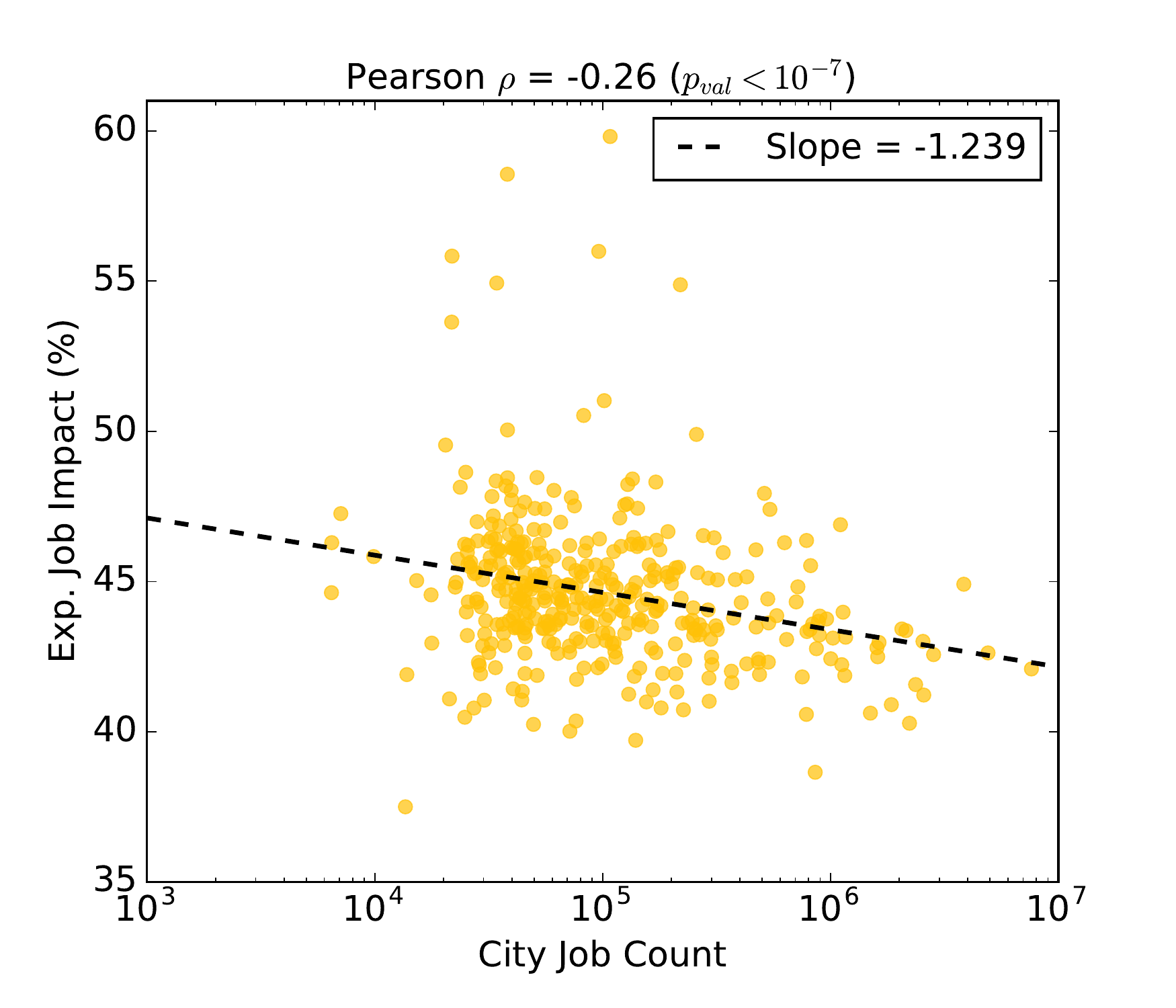}
		\put(50,150){\fbox{\small A}}
	\end{overpic}
	\begin{overpic}[width=.49\textwidth]{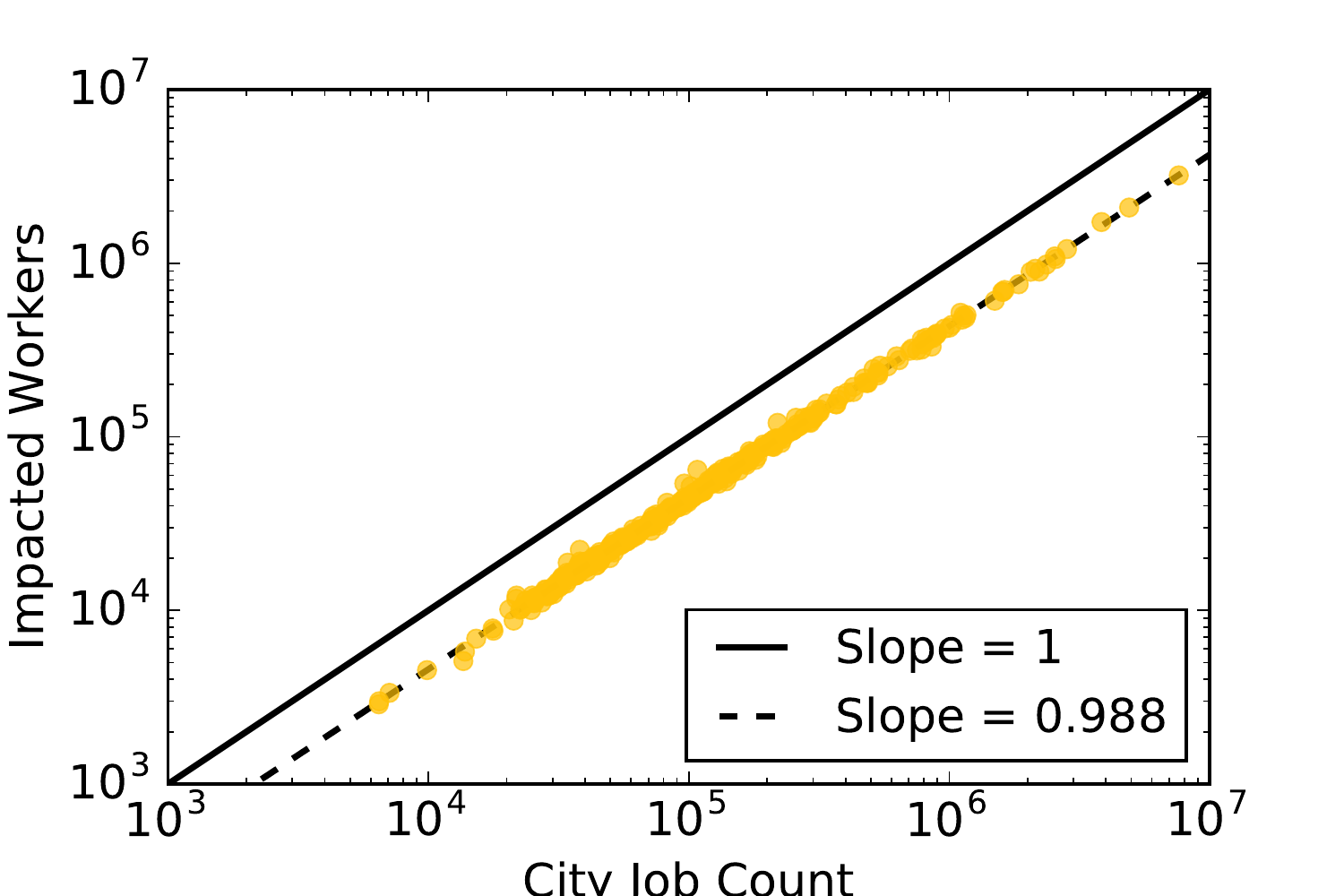}
		\put(50,120){\fbox{\small B}}
	\end{overpic}
	\caption{
		Expected job impact from automation decreases with city size using conservative estimates of job loss.
		{\bf (A)} The expected job impact of cities decreases with city size.
		{\bf (B)} The number of displaced workers per city grows slightly sublinearly with city size ($\beta=0.988$).
	}
	\label{OECD}
\end{figure}

The Organization for Economic Co-operation Development (OECD) released alternative estimates for the probability of job automation with a focus on job categories used by OECD countries~\cite{arntz2016risk}. 
Rather than the job-based approach used in \cite{freythe2013}, assessments on the automatability of workplace skills were derived. 
These skill assessments can be used in combination with government data relating the importance of skills to jobs to assess the likelihood of computerization for jobs.
Contrary to the alarming 47\% of jobs at ''high risk of computerization" found by Frey and Osborne, these new probabilities produce a more mild conclusion of only 9\%. 
These job probabilities were derived with OECD job definitions in mind, but collaborations between OECD and U.S. BLS have lead to an official mapping between the two job definitions.
We utilize this mapping to assess the resilience of labor markets in cities as a function of city size in Figure~\ref{OECD}.
Despite the more conservative estimates in \cite{arntz2016risk}, our results remain; we again observe significantly decreased expected job impact in large cities (Fig.~\ref{OECD}A).

\subsection{Expected Job Impact \& Labor Specialization in Cities}
\begin{figure}[t]
	\centering
	\begin{overpic}[width=.49\textwidth]{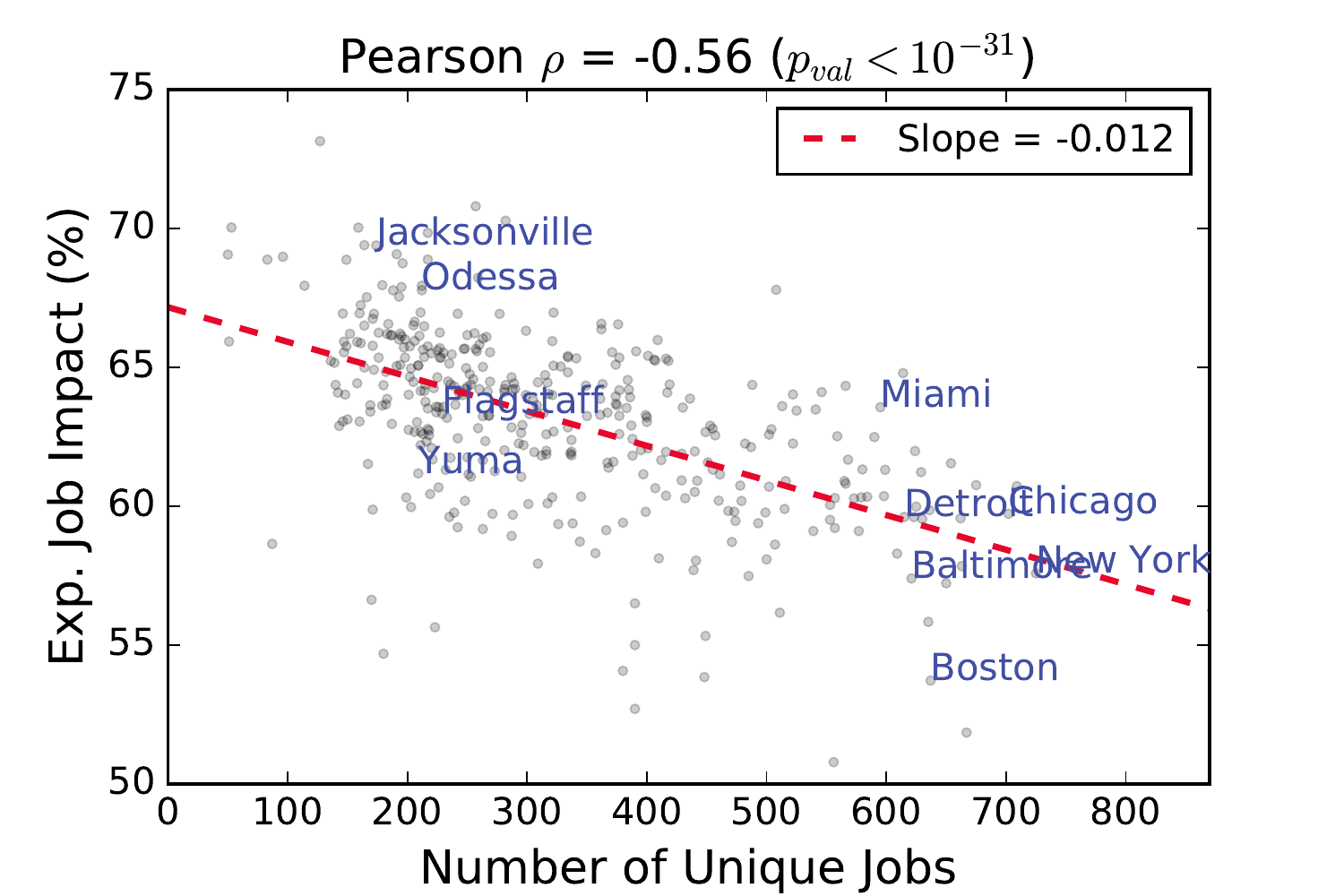}
		\put(35,120){\fbox{\small A}}
	\end{overpic}
	\begin{overpic}[width=.49\textwidth]{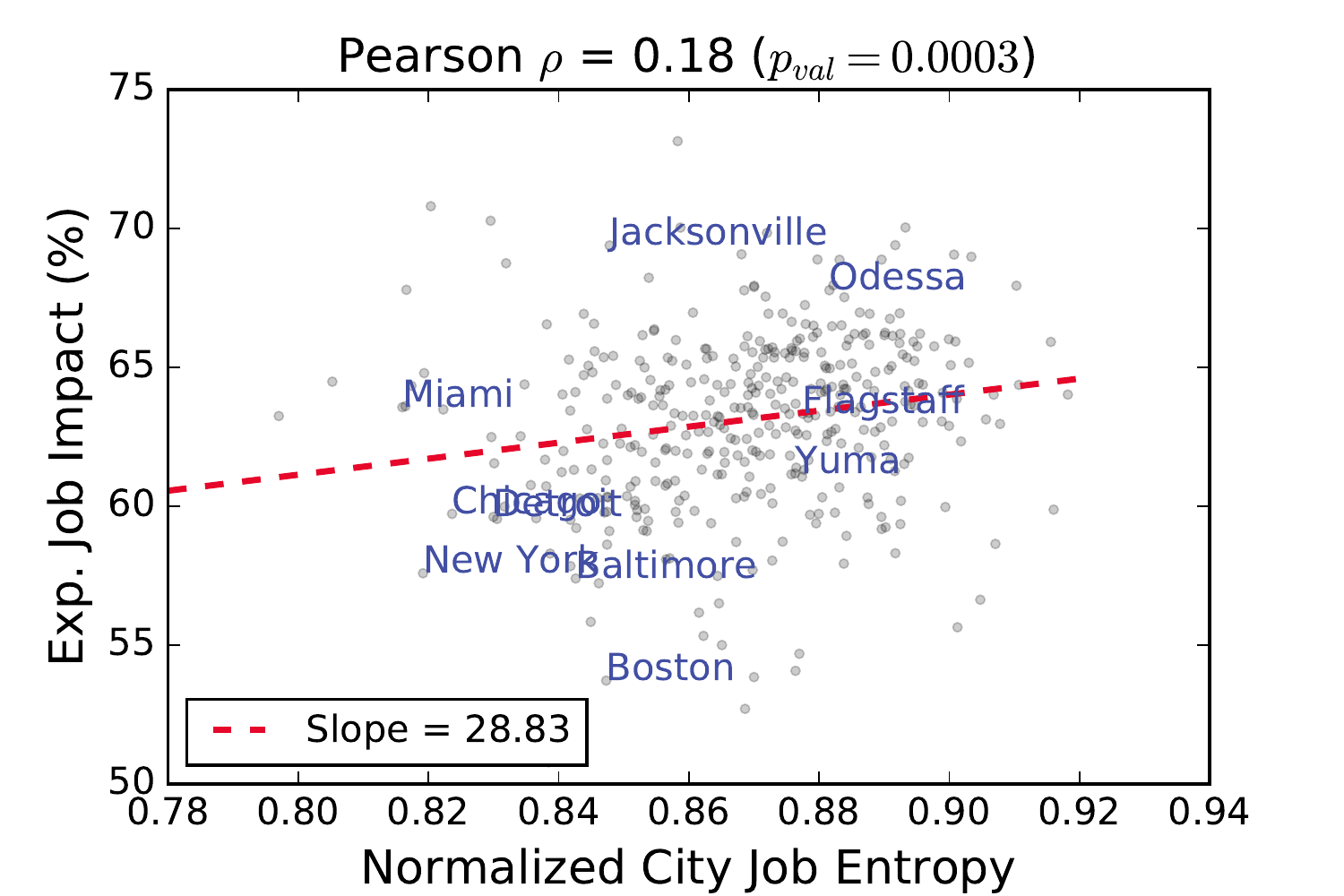}
		\put(35,120){\fbox{\small B}}
	\end{overpic}
	\caption{
	Characterizing the relationship between labor specialization and job impact from automation in cities.
	{\bf (A)} Resilience to job impact is correlated with the number of unique jobs in cities.
	{\bf (B)} Increased labor specialization according to job distributions in cities indicates increased resilience to job impact from automation.
	}
	\label{altJobDisp}
\end{figure}

\indent In Figure~\ref{altJobDisp}, we further characterize the relationship between a city's resilience to job impact from automation and labor specialization.
We provide additional figures in the main text detailing how workplace skills explain the positive correlation we observe between labor specialization and resilience to job impact in cities. 
Here, we demonstrate that resilience to job impact is significantly correlated to the number of unique jobs in a city, and more weakly correlated to the Shannon entropy of job distributions.
This weaker correlation motivates our investigation into workplace skills, in addition to the distribution of jobs, presented in the main text.
\clearpage
\subsection{Explaining Differences in Expected Job Impact}
From equation~\ref{cityJobDisp}, we may observe that both the automatability and the employment share of individual occupations contribute to the expected job impact from automation in a city.
Correspondingly, we measure the difference in expected job impact for cities $m$ and $n$ according to 
\begin{align}
    \begin{split}
        E_m-E_n &= \displaystyle\sum_{j\in Jobs}p_{auto}(j)\cdot\big(share_m(j)-share_n(j)\big)\\
            &= \sum_{j\in Jobs}p_{auto}(j)\cdot\big(share_m(j)-share_n(j)\big)-\sum_{j\in Jobs}E_n\cdot\big(share_m(j)-share_n(j)\big) \\
            &= \sum_{j\in Jobs}\big(p_{auto}(j)-E_n\big)\cdot\big(share_m(j)-share_n(j)\big),
    \end{split}
    \label{occShift}
\end{align}
where we have utilized $\sum E_n\cdot(share_m(j)-share_n(j))=0$.
Here, we let $Jobs$ denote the set of all occupation types across all cities, $p_{auto}(j)$ denotes the probability of computerization of occupation $j$ according to \cite{freythe2013}, and $share_m(j)$ denotes the employment share of occupation $j$ in city $m$.
Equation~\ref{occShift} highlights that occupation $j$'s influence on the difference in expected job impact in cities $m$ and $n$ falls into one of four categories:
\begin{enumerate}
    \item occupation $j$ is relatively resilient to automation (i.e. $(p_{auto}(j)-E_n)>0$) and relatively more abundant in city $m$ (i.e. $(share_m(j)-share_n(j))>0$),
    \item occupation $j$ is relatively susceptible to automation (i.e. $(p_{auto}(j)-E_n)<0$) and relatively less abundant in city $m$ (i.e. $(share_m(j)-share_n(j))<0$),
    \item occupation $j$ is relatively resilient to automation (i.e. $(p_{auto}(j)-E_n)>0$) and relatively less abundant in city $m$ (i.e. $(share_m(j)-share_n(j))<0$), or
    \item occupation $j$ is relatively susceptible to automation (i.e. $(p_{auto}(j)-E_n)<0$) and relatively more abundant in city $m$ (i.e. $(share_m(j)-share_n(j))>0$).
\end{enumerate}
Occupations in categories 1 and 2 effectively increase $E_m-E_n$, while occupations in categories 3 and 4 effectively decrease the difference.

\begin{figure}[!p]
    \centering
    \includegraphics[width=.97\textwidth]{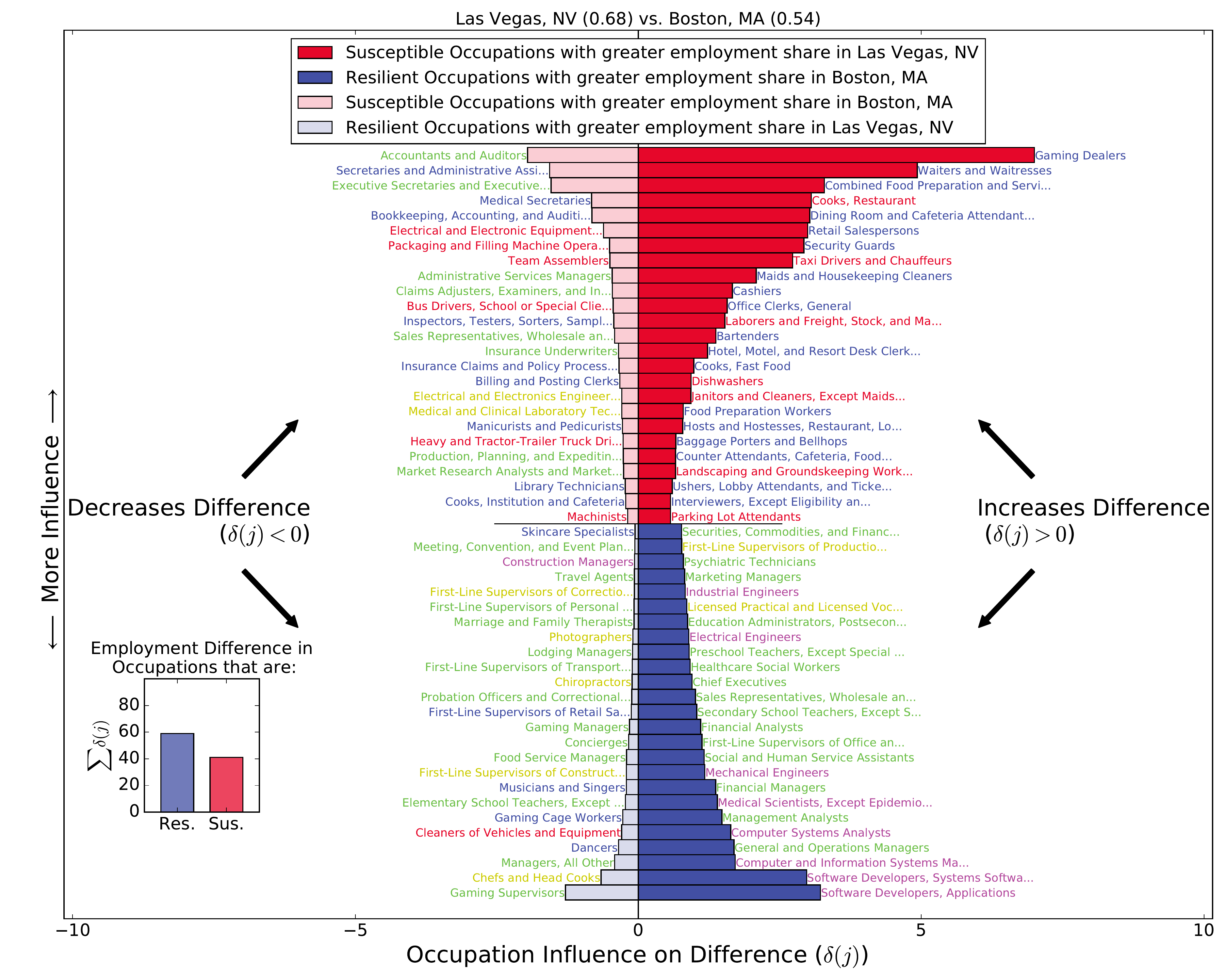}
    \caption{
        An occupation shift explaining the difference in expected job impact for Boston, MA ($E_m=0.54$) compared to Las Vegas, NV ($E_m=0.68$) using equation~\ref{occShift}.
        Each horizontal bar represents $\delta_{(\text{Las Vegas},\text{Boston})}(j)$ of occupation $j$.
        The occupation title is provided next to the corresponding bar and colored according to its job cluster as identified in Figure 4 of the main text.
        Red bars represent occupations with higher risk of computerization compared to Boston's expected job impact.
        Blue bars represent occupations with lower risk of computerization compared to Boston's expected job impact.
        Dark colors represent occupations that effectively increase the difference, while pale colors represent occupations that effectively decrease the difference in expect job impact.
        Bars in each of the quadrants are vertically ordered according to $|\delta_{(\text{Las Vegas},\text{Boston})}(j)|$.
        The inset in the bottom left of the plot summarizes the overall influence of resilient occupations compared to occupations that are at risk of computerization.
    }
    \label{Boston_LasVegas}
\end{figure}

Let 
\begin{equation}
    \delta_{(m,n)}(j) = 100\cdot\frac{\big(p_{auto}(j)-E_n\big)\cdot\big(share_m(j)-share_n(j)\big)}{E_m-E_n}
    \label{diffInfluence}
\end{equation}
denote the percent influence of occupation $j$ on the difference in expected job impact for cities $m$ and $n$.
Figure~\ref{Boston_LasVegas} demonstrates a visualization of equation~\ref{occShift} that we call an ``occupation shift."
Correspondingly, if we add the employment distributions in the 50 largest cities and 50 smallest cities together (respectively), then we can quantify how each occupation contributes to the differential impact of automation on employment in large and small cities.
We present this occupation shift in Figure~\ref{citySizeShift} (also Figure 5 of the main text).

Referring to the job clusters from Figure 5 in the main text, we see that purple occupations and blue occupations contribute the most to the difference in expected job impact, while green and yellow occupation types effectively diminish the difference in both occupation shifts.
However, certain occupations, such occupations of the green job cluster, can both increase and decrease the difference between resilient and susceptible cities.
The occupation shift allows us to understand which occupations explain the overall trend and which occupations go against the overall trend.
If we had only considered occupations that add to the difference (i.e. occupations corresponding to dark colored bars on the right side of the plot), then we may have incorrectly concluded that the differences in relatively susceptible occupations explain the difference we observe in these two examples.
This transparency can help urban policy makers determine how labor shifts in different industries may effect their preparedness for the impact of new technology.

\begin{figure}[!p]
    \centering
    \includegraphics[width=.97\textwidth]{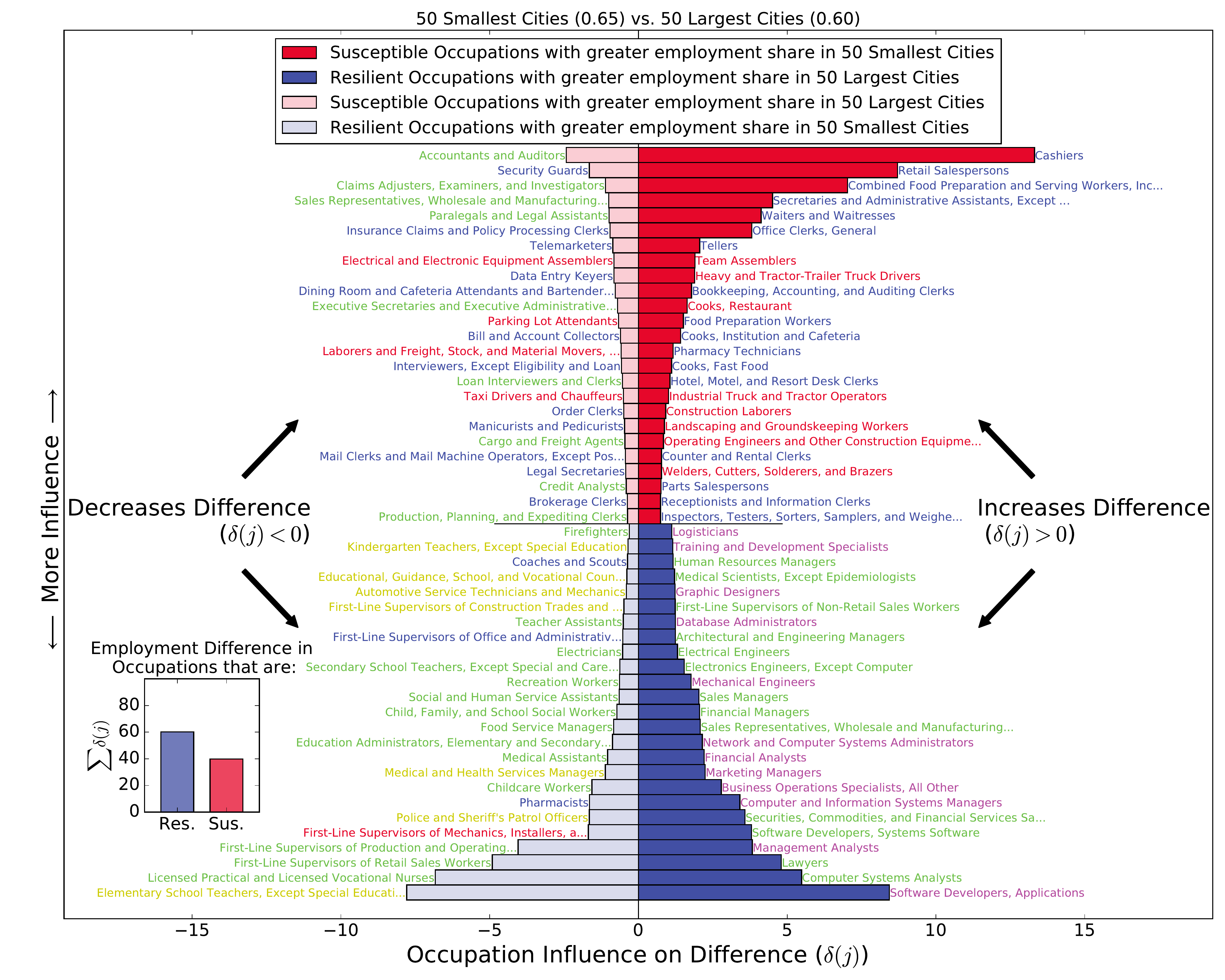}
    \caption{
        An occupation shift explaining the difference in expected job impact for the 50 largest cities ($E_m=0.60$) compared to the 50 smallest cities ($E_m=0.65$) using equation~\ref{occShift}.
        Each horizontal bar represents $\delta_{(\text{Small Cities},\text{Large Cities})}(j)$ of occupation $j$.
        The occupation title is provided next to the corresponding bar and colored according to its job cluster as identified in Figure 5 of the main text.
        Red bars represent occupations with higher risk of computerization compared to the expected job impact in large cities.
        Blue bars represent occupations with lower risk of computerization compared to the expected job impact in large cities.
        Dark colors represent occupations that effectively increase the difference, while pale colors represent occupations that effectively decrease the difference in expect job impact.
        Bars in each of the quadrants are vertically ordered according to $|\delta_{(\text{Small Cities},\text{Large Cities})}(j)|$.
        The inset in the bottom left of the plot summarizes the overall influence of resilient occupations compared to occupations that are at risk of computerization.
    }
    \label{citySizeShift}
\end{figure}

\clearpage
%\section{Multiple Regression for Predicting Expected Job Impact}
\section{Robustness Check of the Linear Regression Model for $E_m$}
\begin{figure}[!ht]
    \centering
    \begin{overpic}[width=.49\textwidth]{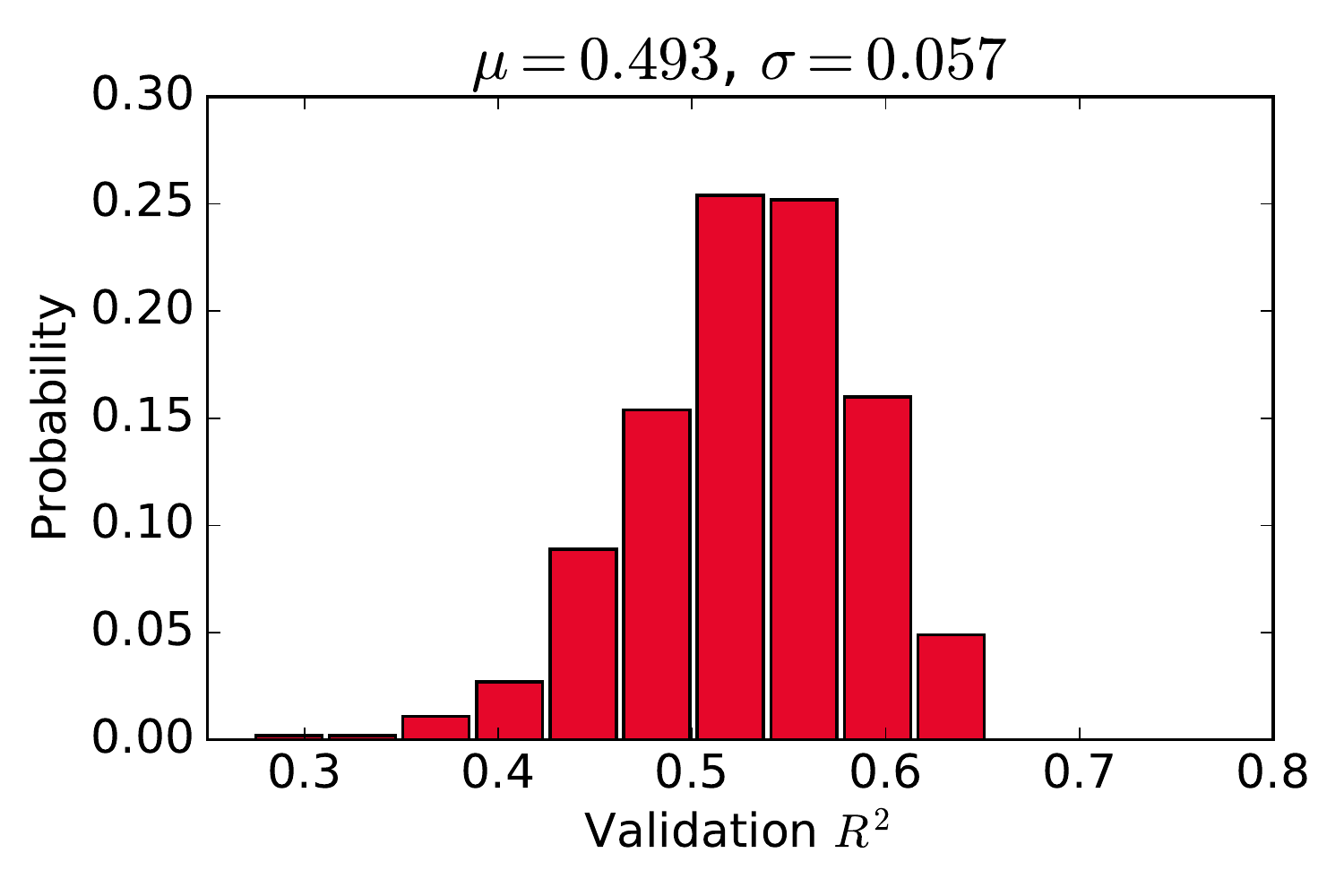}
        \put(40,120){\fbox{\small A}}
    \end{overpic}
    \begin{overpic}[width=.49\textwidth]{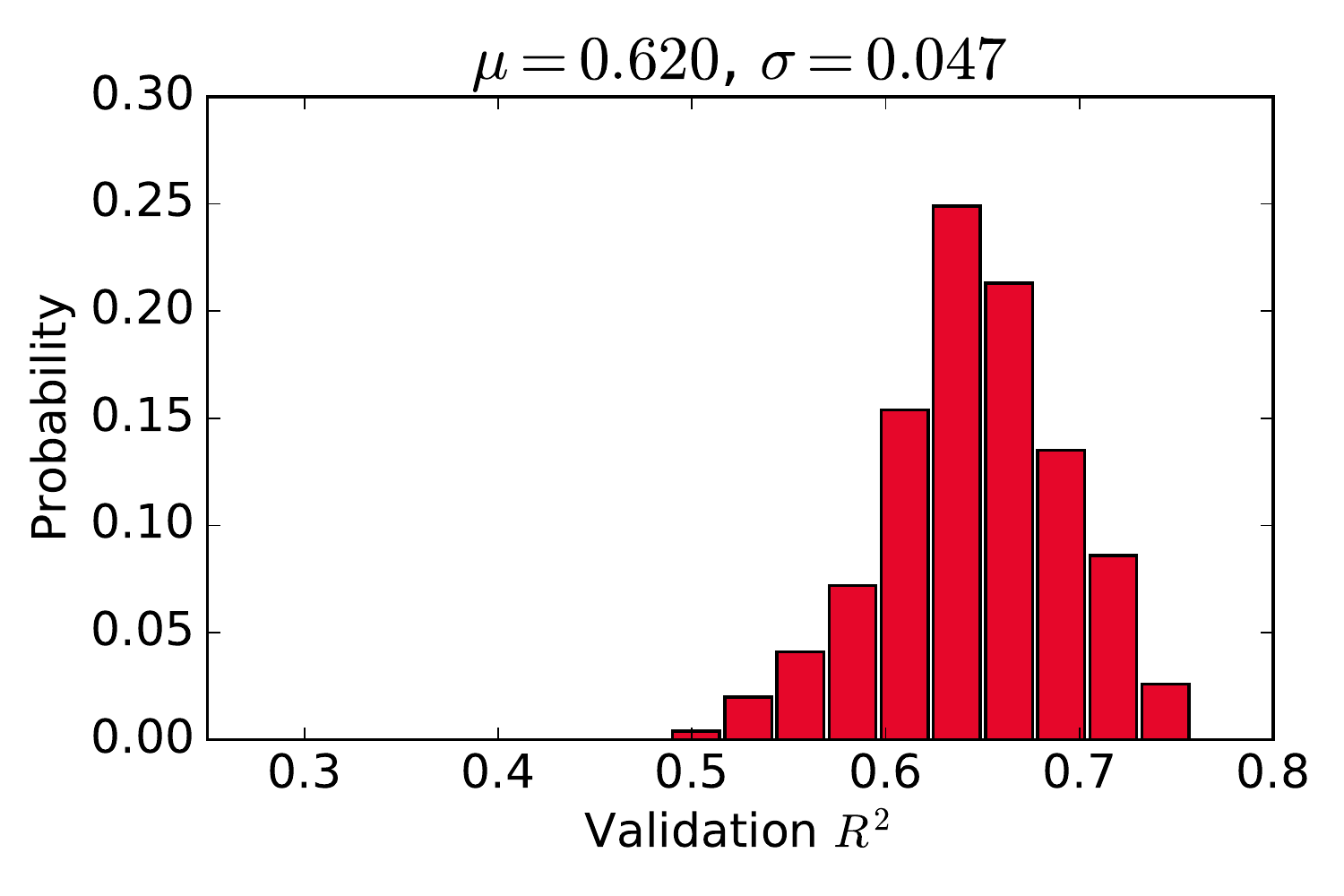}
        \put(40,120){\fbox{\small B}}
    \end{overpic}
    \caption{
        To confirm the validity of the regression models, we perform 1,000 trials where half of the cities are randomly selected without replacement as training data and the remaining cites are used for validation. 
        We under go this process for the regression model using only generic urban indicators {\bf (A)} and the regression model using all variables {\bf (B)}.
        The resulting distributions of variance explained ($R^2$) when the trained models are applied to separate validation data confirms that the full regression model accounts for an additional 10\% of variance on average.
    }
    \label{OLSoverfit}
\end{figure}
%
%\begin{figure}[!ht]
%    \centering
%    \includegraphics[width=.9\textwidth]{figures/MLS_all.pdf}
%    \caption{
%        Comparing the actual and predicted values of expected job displacement using the multiple linear regression model on all variables.
%    }
%    \label{MlsPerformance}
%\end{figure}

%%%%%%%%%%%%%%%%%
\clearpage
\section{Simplifying Jobs \& Skills}
	\label{simplify}
\indent In an effort to clearly identify how jobs contribute to labor specialization in larger cities, we identify aggregate job types based on common workplace skills.
Previous studies, such as \cite{hyejin}, examined the relationship between industry size and city size for various abstractions for industry according to NAICS. 
Here, we are seeking an organic representation of the forces in effect, and so we use K-means clustering based on the raw skill values for each job to identify five clusters of similar jobs (i.e. occupations are instances and the raw \onet importance of each skill are features).
These job groups represent collections of jobs which rely on similar skills for completion.
The BLS jobs comprising each job type are shown in Section~\ref{jobGroups}.
Note that our results and interpretations are consistent for anywhere from three to seven clusters (see \ref{altJobGroups}).
This simplification of the space of jobs allows us to clearly understand which job types are disproportionately emphasized in large cities through the scaling behaviors of these job types.
\\
\begin{figure}[t]
	\centering
	\begin{overpic}[width=.45\textwidth]{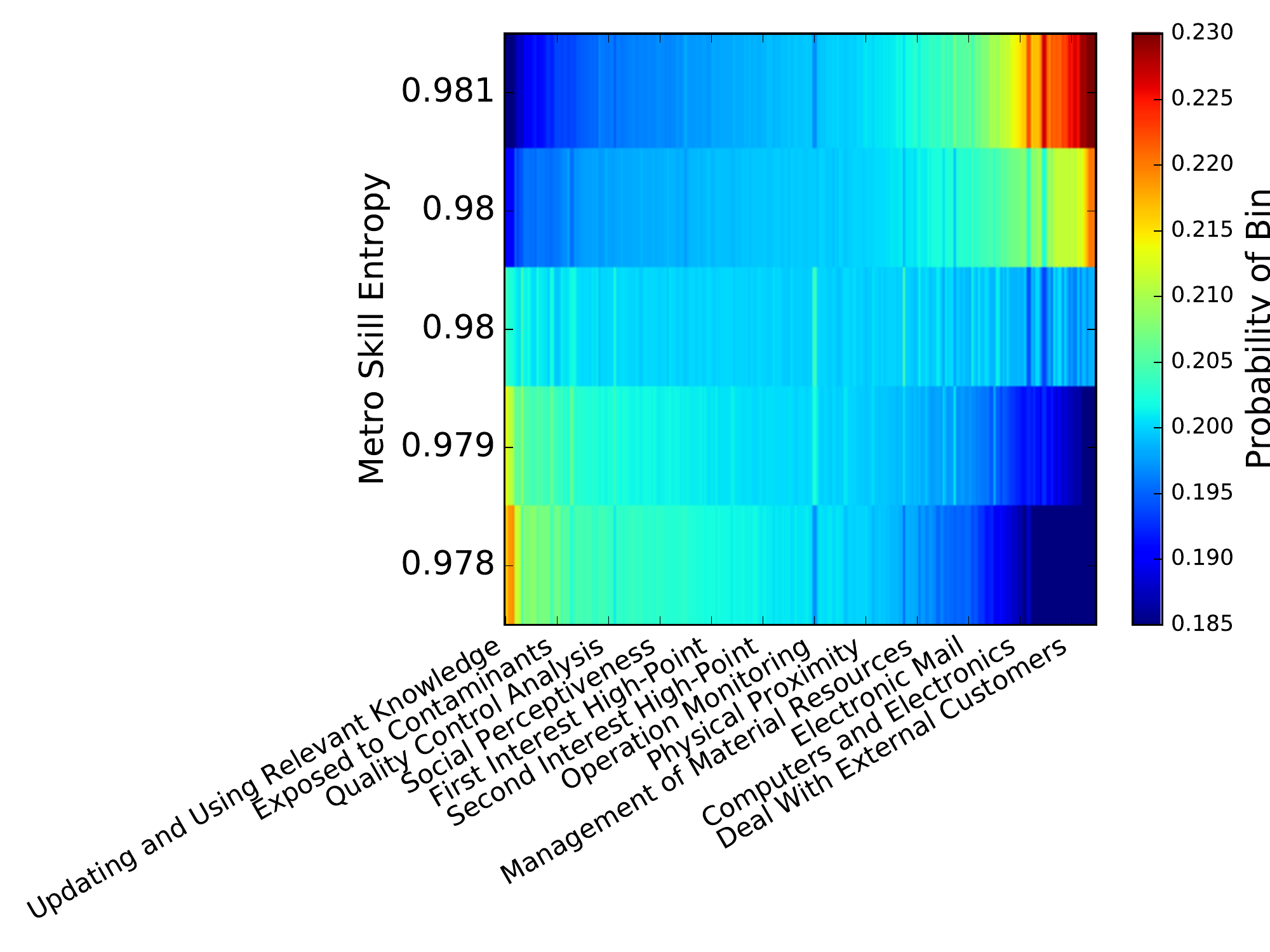}
		\put(20,60){\fbox{\small A}}
	\end{overpic}
	\begin{overpic}[width=.45\textwidth]{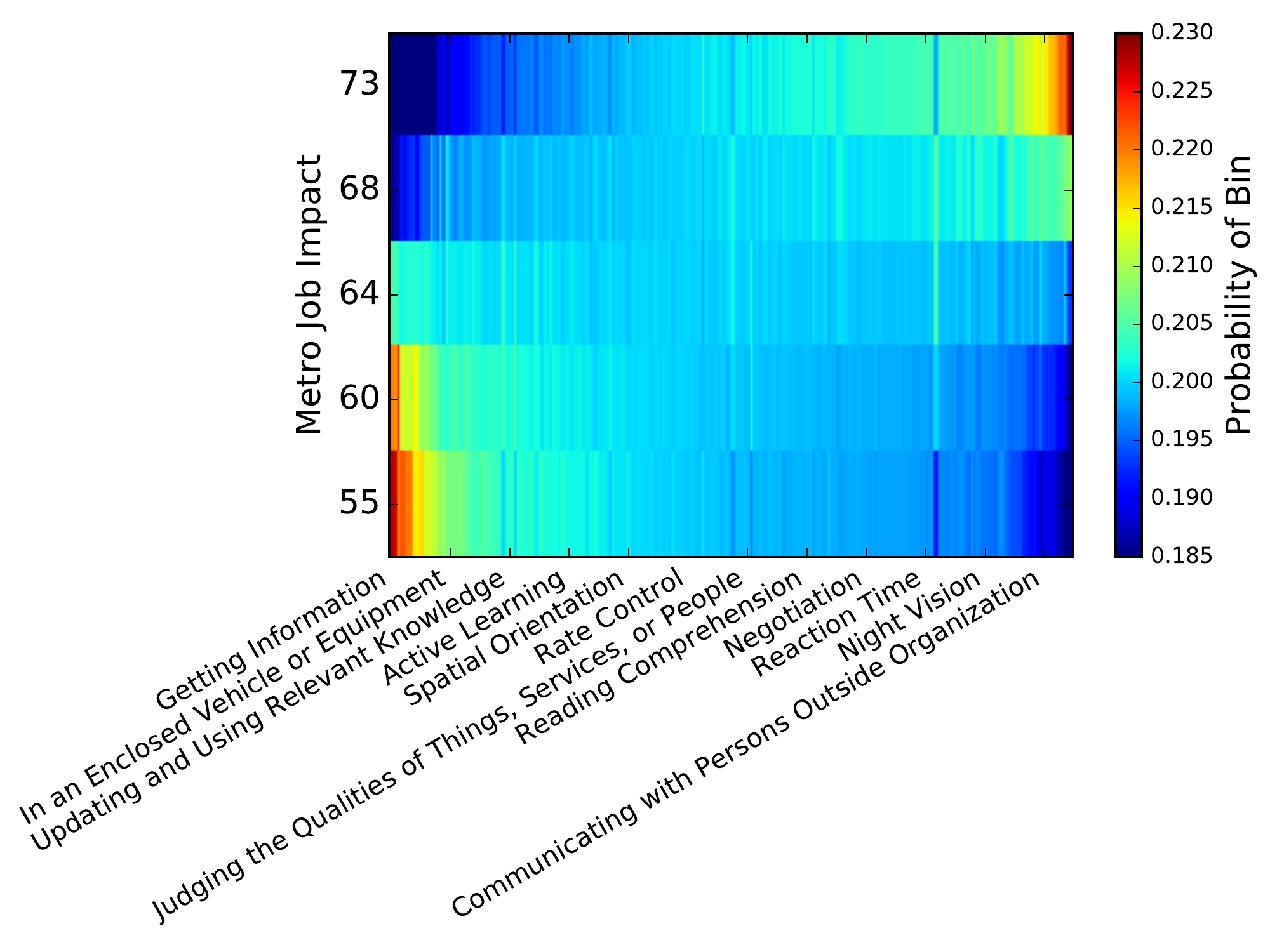}
		\put(20,60){\fbox{\small B}}
	\end{overpic}
	\caption{
	The shape of skills indicating specialization (left) and resilience to expected job displacement (right) in cities is maintained when observing raw \onet skills in place of the aggregate skills used in the main text.
	The colors in each column indicate the probability of a city having the quality on the y-axis given an observation of a labor skill on the x-axis.
	We have labelled a few of the raw \onet skills on the x-axis for reference.	
	}
	\label{rawOnetSkillPlots}
\end{figure}
\indent We also seek to explain our results on the basis of workplace skills.
To this end, we measure the correlation of raw skill values across all BLS jobs for each pair of \onet skills and employ K-means clustering to identify ten groups of co-occurring skills (i.e. workplace skills are instances and the Pearson correlation of the raw \onet importance of that skill to the importance of each other skill are the features).
The complete lists of raw \onet skills comprising each skill type are presented in Section~\ref{aggSkills}.
We summarize the skills comprising each skill type with the groups' titles.
This simplification of the space of skills clarifies how different types of skills explain our results, and trends that we present using these aggregated skill groups are apparent when reproduced using the raw \onet skills instead.
%%%%
\clearpage
\subsection{O$^*$NET Task Groups}

\begin{figure}[!ht]
    \centering
    \begin{overpic}[width=.32\textwidth]{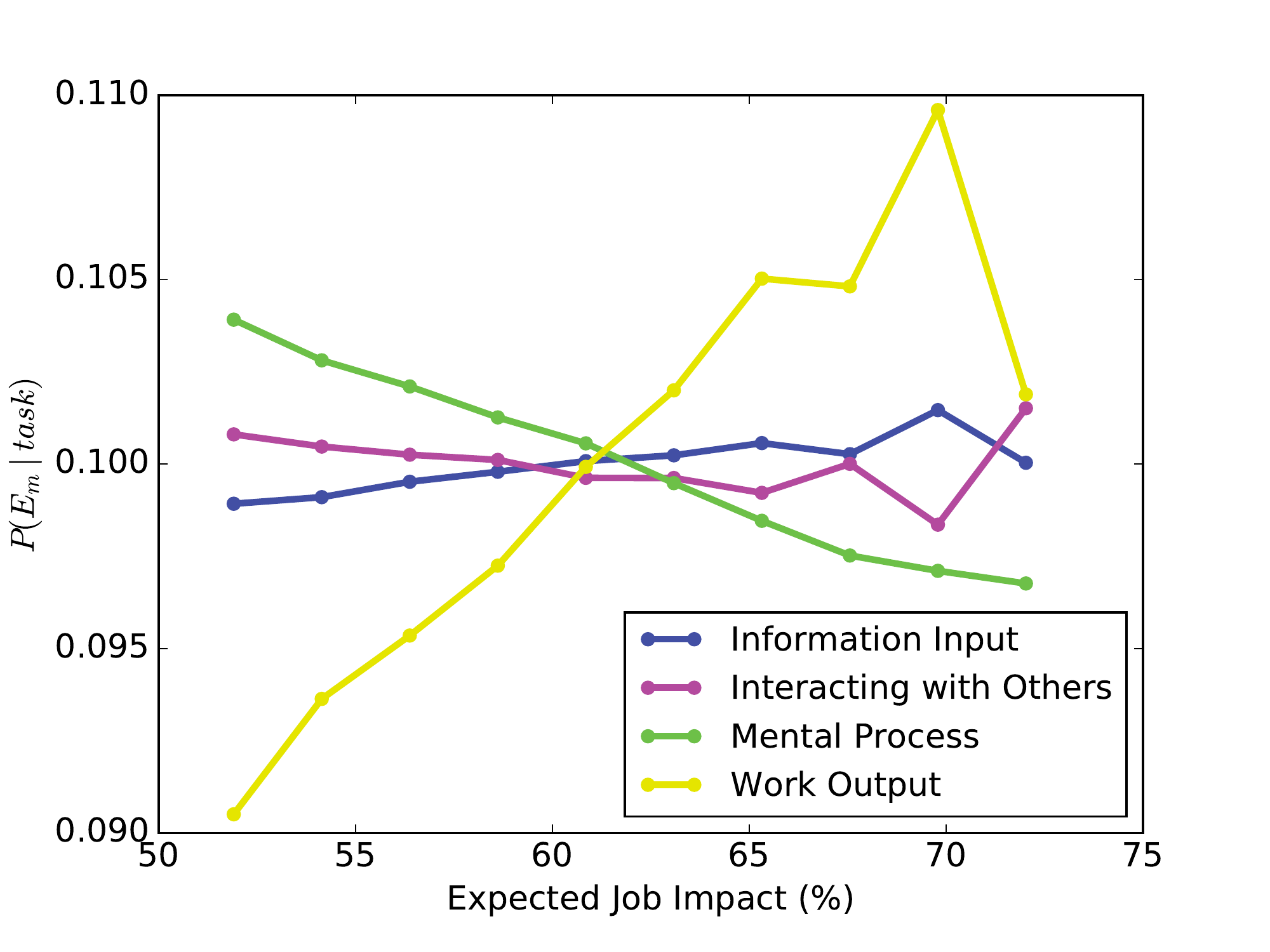}
        \put(30,90){\fbox{\small A}}
    \end{overpic}
     \begin{overpic}[width=.32\textwidth]{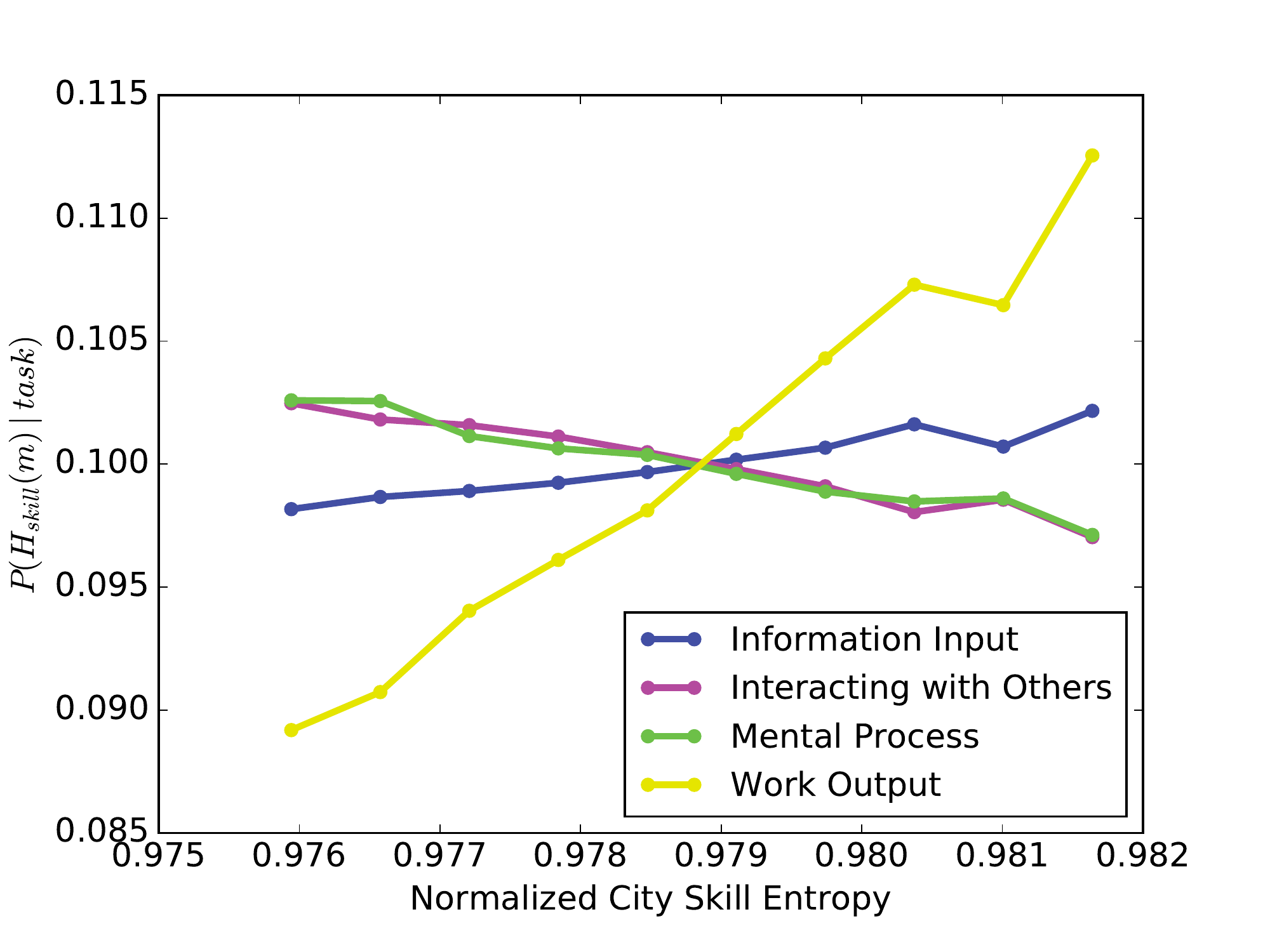}
        \put(30,90){\fbox{\small B}}
    \end{overpic}
     \begin{overpic}[width=.32\textwidth]{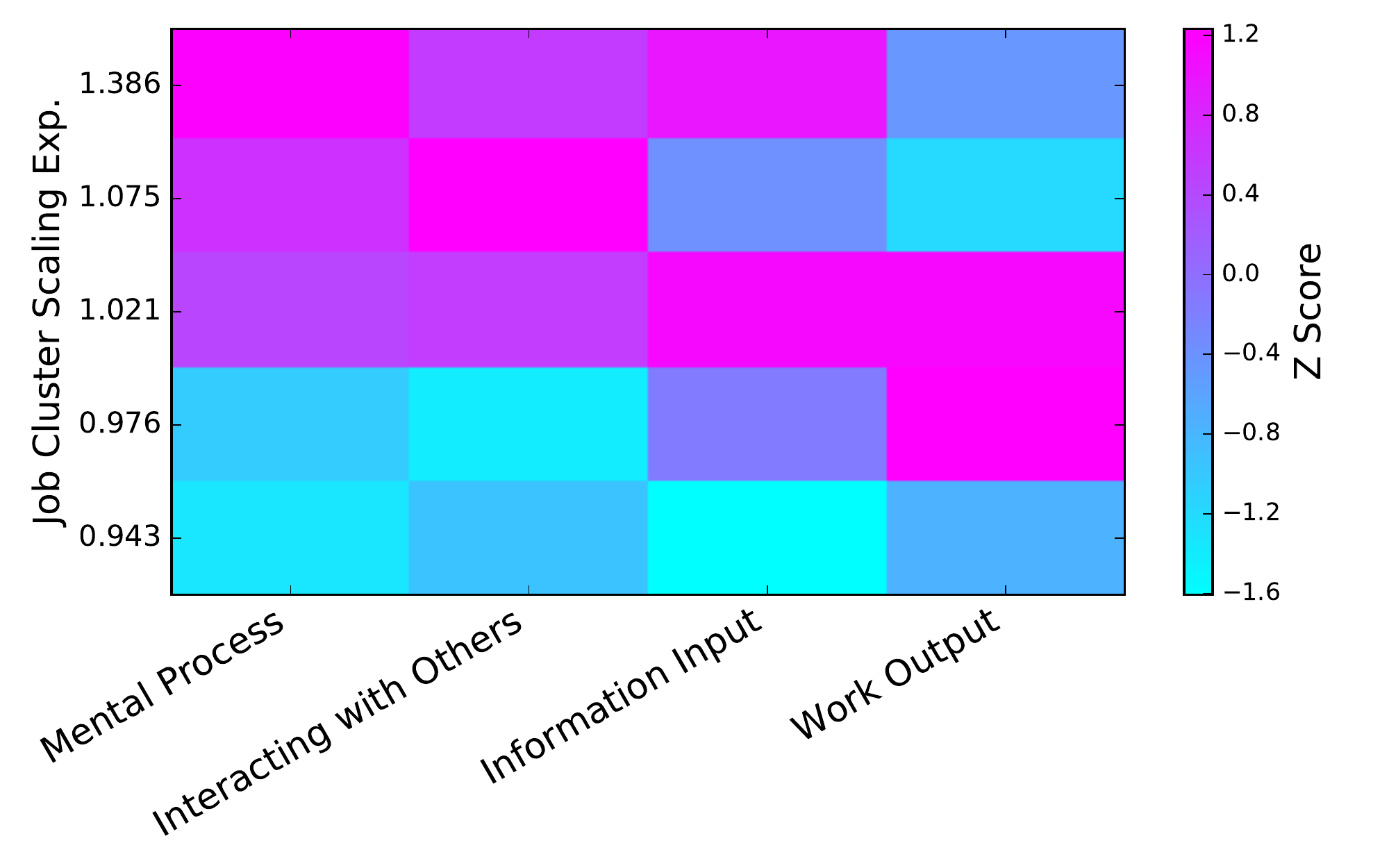}
        \put(-1,90){\fbox{\small C}}
    \end{overpic}
    \caption{
    The relationships between \onet tasks expected job impact from automation, labor specialization, and the scaling of job types.
    {\bf (A)} We bin cities according to their expected job impact from automation
(x-axis). For each task (legend), we normalize the importance of that task across bins
to a probability $P(E_m \mid task)$ representing how strongly that task indicates each level of job
displacement (y-axis).
    {\bf (B)} We bin cities according to their skill specialization (x-
axis) and sum the importance of each task for each bin. For each task (leg-
end), we normalize the importance of that task across bins to a probability $P(H_{skill}(m) \mid task)$
representing how strongly that task indicates each level of specialization (y-axis).
    {\bf (C)} By summing the importance of each task to each job type, we
assess how strongly a task indicates a scaling relationship according to its z score.
For a given task, z scores are calculated according to the distribution of importance
across job clusters.
    }
    \label{taskResults}
\end{figure}

An alternative simplification of the raw \onet skills is the \onet Task Groups, which represent collections of similar work activities.
We provide the definitions for these task groups in Table~\ref{taskDefs}.
These task groups have been used to investigate the task connectivity of urban labor markets in relation to employment growth~\cite{kok2014cities}.
In Figure~\ref{taskResults}, we use these groups as alternative skill aggregations and assess which tasks indicate resilience to job displacement from automation in cities (Fig.~\ref{taskResults}A), which tasks indicate occupational specialization in cities (Fig.~\ref{taskResults}B), and which tasks indicate superlinear scaling of job types (Fig.~\ref{taskResults}C).
We find that Mental Process tasks are indicative of increased specialization in cities, increased resilience to job displacement in cities, and superlinear scaling of job types.
On the other hand, Work Output tasks, which focuses on physical skills, indicate less specialization in cities, less resilience to job displacement in cities, and linear or sublinear scaling of job types.
These findings are in agreement with the results in the main text.

\begin{table}[!ht]
    \centering
    \begin{tabular}[b]{|c|c|c|}
        \hline
        \onet Task & Job Impact & Log$_{10}$ City Size  \\ 
        Group& Corr. & Corr. \\ \hline
        Mental Process & -0.86 ($<10^{-113}$) & 0.67 ($<10^{-49}$) \\ \hline
        Interacting with Others & -0.46 ($<10^{-20}$) & 0.13 (0.01) \\ \hline
        Information Input & -0.082 (0.11) & 0.34 ($<10^{-11}$) \\ \hline
        Work Output & 0.69 ($<10^{-53}$) & -0.37 ($<10^{-12}$) \\ \hline
    \end{tabular}
    \caption{
    Summarizing the relationship between tasks, job impact from automation, and city
size. In the middle (right) column, we present the Pearson correlation of the proportion of each
task to the expected job impact (log 10 city size). We provide the associated
p-values in parentheses
    }
    \label{taskCorrs}
\end{table}

\begin{table}[!ht]
    \centering
    \begin{tabu}{|c|X|}
    \hline
        {\bf Task Group} & {\bf \onet Skills} \\ \hline
        Information Input & Getting Information, Monitor Processes, Materials, or Surroundings, Identifying Objects, Actions, and Events, Inspecting Equipment, Structures, or Material, Estimating the Quantifiable Characteristics of Products, Events, or Information \\ \hline
        Mental Process & Judging the Qualities of Things, Services, or People, Processing Information, Evaluating Information to Determine Compliance with Standards, Analyzing Data or Information, Making Decisions and Solving Problems, Thinking Creatively, Updating and Using Relevant Knowledge, Developing Objectives and Strategies, Scheduling Work and Activities, Organizing, Planning, and Prioritizing Work \\ \hline
        Work Output & Performing General Physical Activities, Handling and Moving Objects, Controlling Machines and Processes, Operating Vehicles, Mechanized Devices, or Equipment, Interacting With Computers, Drafting, Laying Out, and Specifying Technical Devices. Parts. and Equipment, Repairing and Maintaining Mechanical Equipment, Repairing and Maintaining Electronic Equipment, Documenting or Recording Information \\ \hline
        Interacting with Others & Interpreting the Meaning of Information for Others, Communicating with Supervisors, Peers, or Subordinates, Communicating with Persons Outside Organization, Establishing and Maintaining Interpersonal Relationships, Assisting and Caring for Others, Selling or Influencing Others, Resolving Conflicts and Negotiating with Others, Performing for or Working Directly with the Public, Coordinating the Work and Activities of Others, Developing and Building Teams, Training and Teaching Others, Guiding, Directing, and Motivating Subordinates, Coaching and Developing Others, Provide Consultation and Advice to Others, Performing Administrative Activities, Staffing Organizational Units, Monitoring and Controlling Resources \\ \hline
    \end{tabu}
    \caption{
    The \onet skills comprising each  Task Group.
    }
    \label{taskDefs}
\end{table}

\subsection{The Routineness of Tasks}

\begin{figure}[!ht]
    \centering
    \begin{overpic}[width=.32\textwidth]{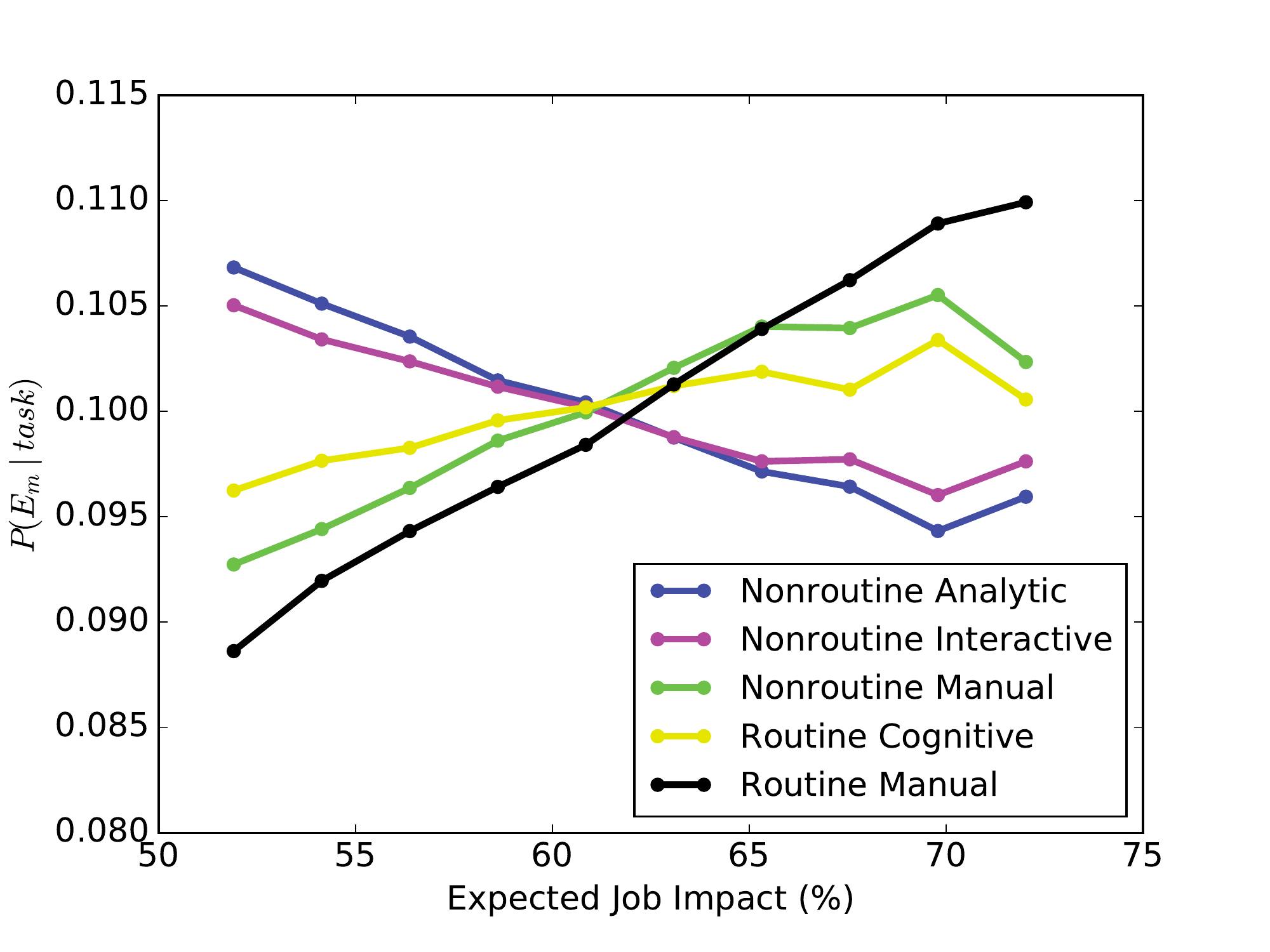}
        \put(30,90){\fbox{\small A}}
    \end{overpic}
     \begin{overpic}[width=.32\textwidth]{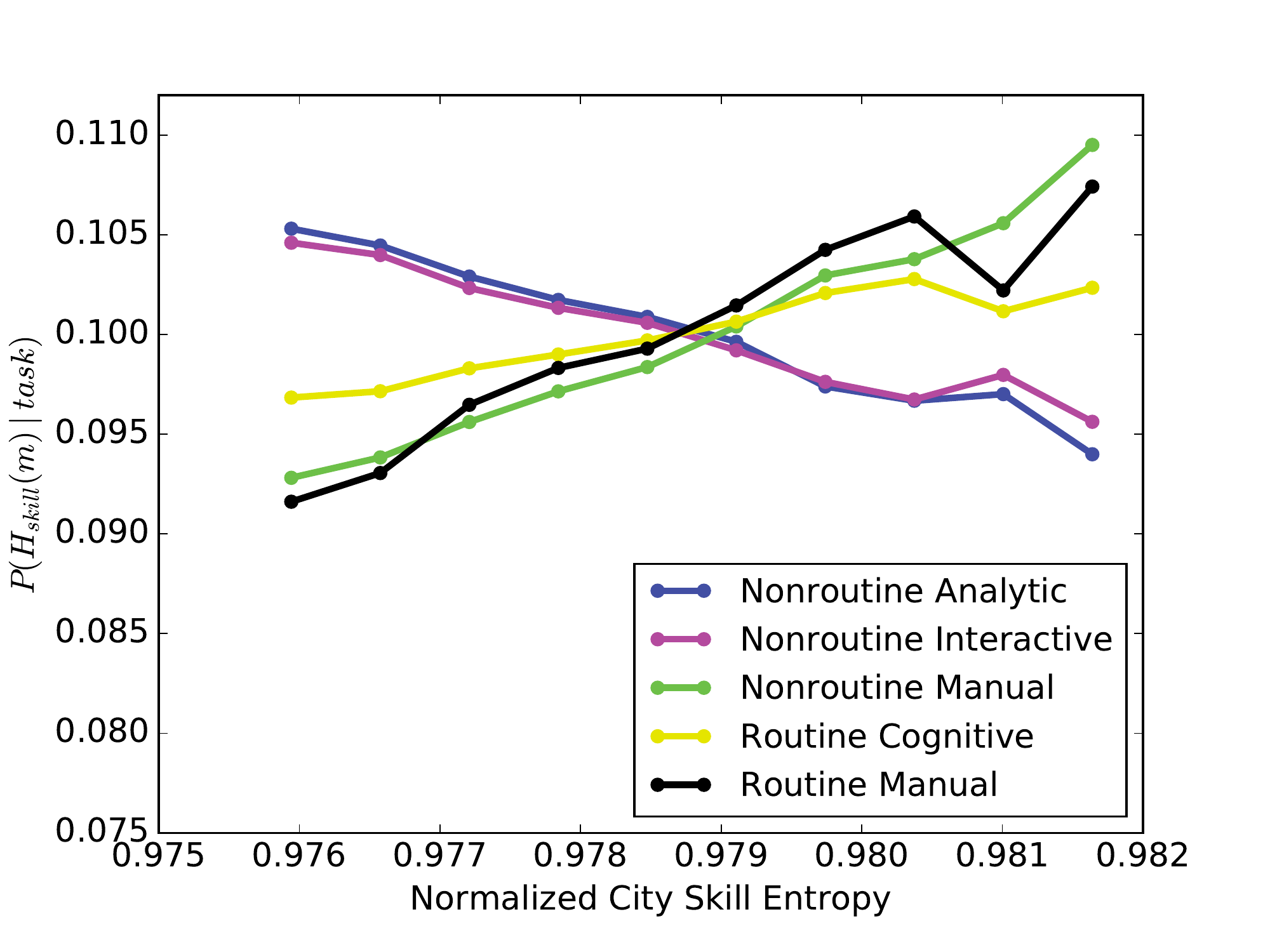}
        \put(30,90){\fbox{\small B}}
    \end{overpic}
     \begin{overpic}[width=.32\textwidth]{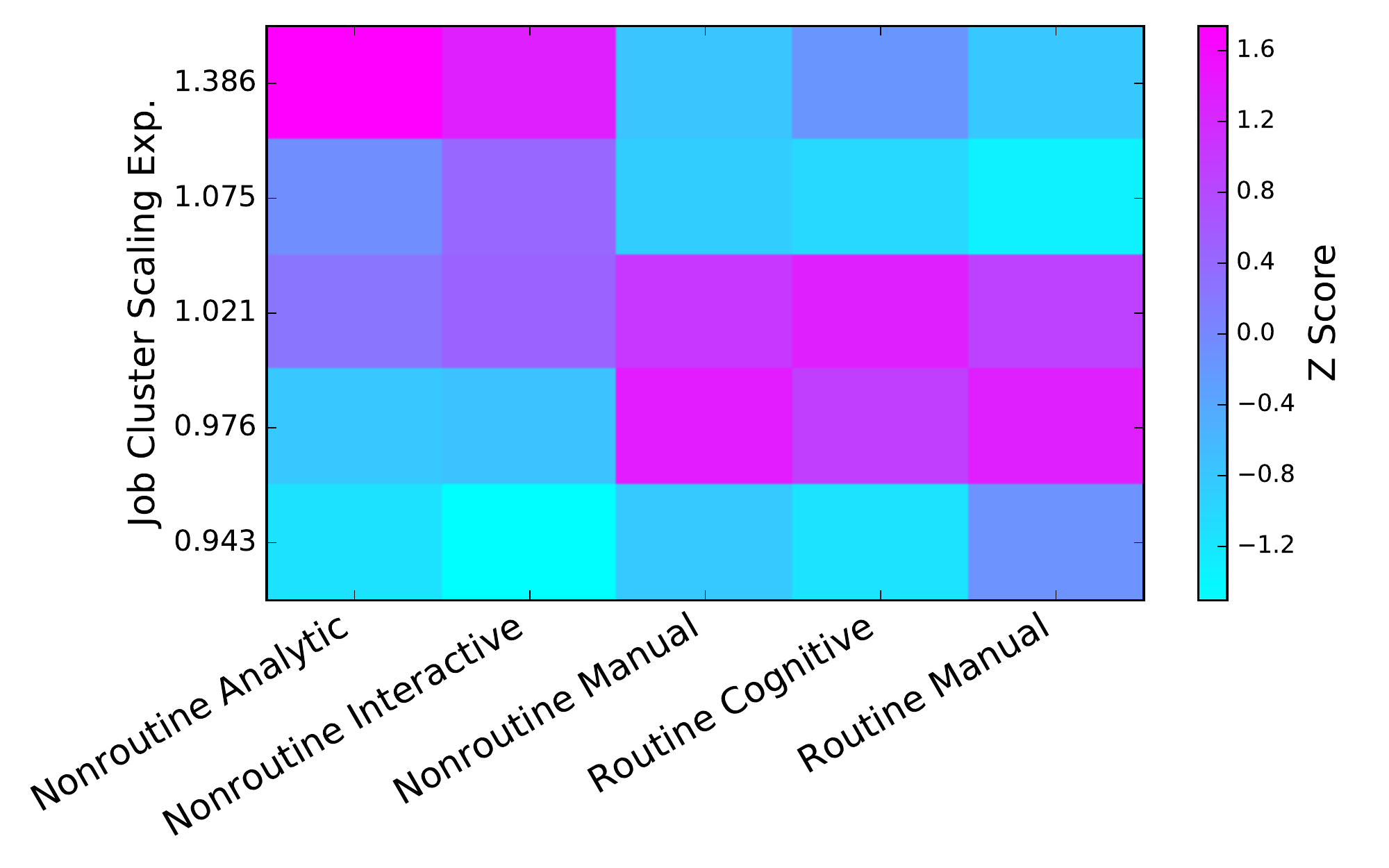}
        \put(-1,90){\fbox{\small C}}
    \end{overpic}
    \caption{
    The relationships between \onet tasks expected job impact from automation, labor specialization, and the scaling of job types.
    {\bf (A)} We bin cities according to their expected job impact from automation
(x-axis). For each task (legend), we normalize the importance of that task across bins
to a probability $P(E_m \mid task)$ representing how strongly that task indicates each level of job
displacement (y-axis).
    {\bf (B)} We bin cities according to their skill specialization (x-
axis) and sum the importance of each task for each bin. For each task (leg-
end), we normalize the importance of that task across bins to a probability $P(H_{skill}(m) \mid task)$
representing how strongly that task indicates each level of specialization (y-axis).
    {\bf (C)} By summing the importance of each task to each job type, we
assess how strongly a task indicates a scaling relationship according to its z score.
For a given task, z scores are calculated according to the distribution of importance
across job clusters.
    }
    \label{AutorResults}
\end{figure}

Autor et al.~\cite{david2001skill,autor2013changing} identify workplace tasks according to their type and how routine the task is.
They find that non-routine tasks are becoming increasingly important to workers relative to routine tasks.
We provide the definitions for these task groups in Table~\ref{AutorDefs}.
In Figure~\ref{AutorResults}, we use these groups as alternative skill aggregations and assess which tasks indicate resilience to job impact from automation in cities (Fig.~\ref{AutorResults}A), which tasks indicate occupational specialization in cities (Fig.~\ref{AutorResults}B), and which tasks indicate superlinear scaling of job types (Fig.~\ref{AutorResults}C).
We find that all non-routine tasks are indicative of increased specialization in cities and increased resilience to job impact in cities.
Non-routine analytic tasks and non-routine interactive tasks are indicative of superlinear scaling of job types, while non-routine manual tasks indicate linear or sublinear scaling of job types.
Routine tasks indicate less specialization in cities, less resilience to job impact in cities, and linear or sublinear scaling of job types.
These findings are in agreement with the results in the main text.

\begin{table}[!ht]
    \centering
    \begin{tabular}[b]{|c|c|c|}
        \hline
        \onet Task & Job Impact & Log$_{10}$ City Size  \\ 
        Type & Corr. & Corr. \\ \hline
        Non-routine Analytic & -0.79 ($<10^{-80}$) & 0.50 ($<10^{-24}$) \\ \hline
        Non-routine Interactive & -0.73 ($<10^{-62}$) & 0.52 ($<10^{26}$) \\ \hline
        Routine Cognitive & 0.47 ($<10^{-21}$) & -0.14 (0.005) \\ \hline
        Non-routine Manual & 0.64 ($<10^{-43}$) & -0.30 ($<10^{-8}$) \\ \hline
        Routine Manual & 0.83 ($<10^{-97}$) & -0.49 ($<10^{-23}$) \\ \hline
    \end{tabular}
    \caption{
    Summarizing the relationship between tasks, job impact, and city
size. In the middle (right) column, we present the Pearson correlation of the proportion of each
task to the expected job impact (log 10 city size). We provide the associated
p-values in parentheses
    }
    \label{AutorCorrs}
\end{table}

\begin{table}[!ht]
    \centering
    \begin{tabu}{|c|X|}
    \hline
        {\bf Task Type} & {\bf \onet Skills} \\ \hline
        Non-routine Analytic &
        Mathematical Reasoning, Mathematics, Deductive Reasoning, Number Facility, Physics, Programming \\ \hline
        Non-routine Interactive &
        Design, Administration and Management, Economics and Accounting, Equipment Selection, Estimating the Quantifiable Characteristics of Products, Events, or Information, Importance of Being Exact or Accurate, Management of Financial Resources, Management of Material Resources, Management of Personnel Resources, Organizing, Planning, and Prioritizing Work, Personnel and Human Resources, Quality Control Analysis, Sales and Marketing, Scheduling Work and Activities, Technology Design, Visualization \\ \hline 
        Routine Cognitive &
        Consequence of Error, Control Precision, Controlling Machines and Processes, Documenting/Recording Information, Evaluating Information to Determine Compliance with Standards, Inspecting Equipment, Structures, or Material, Operation and Control, Quality Control Analysis \\ \hline
        Routine Manual &
        Finger Dexterity, Manual Dexterity, Arm-Hand Steadiness, Wrist-Finger Speed \\ \hline
        Non-routine Manual &
        Reaction Time, Response Orientation, Cramped Work Space, Awkward Positions, Dynamic Flexibility, Spatial Orientation, Transportation, Coordination \\ \hline
    \end{tabu}
    \caption{
    The \onet skills comprising each Task Type.
    }
    \label{AutorDefs}
\end{table}

\clearpage
\section{Data Tables}
	\label{DataTables}
\subsection{Cities Ordered by Expected Job Impact from Automation}
	\label{orderedCities}
\begin{longtable}{|c|c|c|}
	%\caption[Cities ordered by expected job displacement from automation.]{Cities ordered by expected job displacement from automation.}
	%\label{jobDispTable} \\
	\hline
	\bf Rank & \bf Metro. Area & \bf Exp. Job Impact (\%) \\ \hline
1 & San Jose-Sunnyvale-Santa Clara, CA & 50.79 \\ \hline 
2 & Washington-Arlington-Alexandria, DC-VA-MD-WV & 51.85 \\ \hline 
3 & Trenton-Ewing, NJ & 52.71 \\ \hline 
4 & Boston-Cambridge-Quincy, MA-NH & 53.72 \\ \hline 
5 & Durham-Chapel Hill, NC & 53.85 \\ \hline 
6 & Boulder, CO & 54.07 \\ \hline 
7 & Warner Robins, GA & 54.69 \\ \hline 
8 & Huntsville, AL & 55.00 \\ \hline 
9 & Bridgeport-Stamford-Norwalk, CT & 55.33 \\ \hline 
10 & Ithaca, NY & 55.64 \\ \hline 
11 & San Francisco-Oakland-Fremont, CA & 55.84 \\ \hline 
12 & Hartford-West Hartford-East Hartford, CT & 56.17 \\ \hline 
13 & Ann Arbor, MI & 56.50 \\ \hline 
14 & Corvallis, OR & 56.63 \\ \hline 
15 & Seattle-Tacoma-Bellevue, WA & 57.23 \\ \hline 
16 & Baltimore-Towson, MD & 57.40 \\ \hline 
17 & Madison, WI & 57.49 \\ \hline 
18 & New York-Northern New Jersey-Long Island, NY-NJ-PA & 57.59 \\ \hline 
19 & New Haven, CT & 57.70 \\ \hline 
20 & Minneapolis-St. Paul-Bloomington, MN-WI & 57.85 \\ \hline 
21 & Charlottesville, VA & 57.93 \\ \hline 
22 & Worcester, MA-CT & 58.04 \\ \hline 
23 & Albany-Schenectady-Troy, NY & 58.09 \\ \hline 
24 & Colorado Springs, CO & 58.12 \\ \hline 
25 & Denver-Aurora-Broomfield, CO & 58.30 \\ \hline 
26 & Burlington-South Burlington, VT & 58.31 \\ \hline 
27 & Raleigh-Cary, NC & 58.62 \\ \hline 
28 & Hinesville-Fort Stewart, GA & 58.64 \\ \hline 
29 & Springfield, MA-CT & 58.71 \\ \hline 
30 & Cedar Rapids, IA & 58.73 \\ \hline 
31 & Rochester, MN & 58.93 \\ \hline 
32 & Sacramento--Arden-Arcade--Roseville, CA & 59.11 \\ \hline 
33 & Richmond, VA & 59.11 \\ \hline 
34 & Tallahassee, FL & 59.14 \\ \hline 
35 & Bremerton-Silverdale, WA & 59.18 \\ \hline 
36 & Austin-Round Rock-San Marcos, TX & 59.22 \\ \hline 
37 & Portsmouth, NH-ME & 59.24 \\ \hline 
38 & Manchester, NH & 59.36 \\ \hline 
39 & Peoria, IL & 59.38 \\ \hline 
40 & Dayton, OH & 59.39 \\ \hline 
41 & Provo-Orem, UT & 59.41 \\ \hline 
42 & Tucson, AZ & 59.47 \\ \hline 
43 & Columbus, OH & 59.51 \\ \hline 
44 & Atlanta-Sandy Springs-Marietta, GA & 59.54 \\ \hline 
45 & Philadelphia-Camden-Wilmington, PA-NJ-DE-MD & 59.57 \\ \hline 
46 & San Diego-Carlsbad-San Marcos, CA & 59.61 \\ \hline 
47 & Waterbury, CT & 59.62 \\ \hline 
48 & Detroit-Warren-Livonia, MI & 59.62 \\ \hline 
49 & Springfield, IL & 59.69 \\ \hline 
50 & Chicago-Joliet-Naperville, IL-IN-WI & 59.73 \\ \hline 
51 & Olympia, WA & 59.73 \\ \hline 
52 & Danbury, CT & 59.77 \\ \hline 
53 & Little Rock-North Little Rock-Conway, AR & 59.77 \\ \hline 
54 & Palm Bay-Melbourne-Titusville, FL & 59.80 \\ \hline 
55 & Albuquerque, NM & 59.80 \\ \hline 
56 & Des Moines-West Des Moines, IA & 59.83 \\ \hline 
57 & Portland-Vancouver-Hillsboro, OR-WA & 59.86 \\ \hline 
58 & Pittsfield, MA & 59.88 \\ \hline 
59 & Rochester, NY & 59.90 \\ \hline 
60 & Leominster-Fitchburg-Gardner, MA & 59.97 \\ \hline 
61 & Phoenix-Mesa-Glendale, AZ & 59.99 \\ \hline 
62 & Providence-Fall River-Warwick, RI-MA & 60.05 \\ \hline 
63 & Topeka, KS & 60.08 \\ \hline 
64 & Norwich-New London, CT-RI & 60.10 \\ \hline 
65 & Columbia, SC & 60.19 \\ \hline 
66 & Jefferson City, MO & 60.19 \\ \hline 
67 & Baton Rouge, LA & 60.20 \\ \hline 
68 & Charlotte-Gastonia-Rock Hill, NC-SC & 60.28 \\ \hline 
69 & Bakersfield-Delano, CA & 60.29 \\ \hline 
70 & Salt Lake City, UT & 60.29 \\ \hline 
71 & Pascagoula, MS & 60.31 \\ \hline 
72 & Virginia Beach-Norfolk-Newport News, VA-NC & 60.32 \\ \hline 
73 & Kennewick-Pasco-Richland, WA & 60.32 \\ \hline 
74 & Milwaukee-Waukesha-West Allis, WI & 60.34 \\ \hline 
75 & Fort Collins-Loveland, CO & 60.35 \\ \hline 
76 & Cincinnati-Middletown, OH-KY-IN & 60.36 \\ \hline 
77 & Lansing-East Lansing, MI & 60.38 \\ \hline 
78 & Bloomington-Normal, IL & 60.44 \\ \hline 
79 & Portland-South Portland-Biddeford, ME & 60.51 \\ \hline 
80 & Anchorage, AK & 60.65 \\ \hline 
81 & Rochester-Dover, NH-ME & 60.68 \\ \hline 
82 & Honolulu, HI & 60.70 \\ \hline 
83 & Los Angeles-Long Beach-Santa Ana, CA & 60.72 \\ \hline 
84 & Syracuse, NY & 60.74 \\ \hline 
85 & Houston-Sugar Land-Baytown, TX & 60.76 \\ \hline 
86 & Cleveland-Elyria-Mentor, OH & 60.82 \\ \hline 
87 & Omaha-Council Bluffs, NE-IA & 60.90 \\ \hline 
88 & Oklahoma City, OK & 60.91 \\ \hline 
89 & Charleston-North Charleston-Summerville, SC & 60.92 \\ \hline 
90 & Jackson, MS & 60.93 \\ \hline 
91 & Champaign-Urbana, IL & 61.06 \\ \hline 
92 & Iowa City, IA & 61.06 \\ \hline 
93 & Akron, OH & 61.14 \\ \hline 
94 & Santa Fe, NM & 61.15 \\ \hline 
95 & Ogden-Clearfield, UT & 61.15 \\ \hline 
96 & Yuma, AZ & 61.18 \\ \hline 
97 & St. Louis, MO-IL & 61.23 \\ \hline 
98 & Bismarck, ND & 61.27 \\ \hline 
99 & Las Cruces, NM & 61.31 \\ \hline 
100 & Kansas City, MO-KS & 61.31 \\ \hline 
101 & Boise City-Nampa, ID & 61.32 \\ \hline 
102 & Indianapolis-Carmel, IN & 61.33 \\ \hline 
103 & Salem, OR & 61.39 \\ \hline 
104 & Cumberland, MD-WV & 61.52 \\ \hline 
105 & Dallas-Fort Worth-Arlington, TX & 61.55 \\ \hline 
106 & Duluth, MN-WI & 61.57 \\ \hline 
107 & Harrisburg-Carlisle, PA & 61.58 \\ \hline 
108 & Santa Barbara-Santa Maria-Goleta, CA & 61.60 \\ \hline 
109 & Poughkeepsie-Newburgh-Middletown, NY & 61.66 \\ \hline 
110 & Athens-Clarke County, GA & 61.66 \\ \hline 
111 & Nashville-Davidson--Murfreesboro--Franklin, TN & 61.68 \\ \hline 
112 & Battle Creek, MI & 61.69 \\ \hline 
113 & New Bedford, MA & 61.75 \\ \hline 
114 & Bangor, ME & 61.76 \\ \hline 
115 & Lincoln, NE & 61.82 \\ \hline 
116 & Columbus, GA-AL & 61.82 \\ \hline 
117 & Gainesville, FL & 61.83 \\ \hline 
118 & Binghamton, NY & 61.87 \\ \hline 
119 & Oxnard-Thousand Oaks-Ventura, CA & 61.89 \\ \hline 
120 & Utica-Rome, NY & 61.90 \\ \hline 
121 & Vallejo-Fairfield, CA & 61.95 \\ \hline 
122 & Chattanooga, TN-GA & 61.96 \\ \hline 
123 & Kalamazoo-Portage, MI & 61.96 \\ \hline 
124 & Lexington-Fayette, KY & 61.97 \\ \hline 
125 & Pittsburgh, PA & 61.99 \\ \hline 
126 & Killeen-Temple-Fort Hood, TX & 61.99 \\ \hline 
127 & Barnstable Town, MA & 62.01 \\ \hline 
128 & Fayetteville-Springdale-Rogers, AR-MO & 62.02 \\ \hline 
129 & Augusta-Richmond County, GA-SC & 62.08 \\ \hline 
130 & Johnstown, PA & 62.10 \\ \hline 
131 & Knoxville, TN & 62.13 \\ \hline 
132 & Pueblo, CO & 62.20 \\ \hline 
133 & Fayetteville, NC & 62.20 \\ \hline 
134 & Allentown-Bethlehem-Easton, PA-NJ & 62.25 \\ \hline 
135 & Buffalo-Niagara Falls, NY & 62.26 \\ \hline 
136 & Santa Cruz-Watsonville, CA & 62.26 \\ \hline 
137 & Cheyenne, WY & 62.34 \\ \hline 
138 & Florence, SC & 62.34 \\ \hline 
139 & Huntington-Ashland, WV-KY-OH & 62.39 \\ \hline 
140 & Santa Rosa-Petaluma, CA & 62.42 \\ \hline 
141 & Saginaw-Saginaw Township North, MI & 62.45 \\ \hline 
142 & Tampa-St. Petersburg-Clearwater, FL & 62.49 \\ \hline 
143 & San Antonio-New Braunfels, TX & 62.52 \\ \hline 
144 & Fresno, CA & 62.56 \\ \hline 
145 & Alexandria, LA & 62.56 \\ \hline 
146 & Tulsa, OK & 62.58 \\ \hline 
147 & Lynchburg, VA & 62.59 \\ \hline 
148 & Eugene-Springfield, OR & 62.60 \\ \hline 
149 & Morgantown, WV & 62.60 \\ \hline 
150 & Hagerstown-Martinsburg, MD-WV & 62.64 \\ \hline 
151 & Wichita, KS & 62.67 \\ \hline 
152 & Merced, CA & 62.68 \\ \hline 
153 & Vineland-Millville-Bridgeton, NJ & 62.68 \\ \hline 
154 & Rockford, IL & 62.69 \\ \hline 
155 & Albany, GA & 62.73 \\ \hline 
156 & Jackson, MI & 62.75 \\ \hline 
157 & Birmingham-Hoover, AL & 62.77 \\ \hline 
158 & Mankato-North Mankato, MN & 62.78 \\ \hline 
159 & Grand Rapids-Wyoming, MI & 62.79 \\ \hline 
160 & Crestview-Fort Walton Beach-Destin, FL & 62.81 \\ \hline 
161 & Salinas, CA & 62.84 \\ \hline 
162 & Oshkosh-Neenah, WI & 62.84 \\ \hline 
163 & Pine Bluff, AR & 62.89 \\ \hline 
164 & Toledo, OH & 62.90 \\ \hline 
165 & Green Bay, WI & 62.91 \\ \hline 
166 & College Station-Bryan, TX & 62.92 \\ \hline 
167 & Fairbanks, AK & 62.96 \\ \hline 
168 & Lewiston-Auburn, ME & 63.03 \\ \hline 
169 & Winston-Salem, NC & 63.04 \\ \hline 
170 & Pocatello, ID & 63.05 \\ \hline 
171 & Madera-Chowchilla, CA & 63.05 \\ \hline 
172 & Rome, GA & 63.13 \\ \hline 
173 & State College, PA & 63.18 \\ \hline 
174 & Evansville, IN-KY & 63.21 \\ \hline 
175 & Johnson City, TN & 63.24 \\ \hline 
176 & McAllen-Edinburg-Mission, TX & 63.25 \\ \hline 
177 & Beaumont-Port Arthur, TX & 63.25 \\ \hline 
178 & Clarksville, TN-KY & 63.26 \\ \hline 
179 & Yakima, WA & 63.27 \\ \hline 
180 & Davenport-Moline-Rock Island, IA-IL & 63.28 \\ \hline 
181 & Fargo, ND-MN & 63.29 \\ \hline 
182 & San Luis Obispo-Paso Robles, CA & 63.32 \\ \hline 
183 & Flagstaff, AZ & 63.33 \\ \hline 
184 & Visalia-Porterville, CA & 63.34 \\ \hline 
185 & St. Cloud, MN & 63.36 \\ \hline 
186 & Reading, PA & 63.37 \\ \hline 
187 & Salisbury, MD & 63.39 \\ \hline 
188 & Springfield, OH & 63.41 \\ \hline 
189 & New Orleans-Metairie-Kenner, LA & 63.44 \\ \hline 
190 & Memphis, TN-MS-AR & 63.48 \\ \hline 
191 & Kingston, NY & 63.52 \\ \hline 
192 & Canton-Massillon, OH & 63.54 \\ \hline 
193 & Spokane, WA & 63.55 \\ \hline 
194 & Miami-Fort Lauderdale-Pompano Beach, FL & 63.56 \\ \hline 
195 & Monroe, LA & 63.57 \\ \hline 
196 & Niles-Benton Harbor, MI & 63.57 \\ \hline 
197 & Jackson, TN & 63.59 \\ \hline 
198 & San Juan-Caguas-Guaynabo, PR & 63.60 \\ \hline 
199 & Lawton, OK & 63.63 \\ \hline 
200 & Flint, MI & 63.63 \\ \hline 
201 & Charleston, WV & 63.63 \\ \hline 
202 & Sherman-Denison, TX & 63.64 \\ \hline 
203 & Blacksburg-Christiansburg-Radford, VA & 63.66 \\ \hline 
204 & Yuba City, CA & 63.67 \\ \hline 
205 & Roanoke, VA & 63.69 \\ \hline 
206 & Manhattan, KS & 63.76 \\ \hline 
207 & Amarillo, TX & 63.81 \\ \hline 
208 & Steubenville-Weirton, OH-WV & 63.86 \\ \hline 
209 & Wilmington, NC & 63.87 \\ \hline 
210 & Greenville-Mauldin-Easley, SC & 63.87 \\ \hline 
211 & Pensacola-Ferry Pass-Brent, FL & 63.92 \\ \hline 
212 & Corpus Christi, TX & 63.96 \\ \hline 
213 & Tyler, TX & 64.00 \\ \hline 
214 & Kingsport-Bristol-Bristol, TN-VA & 64.00 \\ \hline 
215 & Redding, CA & 64.00 \\ \hline 
216 & Ames, IA & 64.01 \\ \hline 
217 & Carson City, NV & 64.02 \\ \hline 
218 & Jacksonville, FL & 64.02 \\ \hline 
219 & Appleton, WI & 64.05 \\ \hline 
220 & Decatur, IL & 64.09 \\ \hline 
221 & Wheeling, WV-OH & 64.09 \\ \hline 
222 & Scranton--Wilkes-Barre, PA & 64.09 \\ \hline 
223 & Chico, CA & 64.10 \\ \hline 
224 & Louisville-Jefferson County, KY-IN & 64.11 \\ \hline 
225 & Macon, GA & 64.11 \\ \hline 
226 & Eau Claire, WI & 64.13 \\ \hline 
227 & Decatur, AL & 64.15 \\ \hline 
228 & Idaho Falls, ID & 64.15 \\ \hline 
229 & Shreveport-Bossier City, LA & 64.18 \\ \hline 
230 & Springfield, MO & 64.20 \\ \hline 
231 & Medford, OR & 64.20 \\ \hline 
232 & Bloomington, IN & 64.22 \\ \hline 
233 & La Crosse, WI-MN & 64.22 \\ \hline 
234 & Billings, MT & 64.22 \\ \hline 
235 & Bellingham, WA & 64.23 \\ \hline 
236 & Glens Falls, NY & 64.25 \\ \hline 
237 & Missoula, MT & 64.30 \\ \hline 
238 & Coeur d'Alene, ID & 64.32 \\ \hline 
239 & Orlando-Kissimmee-Sanford, FL & 64.33 \\ \hline 
240 & South Bend-Mishawaka, IN-MI & 64.34 \\ \hline 
241 & Ponce, PR & 64.36 \\ \hline 
242 & Holland-Grand Haven, MI & 64.37 \\ \hline 
243 & Greensboro-High Point, NC & 64.37 \\ \hline 
244 & Lewiston, ID-WA & 64.37 \\ \hline 
245 & Mansfield, OH & 64.38 \\ \hline 
246 & Racine, WI & 64.38 \\ \hline 
247 & El Paso, TX & 64.39 \\ \hline 
248 & Bowling Green, KY & 64.39 \\ \hline 
249 & York-Hanover, PA & 64.40 \\ \hline 
250 & Waterloo-Cedar Falls, IA & 64.42 \\ \hline 
251 & Cleveland, TN & 64.43 \\ \hline 
252 & Gulfport-Biloxi, MS & 64.45 \\ \hline 
253 & Port St. Lucie, FL & 64.46 \\ \hline 
254 & Elizabethtown, KY & 64.48 \\ \hline 
255 & Brownsville-Harlingen, TX & 64.48 \\ \hline 
256 & Abilene, TX & 64.49 \\ \hline 
257 & Stockton, CA & 64.54 \\ \hline 
258 & Greeley, CO & 64.57 \\ \hline 
259 & Rocky Mount, NC & 64.64 \\ \hline 
260 & Longview, TX & 64.64 \\ \hline 
261 & Lima, OH & 64.66 \\ \hline 
262 & Spartanburg, SC & 64.72 \\ \hline 
263 & Parkersburg-Marietta-Vienna, WV-OH & 64.76 \\ \hline 
264 & Riverside-San Bernardino-Ontario, CA & 64.79 \\ \hline 
265 & Lubbock, TX & 64.82 \\ \hline 
266 & Lawrence, KS & 64.84 \\ \hline 
267 & Kankakee-Bradley, IL & 64.91 \\ \hline 
268 & Wichita Falls, TX & 64.96 \\ \hline 
269 & Terre Haute, IN & 64.97 \\ \hline 
270 & El Centro, CA & 65.00 \\ \hline 
271 & Greenville, NC & 65.02 \\ \hline 
272 & Erie, PA & 65.03 \\ \hline 
273 & Victoria, TX & 65.05 \\ \hline 
274 & Anderson, SC & 65.06 \\ \hline 
275 & Atlantic City-Hammonton, NJ & 65.06 \\ \hline 
276 & Mount Vernon-Anacortes, WA & 65.08 \\ \hline 
277 & Farmington, NM & 65.09 \\ \hline 
278 & Mobile, AL & 65.10 \\ \hline 
279 & Prescott, AZ & 65.14 \\ \hline 
280 & Hanford-Corcoran, CA & 65.16 \\ \hline 
281 & Sumter, SC & 65.23 \\ \hline 
282 & Lancaster, PA & 65.24 \\ \hline 
283 & Asheville, NC & 65.24 \\ \hline 
284 & North Port-Bradenton-Sarasota, FL & 65.28 \\ \hline 
285 & Fort Wayne, IN & 65.31 \\ \hline 
286 & Fort Smith, AR-OK & 65.33 \\ \hline 
287 & Janesville, WI & 65.35 \\ \hline 
288 & Montgomery, AL & 65.35 \\ \hline 
289 & Anderson, IN & 65.35 \\ \hline 
290 & Winchester, VA-WV & 65.35 \\ \hline 
291 & Lake Havasu City - Kingman, AZ & 65.35 \\ \hline 
292 & Savannah, GA & 65.36 \\ \hline 
293 & Altoona, PA & 65.37 \\ \hline 
294 & Modesto, CA & 65.40 \\ \hline 
295 & Youngstown-Warren-Boardman, OH-PA & 65.41 \\ \hline 
296 & Sheboygan, WI & 65.47 \\ \hline 
297 & Grand Forks, ND-MN & 65.51 \\ \hline 
298 & Joplin, MO & 65.52 \\ \hline 
299 & Bay City, MI & 65.54 \\ \hline 
300 & Columbia, MO & 65.54 \\ \hline 
301 & Sioux Falls, SD & 65.55 \\ \hline 
302 & Lakeland-Winter Haven, FL & 65.58 \\ \hline 
303 & Bend, OR & 65.58 \\ \hline 
304 & Muskegon-Norton Shores, MI & 65.61 \\ \hline 
305 & St. George, UT & 65.64 \\ \hline 
306 & Panama City-Lynn Haven-Panama City Beach, FL & 65.66 \\ \hline 
307 & Houma-Bayou Cane-Thibodaux, LA & 65.67 \\ \hline 
308 & Midland, TX & 65.67 \\ \hline 
309 & Dover, DE & 65.69 \\ \hline 
310 & Texarkana-Texarkana, TX-AR & 65.72 \\ \hline 
311 & Logan, UT-ID & 65.75 \\ \hline 
312 & Aguadilla-Isabela-San Sebastian, PR & 65.75 \\ \hline 
313 & Casper, WY & 65.76 \\ \hline 
314 & Goldsboro, NC & 65.78 \\ \hline 
315 & Sioux City, IA-NE-SD & 65.82 \\ \hline 
316 & Brunswick, GA & 65.87 \\ \hline 
317 & Longview, WA & 65.92 \\ \hline 
318 & Monroe, MI & 65.93 \\ \hline 
319 & Guayama, PR & 65.93 \\ \hline 
320 & Waco, TX & 65.94 \\ \hline 
321 & Hattiesburg, MS & 65.95 \\ \hline 
322 & Reno-Sparks, NV & 65.98 \\ \hline 
323 & Muncie, IN & 66.01 \\ \hline 
324 & Fond du Lac, WI & 66.01 \\ \hline 
325 & Lake Charles, LA & 66.03 \\ \hline 
326 & Rapid City, SD & 66.10 \\ \hline 
327 & Valdosta, GA & 66.12 \\ \hline 
328 & Dubuque, IA & 66.13 \\ \hline 
329 & Wenatchee-East Wenatchee, WA & 66.16 \\ \hline 
330 & Lafayette, IN & 66.16 \\ \hline 
331 & St. Joseph, MO-KS & 66.16 \\ \hline 
332 & Morristown, TN & 66.20 \\ \hline 
333 & Sandusky, OH & 66.21 \\ \hline 
334 & Owensboro, KY & 66.21 \\ \hline 
335 & Wausau, WI & 66.23 \\ \hline 
336 & Elmira, NY & 66.25 \\ \hline 
337 & Grand Junction, CO & 66.25 \\ \hline 
338 & Tuscaloosa, AL & 66.32 \\ \hline 
339 & Hickory-Lenoir-Morganton, NC & 66.37 \\ \hline 
340 & Jonesboro, AR & 66.48 \\ \hline 
341 & Danville, VA & 66.50 \\ \hline 
342 & Florence-Muscle Shoals, AL & 66.51 \\ \hline 
343 & Cape Coral-Fort Myers, FL & 66.55 \\ \hline 
344 & Cape Girardeau-Jackson, MO-IL & 66.56 \\ \hline 
345 & Deltona-Daytona Beach-Ormond Beach, FL & 66.57 \\ \hline 
346 & Dothan, AL & 66.64 \\ \hline 
347 & Great Falls, MT & 66.75 \\ \hline 
348 & Naples-Marco Island, FL & 66.93 \\ \hline 
349 & Columbus, IN & 66.93 \\ \hline 
350 & Gainesville, GA & 66.93 \\ \hline 
351 & Kokomo, IN & 66.94 \\ \hline 
352 & Hot Springs, AR & 66.95 \\ \hline 
353 & Lafayette, LA & 66.97 \\ \hline 
354 & Williamsport, PA & 66.97 \\ \hline 
355 & Ocean City, NJ & 67.24 \\ \hline 
356 & Anniston-Oxford, AL & 67.53 \\ \hline 
357 & Sebastian-Vero Beach, FL & 67.55 \\ \hline 
358 & Auburn-Opelika, AL & 67.77 \\ \hline 
359 & Odessa, TX & 67.78 \\ \hline 
360 & Las Vegas-Paradise, NV & 67.79 \\ \hline 
361 & Lebanon, PA & 67.89 \\ \hline 
362 & Burlington, NC & 67.94 \\ \hline 
363 & Danville, IL & 67.94 \\ \hline 
364 & San Angelo, TX & 67.96 \\ \hline 
365 & Ocala, FL & 68.23 \\ \hline 
366 & Laredo, TX & 68.75 \\ \hline 
367 & Gadsden, AL & 68.87 \\ \hline 
368 & San German-Cabo Rojo, PR & 68.88 \\ \hline 
369 & Napa, CA & 68.88 \\ \hline 
370 & Palm Coast, FL & 68.98 \\ \hline 
371 & Yauco, PR & 69.06 \\ \hline 
372 & Dalton, GA & 69.07 \\ \hline 
373 & Jacksonville, NC & 69.39 \\ \hline 
374 & Michigan City-La Porte, IN & 69.40 \\ \hline 
375 & Harrisonburg, VA & 69.84 \\ \hline 
376 & Punta Gorda, FL & 70.03 \\ \hline 
377 & Fajardo, PR & 70.04 \\ \hline 
378 & Elkhart-Goshen, IN & 70.28 \\ \hline 
379 & Myrtle Beach-North Myrtle Beach-Conway, SC & 70.80 \\ \hline 
380 & Mayaguez, PR & 73.14 \\ \hline 

\end{longtable}
\subsection{Relating City Trends to BLS Jobs}
	\label{jobTableSection}
\indent We present BLS jobs ordered by decreasing skill specialization in Table~\ref{jobTableSection}.
We also provide the scaling exponent of each BLS job, along with the Pearson correlation of the relative abundance of each job to the expected job impact from automation (discussed below) across cities.
p-values for the correlations are presented in parentheses.
\begin{longtabu} to \textwidth {|c|X|c|c|c|c|}
	\hline
	\bf Rank &\bf Job Title & \bf $H_j$ & \bf $\beta$ & \bf Corr. to Job Impact\\ \hline 
\small 1 &\small  Statisticians &\small  0.949 &\small  0.748 &\small  -0.434 (0) \\ \hline
\small 2 &\small  Telemarketers &\small  0.951 &\small  0.955 &\small  0.233 (0) \\ \hline
\small 3 &\small  Securities, Commodities, and Financial Services Sales Agents &\small  0.955 &\small  1.128 &\small  -0.265 (0) \\ \hline
\small 4 &\small  Loan Interviewers and Clerks &\small  0.955 &\small  1.005 &\small  0.060 ($<10^{-54}$) \\ \hline
\small 5 &\small  Actuaries &\small  0.956 &\small  0.756 &\small  -0.193 ($<10^{-103}$) \\ \hline
\small 6 &\small  Court, Municipal, and License Clerks &\small  0.956 &\small  0.762 &\small  0.176 (0) \\ \hline
\small 7 &\small  Court Reporters &\small  0.957 &\small  0.637 &\small  0.140 ($<10^{-65}$) \\ \hline
\small 8 &\small  Credit Counselors &\small  0.957 &\small  0.825 &\small  0.220 ($<10^{-230}$) \\ \hline
\small 9 &\small  Medical Transcriptionists &\small  0.957 &\small  0.696 &\small  0.341 (0) \\ \hline
\small 10 &\small  Financial Managers &\small  0.958 &\small  1.103 &\small  -0.493 (0) \\ \hline
\small 11 &\small  Training and Development Specialists &\small  0.958 &\small  1.054 &\small  -0.316 (0) \\ \hline
\small 12 &\small  Billing and Posting Clerks &\small  0.958 &\small  0.972 &\small  0.145 (0) \\ \hline
\small 13 &\small  Credit Authorizers, Checkers, and Clerks &\small  0.958 &\small  0.792 &\small  0.192 ($<10^{-239}$) \\ \hline
\small 14 &\small  Legal Secretaries &\small  0.958 &\small  0.990 &\small  0.053 ($<10^{-41}$) \\ \hline
\small 15 &\small  Clinical, Counseling, and School Psychologists &\small  0.958 &\small  0.861 &\small  -0.087 ($<10^{-103}$) \\ \hline
\small 16 &\small  Operations Research Analysts &\small  0.959 &\small  0.985 &\small  -0.282 (0) \\ \hline
\small 17 &\small  Eligibility Interviewers, Government Programs &\small  0.959 &\small  0.821 &\small  0.119 ($<10^{-180}$) \\ \hline
\small 18 &\small  Bookkeeping, Accounting, and Auditing Clerks &\small  0.959 &\small  0.958 &\small  0.098 ($<10^{-183}$) \\ \hline
\small 19 &\small  Demonstrators and Product Promoters &\small  0.959 &\small  0.718 &\small  0.283 (0) \\ \hline
\small 20 &\small  Marriage and Family Therapists &\small  0.959 &\small  0.610 &\small  0.280 (0) \\ \hline
\small 21 &\small  Financial Examiners &\small  0.959 &\small  0.910 &\small  0.006 ($<10^{0}$) \\ \hline
\small 22 &\small  Insurance Claims and Policy Processing Clerks &\small  0.959 &\small  1.031 &\small  -0.055 ($<10^{-33}$) \\ \hline
\small 23 &\small  Judges, Magistrate Judges, and Magistrates &\small  0.959 &\small  0.601 &\small  0.327 (0) \\ \hline
\small 24 &\small  Payroll and Timekeeping Clerks &\small  0.960 &\small  0.949 &\small  0.135 (0) \\ \hline
\small 25 &\small  Accountants and Auditors &\small  0.960 &\small  1.111 &\small  -0.458 (0) \\ \hline
\small 26 &\small  Cost Estimators &\small  0.960 &\small  0.986 &\small  0.059 ($<10^{-61}$) \\ \hline
\small 27 &\small  Administrative Law Judges, Adjudicators, and Hearing Officers &\small  0.960 &\small  0.572 &\small  0.132 ($<10^{-63}$) \\ \hline
\small 28 &\small  Word Processors and Typists &\small  0.960 &\small  0.600 &\small  0.226 (0) \\ \hline
\small 29 &\small  Budget Analysts &\small  0.960 &\small  0.795 &\small  -0.123 ($<10^{-128}$) \\ \hline
\small 30 &\small  Paralegals and Legal Assistants &\small  0.960 &\small  1.093 &\small  -0.213 (0) \\ \hline
\small 31 &\small  Office Clerks, General &\small  0.960 &\small  0.934 &\small  0.196 (0) \\ \hline
\small 32 &\small  Computer Programmers &\small  0.960 &\small  1.168 &\small  -0.501 (0) \\ \hline
\small 33 &\small  Procurement Clerks &\small  0.961 &\small  0.845 &\small  0.142 ($<10^{-261}$) \\ \hline
\small 34 &\small  Crossing Guards &\small  0.961 &\small  0.738 &\small  0.277 (0) \\ \hline
\small 35 &\small  Executive Secretaries and Executive Administrative Assistants &\small  0.961 &\small  1.093 &\small  -0.341 (0) \\ \hline
\small 36 &\small  Real Estate Brokers &\small  0.961 &\small  0.706 &\small  0.249 ($<10^{-321}$) \\ \hline
\small 37 &\small  Insurance Sales Agents &\small  0.961 &\small  1.021 &\small  0.051 ($<10^{-47}$) \\ \hline
\small 38 &\small  Mental Health Counselors &\small  0.961 &\small  0.814 &\small  -0.008 ($<10^{0}$) \\ \hline
\small 39 &\small  Loan Officers &\small  0.961 &\small  1.033 &\small  -0.012 ($<10^{-2}$) \\ \hline
\small 40 &\small  Lawyers &\small  0.961 &\small  1.226 &\small  -0.473 (0) \\ \hline
\small 41 &\small  Financial Analysts &\small  0.962 &\small  1.270 &\small  -0.529 (0) \\ \hline
\small 42 &\small  Travel Agents &\small  0.962 &\small  1.056 &\small  0.042 ($<10^{-13}$) \\ \hline
\small 43 &\small  Statistical Assistants &\small  0.962 &\small  0.402 &\small  0.313 ($<10^{-274}$) \\ \hline
\small 44 &\small  Health Educators &\small  0.962 &\small  0.793 &\small  -0.084 ($<10^{-73}$) \\ \hline
\small 45 &\small  Title Examiners, Abstractors, and Searchers &\small  0.962 &\small  0.806 &\small  0.313 (0) \\ \hline
\small 46 &\small  Software Developers, Applications &\small  0.963 &\small  1.304 &\small  -0.663 (0) \\ \hline
\small 47 &\small  Compensation, Benefits, and Job Analysis Specialists &\small  0.963 &\small  1.037 &\small  -0.391 (0) \\ \hline
\small 48 &\small  Medical Records and Health Information Technicians &\small  0.963 &\small  0.893 &\small  0.073 ($<10^{-92}$) \\ \hline
\small 49 &\small  Educational, Guidance, School, and Vocational Counselors &\small  0.963 &\small  0.919 &\small  -0.097 ($<10^{-170}$) \\ \hline
\small 50 &\small  Personal Financial Advisors &\small  0.963 &\small  1.132 &\small  -0.356 (0) \\ \hline
\small 51 &\small  Public Relations and Fundraising Managers &\small  0.963 &\small  0.909 &\small  -0.405 (0) \\ \hline
\small 52 &\small  Healthcare Social Workers &\small  0.964 &\small  0.885 &\small  -0.053 ($<10^{-46}$) \\ \hline
\small 53 &\small  Psychologists, All Other &\small  0.964 &\small  0.552 &\small  0.194 ($<10^{-102}$) \\ \hline
\small 54 &\small  Bill and Account Collectors &\small  0.964 &\small  1.129 &\small  0.028 ($<10^{-13}$) \\ \hline
\small 55 &\small  Market Research Analysts and Marketing Specialists &\small  0.964 &\small  1.237 &\small  -0.575 (0) \\ \hline
\small 56 &\small  Compensation and Benefits Managers &\small  0.964 &\small  0.877 &\small  -0.449 (0) \\ \hline
\small 57 &\small  Software Developers, Systems Software &\small  0.964 &\small  1.233 &\small  -0.612 (0) \\ \hline
\small 58 &\small  Human Resources Assistants, Except Payroll and Timekeeping &\small  0.964 &\small  0.937 &\small  0.026 ($<10^{-11}$) \\ \hline
\small 59 &\small  Elementary School Teachers, Except Special Education &\small  0.964 &\small  0.855 &\small  0.226 (0) \\ \hline
\small 60 &\small  Human Resources Managers &\small  0.964 &\small  1.025 &\small  -0.348 (0) \\ \hline
\small 61 &\small  Managers, All Other &\small  0.964 &\small  1.040 &\small  -0.298 (0) \\ \hline
\small 62 &\small  Technical Writers &\small  0.964 &\small  0.888 &\small  -0.382 (0) \\ \hline
\small 63 &\small  Library Technicians &\small  0.964 &\small  0.765 &\small  -0.003 ($<10^{0}$) \\ \hline
\small 64 &\small  Speech-Language Pathologists &\small  0.964 &\small  0.884 &\small  0.080 ($<10^{-105}$) \\ \hline
\small 65 &\small  Photographic Process Workers and Processing Machine Operators &\small  0.964 &\small  0.695 &\small  0.148 ($<10^{-107}$) \\ \hline
\small 66 &\small  Chief Executives &\small  0.964 &\small  0.977 &\small  -0.072 ($<10^{-88}$) \\ \hline
\small 67 &\small  Credit Analysts &\small  0.964 &\small  1.039 &\small  -0.064 ($<10^{-31}$) \\ \hline
\small 68 &\small  Receptionists and Information Clerks &\small  0.965 &\small  0.955 &\small  0.136 (0) \\ \hline
\small 69 &\small  Tax Examiners and Collectors, and Revenue Agents &\small  0.965 &\small  0.843 &\small  -0.054 ($<10^{-24}$) \\ \hline
\small 70 &\small  Education Administrators, Postsecondary &\small  0.965 &\small  0.792 &\small  -0.000 ($<10^{0}$) \\ \hline
\small 71 &\small  Switchboard Operators, Including Answering Service &\small  0.965 &\small  0.938 &\small  0.247 (0) \\ \hline
\small 72 &\small  Financial Specialists, All Other &\small  0.965 &\small  0.966 &\small  -0.258 (0) \\ \hline
\small 73 &\small  Insurance Underwriters &\small  0.965 &\small  0.887 &\small  -0.053 ($<10^{-18}$) \\ \hline
\small 74 &\small  Secretaries and Administrative Assistants, Except Legal, Medical, and Executive &\small  0.965 &\small  0.921 &\small  0.221 (0) \\ \hline
\small 75 &\small  Advertising Sales Agents &\small  0.965 &\small  0.991 &\small  0.066 ($<10^{-64}$) \\ \hline
\small 76 &\small  Security Guards &\small  0.965 &\small  1.161 &\small  0.053 ($<10^{-51}$) \\ \hline
\small 77 &\small  Producers and Directors &\small  0.965 &\small  1.057 &\small  -0.171 ($<10^{-267}$) \\ \hline
\small 78 &\small  Claims Adjusters, Examiners, and Investigators &\small  0.965 &\small  1.137 &\small  -0.104 ($<10^{-134}$) \\ \hline
\small 79 &\small  Brokerage Clerks &\small  0.965 &\small  1.038 &\small  -0.082 ($<10^{-42}$) \\ \hline
\small 80 &\small  First-Line Supervisors of Non-Retail Sales Workers &\small  0.965 &\small  1.088 &\small  -0.075 ($<10^{-99}$) \\ \hline
\small 81 &\small  Interviewers, Except Eligibility and Loan &\small  0.965 &\small  0.934 &\small  0.038 ($<10^{-17}$) \\ \hline
\small 82 &\small  Merchandise Displayers and Window Trimmers &\small  0.965 &\small  0.870 &\small  0.218 (0) \\ \hline
\small 83 &\small  Molding, Coremaking, and Casting Machine Setters, Operators, and Tenders, Metal and Plastic &\small  0.965 &\small  0.624 &\small  0.288 (0) \\ \hline
\small 84 &\small  New Accounts Clerks &\small  0.966 &\small  0.701 &\small  0.396 (0) \\ \hline
\small 85 &\small  Transportation, Storage, and Distribution Managers &\small  0.966 &\small  0.961 &\small  0.140 ($<10^{-269}$) \\ \hline
\small 86 &\small  Marketing Managers &\small  0.966 &\small  1.163 &\small  -0.549 (0) \\ \hline
\small 87 &\small  Public Relations Specialists &\small  0.966 &\small  1.052 &\small  -0.398 (0) \\ \hline
\small 88 &\small  Education Administrators, Elementary and Secondary School &\small  0.966 &\small  0.885 &\small  0.086 ($<10^{-135}$) \\ \hline
\small 89 &\small  Kindergarten Teachers, Except Special Education &\small  0.966 &\small  0.841 &\small  0.217 (0) \\ \hline
\small 90 &\small  Writers and Authors &\small  0.966 &\small  0.911 &\small  -0.384 (0) \\ \hline
\small 91 &\small  Police, Fire, and Ambulance Dispatchers &\small  0.966 &\small  0.781 &\small  0.246 (0) \\ \hline
\small 92 &\small  Massage Therapists &\small  0.966 &\small  0.852 &\small  0.163 ($<10^{-244}$) \\ \hline
\small 93 &\small  Sales Representatives, Wholesale and Manufacturing, Except Technical and Scientific Products &\small  0.966 &\small  1.082 &\small  0.155 (0) \\ \hline
\small 94 &\small  Ushers, Lobby Attendants, and Ticket Takers &\small  0.966 &\small  1.084 &\small  0.114 ($<10^{-88}$) \\ \hline
\small 95 &\small  Music Directors and Composers &\small  0.966 &\small  0.555 &\small  0.297 (0) \\ \hline
\small 96 &\small  Probation Officers and Correctional Treatment Specialists &\small  0.966 &\small  0.769 &\small  0.187 ($<10^{-308}$) \\ \hline
\small 97 &\small  Property, Real Estate, and Community Association Managers &\small  0.966 &\small  0.986 &\small  0.173 (0) \\ \hline
\small 98 &\small  Radio and Television Announcers &\small  0.966 &\small  0.557 &\small  0.417 (0) \\ \hline
\small 99 &\small  Health Diagnosing and Treating Practitioners, All Other &\small  0.966 &\small  0.786 &\small  -0.155 ($<10^{-105}$) \\ \hline
\small 100 &\small  Editors &\small  0.966 &\small  0.956 &\small  -0.285 (0) \\ \hline
\small 101 &\small  Database Administrators &\small  0.967 &\small  1.087 &\small  -0.229 (0) \\ \hline
\small 102 &\small  Order Clerks &\small  0.967 &\small  1.001 &\small  0.104 ($<10^{-157}$) \\ \hline
\small 103 &\small  Business Operations Specialists, All Other &\small  0.967 &\small  1.117 &\small  -0.468 (0) \\ \hline
\small 104 &\small  Management Analysts &\small  0.967 &\small  1.205 &\small  -0.460 (0) \\ \hline
\small 105 &\small  Mechanical Drafters &\small  0.967 &\small  0.751 &\small  0.347 (0) \\ \hline
\small 106 &\small  Mental Health and Substance Abuse Social Workers &\small  0.967 &\small  0.749 &\small  0.098 ($<10^{-118}$) \\ \hline
\small 107 &\small  Advertising and Promotions Managers &\small  0.967 &\small  0.940 &\small  -0.155 ($<10^{-121}$) \\ \hline
\small 108 &\small  Cartographers and Photogrammetrists &\small  0.967 &\small  0.446 &\small  -0.146 ($<10^{-68}$) \\ \hline
\small 109 &\small  Data Entry Keyers &\small  0.967 &\small  1.076 &\small  -0.090 ($<10^{-123}$) \\ \hline
\small 110 &\small  Medical Secretaries &\small  0.967 &\small  0.906 &\small  0.114 ($<10^{-241}$) \\ \hline
\small 111 &\small  Computer Hardware Engineers &\small  0.967 &\small  0.689 &\small  -0.355 (0) \\ \hline
\small 112 &\small  Rehabilitation Counselors &\small  0.967 &\small  0.770 &\small  0.010 ($<10^{0}$) \\ \hline
\small 113 &\small  Instructional Coordinators &\small  0.967 &\small  0.884 &\small  -0.119 ($<10^{-209}$) \\ \hline
\small 114 &\small  Cargo and Freight Agents &\small  0.967 &\small  0.869 &\small  0.279 (0) \\ \hline
\small 115 &\small  Stock Clerks and Order Fillers &\small  0.967 &\small  0.936 &\small  0.307 (0) \\ \hline
\small 116 &\small  Physical Scientists, All Other &\small  0.967 &\small  0.466 &\small  -0.231 ($<10^{-137}$) \\ \hline
\small 117 &\small  Directors, Religious Activities and Education &\small  0.967 &\small  0.489 &\small  0.243 ($<10^{-206}$) \\ \hline
\small 118 &\small  Secondary School Teachers, Except Special and Career/Technical Education &\small  0.967 &\small  0.918 &\small  0.103 ($<10^{-175}$) \\ \hline
\small 119 &\small  Special Education Teachers, Secondary School &\small  0.967 &\small  0.850 &\small  0.064 ($<10^{-52}$) \\ \hline
\small 120 &\small  Education Administrators, Preschool and Childcare Center/Program &\small  0.967 &\small  0.819 &\small  -0.227 (0) \\ \hline
\small 121 &\small  Parts Salespersons &\small  0.968 &\small  0.847 &\small  0.384 (0) \\ \hline
\small 122 &\small  Weighers, Measurers, Checkers, and Samplers, Recordkeeping &\small  0.968 &\small  0.749 &\small  0.350 (0) \\ \hline
\small 123 &\small  Musicians and Singers &\small  0.968 &\small  0.738 &\small  0.077 ($<10^{-25}$) \\ \hline
\small 124 &\small  Social and Community Service Managers &\small  0.968 &\small  0.835 &\small  -0.153 (0) \\ \hline
\small 125 &\small  Training and Development Managers &\small  0.968 &\small  0.894 &\small  -0.233 ($<10^{-313}$) \\ \hline
\small 126 &\small  Social Scientists and Related Workers, All Other &\small  0.968 &\small  0.536 &\small  -0.182 ($<10^{-144}$) \\ \hline
\small 127 &\small  Interpreters and Translators &\small  0.968 &\small  0.772 &\small  0.139 ($<10^{-138}$) \\ \hline
\small 128 &\small  Meeting, Convention, and Event Planners &\small  0.968 &\small  0.933 &\small  -0.113 ($<10^{-131}$) \\ \hline
\small 129 &\small  Child, Family, and School Social Workers &\small  0.968 &\small  0.883 &\small  0.011 ($<10^{-2}$) \\ \hline
\small 130 &\small  Bailiffs &\small  0.968 &\small  0.409 &\small  0.397 (0) \\ \hline
\small 131 &\small  Mail Clerks and Mail Machine Operators, Except Postal Service &\small  0.968 &\small  0.952 &\small  -0.023 ($<10^{-5}$) \\ \hline
\small 132 &\small  Biological Scientists, All Other &\small  0.968 &\small  0.534 &\small  -0.082 ($<10^{-27}$) \\ \hline
\small 133 &\small  Special Education Teachers, Middle School &\small  0.968 &\small  0.820 &\small  0.120 ($<10^{-148}$) \\ \hline
\small 134 &\small  Appraisers and Assessors of Real Estate &\small  0.968 &\small  0.765 &\small  0.361 (0) \\ \hline
\small 135 &\small  Tellers &\small  0.968 &\small  0.850 &\small  0.458 (0) \\ \hline
\small 136 &\small  Teacher Assistants &\small  0.968 &\small  0.894 &\small  -0.076 ($<10^{-109}$) \\ \hline
\small 137 &\small  Network and Computer Systems Administrators &\small  0.968 &\small  1.149 &\small  -0.564 (0) \\ \hline
\small 138 &\small  First-Line Supervisors of Office and Administrative Support Workers &\small  0.968 &\small  0.989 &\small  -0.028 ($<10^{-15}$) \\ \hline
\small 139 &\small  Manicurists and Pedicurists &\small  0.968 &\small  0.942 &\small  -0.213 ($<10^{-168}$) \\ \hline
\small 140 &\small  Concierges &\small  0.968 &\small  0.861 &\small  0.291 ($<10^{-292}$) \\ \hline
\small 141 &\small  Food and Tobacco Roasting, Baking, and Drying Machine Operators and Tenders &\small  0.968 &\small  0.384 &\small  0.364 (0) \\ \hline
\small 142 &\small  Print Binding and Finishing Workers &\small  0.968 &\small  0.667 &\small  0.306 (0) \\ \hline
\small 143 &\small  Sales Managers &\small  0.968 &\small  1.126 &\small  -0.267 (0) \\ \hline
\small 144 &\small  Curators &\small  0.968 &\small  0.518 &\small  -0.147 ($<10^{-72}$) \\ \hline
\small 145 &\small  Reporters and Correspondents &\small  0.969 &\small  0.762 &\small  0.061 ($<10^{-26}$) \\ \hline
\small 146 &\small  Waiters and Waitresses &\small  0.969 &\small  0.948 &\small  0.306 (0) \\ \hline
\small 147 &\small  Pharmacists &\small  0.969 &\small  0.921 &\small  0.179 (0) \\ \hline
\small 148 &\small  Bus Drivers, School or Special Client &\small  0.969 &\small  0.887 &\small  0.151 (0) \\ \hline
\small 149 &\small  Locker Room, Coatroom, and Dressing Room Attendants &\small  0.969 &\small  0.689 &\small  0.403 (0) \\ \hline
\small 150 &\small  Social and Human Service Assistants &\small  0.969 &\small  0.901 &\small  -0.105 ($<10^{-192}$) \\ \hline
\small 151 &\small  Graphic Designers &\small  0.969 &\small  1.063 &\small  -0.304 (0) \\ \hline
\small 152 &\small  Furnace, Kiln, Oven, Drier, and Kettle Operators and Tenders &\small  0.969 &\small  0.433 &\small  0.368 (0) \\ \hline
\small 153 &\small  Furniture Finishers &\small  0.969 &\small  0.276 &\small  0.416 (0) \\ \hline
\small 154 &\small  Clergy &\small  0.969 &\small  0.693 &\small  0.072 ($<10^{-45}$) \\ \hline
\small 155 &\small  Purchasing Managers &\small  0.969 &\small  0.924 &\small  -0.119 ($<10^{-146}$) \\ \hline
\small 156 &\small  Adult Basic and Secondary Education and Literacy Teachers and Instructors &\small  0.969 &\small  0.569 &\small  0.187 ($<10^{-241}$) \\ \hline
\small 157 &\small  Reservation and Transportation Ticket Agents and Travel Clerks &\small  0.969 &\small  1.078 &\small  0.187 ($<10^{-178}$) \\ \hline
\small 158 &\small  Skincare Specialists &\small  0.969 &\small  0.798 &\small  0.304 (0) \\ \hline
\small 159 &\small  Art Directors &\small  0.969 &\small  0.974 &\small  -0.186 ($<10^{-176}$) \\ \hline
\small 160 &\small  Customer Service Representatives &\small  0.969 &\small  1.069 &\small  -0.115 ($<10^{-249}$) \\ \hline
\small 161 &\small  Sales Representatives, Wholesale and Manufacturing, Technical and Scientific Products &\small  0.969 &\small  1.189 &\small  -0.409 (0) \\ \hline
\small 162 &\small  Team Assemblers &\small  0.969 &\small  0.882 &\small  0.306 (0) \\ \hline
\small 163 &\small  Computer and Information Systems Managers &\small  0.969 &\small  1.219 &\small  -0.718 (0) \\ \hline
\small 164 &\small  Hotel, Motel, and Resort Desk Clerks &\small  0.969 &\small  0.788 &\small  0.313 (0) \\ \hline
\small 165 &\small  Aerospace Engineers &\small  0.969 &\small  0.478 &\small  -0.236 ($<10^{-164}$) \\ \hline
\small 166 &\small  Health and Safety Engineers, Except Mining Safety Engineers and Inspectors &\small  0.969 &\small  0.634 &\small  0.133 ($<10^{-94}$) \\ \hline
\small 167 &\small  Tax Preparers &\small  0.969 &\small  0.845 &\small  0.214 (0) \\ \hline
\small 168 &\small  Private Detectives and Investigators &\small  0.969 &\small  0.555 &\small  0.194 ($<10^{-107}$) \\ \hline
\small 169 &\small  Computer and Information Research Scientists &\small  0.969 &\small  0.440 &\small  -0.343 ($<10^{-320}$) \\ \hline
\small 170 &\small  Civil Engineering Technicians &\small  0.970 &\small  0.721 &\small  0.148 ($<10^{-256}$) \\ \hline
\small 171 &\small  Printing Press Operators &\small  0.970 &\small  0.930 &\small  0.126 ($<10^{-243}$) \\ \hline
\small 172 &\small  Biological Technicians &\small  0.970 &\small  0.632 &\small  -0.081 ($<10^{-44}$) \\ \hline
\small 173 &\small  Milling and Planing Machine Setters, Operators, and Tenders, Metal and Plastic &\small  0.970 &\small  0.538 &\small  0.315 (0) \\ \hline
\small 174 &\small  Electrical and Electronics Drafters &\small  0.970 &\small  0.798 &\small  0.001 ($<10^{0}$) \\ \hline
\small 175 &\small  Medical Scientists, Except Epidemiologists &\small  0.970 &\small  0.937 &\small  -0.475 (0) \\ \hline
\small 176 &\small  Insurance Appraisers, Auto Damage &\small  0.970 &\small  0.490 &\small  0.332 ($<10^{-297}$) \\ \hline
\small 177 &\small  Real Estate Sales Agents &\small  0.970 &\small  0.929 &\small  0.247 (0) \\ \hline
\small 178 &\small  Library Assistants, Clerical &\small  0.970 &\small  0.777 &\small  0.067 ($<10^{-60}$) \\ \hline
\small 179 &\small  Environmental Scientists and Specialists, Including Health &\small  0.970 &\small  0.796 &\small  -0.131 ($<10^{-200}$) \\ \hline
\small 180 &\small  Counter Attendants, Cafeteria, Food Concession, and Coffee Shop &\small  0.970 &\small  0.958 &\small  0.099 ($<10^{-137}$) \\ \hline
\small 181 &\small  Civil Engineers &\small  0.970 &\small  1.039 &\small  -0.233 (0) \\ \hline
\small 182 &\small  Librarians &\small  0.970 &\small  0.873 &\small  -0.031 ($<10^{-16}$) \\ \hline
\small 183 &\small  Construction Managers &\small  0.970 &\small  1.027 &\small  -0.002 ($<10^{0}$) \\ \hline
\small 184 &\small  Mechanical Engineers &\small  0.970 &\small  0.971 &\small  -0.166 (0) \\ \hline
\small 185 &\small  Commercial and Industrial Designers &\small  0.970 &\small  0.644 &\small  0.224 ($<10^{-251}$) \\ \hline
\small 186 &\small  Hairdressers, Hairstylists, and Cosmetologists &\small  0.970 &\small  0.952 &\small  0.013 ($<10^{-2}$) \\ \hline
\small 187 &\small  Recreational Therapists &\small  0.970 &\small  0.485 &\small  0.222 ($<10^{-260}$) \\ \hline
\small 188 &\small  Couriers and Messengers &\small  0.970 &\small  0.834 &\small  0.236 (0) \\ \hline
\small 189 &\small  Helpers--Production Workers &\small  0.970 &\small  0.864 &\small  0.307 (0) \\ \hline
\small 190 &\small  Lathe and Turning Machine Tool Setters, Operators, and Tenders, Metal and Plastic &\small  0.970 &\small  0.503 &\small  0.317 (0) \\ \hline
\small 191 &\small  Residential Advisors &\small  0.970 &\small  0.692 &\small  0.209 (0) \\ \hline
\small 192 &\small  Psychiatric Technicians &\small  0.970 &\small  0.412 &\small  0.206 ($<10^{-182}$) \\ \hline
\small 193 &\small  Film and Video Editors &\small  0.971 &\small  0.916 &\small  0.089 ($<10^{-22}$) \\ \hline
\small 194 &\small  Substance Abuse and Behavioral Disorder Counselors &\small  0.971 &\small  0.726 &\small  0.125 ($<10^{-180}$) \\ \hline
\small 195 &\small  Logisticians &\small  0.971 &\small  0.936 &\small  -0.237 (0) \\ \hline
\small 196 &\small  Architectural and Civil Drafters &\small  0.971 &\small  0.934 &\small  0.022 ($<10^{-5}$) \\ \hline
\small 197 &\small  Chemical Technicians &\small  0.971 &\small  0.760 &\small  0.155 ($<10^{-221}$) \\ \hline
\small 198 &\small  Biomedical Engineers &\small  0.971 &\small  0.645 &\small  -0.565 (0) \\ \hline
\small 199 &\small  Wholesale and Retail Buyers, Except Farm Products &\small  0.971 &\small  0.986 &\small  -0.048 ($<10^{-27}$) \\ \hline
\small 200 &\small  Computer Systems Analysts &\small  0.971 &\small  1.277 &\small  -0.584 (0) \\ \hline
\small 201 &\small  Engineers, All Other &\small  0.971 &\small  0.894 &\small  -0.262 (0) \\ \hline
\small 202 &\small  Chemical Engineers &\small  0.971 &\small  0.524 &\small  0.207 ($<10^{-215}$) \\ \hline
\small 203 &\small  Occupational Therapists &\small  0.971 &\small  0.879 &\small  0.037 ($<10^{-21}$) \\ \hline
\small 204 &\small  Pharmacy Aides &\small  0.971 &\small  0.627 &\small  0.399 (0) \\ \hline
\small 205 &\small  Middle School Teachers, Except Special and Career/Technical Education &\small  0.971 &\small  0.926 &\small  0.138 ($<10^{-302}$) \\ \hline
\small 206 &\small  Purchasing Agents, Except Wholesale, Retail, and Farm Products &\small  0.971 &\small  1.047 &\small  -0.296 (0) \\ \hline
\small 207 &\small  Computer Operators &\small  0.971 &\small  0.858 &\small  0.041 ($<10^{-14}$) \\ \hline
\small 208 &\small  Postmasters and Mail Superintendents &\small  0.971 &\small  0.348 &\small  0.344 (0) \\ \hline
\small 209 &\small  Fitness Trainers and Aerobics Instructors &\small  0.971 &\small  0.952 &\small  -0.204 (0) \\ \hline
\small 210 &\small  Administrative Services Managers &\small  0.971 &\small  1.020 &\small  -0.288 (0) \\ \hline
\small 211 &\small  Medical Assistants &\small  0.971 &\small  0.951 &\small  0.124 ($<10^{-287}$) \\ \hline
\small 212 &\small  Urban and Regional Planners &\small  0.971 &\small  0.670 &\small  0.031 ($<10^{-7}$) \\ \hline
\small 213 &\small  Medical and Health Services Managers &\small  0.971 &\small  0.924 &\small  -0.178 (0) \\ \hline
\small 214 &\small  Natural Sciences Managers &\small  0.971 &\small  0.711 &\small  -0.371 (0) \\ \hline
\small 215 &\small  Preschool Teachers, Except Special Education &\small  0.972 &\small  0.956 &\small  -0.145 (0) \\ \hline
\small 216 &\small  Dietitians and Nutritionists &\small  0.972 &\small  0.844 &\small  0.031 ($<10^{-11}$) \\ \hline
\small 217 &\small  Production, Planning, and Expediting Clerks &\small  0.972 &\small  1.022 &\small  0.002 ($<10^{0}$) \\ \hline
\small 218 &\small  Lifeguards, Ski Patrol, and Other Recreational Protective Service Workers &\small  0.972 &\small  0.908 &\small  0.151 ($<10^{-241}$) \\ \hline
\small 219 &\small  Floral Designers &\small  0.972 &\small  0.751 &\small  0.296 (0) \\ \hline
\small 220 &\small  Postal Service Mail Carriers &\small  0.972 &\small  0.891 &\small  0.253 (0) \\ \hline
\small 221 &\small  Multimedia Artists and Animators &\small  0.972 &\small  1.012 &\small  -0.244 ($<10^{-196}$) \\ \hline
\small 222 &\small  Architectural and Engineering Managers &\small  0.972 &\small  1.039 &\small  -0.390 (0) \\ \hline
\small 223 &\small  Chemical Equipment Operators and Tenders &\small  0.972 &\small  0.508 &\small  0.305 (0) \\ \hline
\small 224 &\small  Soil and Plant Scientists &\small  0.972 &\small  0.166 &\small  0.116 ($<10^{-38}$) \\ \hline
\small 225 &\small  Sewing Machine Operators &\small  0.972 &\small  0.743 &\small  0.277 (0) \\ \hline
\small 226 &\small  Chemists &\small  0.972 &\small  0.881 &\small  -0.200 (0) \\ \hline
\small 227 &\small  Psychiatric Aides &\small  0.972 &\small  0.322 &\small  0.331 (0) \\ \hline
\small 228 &\small  Inspectors, Testers, Sorters, Samplers, and Weighers &\small  0.972 &\small  0.906 &\small  0.267 (0) \\ \hline
\small 229 &\small  Tour Guides and Escorts &\small  0.972 &\small  0.551 &\small  0.363 (0) \\ \hline
\small 230 &\small  Maids and Housekeeping Cleaners &\small  0.972 &\small  0.907 &\small  0.260 (0) \\ \hline
\small 231 &\small  Compliance Officers &\small  0.972 &\small  1.039 &\small  -0.111 ($<10^{-217}$) \\ \hline
\small 232 &\small  Microbiologists &\small  0.972 &\small  0.531 &\small  -0.128 ($<10^{-47}$) \\ \hline
\small 233 &\small  Woodworking Machine Setters, Operators, and Tenders, Except Sawing &\small  0.972 &\small  0.410 &\small  0.492 (0) \\ \hline
\small 234 &\small  Funeral Attendants &\small  0.973 &\small  0.509 &\small  0.452 (0) \\ \hline
\small 235 &\small  Sawing Machine Setters, Operators, and Tenders, Wood &\small  0.973 &\small  0.320 &\small  0.472 (0) \\ \hline
\small 236 &\small  Paper Goods Machine Setters, Operators, and Tenders &\small  0.973 &\small  0.514 &\small  0.251 (0) \\ \hline
\small 237 &\small  Cutting and Slicing Machine Setters, Operators, and Tenders &\small  0.973 &\small  0.514 &\small  0.411 (0) \\ \hline
\small 238 &\small  Helpers--Electricians &\small  0.973 &\small  0.640 &\small  0.441 (0) \\ \hline
\small 239 &\small  Electrical and Electronic Equipment Assemblers &\small  0.973 &\small  0.914 &\small  -0.030 ($<10^{-8}$) \\ \hline
\small 240 &\small  Postal Service Mail Sorters, Processors, and Processing Machine Operators &\small  0.973 &\small  0.975 &\small  -0.023 ($<10^{-5}$) \\ \hline
\small 241 &\small  Interior Designers &\small  0.973 &\small  1.038 &\small  0.080 ($<10^{-40}$) \\ \hline
\small 242 &\small  Physical Therapists &\small  0.973 &\small  0.934 &\small  -0.057 ($<10^{-56}$) \\ \hline
\small 243 &\small  Industrial Engineering Technicians &\small  0.973 &\small  0.600 &\small  0.136 ($<10^{-155}$) \\ \hline
\small 244 &\small  First-Line Supervisors of Correctional Officers &\small  0.973 &\small  0.456 &\small  0.195 ($<10^{-236}$) \\ \hline
\small 245 &\small  Hosts and Hostesses, Restaurant, Lounge, and Coffee Shop &\small  0.973 &\small  0.946 &\small  0.246 (0) \\ \hline
\small 246 &\small  Buyers and Purchasing Agents, Farm Products &\small  0.973 &\small  0.370 &\small  0.335 (0) \\ \hline
\small 247 &\small  First-Line Supervisors of Personal Service Workers &\small  0.973 &\small  0.913 &\small  0.090 ($<10^{-142}$) \\ \hline
\small 248 &\small  Amusement and Recreation Attendants &\small  0.973 &\small  0.979 &\small  0.232 (0) \\ \hline
\small 249 &\small  Coaches and Scouts &\small  0.973 &\small  0.876 &\small  -0.014 ($<10^{-2}$) \\ \hline
\small 250 &\small  Self-Enrichment Education Teachers &\small  0.973 &\small  0.948 &\small  -0.213 (0) \\ \hline
\small 251 &\small  Medical and Clinical Laboratory Technologists &\small  0.973 &\small  0.905 &\small  0.237 (0) \\ \hline
\small 252 &\small  Pharmacy Technicians &\small  0.973 &\small  0.882 &\small  0.310 (0) \\ \hline
\small 253 &\small  Food Scientists and Technologists &\small  0.973 &\small  0.373 &\small  0.268 ($<10^{-216}$) \\ \hline
\small 254 &\small  Painting, Coating, and Decorating Workers &\small  0.973 &\small  0.570 &\small  0.519 (0) \\ \hline
\small 255 &\small  Sales Engineers &\small  0.973 &\small  1.004 &\small  -0.442 (0) \\ \hline
\small 256 &\small  Food Servers, Nonrestaurant &\small  0.973 &\small  0.912 &\small  0.064 ($<10^{-60}$) \\ \hline
\small 257 &\small  Geoscientists, Except Hydrologists and Geographers &\small  0.974 &\small  0.515 &\small  0.320 (0) \\ \hline
\small 258 &\small  Electronics Engineers, Except Computer &\small  0.974 &\small  1.002 &\small  -0.350 (0) \\ \hline
\small 259 &\small  Optometrists &\small  0.974 &\small  0.813 &\small  0.175 ($<10^{-216}$) \\ \hline
\small 260 &\small  Occupational Therapy Assistants &\small  0.974 &\small  0.583 &\small  0.323 (0) \\ \hline
\small 261 &\small  Personal Care Aides &\small  0.974 &\small  0.877 &\small  0.100 ($<10^{-169}$) \\ \hline
\small 262 &\small  Postal Service Clerks &\small  0.974 &\small  0.814 &\small  0.219 (0) \\ \hline
\small 263 &\small  Painters, Transportation Equipment &\small  0.974 &\small  0.701 &\small  0.350 (0) \\ \hline
\small 264 &\small  Food Batchmakers &\small  0.974 &\small  0.663 &\small  0.311 (0) \\ \hline
\small 265 &\small  Welding, Soldering, and Brazing Machine Setters, Operators, and Tenders &\small  0.974 &\small  0.443 &\small  0.240 (0) \\ \hline
\small 266 &\small  Dining Room and Cafeteria Attendants and Bartender Helpers &\small  0.974 &\small  1.019 &\small  0.139 ($<10^{-312}$) \\ \hline
\small 267 &\small  Environmental Engineers &\small  0.974 &\small  0.780 &\small  -0.154 ($<10^{-199}$) \\ \hline
\small 268 &\small  Industrial Engineers &\small  0.974 &\small  0.899 &\small  0.080 ($<10^{-94}$) \\ \hline
\small 269 &\small  Electrical Engineers &\small  0.974 &\small  1.020 &\small  -0.403 (0) \\ \hline
\small 270 &\small  Industrial Production Managers &\small  0.974 &\small  0.840 &\small  0.256 (0) \\ \hline
\small 271 &\small  Pressers, Textile, Garment, and Related Materials &\small  0.974 &\small  0.749 &\small  0.351 (0) \\ \hline
\small 272 &\small  Materials Engineers &\small  0.974 &\small  0.679 &\small  0.037 ($<10^{-5}$) \\ \hline
\small 273 &\small  Extruding, Forming, Pressing, and Compacting Machine Setters, Operators, and Tenders &\small  0.974 &\small  0.540 &\small  0.329 (0) \\ \hline
\small 274 &\small  Plating and Coating Machine Setters, Operators, and Tenders, Metal and Plastic &\small  0.974 &\small  0.487 &\small  0.382 (0) \\ \hline
\small 275 &\small  Cooks, Institution and Cafeteria &\small  0.974 &\small  0.776 &\small  0.290 (0) \\ \hline
\small 276 &\small  Bartenders &\small  0.974 &\small  0.932 &\small  0.118 ($<10^{-253}$) \\ \hline
\small 277 &\small  General and Operations Managers &\small  0.974 &\small  1.024 &\small  -0.269 (0) \\ \hline
\small 278 &\small  Dispatchers, Except Police, Fire, and Ambulance &\small  0.974 &\small  0.991 &\small  0.225 (0) \\ \hline
\small 279 &\small  Cutting, Punching, and Press Machine Setters, Operators, and Tenders, Metal and Plastic &\small  0.974 &\small  0.712 &\small  0.355 (0) \\ \hline
\small 280 &\small  Licensed Practical and Licensed Vocational Nurses &\small  0.974 &\small  0.845 &\small  0.293 (0) \\ \hline
\small 281 &\small  First-Line Supervisors of Food Preparation and Serving Workers &\small  0.974 &\small  0.881 &\small  0.363 (0) \\ \hline
\small 282 &\small  Tool and Die Makers &\small  0.974 &\small  0.531 &\small  0.256 (0) \\ \hline
\small 283 &\small  Landscape Architects &\small  0.974 &\small  0.680 &\small  0.087 ($<10^{-21}$) \\ \hline
\small 284 &\small  Lodging Managers &\small  0.975 &\small  0.593 &\small  0.397 (0) \\ \hline
\small 285 &\small  Physician Assistants &\small  0.975 &\small  0.816 &\small  0.031 ($<10^{-12}$) \\ \hline
\small 286 &\small  Architects, Except Landscape and Naval &\small  0.975 &\small  1.034 &\small  -0.288 (0) \\ \hline
\small 287 &\small  Counter and Rental Clerks &\small  0.975 &\small  0.933 &\small  0.215 (0) \\ \hline
\small 288 &\small  Baggage Porters and Bellhops &\small  0.975 &\small  0.783 &\small  0.439 (0) \\ \hline
\small 289 &\small  Food Cooking Machine Operators and Tenders &\small  0.975 &\small  0.359 &\small  0.486 (0) \\ \hline
\small 290 &\small  Computer, Automated Teller, and Office Machine Repairers &\small  0.975 &\small  0.982 &\small  0.028 ($<10^{-9}$) \\ \hline
\small 291 &\small  Cashiers &\small  0.975 &\small  0.854 &\small  0.491 (0) \\ \hline
\small 292 &\small  Electrical and Electronics Engineering Technicians &\small  0.975 &\small  0.958 &\small  -0.310 (0) \\ \hline
\small 293 &\small  Zoologists and Wildlife Biologists &\small  0.975 &\small  0.299 &\small  0.147 ($<10^{-82}$) \\ \hline
\small 294 &\small  Nonfarm Animal Caretakers &\small  0.975 &\small  0.960 &\small  0.048 ($<10^{-37}$) \\ \hline
\small 295 &\small  First-Line Supervisors of Transportation and Material-Moving Machine and Vehicle Operators &\small  0.975 &\small  0.949 &\small  0.319 (0) \\ \hline
\small 296 &\small  Dental Hygienists &\small  0.975 &\small  0.898 &\small  0.103 ($<10^{-184}$) \\ \hline
\small 297 &\small  Dental Assistants &\small  0.975 &\small  0.920 &\small  0.017 ($<10^{-5}$) \\ \hline
\small 298 &\small  Ophthalmic Laboratory Technicians &\small  0.975 &\small  0.632 &\small  0.152 ($<10^{-95}$) \\ \hline
\small 299 &\small  Retail Salespersons &\small  0.975 &\small  0.904 &\small  0.456 (0) \\ \hline
\small 300 &\small  Chiropractors &\small  0.975 &\small  0.727 &\small  0.295 (0) \\ \hline
\small 301 &\small  Coin, Vending, and Amusement Machine Servicers and Repairers &\small  0.975 &\small  0.635 &\small  0.400 (0) \\ \hline
\small 302 &\small  Upholsterers &\small  0.975 &\small  0.459 &\small  0.345 (0) \\ \hline
\small 303 &\small  Motorcycle Mechanics &\small  0.975 &\small  0.429 &\small  0.490 (0) \\ \hline
\small 304 &\small  Hazardous Materials Removal Workers &\small  0.975 &\small  0.712 &\small  0.319 (0) \\ \hline
\small 305 &\small  Grinding, Lapping, Polishing, and Buffing Machine Tool Setters, Operators, and Tenders, Metal and Plastic &\small  0.975 &\small  0.624 &\small  0.335 (0) \\ \hline
\small 306 &\small  Dentists, General &\small  0.975 &\small  0.900 &\small  0.074 ($<10^{-73}$) \\ \hline
\small 307 &\small  Photographers &\small  0.975 &\small  0.929 &\small  0.203 (0) \\ \hline
\small 308 &\small  Jewelers and Precious Stone and Metal Workers &\small  0.975 &\small  0.743 &\small  0.187 ($<10^{-108}$) \\ \hline
\small 309 &\small  Maintenance Workers, Machinery &\small  0.975 &\small  0.685 &\small  0.412 (0) \\ \hline
\small 310 &\small  Stationary Engineers and Boiler Operators &\small  0.975 &\small  0.572 &\small  0.205 ($<10^{-253}$) \\ \hline
\small 311 &\small  Cabinetmakers and Bench Carpenters &\small  0.976 &\small  0.726 &\small  0.329 (0) \\ \hline
\small 312 &\small  Medical Equipment Preparers &\small  0.976 &\small  0.766 &\small  0.182 ($<10^{-291}$) \\ \hline
\small 313 &\small  Coating, Painting, and Spraying Machine Setters, Operators, and Tenders &\small  0.976 &\small  0.657 &\small  0.398 (0) \\ \hline
\small 314 &\small  Butchers and Meat Cutters &\small  0.976 &\small  0.844 &\small  0.307 (0) \\ \hline
\small 315 &\small  Driver/Sales Workers &\small  0.976 &\small  0.943 &\small  0.204 (0) \\ \hline
\small 316 &\small  Aircraft Mechanics and Service Technicians &\small  0.976 &\small  0.812 &\small  -0.034 ($<10^{-9}$) \\ \hline
\small 317 &\small  Glaziers &\small  0.976 &\small  0.728 &\small  0.249 ($<10^{-305}$) \\ \hline
\small 318 &\small  Home Health Aides &\small  0.976 &\small  0.937 &\small  0.018 ($<10^{-4}$) \\ \hline
\small 319 &\small  Power Plant Operators &\small  0.976 &\small  0.549 &\small  0.364 (0) \\ \hline
\small 320 &\small  Electromechanical Equipment Assemblers &\small  0.976 &\small  0.625 &\small  0.103 ($<10^{-41}$) \\ \hline
\small 321 &\small  Broadcast Technicians &\small  0.976 &\small  0.713 &\small  0.309 (0) \\ \hline
\small 322 &\small  Prepress Technicians and Workers &\small  0.976 &\small  0.742 &\small  0.199 ($<10^{-280}$) \\ \hline
\small 323 &\small  Tile and Marble Setters &\small  0.976 &\small  0.646 &\small  0.378 (0) \\ \hline
\small 324 &\small  Excavating and Loading Machine and Dragline Operators &\small  0.976 &\small  0.423 &\small  0.335 (0) \\ \hline
\small 325 &\small  Veterinarians &\small  0.976 &\small  0.813 &\small  0.132 ($<10^{-228}$) \\ \hline
\small 326 &\small  Shipping, Receiving, and Traffic Clerks &\small  0.976 &\small  1.026 &\small  0.181 (0) \\ \hline
\small 327 &\small  Cooks, Fast Food &\small  0.976 &\small  0.763 &\small  0.395 (0) \\ \hline
\small 328 &\small  Logging Equipment Operators &\small  0.976 &\small  0.104 &\small  0.150 ($<10^{-65}$) \\ \hline
\small 329 &\small  Cardiovascular Technologists and Technicians &\small  0.976 &\small  0.807 &\small  0.350 (0) \\ \hline
\small 330 &\small  Packaging and Filling Machine Operators and Tenders &\small  0.976 &\small  0.805 &\small  0.273 (0) \\ \hline
\small 331 &\small  Helpers--Brickmasons, Blockmasons, Stonemasons, and Tile and Marble Setters &\small  0.976 &\small  0.510 &\small  0.374 (0) \\ \hline
\small 332 &\small  Childcare Workers &\small  0.976 &\small  0.945 &\small  0.003 ($<10^{0}$) \\ \hline
\small 333 &\small  Rolling Machine Setters, Operators, and Tenders, Metal and Plastic &\small  0.976 &\small  0.422 &\small  0.446 (0) \\ \hline
\small 334 &\small  Home Appliance Repairers &\small  0.976 &\small  0.766 &\small  0.426 (0) \\ \hline
\small 335 &\small  Helpers--Pipelayers, Plumbers, Pipefitters, and Steamfitters &\small  0.976 &\small  0.640 &\small  0.247 (0) \\ \hline
\small 336 &\small  Packers and Packagers, Hand &\small  0.976 &\small  0.969 &\small  0.176 (0) \\ \hline
\small 337 &\small  Farm Equipment Mechanics and Service Technicians &\small  0.976 &\small  0.164 &\small  0.425 (0) \\ \hline
\small 338 &\small  Emergency Management Directors &\small  0.976 &\small  0.385 &\small  0.265 ($<10^{-182}$) \\ \hline
\small 339 &\small  Computer Numerically Controlled Machine Tool Programmers, Metal and Plastic &\small  0.976 &\small  0.534 &\small  0.236 ($<10^{-300}$) \\ \hline
\small 340 &\small  Opticians, Dispensing &\small  0.976 &\small  0.798 &\small  0.319 (0) \\ \hline
\small 341 &\small  Laborers and Freight, Stock, and Material Movers, Hand &\small  0.976 &\small  1.015 &\small  0.262 (0) \\ \hline
\small 342 &\small  Crane and Tower Operators &\small  0.976 &\small  0.509 &\small  0.323 (0) \\ \hline
\small 343 &\small  Animal Control Workers &\small  0.977 &\small  0.491 &\small  0.296 (0) \\ \hline
\small 344 &\small  Grinding and Polishing Workers, Hand &\small  0.977 &\small  0.499 &\small  0.392 (0) \\ \hline
\small 345 &\small  Food Preparation Workers &\small  0.977 &\small  0.914 &\small  0.219 (0) \\ \hline
\small 346 &\small  Meter Readers, Utilities &\small  0.977 &\small  0.605 &\small  0.320 (0) \\ \hline
\small 347 &\small  Pipelayers &\small  0.977 &\small  0.595 &\small  0.471 (0) \\ \hline
\small 348 &\small  Audio and Video Equipment Technicians &\small  0.977 &\small  0.974 &\small  0.072 ($<10^{-36}$) \\ \hline
\small 349 &\small  Police and Sheriff's Patrol Officers &\small  0.977 &\small  0.918 &\small  0.137 (0) \\ \hline
\small 350 &\small  First-Line Supervisors of Retail Sales Workers &\small  0.977 &\small  0.868 &\small  0.479 (0) \\ \hline
\small 351 &\small  Helpers--Installation, Maintenance, and Repair Workers &\small  0.977 &\small  0.855 &\small  0.344 (0) \\ \hline
\small 352 &\small  Occupational Health and Safety Specialists &\small  0.977 &\small  0.786 &\small  0.077 ($<10^{-71}$) \\ \hline
\small 353 &\small  Medical and Clinical Laboratory Technicians &\small  0.977 &\small  0.953 &\small  0.027 ($<10^{-9}$) \\ \hline
\small 354 &\small  Correctional Officers and Jailers &\small  0.977 &\small  0.743 &\small  0.102 ($<10^{-107}$) \\ \hline
\small 355 &\small  Drywall and Ceiling Tile Installers &\small  0.977 &\small  0.801 &\small  0.198 ($<10^{-309}$) \\ \hline
\small 356 &\small  Engineering Technicians, Except Drafters, All Other &\small  0.977 &\small  0.724 &\small  -0.013 ($<10^{0}$) \\ \hline
\small 357 &\small  Cooks, Short Order &\small  0.977 &\small  0.759 &\small  0.078 ($<10^{-59}$) \\ \hline
\small 358 &\small  Athletic Trainers &\small  0.977 &\small  0.662 &\small  0.145 ($<10^{-120}$) \\ \hline
\small 359 &\small  Veterinary Technologists and Technicians &\small  0.977 &\small  0.843 &\small  0.042 ($<10^{-21}$) \\ \hline
\small 360 &\small  Taxi Drivers and Chauffeurs &\small  0.977 &\small  0.958 &\small  -0.014 ($<10^{-2}$) \\ \hline
\small 361 &\small  Detectives and Criminal Investigators &\small  0.977 &\small  0.832 &\small  0.109 ($<10^{-150}$) \\ \hline
\small 362 &\small  Agricultural and Food Science Technicians &\small  0.977 &\small  0.245 &\small  0.353 (0) \\ \hline
\small 363 &\small  Physical Therapist Aides &\small  0.977 &\small  0.702 &\small  0.294 (0) \\ \hline
\small 364 &\small  Life, Physical, and Social Science Technicians, All Other &\small  0.977 &\small  0.780 &\small  -0.109 ($<10^{-92}$) \\ \hline
\small 365 &\small  Veterinary Assistants and Laboratory Animal Caretakers &\small  0.977 &\small  0.764 &\small  0.103 ($<10^{-108}$) \\ \hline
\small 366 &\small  First-Line Supervisors of Production and Operating Workers &\small  0.977 &\small  0.856 &\small  0.350 (0) \\ \hline
\small 367 &\small  Structural Metal Fabricators and Fitters &\small  0.977 &\small  0.660 &\small  0.378 (0) \\ \hline
\small 368 &\small  Drilling and Boring Machine Tool Setters, Operators, and Tenders, Metal and Plastic &\small  0.977 &\small  0.482 &\small  0.576 (0) \\ \hline
\small 369 &\small  Construction and Building Inspectors &\small  0.977 &\small  0.875 &\small  0.081 ($<10^{-88}$) \\ \hline
\small 370 &\small  Nuclear Medicine Technologists &\small  0.977 &\small  0.685 &\small  0.332 (0) \\ \hline
\small 371 &\small  Electrical Power-Line Installers and Repairers &\small  0.977 &\small  0.723 &\small  0.247 (0) \\ \hline
\small 372 &\small  Dental Laboratory Technicians &\small  0.977 &\small  0.722 &\small  0.298 (0) \\ \hline
\small 373 &\small  Medical Equipment Repairers &\small  0.977 &\small  0.794 &\small  0.207 ($<10^{-322}$) \\ \hline
\small 374 &\small  Physical Therapist Assistants &\small  0.977 &\small  0.742 &\small  0.361 (0) \\ \hline
\small 375 &\small  Locksmiths and Safe Repairers &\small  0.977 &\small  0.669 &\small  0.406 (0) \\ \hline
\small 376 &\small  Respiratory Therapists &\small  0.977 &\small  0.865 &\small  0.282 (0) \\ \hline
\small 377 &\small  Security and Fire Alarm Systems Installers &\small  0.977 &\small  0.885 &\small  0.178 ($<10^{-230}$) \\ \hline
\small 378 &\small  Machine Feeders and Offbearers &\small  0.977 &\small  0.628 &\small  0.281 (0) \\ \hline
\small 379 &\small  Bakers &\small  0.977 &\small  0.924 &\small  0.168 (0) \\ \hline
\small 380 &\small  Parking Lot Attendants &\small  0.977 &\small  1.174 &\small  0.063 ($<10^{-29}$) \\ \hline
\small 381 &\small  Recreation Workers &\small  0.978 &\small  0.940 &\small  0.003 ($<10^{0}$) \\ \hline
\small 382 &\small  Landscaping and Groundskeeping Workers &\small  0.978 &\small  0.972 &\small  0.163 (0) \\ \hline
\small 383 &\small  Career/Technical Education Teachers, Secondary School &\small  0.978 &\small  0.657 &\small  0.367 (0) \\ \hline
\small 384 &\small  Surveyors &\small  0.978 &\small  0.690 &\small  0.304 (0) \\ \hline
\small 385 &\small  Office Machine Operators, Except Computer &\small  0.978 &\small  0.944 &\small  -0.002 ($<10^{0}$) \\ \hline
\small 386 &\small  Commercial Pilots &\small  0.978 &\small  0.586 &\small  0.249 (0) \\ \hline
\small 387 &\small  Engine and Other Machine Assemblers &\small  0.978 &\small  0.322 &\small  0.109 ($<10^{-30}$) \\ \hline
\small 388 &\small  Bus Drivers, Transit and Intercity &\small  0.978 &\small  0.938 &\small  0.034 ($<10^{-7}$) \\ \hline
\small 389 &\small  Industrial Truck and Tractor Operators &\small  0.978 &\small  0.917 &\small  0.345 (0) \\ \hline
\small 390 &\small  Pesticide Handlers, Sprayers, and Applicators, Vegetation &\small  0.978 &\small  0.293 &\small  0.523 (0) \\ \hline
\small 391 &\small  Roofers &\small  0.978 &\small  0.795 &\small  0.382 (0) \\ \hline
\small 392 &\small  Motorboat Mechanics and Service Technicians &\small  0.978 &\small  0.380 &\small  0.490 (0) \\ \hline
\small 393 &\small  Light Truck or Delivery Services Drivers &\small  0.978 &\small  0.964 &\small  0.174 (0) \\ \hline
\small 394 &\small  Operating Engineers and Other Construction Equipment Operators &\small  0.978 &\small  0.853 &\small  0.221 (0) \\ \hline
\small 395 &\small  Mixing and Blending Machine Setters, Operators, and Tenders &\small  0.978 &\small  0.747 &\small  0.317 (0) \\ \hline
\small 396 &\small  Carpet Installers &\small  0.978 &\small  0.644 &\small  0.244 ($<10^{-162}$) \\ \hline
\small 397 &\small  Industrial Machinery Mechanics &\small  0.978 &\small  0.773 &\small  0.382 (0) \\ \hline
\small 398 &\small  Brickmasons and Blockmasons &\small  0.978 &\small  0.775 &\small  0.266 (0) \\ \hline
\small 399 &\small  Surveying and Mapping Technicians &\small  0.978 &\small  0.688 &\small  0.262 (0) \\ \hline
\small 400 &\small  Agricultural Inspectors &\small  0.978 &\small  0.137 &\small  0.520 (0) \\ \hline
\small 401 &\small  First-Line Supervisors of Housekeeping and Janitorial Workers &\small  0.978 &\small  0.938 &\small  0.088 ($<10^{-140}$) \\ \hline
\small 402 &\small  Janitors and Cleaners, Except Maids and Housekeeping Cleaners &\small  0.978 &\small  0.940 &\small  0.114 ($<10^{-247}$) \\ \hline
\small 403 &\small  Welders, Cutters, Solderers, and Brazers &\small  0.978 &\small  0.812 &\small  0.247 (0) \\ \hline
\small 404 &\small  First-Line Supervisors of Police and Detectives &\small  0.978 &\small  0.791 &\small  0.114 ($<10^{-201}$) \\ \hline
\small 405 &\small  Animal Trainers &\small  0.978 &\small  0.437 &\small  0.420 (0) \\ \hline
\small 406 &\small  Radiation Therapists &\small  0.978 &\small  0.610 &\small  0.372 (0) \\ \hline
\small 407 &\small  First-Line Supervisors of Landscaping, Lawn Service, and Groundskeeping Workers &\small  0.978 &\small  0.902 &\small  0.226 (0) \\ \hline
\small 408 &\small  Computer-Controlled Machine Tool Operators, Metal and Plastic &\small  0.978 &\small  0.649 &\small  0.281 (0) \\ \hline
\small 409 &\small  Fire Inspectors and Investigators &\small  0.979 &\small  0.468 &\small  0.207 ($<10^{-145}$) \\ \hline
\small 410 &\small  Surgical Technologists &\small  0.979 &\small  0.796 &\small  0.350 (0) \\ \hline
\small 411 &\small  Sheet Metal Workers &\small  0.979 &\small  0.870 &\small  0.167 (0) \\ \hline
\small 412 &\small  Camera Operators, Television, Video, and Motion Picture &\small  0.979 &\small  0.860 &\small  0.126 ($<10^{-60}$) \\ \hline
\small 413 &\small  Electrical and Electronics Repairers, Powerhouse, Substation, and Relay &\small  0.979 &\small  0.448 &\small  0.204 ($<10^{-125}$) \\ \hline
\small 414 &\small  Environmental Science and Protection Technicians, Including Health &\small  0.979 &\small  0.717 &\small  0.206 ($<10^{-315}$) \\ \hline
\small 415 &\small  Pest Control Workers &\small  0.979 &\small  0.734 &\small  0.435 (0) \\ \hline
\small 416 &\small  File Clerks &\small  0.979 &\small  0.965 &\small  0.087 ($<10^{-114}$) \\ \hline
\small 417 &\small  Combined Food Preparation and Serving Workers, Including Fast Food &\small  0.979 &\small  0.909 &\small  0.281 (0) \\ \hline
\small 418 &\small  Conservation Scientists &\small  0.979 &\small  0.255 &\small  0.092 ($<10^{-37}$) \\ \hline
\small 419 &\small  Millwrights &\small  0.979 &\small  0.466 &\small  0.295 (0) \\ \hline
\small 420 &\small  Dietetic Technicians &\small  0.979 &\small  0.574 &\small  0.311 (0) \\ \hline
\small 421 &\small  Structural Iron and Steel Workers &\small  0.979 &\small  0.713 &\small  0.339 (0) \\ \hline
\small 422 &\small  Mechanical Engineering Technicians &\small  0.979 &\small  0.751 &\small  0.037 ($<10^{-9}$) \\ \hline
\small 423 &\small  Molders, Shapers, and Casters, Except Metal and Plastic &\small  0.979 &\small  0.558 &\small  0.386 (0) \\ \hline
\small 424 &\small  Conveyor Operators and Tenders &\small  0.979 &\small  0.446 &\small  0.476 (0) \\ \hline
\small 425 &\small  Transportation Inspectors &\small  0.979 &\small  0.634 &\small  0.130 ($<10^{-55}$) \\ \hline
\small 426 &\small  Forensic Science Technicians &\small  0.979 &\small  0.498 &\small  0.156 ($<10^{-83}$) \\ \hline
\small 427 &\small  Diagnostic Medical Sonographers &\small  0.979 &\small  0.817 &\small  0.198 (0) \\ \hline
\small 428 &\small  Cleaners of Vehicles and Equipment &\small  0.979 &\small  0.930 &\small  0.313 (0) \\ \hline
\small 429 &\small  Cement Masons and Concrete Finishers &\small  0.979 &\small  0.847 &\small  0.277 (0) \\ \hline
\small 430 &\small  Machinists &\small  0.979 &\small  0.877 &\small  0.194 (0) \\ \hline
\small 431 &\small  Dishwashers &\small  0.979 &\small  0.945 &\small  0.133 (0) \\ \hline
\small 432 &\small  Multiple Machine Tool Setters, Operators, and Tenders, Metal and Plastic &\small  0.979 &\small  0.460 &\small  0.317 (0) \\ \hline
\small 433 &\small  Chefs and Head Cooks &\small  0.979 &\small  0.931 &\small  0.072 ($<10^{-65}$) \\ \hline
\small 434 &\small  Outdoor Power Equipment and Other Small Engine Mechanics &\small  0.979 &\small  0.532 &\small  0.532 (0) \\ \hline
\small 435 &\small  Insulation Workers, Mechanical &\small  0.979 &\small  0.414 &\small  0.326 (0) \\ \hline
\small 436 &\small  Heavy and Tractor-Trailer Truck Drivers &\small  0.979 &\small  0.901 &\small  0.380 (0) \\ \hline
\small 437 &\small  Electric Motor, Power Tool, and Related Repairers &\small  0.979 &\small  0.479 &\small  0.443 (0) \\ \hline
\small 438 &\small  Food Service Managers &\small  0.980 &\small  0.906 &\small  0.048 ($<10^{-41}$) \\ \hline
\small 439 &\small  Automotive and Watercraft Service Attendants &\small  0.980 &\small  0.762 &\small  0.155 ($<10^{-280}$) \\ \hline
\small 440 &\small  Refuse and Recyclable Material Collectors &\small  0.980 &\small  0.763 &\small  0.329 (0) \\ \hline
\small 441 &\small  Automotive Service Technicians and Mechanics &\small  0.980 &\small  0.888 &\small  0.324 (0) \\ \hline
\small 442 &\small  Electrical and Electronics Repairers, Commercial and Industrial Equipment &\small  0.980 &\small  0.715 &\small  -0.076 ($<10^{-57}$) \\ \hline
\small 443 &\small  Occupational Health and Safety Technicians &\small  0.980 &\small  0.420 &\small  0.436 (0) \\ \hline
\small 444 &\small  Bus and Truck Mechanics and Diesel Engine Specialists &\small  0.980 &\small  0.895 &\small  0.286 (0) \\ \hline
\small 445 &\small  Paving, Surfacing, and Tamping Equipment Operators &\small  0.980 &\small  0.635 &\small  0.376 (0) \\ \hline
\small 446 &\small  Water and Wastewater Treatment Plant and System Operators &\small  0.980 &\small  0.721 &\small  0.312 (0) \\ \hline
\small 447 &\small  First-Line Supervisors of Construction Trades and Extraction Workers &\small  0.980 &\small  0.955 &\small  0.155 (0) \\ \hline
\small 448 &\small  Telecommunications Line Installers and Repairers &\small  0.980 &\small  0.854 &\small  0.256 (0) \\ \hline
\small 449 &\small  Laundry and Dry-Cleaning Workers &\small  0.980 &\small  0.921 &\small  0.239 (0) \\ \hline
\small 450 &\small  Chemical Plant and System Operators &\small  0.980 &\small  0.099 &\small  0.412 (0) \\ \hline
\small 451 &\small  Environmental Engineering Technicians &\small  0.981 &\small  0.589 &\small  0.287 (0) \\ \hline
\small 452 &\small  Crushing, Grinding, and Polishing Machine Setters, Operators, and Tenders &\small  0.981 &\small  0.435 &\small  0.425 (0) \\ \hline
\small 453 &\small  Mobile Heavy Equipment Mechanics, Except Engines &\small  0.981 &\small  0.780 &\small  0.203 (0) \\ \hline
\small 454 &\small  Cleaning, Washing, and Metal Pickling Equipment Operators and Tenders &\small  0.981 &\small  0.349 &\small  0.255 ($<10^{-234}$) \\ \hline
\small 455 &\small  Heating, Air Conditioning, and Refrigeration Mechanics and Installers &\small  0.981 &\small  0.916 &\small  0.208 (0) \\ \hline
\small 456 &\small  Separating, Filtering, Clarifying, Precipitating, and Still Machine Setters, Operators, and Tenders &\small  0.981 &\small  0.366 &\small  0.312 (0) \\ \hline
\small 457 &\small  Plumbers, Pipefitters, and Steamfitters &\small  0.981 &\small  0.960 &\small  0.037 ($<10^{-24}$) \\ \hline
\small 458 &\small  Heat Treating Equipment Setters, Operators, and Tenders, Metal and Plastic &\small  0.981 &\small  0.415 &\small  0.366 (0) \\ \hline
\small 459 &\small  Helpers--Carpenters &\small  0.981 &\small  0.576 &\small  0.297 (0) \\ \hline
\small 460 &\small  Tire Repairers and Changers &\small  0.981 &\small  0.698 &\small  0.451 (0) \\ \hline
\small 461 &\small  Electricians &\small  0.981 &\small  0.964 &\small  0.062 ($<10^{-70}$) \\ \hline
\small 462 &\small  Automotive Body and Related Repairers &\small  0.981 &\small  0.890 &\small  0.293 (0) \\ \hline
\small 463 &\small  Highway Maintenance Workers &\small  0.981 &\small  0.587 &\small  0.212 (0) \\ \hline
\small 464 &\small  Electronic Home Entertainment Equipment Installers and Repairers &\small  0.981 &\small  0.571 &\small  0.401 (0) \\ \hline
\small 465 &\small  Meat, Poultry, and Fish Cutters and Trimmers &\small  0.981 &\small  0.528 &\small  0.352 (0) \\ \hline
\small 466 &\small  Cooks, Restaurant &\small  0.982 &\small  0.964 &\small  0.197 (0) \\ \hline
\small 467 &\small  Firefighters &\small  0.982 &\small  0.851 &\small  0.207 (0) \\ \hline
\small 468 &\small  Control and Valve Installers and Repairers, Except Mechanical Door &\small  0.982 &\small  0.584 &\small  0.415 (0) \\ \hline
\small 469 &\small  Painters, Construction and Maintenance &\small  0.982 &\small  0.969 &\small  0.054 ($<10^{-49}$) \\ \hline
\small 470 &\small  Maintenance and Repair Workers, General &\small  0.982 &\small  0.908 &\small  0.364 (0) \\ \hline
\small 471 &\small  First-Line Supervisors of Mechanics, Installers, and Repairers &\small  0.982 &\small  0.911 &\small  0.181 (0) \\ \hline
\small 472 &\small  Telecommunications Equipment Installers and Repairers, Except Line Installers &\small  0.982 &\small  0.963 &\small  0.062 ($<10^{-55}$) \\ \hline
\small 473 &\small  Septic Tank Servicers and Sewer Pipe Cleaners &\small  0.982 &\small  0.522 &\small  0.475 (0) \\ \hline
\small 474 &\small  First-Line Supervisors of Farming, Fishing, and Forestry Workers &\small  0.982 &\small  0.035 &\small  0.321 (0) \\ \hline
\small 475 &\small  First-Line Supervisors of Fire Fighting and Prevention Workers &\small  0.982 &\small  0.734 &\small  0.205 (0) \\ \hline
\small 476 &\small  Carpenters &\small  0.983 &\small  0.957 &\small  0.115 ($<10^{-250}$) \\ \hline
\small 477 &\small  Emergency Medical Technicians and Paramedics &\small  0.983 &\small  0.807 &\small  0.217 (0) \\ \hline
\small 478 &\small  Construction Laborers &\small  0.983 &\small  0.943 &\small  0.230 (0) \\ \hline
\small 479 &\small  Extruding and Drawing Machine Setters, Operators, and Tenders, Metal and Plastic &\small  0.984 &\small  0.463 &\small  0.429 (0) \\ \hline
\small 480 &\small  Insulation Workers, Floor, Ceiling, and Wall &\small  0.985 &\small  0.499 &\small  0.289 ($<10^{-218}$) \\ \hline
\small 481 &\small  Forest and Conservation Technicians &\small  0.985 &\small  0.058 &\small  0.376 (0) \\ \hline

\end{longtabu}
\subsection{Job Groups}
	\label{jobGroups}
\indent The \onet skills database allows us to identify how important each of 230 workplace skills is to completing each of the BLS jobs. 
We use K-means clustering to group jobs into five groups according to the skills required to perform those jobs.
The complete list of BLS jobs comprising each job group is presented in the table below.
Our interpretation about the scaling behaviors of jobs, and how aggregate skills indicate those scaling behaviors, is the same if we use anywhere between three and seven job groups instead of five while computing the K-means clustering algorithm. 
\begin{longtabu} to \textwidth {|c|X|}
	\hline
	\cellcolor[gray]{0.8}\bf Group ($\beta$)& \bf BLS Jobs \\ \hline
	\scriptsize\cellcolor[gray]{0.8} \textcolor{myMagenta}{Purple (1.39)} &\scriptsize Aerospace Engineers, Agricultural Engineers, Animal Scientists, Anthropologists and Archeologists, Architects, Except Landscape and Naval, Architectural and Civil Drafters, Architectural and Engineering Managers, Astronomers, Atmospheric and Space Scientists, Biochemists and Biophysicists, Biological Scientists, All Other, Biomedical Engineers, Cartographers and Photogrammetrists, Chemical Engineers, Chemists, Civil Engineering Technicians, Civil Engineers, Commercial and Industrial Designers, Computer Hardware Engineers, Computer Programmers, Computer Systems Analysts, Computer and Information Research Scientists, Computer and Information Systems Managers, Construction Managers, Database Administrators, Electrical Engineers, Electrical and Electronics Drafters, Electronics Engineers, Except Computer, Engineers, All Other, Environmental Engineers, Environmental Scientists and Specialists, Including Health, Food Scientists and Technologists, Geographers, Geoscientists, Except Hydrologists and Geographers, Health and Safety Engineers, Except Mining Safety Engineers and Inspectors, Hydrologists, Industrial Engineering Technicians, Industrial Engineers, Landscape Architects, Logisticians, Marine Engineers and Naval Architects, Materials Engineers, Materials Scientists, Mathematical Technicians, Mathematicians, Mechanical Drafters, Mechanical Engineers, Medical Scientists, Except Epidemiologists, Microbiologists, Mining and Geological Engineers, Including Mining Safety Engineers, Multimedia Artists and Animators, Natural Sciences Managers, Network and Computer Systems Administrators, Nuclear Engineers, Occupational Health and Safety Specialists, Operations Research Analysts, Petroleum Engineers, Physical Scientists, All Other, Physicists, Sales Engineers, Set and Exhibit Designers, Social Scientists and Related Workers, All Other, Software Developers, Applications, Software Developers, Systems Software, Soil and Plant Scientists, Statistical Assistants, Statisticians, Technical Writers\\ \hline 
\scriptsize\cellcolor[gray]{0.8} \textcolor{myGreen}{Green (1.08)} &\scriptsize Accountants and Auditors, Actuaries, Administrative Law Judges, Adjudicators, and Hearing Officers, Administrative Services Managers, Adult Basic and Secondary Education and Literacy Teachers and Instructors, Advertising Sales Agents, Advertising and Promotions Managers, Agents and Business Managers of Artists, Performers, and Athletes, Air Traffic Controllers, Appraisers and Assessors of Real Estate, Arbitrators, Mediators, and Conciliators, Archivists, Art Directors, Audiologists, Broadcast News Analysts, Budget Analysts, Business Operations Specialists, All Other, Buyers and Purchasing Agents, Farm Products, Career/Technical Education Teachers, Middle School, Cargo and Freight Agents, Chief Executives, Child, Family, and School Social Workers, Choreographers, Claims Adjusters, Examiners, and Investigators, Clergy, Clinical, Counseling, and School Psychologists, Coaches and Scouts, Compensation and Benefits Managers, Compensation, Benefits, and Job Analysis Specialists, Compliance Officers, Concierges, Cost Estimators, Credit Analysts, Credit Authorizers, Checkers, and Clerks, Credit Counselors, Curators, Customer Service Representatives, Dietitians and Nutritionists, Directors, Religious Activities and Education, Dispatchers, Except Police, Fire, and Ambulance, Economists, Editors, Education Administrators, Elementary and Secondary School, Education Administrators, Postsecondary, Education Administrators, Preschool and Childcare Center/Program, Educational, Guidance, School, and Vocational Counselors, Elementary School Teachers, Except Special Education, Eligibility Interviewers, Government Programs, Emergency Management Directors, Epidemiologists, Executive Secretaries and Executive Administrative Assistants, Farm and Home Management Advisors, Fashion Designers, Film and Video Editors, Financial Analysts, Financial Examiners, Financial Managers, Financial Specialists, All Other, First-Line Supervisors of Non-Retail Sales Workers, First-Line Supervisors of Office and Administrative Support Workers, First-Line Supervisors of Personal Service Workers, First-Line Supervisors of Transportation and Material-Moving Machine and Vehicle Operators, Food Service Managers, Gaming Managers, Gaming Supervisors, General and Operations Managers, Graphic Designers, Health Diagnosing and Treating Practitioners, All Other, Health Educators, Healthcare Social Workers, Historians, Human Resources Assistants, Except Payroll and Timekeeping, Human Resources Managers, Industrial-Organizational Psychologists, Instructional Coordinators, Insurance Sales Agents, Insurance Underwriters, Interior Designers, Interpreters and Translators, Judges, Magistrate Judges, and Magistrates, Judicial Law Clerks, Kindergarten Teachers, Except Special Education, Lawyers, Librarians, Loan Interviewers and Clerks, Loan Officers, Lodging Managers, Management Analysts, Managers, All Other, Market Research Analysts and Marketing Specialists, Marketing Managers, Marriage and Family Therapists, Medical Assistants, Medical and Health Services Managers, Meeting, Convention, and Event Planners, Mental Health Counselors, Mental Health and Substance Abuse Social Workers, Middle School Teachers, Except Special and Career/Technical Education, Music Directors and Composers, New Accounts Clerks, Occupational Therapists, Opticians, Dispensing, Paralegals and Legal Assistants, Personal Financial Advisors, Pharmacists, Police, Fire, and Ambulance Dispatchers, Political Scientists, Postmasters and Mail Superintendents, Preschool Teachers, Except Special Education, Private Detectives and Investigators, Probation Officers and Correctional Treatment Specialists, Procurement Clerks, Producers and Directors, Production, Planning, and Expediting Clerks, Property, Real Estate, and Community Association Managers, Psychiatric Technicians, Psychologists, All Other, Public Relations Specialists, Public Relations and Fundraising Managers, Purchasing Agents, Except Wholesale, Retail, and Farm Products, Purchasing Managers, Radio and Television Announcers, Real Estate Brokers, Real Estate Sales Agents, Recreation Workers, Recreational Therapists, Rehabilitation Counselors, Reporters and Correspondents, Residential Advisors, Sales Managers, Sales Representatives, Wholesale and Manufacturing, Except Technical and Scientific Products, Sales Representatives, Wholesale and Manufacturing, Technical and Scientific Products, Secondary School Teachers, Except Special and Career/Technical Education, Securities, Commodities, and Financial Services Sales Agents, Social Science Research Assistants, Social and Community Service Managers, Social and Human Service Assistants, Sociologists, Special Education Teachers, Middle School, Special Education Teachers, Secondary School, Speech-Language Pathologists, Substance Abuse and Behavioral Disorder Counselors, Survey Researchers, Tax Examiners and Collectors, and Revenue Agents, Tax Preparers, Training and Development Managers, Training and Development Specialists, Transportation, Storage, and Distribution Managers, Travel Agents, Travel Guides, Urban and Regional Planners, Wholesale and Retail Buyers, Except Farm Products, Writers and Authors\\ \hline 
\scriptsize\cellcolor[gray]{0.8} \textcolor{myYellow}{Yellow (1.02)} &\scriptsize Aerospace Engineering and Operations Technicians, Agricultural Inspectors, Agricultural and Food Science Technicians, Aircraft Cargo Handling Supervisors, Aircraft Mechanics and Service Technicians, Airfield Operations Specialists, Airline Pilots, Copilots, and Flight Engineers, Animal Control Workers, Animal Trainers, Athletic Trainers, Audio and Video Equipment Technicians, Audio-Visual and Multimedia Collections Specialists, Avionics Technicians, Biological Technicians, Broadcast Technicians, Captains, Mates, and Pilots of Water Vessels, Cardiovascular Technologists and Technicians, Career/Technical Education Teachers, Secondary School, Chefs and Head Cooks, Chemical Technicians, Chiropractors, Commercial Divers, Commercial Pilots, Computer Numerically Controlled Machine Tool Programmers, Metal and Plastic, Computer Operators, Computer, Automated Teller, and Office Machine Repairers, Conservation Scientists, Construction and Building Inspectors, Correctional Officers and Jailers, Dental Assistants, Dental Hygienists, Dental Laboratory Technicians, Dentists, General, Desktop Publishers, Detectives and Criminal Investigators, Diagnostic Medical Sonographers, Dietetic Technicians, Electrical Power-Line Installers and Repairers, Electrical and Electronics Engineering Technicians, Electrical and Electronics Repairers, Commercial and Industrial Equipment, Electrical and Electronics Repairers, Powerhouse, Substation, and Relay, Electricians, Electro-Mechanical Technicians, Electronic Equipment Installers and Repairers, Motor Vehicles, Electronic Home Entertainment Equipment Installers and Repairers, Elevator Installers and Repairers, Embalmers, Emergency Medical Technicians and Paramedics, Engineering Technicians, Except Drafters, All Other, Environmental Engineering Technicians, Environmental Science and Protection Technicians, Including Health, Explosives Workers, Ordnance Handling Experts, and Blasters, Fabric and Apparel Patternmakers, Farmers, Ranchers, and Other Agricultural Managers, Fire Inspectors and Investigators, Firefighters, First-Line Supervisors of Construction Trades and Extraction Workers, First-Line Supervisors of Correctional Officers, First-Line Supervisors of Farming, Fishing, and Forestry Workers, First-Line Supervisors of Fire Fighting and Prevention Workers, First-Line Supervisors of Landscaping, Lawn Service, and Groundskeeping Workers, First-Line Supervisors of Mechanics, Installers, and Repairers, First-Line Supervisors of Police and Detectives, First-Line Supervisors of Production and Operating Workers, Fish and Game Wardens, Forensic Science Technicians, Forest Fire Inspectors and Prevention Specialists, Forest and Conservation Technicians, Forest and Conservation Workers, Foresters, Gaming Surveillance Officers and Gaming Investigators, Geological and Petroleum Technicians, Hazardous Materials Removal Workers, Heating, Air Conditioning, and Refrigeration Mechanics and Installers, Industrial Production Managers, Licensed Practical and Licensed Vocational Nurses, Life, Physical, and Social Science Technicians, All Other, Manufactured Building and Mobile Home Installers, Mechanical Engineering Technicians, Medical Appliance Technicians, Medical Equipment Preparers, Medical Equipment Repairers, Medical and Clinical Laboratory Technicians, Medical and Clinical Laboratory Technologists, Museum Technicians and Conservators, Nuclear Medicine Technologists, Nuclear Power Reactor Operators, Nuclear Technicians, Occupational Health and Safety Technicians, Occupational Therapy Assistants, Optometrists, Oral and Maxillofacial Surgeons, Orthodontists, Orthotists and Prosthetists, Pest Control Workers, Photographers, Physical Therapist Assistants, Physical Therapists, Physician Assistants, Podiatrists, Police and Sheriff's Patrol Officers, Power Distributors and Dispatchers, Radiation Therapists, Radio, Cellular, and Tower Equipment Installers and Repairers, Railroad Conductors and Yardmasters, Respiratory Therapists, Respiratory Therapy Technicians, Service Unit Operators, Oil, Gas, and Mining, Ship Engineers, Sound Engineering Technicians, Stationary Engineers and Boiler Operators, Surgical Technologists, Surveying and Mapping Technicians, Surveyors, Telecommunications Equipment Installers and Repairers, Except Line Installers, Traffic Technicians, Transit and Railroad Police, Transportation Attendants, Except Flight Attendants, Transportation Inspectors, Veterinarians, Veterinary Assistants and Laboratory Animal Caretakers, Veterinary Technologists and Technicians, Water and Wastewater Treatment Plant and System Operators, Zoologists and Wildlife Biologists\\ \hline 
\scriptsize\cellcolor[gray]{0.8} \textcolor{myRed}{Red (0.98)} &\scriptsize Adhesive Bonding Machine Operators and Tenders, Aircraft Structure, Surfaces, Rigging, and Systems Assemblers, Ambulance Drivers and Attendants, Except Emergency Medical Technicians, Animal Breeders, Automotive Body and Related Repairers, Automotive Glass Installers and Repairers, Automotive Service Technicians and Mechanics, Automotive and Watercraft Service Attendants, Bakers, Bicycle Repairers, Boilermakers, Brickmasons and Blockmasons, Bridge and Lock Tenders, Bus Drivers, School or Special Client, Bus Drivers, Transit and Intercity, Bus and Truck Mechanics and Diesel Engine Specialists, Butchers and Meat Cutters, Cabinetmakers and Bench Carpenters, Camera Operators, Television, Video, and Motion Picture, Camera and Photographic Equipment Repairers, Carpenters, Carpet Installers, Cement Masons and Concrete Finishers, Chemical Equipment Operators and Tenders, Chemical Plant and System Operators, Cleaners of Vehicles and Equipment, Cleaning, Washing, and Metal Pickling Equipment Operators and Tenders, Coating, Painting, and Spraying Machine Setters, Operators, and Tenders, Coil Winders, Tapers, and Finishers, Coin, Vending, and Amusement Machine Servicers and Repairers, Computer-Controlled Machine Tool Operators, Metal and Plastic, Construction Laborers, Continuous Mining Machine Operators, Control and Valve Installers and Repairers, Except Mechanical Door, Conveyor Operators and Tenders, Cooks, Restaurant, Cooling and Freezing Equipment Operators and Tenders, Couriers and Messengers, Craft Artists, Crane and Tower Operators, Crushing, Grinding, and Polishing Machine Setters, Operators, and Tenders, Cutters and Trimmers, Hand, Cutting and Slicing Machine Setters, Operators, and Tenders, Cutting, Punching, and Press Machine Setters, Operators, and Tenders, Metal and Plastic, Derrick Operators, Oil and Gas, Dishwashers, Drilling and Boring Machine Tool Setters, Operators, and Tenders, Metal and Plastic, Drywall and Ceiling Tile Installers, Earth Drillers, Except Oil and Gas, Electric Motor, Power Tool, and Related Repairers, Electrical and Electronic Equipment Assemblers, Electrical and Electronics Installers and Repairers, Transportation Equipment, Electromechanical Equipment Assemblers, Engine and Other Machine Assemblers, Etchers and Engravers, Excavating and Loading Machine and Dragline Operators, Extruding and Drawing Machine Setters, Operators, and Tenders, Metal and Plastic, Extruding and Forming Machine Setters, Operators, and Tenders, Synthetic and Glass Fibers, Extruding, Forming, Pressing, and Compacting Machine Setters, Operators, and Tenders, Fabric Menders, Except Garment, Fallers, Farm Equipment Mechanics and Service Technicians, Fence Erectors, Fiberglass Laminators and Fabricators, Fishers and Related Fishing Workers, Floor Layers, Except Carpet, Wood, and Hard Tiles, Floor Sanders and Finishers, Food Batchmakers, Food Cooking Machine Operators and Tenders, Food and Tobacco Roasting, Baking, and Drying Machine Operators and Tenders, Forging Machine Setters, Operators, and Tenders, Metal and Plastic, Foundry Mold and Coremakers, Furnace, Kiln, Oven, Drier, and Kettle Operators and Tenders, Furniture Finishers, Gas Compressor and Gas Pumping Station Operators, Gas Plant Operators, Glaziers, Grinding and Polishing Workers, Hand, Grinding, Lapping, Polishing, and Buffing Machine Tool Setters, Operators, and Tenders, Metal and Plastic, Heat Treating Equipment Setters, Operators, and Tenders, Metal and Plastic, Heavy and Tractor-Trailer Truck Drivers, Helpers--Brickmasons, Blockmasons, Stonemasons, and Tile and Marble Setters, Helpers--Carpenters, Helpers--Electricians, Helpers--Extraction Workers, Helpers--Installation, Maintenance, and Repair Workers, Helpers--Painters, Paperhangers, Plasterers, and Stucco Masons, Helpers--Pipelayers, Plumbers, Pipefitters, and Steamfitters, Helpers--Production Workers, Helpers--Roofers, Highway Maintenance Workers, Hoist and Winch Operators, Home Appliance Repairers, Industrial Machinery Mechanics, Industrial Truck and Tractor Operators, Insulation Workers, Floor, Ceiling, and Wall, Insulation Workers, Mechanical, Janitors and Cleaners, Except Maids and Housekeeping Cleaners, Jewelers and Precious Stone and Metal Workers, Laborers and Freight, Stock, and Material Movers, Hand, Landscaping and Groundskeeping Workers, Lathe and Turning Machine Tool Setters, Operators, and Tenders, Metal and Plastic, Laundry and Dry-Cleaning Workers, Layout Workers, Metal and Plastic, Light Truck or Delivery Services Drivers, Locksmiths and Safe Repairers, Locomotive Engineers, Log Graders and Scalers, Logging Equipment Operators, Machine Feeders and Offbearers, Machinists, Maintenance Workers, Machinery, Maintenance and Repair Workers, General, Meat, Poultry, and Fish Cutters and Trimmers, Mechanical Door Repairers, Metal-Refining Furnace Operators and Tenders, Meter Readers, Utilities, Milling and Planing Machine Setters, Operators, and Tenders, Metal and Plastic, Millwrights, Mine Cutting and Channeling Machine Operators, Mine Shuttle Car Operators, Mixing and Blending Machine Setters, Operators, and Tenders, Mobile Heavy Equipment Mechanics, Except Engines, Model Makers, Metal and Plastic, Model Makers, Wood, Molders, Shapers, and Casters, Except Metal and Plastic, Molding, Coremaking, and Casting Machine Setters, Operators, and Tenders, Metal and Plastic, Motion Picture Projectionists, Motorboat Mechanics and Service Technicians, Motorboat Operators, Motorcycle Mechanics, Multiple Machine Tool Setters, Operators, and Tenders, Metal and Plastic, Musical Instrument Repairers and Tuners, Office Machine Operators, Except Computer, Operating Engineers and Other Construction Equipment Operators, Ophthalmic Laboratory Technicians, Outdoor Power Equipment and Other Small Engine Mechanics, Packaging and Filling Machine Operators and Tenders, Packers and Packagers, Hand, Painters, Construction and Maintenance, Painters, Transportation Equipment, Painting, Coating, and Decorating Workers, Paper Goods Machine Setters, Operators, and Tenders, Paperhangers, Parking Lot Attendants, Patternmakers, Metal and Plastic, Patternmakers, Wood, Paving, Surfacing, and Tamping Equipment Operators, Pesticide Handlers, Sprayers, and Applicators, Vegetation, Petroleum Pump System Operators, Refinery Operators, and Gaugers, Pile-Driver Operators, Pipelayers, Plant and System Operators, All Other, Plasterers and Stucco Masons, Plating and Coating Machine Setters, Operators, and Tenders, Metal and Plastic, Plumbers, Pipefitters, and Steamfitters, Pourers and Casters, Metal, Power Plant Operators, Pressers, Textile, Garment, and Related Materials, Print Binding and Finishing Workers, Printing Press Operators, Pump Operators, Except Wellhead Pumpers, Rail Car Repairers, Rail Yard Engineers, Dinkey Operators, and Hostlers, Rail-Track Laying and Maintenance Equipment Operators, Railroad Brake, Signal, and Switch Operators, Recreational Vehicle Service Technicians, Refractory Materials Repairers, Except Brickmasons, Refuse and Recyclable Material Collectors, Reinforcing Iron and Rebar Workers, Riggers, Rock Splitters, Quarry, Rolling Machine Setters, Operators, and Tenders, Metal and Plastic, Roof Bolters, Mining, Roofers, Rotary Drill Operators, Oil and Gas, Roustabouts, Oil and Gas, Sailors and Marine Oilers, Sawing Machine Setters, Operators, and Tenders, Wood, Security and Fire Alarm Systems Installers, Semiconductor Processors, Separating, Filtering, Clarifying, Precipitating, and Still Machine Setters, Operators, and Tenders, Septic Tank Servicers and Sewer Pipe Cleaners, Sewing Machine Operators, Sheet Metal Workers, Shoe Machine Operators and Tenders, Shoe and Leather Workers and Repairers, Signal and Track Switch Repairers, Slaughterers and Meat Packers, Stonemasons, Structural Iron and Steel Workers, Structural Metal Fabricators and Fitters, Subway and Streetcar Operators, Tailors, Dressmakers, and Custom Sewers, Tank Car, Truck, and Ship Loaders, Tapers, Taxi Drivers and Chauffeurs, Team Assemblers, Telecommunications Line Installers and Repairers, Terrazzo Workers and Finishers, Textile Bleaching and Dyeing Machine Operators and Tenders, Textile Cutting Machine Setters, Operators, and Tenders, Textile Knitting and Weaving Machine Setters, Operators, and Tenders, Textile Winding, Twisting, and Drawing Out Machine Setters, Operators, and Tenders, Tile and Marble Setters, Timing Device Assemblers and Adjusters, Tire Builders, Tire Repairers and Changers, Tool Grinders, Filers, and Sharpeners, Tool and Die Makers, Tree Trimmers and Pruners, Upholsterers, Watch Repairers, Welders, Cutters, Solderers, and Brazers, Welding, Soldering, and Brazing Machine Setters, Operators, and Tenders, Wellhead Pumpers, Woodworking Machine Setters, Operators, and Tenders, Except Sawing\\ \hline 
\scriptsize\cellcolor[gray]{0.8} \textcolor{myBlue}{Blue (0.94)} &\scriptsize Actors, Amusement and Recreation Attendants, Athletes and Sports Competitors, Baggage Porters and Bellhops, Bailiffs, Barbers, Bartenders, Bill and Account Collectors, Billing and Posting Clerks, Bookkeeping, Accounting, and Auditing Clerks, Brokerage Clerks, Cashiers, Childcare Workers, Combined Food Preparation and Serving Workers, Including Fast Food, Cooks, Fast Food, Cooks, Institution and Cafeteria, Cooks, Short Order, Correspondence Clerks, Costume Attendants, Counter Attendants, Cafeteria, Food Concession, and Coffee Shop, Counter and Rental Clerks, Court Reporters, Court, Municipal, and License Clerks, Crossing Guards, Dancers, Data Entry Keyers, Demonstrators and Product Promoters, Dining Room and Cafeteria Attendants and Bartender Helpers, Door-to-Door Sales Workers, News and Street Vendors, and Related Workers, Driver/Sales Workers, Farm Labor Contractors, File Clerks, Fine Artists, Including Painters, Sculptors, and Illustrators, First-Line Supervisors of Food Preparation and Serving Workers, First-Line Supervisors of Housekeeping and Janitorial Workers, First-Line Supervisors of Retail Sales Workers, Fitness Trainers and Aerobics Instructors, Flight Attendants, Floral Designers, Food Preparation Workers, Food Servers, Nonrestaurant, Funeral Attendants, Gaming Cage Workers, Gaming Change Persons and Booth Cashiers, Gaming Dealers, Gaming and Sports Book Writers and Runners, Graders and Sorters, Agricultural Products, Hairdressers, Hairstylists, and Cosmetologists, Home Health Aides, Hosts and Hostesses, Restaurant, Lounge, and Coffee Shop, Hotel, Motel, and Resort Desk Clerks, Inspectors, Testers, Sorters, Samplers, and Weighers, Insurance Appraisers, Auto Damage, Insurance Claims and Policy Processing Clerks, Interviewers, Except Eligibility and Loan, Legal Secretaries, Library Assistants, Clerical, Library Technicians, Lifeguards, Ski Patrol, and Other Recreational Protective Service Workers, Locker Room, Coatroom, and Dressing Room Attendants, Maids and Housekeeping Cleaners, Mail Clerks and Mail Machine Operators, Except Postal Service, Makeup Artists, Theatrical and Performance, Manicurists and Pedicurists, Massage Therapists, Medical Records and Health Information Technicians, Medical Secretaries, Medical Transcriptionists, Merchandise Displayers and Window Trimmers, Models, Musicians and Singers, Nonfarm Animal Caretakers, Occupational Therapy Aides, Office Clerks, General, Order Clerks, Parking Enforcement Workers, Parts Salespersons, Payroll and Timekeeping Clerks, Personal Care Aides, Pharmacy Aides, Pharmacy Technicians, Photographic Process Workers and Processing Machine Operators, Physical Therapist Aides, Postal Service Clerks, Postal Service Mail Carriers, Postal Service Mail Sorters, Processors, and Processing Machine Operators, Prepress Technicians and Workers, Proofreaders and Copy Markers, Psychiatric Aides, Public Address System and Other Announcers, Radio Operators, Receptionists and Information Clerks, Reservation and Transportation Ticket Agents and Travel Clerks, Retail Salespersons, Secretaries and Administrative Assistants, Except Legal, Medical, and Executive, Security Guards, Self-Enrichment Education Teachers, Sewers, Hand, Shampooers, Shipping, Receiving, and Traffic Clerks, Skincare Specialists, Slot Supervisors, Stock Clerks and Order Fillers, Switchboard Operators, Including Answering Service, Teacher Assistants, Telemarketers, Telephone Operators, Tellers, Title Examiners, Abstractors, and Searchers, Tour Guides and Escorts, Umpires, Referees, and Other Sports Officials, Ushers, Lobby Attendants, and Ticket Takers, Waiters and Waitresses, Weighers, Measurers, Checkers, and Samplers, Recordkeeping, Word Processors and Typists\\ \hline 

\end{longtabu}
\subsubsection{Alternative Job Groups using K-means}
\label{altJobGroups}
\indent We demonstrate that our choice to focus on five groups of jobs according to skills produces results that are consistent for several alternative numbers of groups.
Using K-means to identify between three and seven job groups continues demonstrate that computational/analytical and managerial skill are more indicative of super linear job growth, while physical skills are more indicative of linear or sub linear job growth.
Likewise, our conclusions relating job scaling to expected job impact by comparing skills hold as well.
\\
\\
\newcounter{numG}
\forloop{numG}{3}{\value{numG} < 8}{
	\begin{center}	
		K-means clustering of similar jobs ($k=\arabic{numG}$)\\
		\includegraphics[width=.45\textwidth]{figures/scalingJobClustersSM_\arabic{numG}.pdf}
		\includegraphics[width=.45\textwidth]{figures/jobClusterAggSkill_\arabic{numG}.pdf}\\
	\end{center}
    }

%\subsection{Alternative Job Grouping (K means with 3)}
%\label{altJobGroups3}
%\begin{center}
%	\includegraphics[width=.45\textwidth]{journalFigures/scalingJobClusters_3.pdf}
%	\includegraphics[width=.45\textwidth]{journalFigures/jobClusterAggSkill_3.pdf}
%\end{center}
%\begin{longtabu} to \textwidth {|c|X|}
%	\hline
%	\cellcolor[gray]{0.8}\bf Group ($\beta$)& \bf BLS Jobs \\ \hline
%	\input{jobGroups_3.tex}
%\end{longtabu}
%\subsection{Alternative Job Grouping (K means with 7)}
%\label{altJobGroups7}
%\begin{center}
%	\includegraphics[width=.45\textwidth]{journalFigures/scalingJobClusters_7.pdf}
%	\includegraphics[width=.45\textwidth]{journalFigures/jobClusterAggSkill_7.pdf}
%\end{center}
%\begin{longtabu} to \textwidth {|c|X|}
%	\hline
%	\cellcolor[gray]{0.8}\bf Group ($\beta$)& \bf BLS Jobs \\ \hline
%	\input{jobGroups_7.tex}
%\end{longtabu}
\subsubsection{Stability Testing for Job Groups}
We want to test the stability of the scaling results we observe when using five job clusters obtained from k-means clustering.
In particular, how robust to sub-sampling is our observation that one job clusters scales faster than the rest?
For a single trial, we sub-sample from the complete list of BLS occupations (percent indicated in plot titles) to obtain a matrix where each row represents a single occupation which was sub-sampled and each column represents the raw O*NET importance of a skill to each occupation. We apply k-means to this occupation-skill matrix (i.e. occupations are instances and skills are features) to obtain five occupation clusters (note: examination of prescribing between three and seven clusters is discussed in the SM). We then measure the scaling exponent ($\beta$) of each occupation cluster and rank the occupation clusters according to scaling exponent (rank indicated by color in plots). We perform 100 independent trials for each sub-sampling proportion in $\{10\%, 20\%, \dots, 90\%, 100\%\}$ and plot the resulting scaling exponent distributions. In good agreement with our original findings, we find that indeed one occupation cluster (indicated in purple) tends to grow much faster than the other occupation clusters despite varying sub-sampling of occupations.
\begin{figure}[!b]
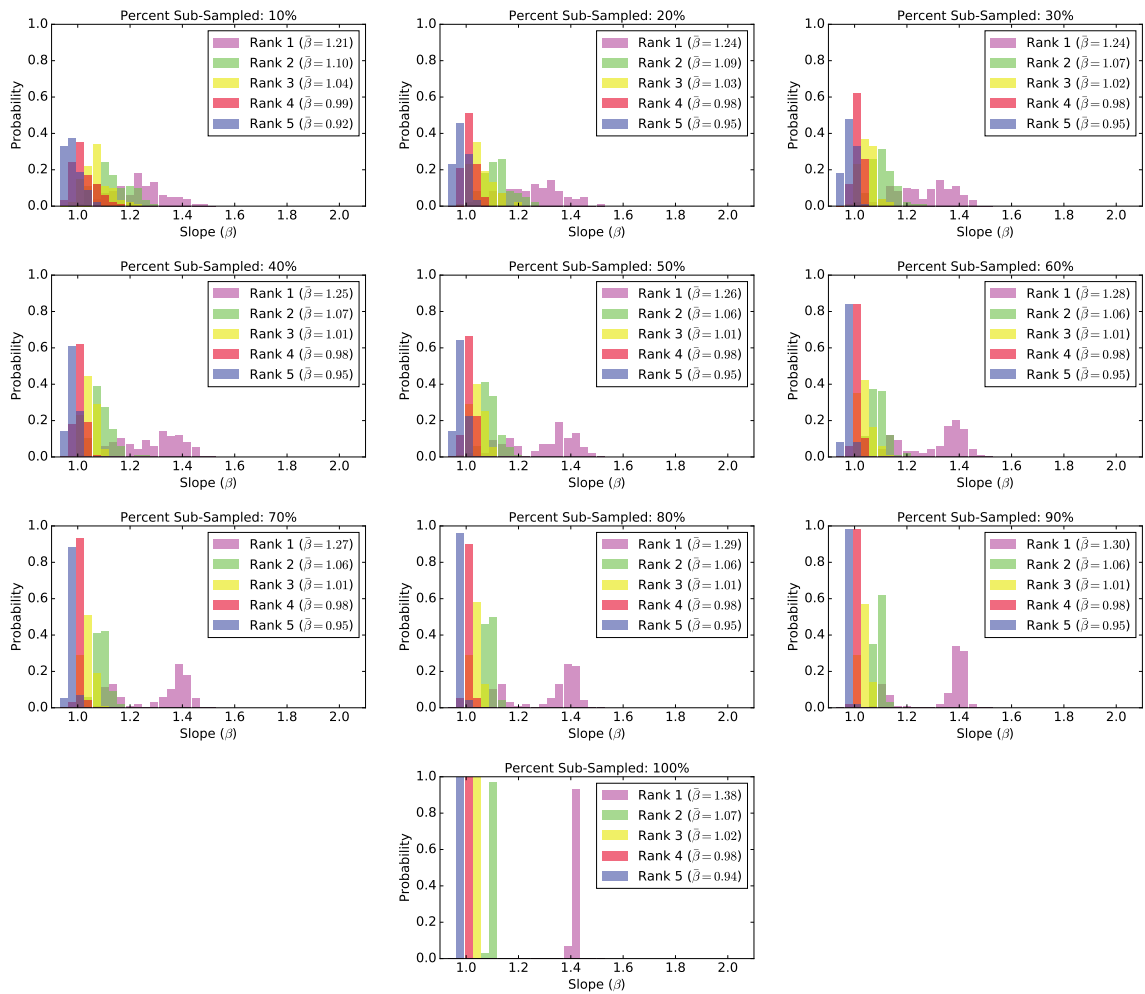

    \centering
    \forloop{numG}{0}{\value{numG} < 9}{
		\includegraphics[width=.3\textwidth]{figures/bootstrapJobScaling_\arabic{numG}.pdf}
		\addtocounter{numG}{1}
		\includegraphics[width=.3\textwidth]{figures/bootstrapJobScaling_\arabic{numG}.pdf}
		\addtocounter{numG}{1}
		\includegraphics[width=.3\textwidth]{figures/bootstrapJobScaling_\arabic{numG}.pdf}
	}
	\includegraphics[width=.3\textwidth]{figures/bootstrapJobScaling_\arabic{numG}.pdf}	
    \caption{
        Boot-strapping at various rates of sub-sampling demonstrates the stability of our result that one job cluster scales at a greater rate than the rest when using five clusters obtained from k-means clustering.
    }
    \label{bootstrap}
\end{figure}
\clearpage
\subsubsection{Checking the Statistical Robustness of Job Group Scaling}
Readers who are familiar with the urban scaling literature may be aware of an ongoing debate about the statistical significance of exponent measurements and identification of underlying statistical models to explain that growth.
For example, what model should one assume to test if a trend is significantly superlinear?
Rather than solving this ongoing and important problem, the goal of this study is only to understand the relationship between automation and urbanization.
Our narrative requires only that highly specialized occupations (represented by purple dots in Figure 3A of the main text) exhibit superlinear growth and be notably different from the growth exhibited by other occupations.

\begin{figure}[h]
    \centering
    \includegraphics[width=.32\textwidth]{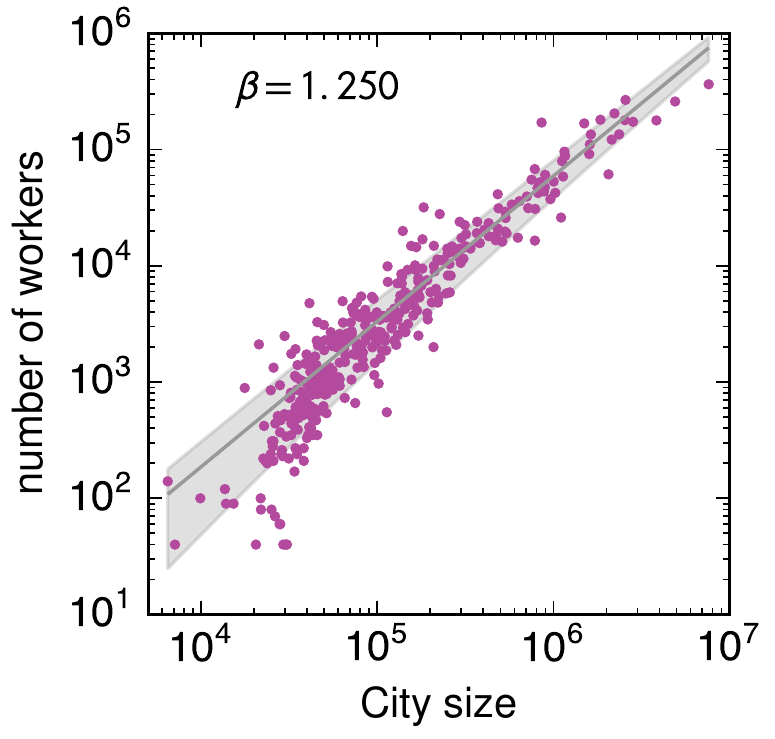}
    \includegraphics[width=.32\textwidth]{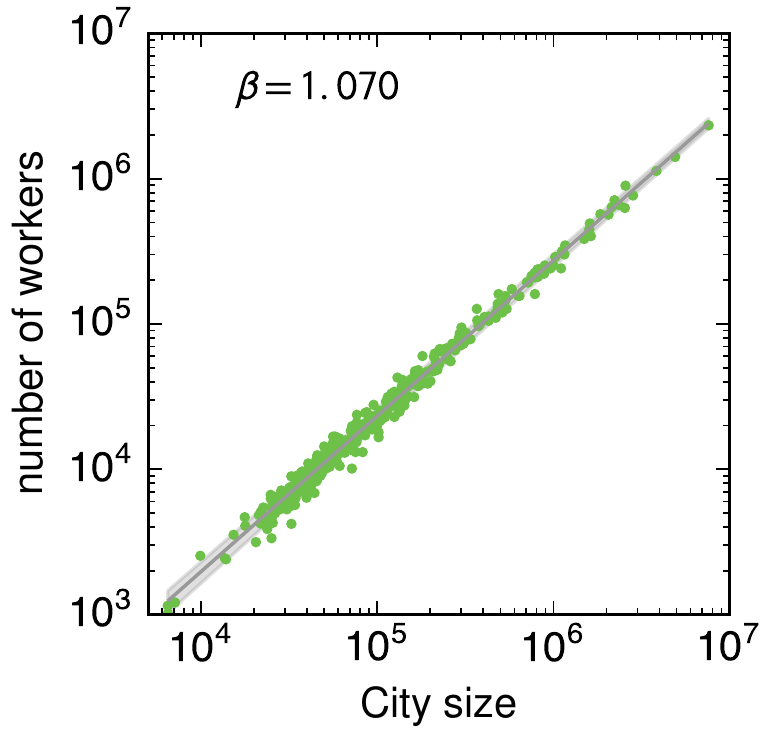}
    \includegraphics[width=.32\textwidth]{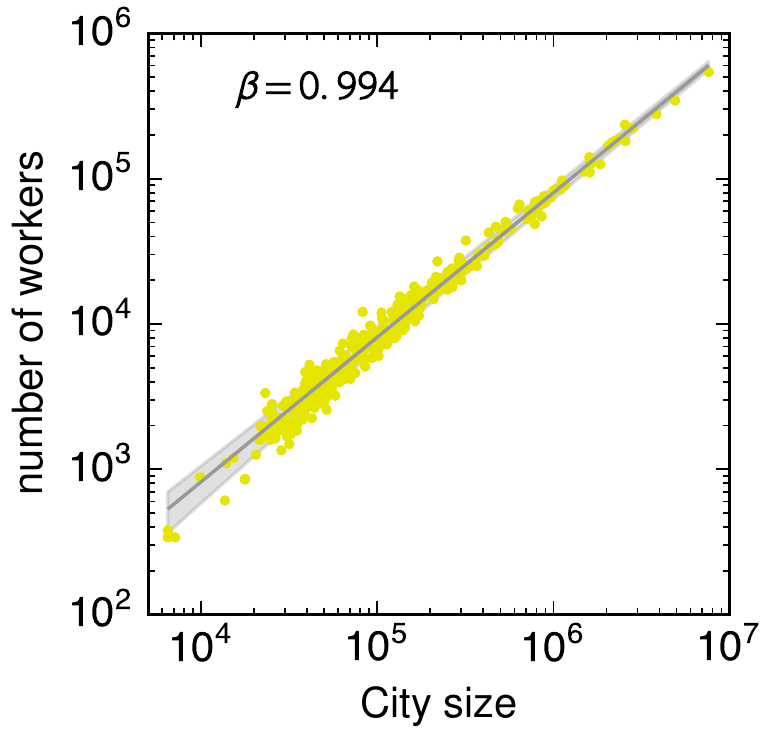}
    \includegraphics[width=.32\textwidth]{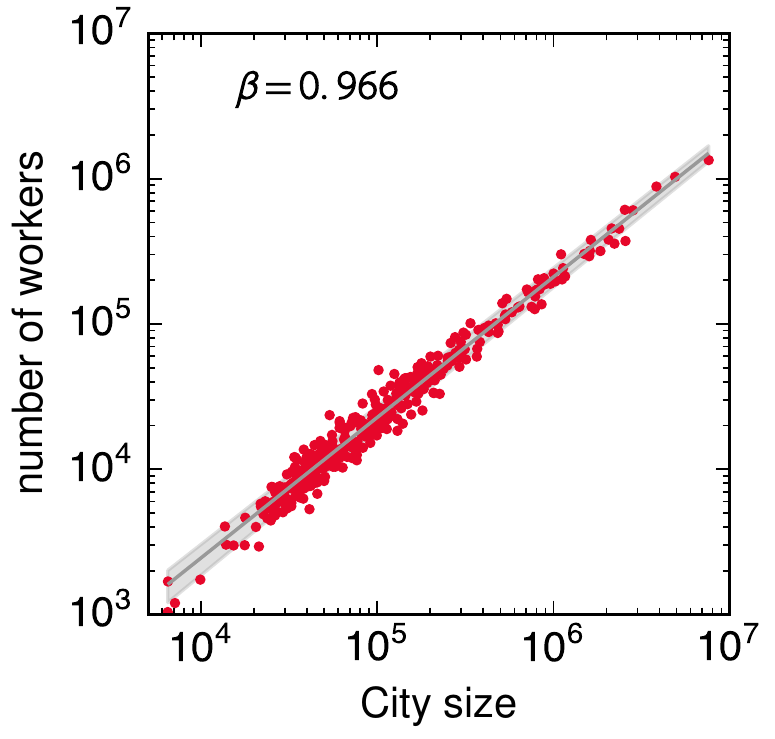}
    \includegraphics[width=.32\textwidth]{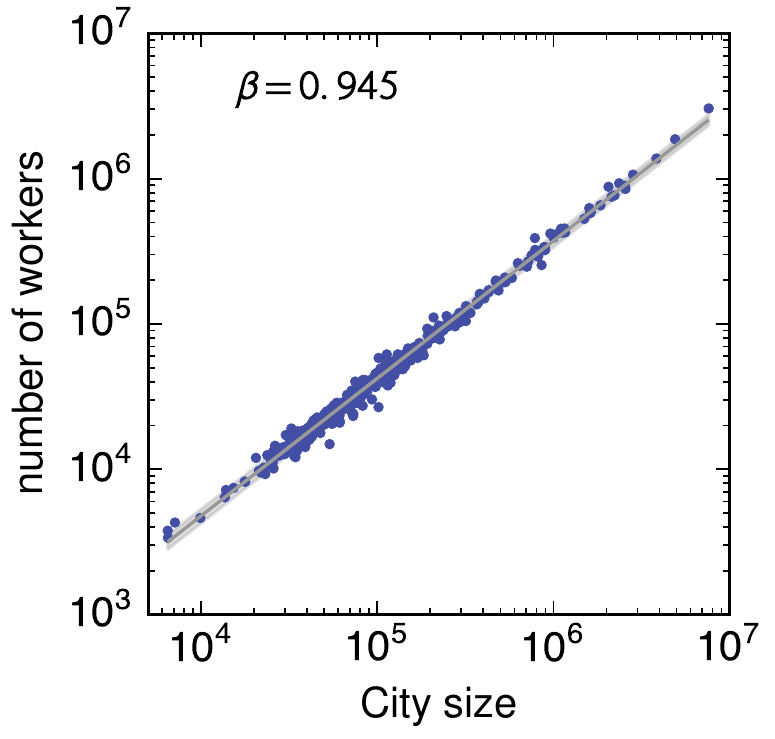}    
    \caption{
        Following the work of Leitao et al., we provide slope estimates along with standard errors for the scaling of each occupation cluster according to unconstrained logarithm model.
    }
    \label{leitaoPlots}
\end{figure}

Recent work by Leitao et al.~\cite{leitao2016scaling} proposes several statistical models that may explain urban scaling trends, and they apply them to a variety of datasets to test the models' ability to explain urban scaling. 
Here, we employ these same models to test if our requirements on the scaling of highly specialized occupations are met according to the five job groups discussed in the main text (i.e. K-means clustering with $k=5$).
As an example, Figure~\ref{leitaoPlots} provides estimates of the scaling exponent along with standard errors for the scaling of each job group according to the unconstrained logarithm model.
Table~\ref{leitao} details the complete analysis in line with the methods in \cite{leitao2016scaling}.
For each model tested, we find that the purple job group, which represents highly specialized occupations, exhibits significantly superlinear scaling, and, furthermore, consistently exhibits faster growth rates than other job groups. 

\begin{table}[h]
    \centering
    \begin{tabular}{c|cc|cc||c}
        &\multicolumn{4}{c||}{city model}&\\ \hline
        & \multicolumn{2}{c|}{lognormal} & \multicolumn{2}{c||}{Gaussian}& person model \\ 
        data& $\delta=2$ & $\delta\in[1,3]$ & $\delta=1$ & $\delta\in[1,2]$& \\ \hline
        purple&1.380 (0.226)$\nearrow$&1.250 (0.079)$\nearrow$&1.289 (0.234)$\nearrow$&1.284 (0.065)$\nearrow$&1.130 (0.102)$\nearrow$ \\ 
        green&1.075 (0.097)$\nearrow$&1.070 (0.044)$\nearrow$&1.065 (0.013)$\nearrow$&1.071 (0.009)$\nearrow^*$&1.061 (0.012)$\nearrow$ \\
        yellow&1.021 (0.060)$\circ^*$&0.994 (0.040)$\rightarrow$&0.980 (0.018)$\searrow$&0.996 (0.014)$\rightarrow$&0.976 (0.017)$\searrow$ \\
        red&0.976 (0.038)$\circ^*$&0.966 (0.013)$\searrow^*$&0.963 (0.022)$\searrow$&0.966 (0.015)$\searrow$&0.965 (0.020)$\searrow$ \\ 
        blue&0.943 (0.009)$\searrow$&0.945 (0.010)$\searrow^*$&0.974 (0.019)$\searrow$&0.945 (0.009)$\searrow$&0.971 (0.018)$\searrow$ \\ \hline
    \end{tabular}
    \caption{
        An analysis of scaling exponents following the work of Leitao et al.
        The entries in the table represent the scaling exponent $\beta$.
        The value obtained through least-squares fitting in log-scale coincides with the value reported in the first column. 
        The error bars were computed with bootstrap. 
        The asterisk indicates that the model has a p-value higher than 0.05.
        If the difference $\Delta BIC$ between the $BIC$ of each model with the same model with a fixed $\beta=1$ is below 0, the model is linear ($\rightarrow$), between zero and six is inconclusive ($\circ$) and higher than six (strong evidence) is super-linear ($\nearrow$)/sublinear ($\searrow$).         
    }
    \label{leitao}
\end{table}

\clearpage
\subsection{Relating City Trends to \onet Skill}
	\label{skillsRelate}
\begin{longtabu} to \textwidth{|X|c|c|c|}
	\hline
	\bf Skill & \bf Corr. to Job Impact & \bf Corr to $H_{skill}(m)$ &\bf Corr. to Log$_{10}$ City Size \\ \hline
\small Thinking Creatively &\small  -0.71 ($<10^{-56}$) &\small  -0.24 ($<10^{-4}$) &\small  0.75 ($<10^{-67}$) \\ \hline 
\small Category Flexibility &\small  -0.70 ($<10^{-56}$) &\small  -0.22 ($<10^{-3}$) &\small  0.63 ($<10^{-41}$) \\ \hline 
\small Inductive Reasoning &\small  -0.70 ($<10^{-56}$) &\small  -0.27 ($<10^{-5}$) &\small  0.62 ($<10^{-40}$) \\ \hline 
\small Deductive Reasoning &\small  -0.70 ($<10^{-55}$) &\small  -0.27 ($<10^{-6}$) &\small  0.57 ($<10^{-32}$) \\ \hline 
\small Active Learning &\small  -0.70 ($<10^{-55}$) &\small  -0.32 ($<10^{-9}$) &\small  0.52 ($<10^{-26}$) \\ \hline 
\small Problem Sensitivity &\small  -0.70 ($<10^{-54}$) &\small  -0.22 ($<10^{-3}$) &\small  0.63 ($<10^{-41}$) \\ \hline 
\small Originality &\small  -0.70 ($<10^{-54}$) &\small  -0.32 ($<10^{-9}$) &\small  0.57 ($<10^{-32}$) \\ \hline 
\small Information Ordering &\small  -0.70 ($<10^{-54}$) &\small  -0.20 ($<10^{-3}$) &\small  0.65 ($<10^{-46}$) \\ \hline 
\small Critical Thinking &\small  -0.69 ($<10^{-54}$) &\small  -0.29 ($<10^{-6}$) &\small  0.61 ($<10^{-38}$) \\ \hline 
\small Complex Problem Solving &\small  -0.69 ($<10^{-53}$) &\small  -0.18 ($<10^{-2}$) &\small  0.69 ($<10^{-52}$) \\ \hline 
\small Interpreting the Meaning of Information for Others &\small  -0.69 ($<10^{-53}$) &\small  -0.27 ($<10^{-6}$) &\small  0.65 ($<10^{-44}$) \\ \hline 
\small Writing &\small  -0.68 ($<10^{-52}$) &\small  -0.40 ($<10^{-14}$) &\small  0.39 ($<10^{-13}$) \\ \hline 
\small Written Comprehension &\small  -0.68 ($<10^{-51}$) &\small  -0.38 ($<10^{-12}$) &\small  0.41 ($<10^{-15}$) \\ \hline 
\small Artistic &\small  -0.68 ($<10^{-51}$) &\small  -0.32 ($<10^{-9}$) &\small  0.79 ($<10^{-81}$) \\ \hline 
\small Reading Comprehension &\small  -0.68 ($<10^{-50}$) &\small  -0.37 ($<10^{-11}$) &\small  0.43 ($<10^{-16}$) \\ \hline 
\small Judgment and Decision Making &\small  -0.68 ($<10^{-50}$) &\small  -0.26 ($<10^{-5}$) &\small  0.52 ($<10^{-25}$) \\ \hline 
\small Science &\small  -0.68 ($<10^{-50}$) &\small  -0.16 ($<10^{-1}$) &\small  0.80 ($<10^{-84}$) \\ \hline 
\small Written Expression &\small  -0.68 ($<10^{-50}$) &\small  -0.40 ($<10^{-14}$) &\small  0.36 ($<10^{-11}$) \\ \hline 
\small Computers and Electronics &\small  -0.67 ($<10^{-49}$) &\small  -0.25 ($<10^{-4}$) &\small  0.57 ($<10^{-33}$) \\ \hline 
\small Fluency of Ideas &\small  -0.67 ($<10^{-49}$) &\small  -0.34 ($<10^{-9}$) &\small  0.47 ($<10^{-20}$) \\ \hline 
\small Analyzing Data or Information &\small  -0.67 ($<10^{-48}$) &\small  -0.21 ($<10^{-3}$) &\small  0.66 ($<10^{-46}$) \\ \hline 
\small Investigative &\small  -0.66 ($<10^{-47}$) &\small  -0.12 ($<10^{0}$) &\small  0.82 ($<10^{-92}$) \\ \hline 
\small Making Decisions and Solving Problems &\small  -0.66 ($<10^{-47}$) &\small  -0.19 ($<10^{-2}$) &\small  0.72 ($<10^{-59}$) \\ \hline 
\small Communications and Media &\small  -0.66 ($<10^{-46}$) &\small  -0.47 ($<10^{-20}$) &\small  0.50 ($<10^{-23}$) \\ \hline 
\small History and Archeology &\small  -0.66 ($<10^{-46}$) &\small  -0.34 ($<10^{-10}$) &\small  0.61 ($<10^{-37}$) \\ \hline 
\small Updating and Using Relevant Knowledge &\small  -0.65 ($<10^{-45}$) &\small  -0.25 ($<10^{-4}$) &\small  0.66 ($<10^{-46}$) \\ \hline 
\small Processing Information &\small  -0.64 ($<10^{-43}$) &\small  -0.23 ($<10^{-4}$) &\small  0.56 ($<10^{-31}$) \\ \hline 
\small Developing Objectives and Strategies &\small  -0.63 ($<10^{-41}$) &\small  -0.20 ($<10^{-3}$) &\small  0.46 ($<10^{-19}$) \\ \hline 
\small Programming &\small  -0.63 ($<10^{-41}$) &\small  -0.14 ($<10^{-1}$) &\small  0.63 ($<10^{-41}$) \\ \hline 
\small Getting Information &\small  -0.63 ($<10^{-40}$) &\small  -0.32 ($<10^{-8}$) &\small  0.57 ($<10^{-31}$) \\ \hline 
\small Spend Time Sitting &\small  -0.63 ($<10^{-40}$) &\small  -0.40 ($<10^{-14}$) &\small  0.51 ($<10^{-24}$) \\ \hline 
\small Operations Analysis &\small  -0.62 ($<10^{-40}$) &\small  -0.14 ($<10^{-1}$) &\small  0.66 ($<10^{-46}$) \\ \hline 
\small Systems Analysis &\small  -0.62 ($<10^{-40}$) &\small  -0.21 ($<10^{-3}$) &\small  0.37 ($<10^{-12}$) \\ \hline 
\small Systems Evaluation &\small  -0.61 ($<10^{-38}$) &\small  -0.19 ($<10^{-2}$) &\small  0.35 ($<10^{-10}$) \\ \hline 
\small Flexibility of Closure &\small  -0.61 ($<10^{-38}$) &\small  -0.07 ($<10^{0}$) &\small  0.83 ($<10^{-97}$) \\ \hline 
\small Fine Arts &\small  -0.61 ($<10^{-37}$) &\small  -0.34 ($<10^{-10}$) &\small  0.79 ($<10^{-81}$) \\ \hline 
\small Near Vision &\small  -0.61 ($<10^{-37}$) &\small  -0.16 ($<10^{-1}$) &\small  0.78 ($<10^{-76}$) \\ \hline 
\small Electronic Mail &\small  -0.61 ($<10^{-37}$) &\small  -0.42 ($<10^{-15}$) &\small  0.25 ($<10^{-5}$) \\ \hline 
\small Documenting/Recording Information &\small  -0.60 ($<10^{-37}$) &\small  -0.21 ($<10^{-3}$) &\small  0.47 ($<10^{-20}$) \\ \hline 
\small Identifying Objects, Actions, and Events &\small  -0.60 ($<10^{-37}$) &\small  -0.24 ($<10^{-4}$) &\small  0.67 ($<10^{-49}$) \\ \hline 
\small Geography &\small  -0.60 ($<10^{-36}$) &\small  -0.26 ($<10^{-5}$) &\small  0.61 ($<10^{-37}$) \\ \hline 
\small Technology Design &\small  -0.60 ($<10^{-36}$) &\small  0.00 ($<10^{0}$) &\small  0.72 ($<10^{-60}$) \\ \hline 
\small Biology &\small  -0.60 ($<10^{-36}$) &\small  -0.32 ($<10^{-8}$) &\small  0.63 ($<10^{-41}$) \\ \hline 
\small Freedom to Make Decisions &\small  -0.60 ($<10^{-36}$) &\small  -0.21 ($<10^{-3}$) &\small  0.75 ($<10^{-68}$) \\ \hline 
\small Speed of Closure &\small  -0.59 ($<10^{-35}$) &\small  -0.13 ($<10^{0}$) &\small  0.54 ($<10^{-28}$) \\ \hline 
\small Scheduling Work and Activities &\small  -0.58 ($<10^{-34}$) &\small  -0.23 ($<10^{-4}$) &\small  0.31 ($<10^{-8}$) \\ \hline 
\small Selective Attention &\small  -0.58 ($<10^{-33}$) &\small  0.03 ($<10^{0}$) &\small  0.84 ($<10^{-101}$) \\ \hline 
\small Education and Training &\small  -0.58 ($<10^{-33}$) &\small  -0.08 ($<10^{0}$) &\small  0.34 ($<10^{-10}$) \\ \hline 
\small Estimating the Quantifiable Characteristics of Products, Events, or Information &\small  -0.58 ($<10^{-33}$) &\small  -0.02 ($<10^{0}$) &\small  0.81 ($<10^{-86}$) \\ \hline 
\small Interacting With Computers &\small  -0.57 ($<10^{-32}$) &\small  -0.30 ($<10^{-7}$) &\small  0.34 ($<10^{-9}$) \\ \hline 
\small Mathematical Reasoning &\small  -0.56 ($<10^{-31}$) &\small  -0.19 ($<10^{-2}$) &\small  0.29 ($<10^{-7}$) \\ \hline 
\small Mathematics &\small  -0.56 ($<10^{-31}$) &\small  -0.04 ($<10^{0}$) &\small  0.52 ($<10^{-25}$) \\ \hline 
\small Judging the Qualities of Things, Services, or People &\small  -0.56 ($<10^{-30}$) &\small  -0.11 ($<10^{0}$) &\small  0.58 ($<10^{-33}$) \\ \hline 
\small English Language &\small  -0.56 ($<10^{-30}$) &\small  -0.43 ($<10^{-16}$) &\small  0.17 ($<10^{-2}$) \\ \hline 
\small Provide Consultation and Advice to Others &\small  -0.55 ($<10^{-29}$) &\small  -0.21 ($<10^{-3}$) &\small  0.27 ($<10^{-6}$) \\ \hline 
\small Monitoring &\small  -0.55 ($<10^{-29}$) &\small  -0.07 ($<10^{0}$) &\small  0.26 ($<10^{-5}$) \\ \hline 
\small Physics &\small  -0.53 ($<10^{-27}$) &\small  0.11 ($<10^{0}$) &\small  0.83 ($<10^{-94}$) \\ \hline 
\small Design &\small  -0.53 ($<10^{-26}$) &\small  0.09 ($<10^{0}$) &\small  0.83 ($<10^{-96}$) \\ \hline 
\small Structured versus Unstructured Work &\small  -0.53 ($<10^{-26}$) &\small  -0.34 ($<10^{-9}$) &\small  0.33 ($<10^{-9}$) \\ \hline 
\small Engineering and Technology &\small  -0.52 ($<10^{-26}$) &\small  0.13 ($<10^{0}$) &\small  0.82 ($<10^{-90}$) \\ \hline 
\small Visualization &\small  -0.52 ($<10^{-25}$) &\small  0.06 ($<10^{0}$) &\small  0.87 ($<10^{-114}$) \\ \hline 
\small Oral Comprehension &\small  -0.52 ($<10^{-25}$) &\small  -0.48 ($<10^{-21}$) &\small  0.07 ($<10^{0}$) \\ \hline 
\small Memorization &\small  -0.51 ($<10^{-24}$) &\small  -0.27 ($<10^{-6}$) &\small  0.08 ($<10^{0}$) \\ \hline 
\small Duration of Typical Work Week &\small  -0.50 ($<10^{-23}$) &\small  0.17 ($<10^{-2}$) &\small  0.67 ($<10^{-49}$) \\ \hline 
\small Level of Competition &\small  -0.50 ($<10^{-23}$) &\small  -0.16 ($<10^{-1}$) &\small  0.79 ($<10^{-80}$) \\ \hline 
\small Oral Expression &\small  -0.49 ($<10^{-22}$) &\small  -0.48 ($<10^{-21}$) &\small  0.03 ($<10^{0}$) \\ \hline 
\small Organizing, Planning, and Prioritizing Work &\small  -0.49 ($<10^{-22}$) &\small  -0.33 ($<10^{-9}$) &\small  0.09 ($<10^{0}$) \\ \hline 
\small Number Facility &\small  -0.48 ($<10^{-21}$) &\small  -0.17 ($<10^{-2}$) &\small  0.21 ($<10^{-3}$) \\ \hline 
\small Active Listening &\small  -0.48 ($<10^{-21}$) &\small  -0.49 ($<10^{-22}$) &\small  0.04 ($<10^{0}$) \\ \hline 
\small Instructing &\small  -0.48 ($<10^{-21}$) &\small  -0.22 ($<10^{-3}$) &\small  -0.03 ($<10^{0}$) \\ \hline 
\small Telecommunications &\small  -0.47 ($<10^{-20}$) &\small  -0.22 ($<10^{-3}$) &\small  0.42 ($<10^{-15}$) \\ \hline 
\small Indoors, Environmentally Controlled &\small  -0.47 ($<10^{-20}$) &\small  -0.54 ($<10^{-28}$) &\small  0.21 ($<10^{-3}$) \\ \hline 
\small Learning Strategies &\small  -0.47 ($<10^{-20}$) &\small  -0.26 ($<10^{-5}$) &\small  -0.06 ($<10^{0}$) \\ \hline 
\small Chemistry &\small  -0.47 ($<10^{-20}$) &\small  0.13 ($<10^{-1}$) &\small  0.78 ($<10^{-75}$) \\ \hline 
\small Far Vision &\small  -0.46 ($<10^{-19}$) &\small  0.08 ($<10^{0}$) &\small  0.71 ($<10^{-57}$) \\ \hline 
\small Monitor Processes, Materials, or Surroundings &\small  -0.46 ($<10^{-19}$) &\small  0.14 ($<10^{-1}$) &\small  0.65 ($<10^{-45}$) \\ \hline 
\small Drafting, Laying Out, and Specifying Technical Devices, Parts, and Equipment &\small  -0.45 ($<10^{-18}$) &\small  0.18 ($<10^{-2}$) &\small  0.80 ($<10^{-84}$) \\ \hline 
\small Perceptual Speed &\small  -0.45 ($<10^{-18}$) &\small  0.16 ($<10^{-1}$) &\small  0.77 ($<10^{-74}$) \\ \hline 
\small Evaluating Information to Determine Compliance with Standards &\small  -0.44 ($<10^{-17}$) &\small  -0.00 ($<10^{0}$) &\small  0.30 ($<10^{-7}$) \\ \hline 
\small Public Speaking &\small  -0.44 ($<10^{-17}$) &\small  -0.31 ($<10^{-8}$) &\small  -0.03 ($<10^{0}$) \\ \hline 
\small Speaking &\small  -0.43 ($<10^{-16}$) &\small  -0.47 ($<10^{-21}$) &\small  -0.07 ($<10^{0}$) \\ \hline 
\small Importance of Being Exact or Accurate &\small  -0.42 ($<10^{-15}$) &\small  -0.07 ($<10^{0}$) &\small  0.80 ($<10^{-84}$) \\ \hline 
\small Visual Color Discrimination &\small  -0.41 ($<10^{-14}$) &\small  0.09 ($<10^{0}$) &\small  0.84 ($<10^{-101}$) \\ \hline 
\small Communicating with Supervisors, Peers, or Subordinates &\small  -0.40 ($<10^{-14}$) &\small  -0.16 ($<10^{-1}$) &\small  -0.01 ($<10^{0}$) \\ \hline 
\small Face-to-Face Discussions &\small  -0.38 ($<10^{-12}$) &\small  -0.11 ($<10^{0}$) &\small  0.34 ($<10^{-9}$) \\ \hline 
\small Letters and Memos &\small  -0.38 ($<10^{-12}$) &\small  -0.37 ($<10^{-12}$) &\small  -0.12 ($<10^{0}$) \\ \hline 
\small Impact of Decisions on Co-workers or Company Results &\small  -0.37 ($<10^{-12}$) &\small  -0.14 ($<10^{-1}$) &\small  0.48 ($<10^{-21}$) \\ \hline 
\small Time Management &\small  -0.37 ($<10^{-11}$) &\small  -0.27 ($<10^{-5}$) &\small  -0.18 ($<10^{-2}$) \\ \hline 
\small Training and Teaching Others &\small  -0.36 ($<10^{-11}$) &\small  -0.04 ($<10^{0}$) &\small  0.02 ($<10^{0}$) \\ \hline 
\small Quality Control Analysis &\small  -0.36 ($<10^{-11}$) &\small  0.25 ($<10^{-5}$) &\small  0.80 ($<10^{-84}$) \\ \hline 
\small Consequence of Error &\small  -0.35 ($<10^{-10}$) &\small  0.16 ($<10^{-1}$) &\small  0.75 ($<10^{-69}$) \\ \hline 
\small Communicating with Persons Outside Organization &\small  -0.35 ($<10^{-10}$) &\small  -0.50 ($<10^{-23}$) &\small  -0.03 ($<10^{0}$) \\ \hline 
\small Repairing and Maintaining Electronic Equipment &\small  -0.35 ($<10^{-10}$) &\small  0.19 ($<10^{-2}$) &\small  0.81 ($<10^{-88}$) \\ \hline 
\small Speech Clarity &\small  -0.35 ($<10^{-10}$) &\small  -0.47 ($<10^{-20}$) &\small  -0.18 ($<10^{-2}$) \\ \hline 
\small Philosophy and Theology &\small  -0.34 ($<10^{-9}$) &\small  -0.35 ($<10^{-10}$) &\small  -0.06 ($<10^{0}$) \\ \hline 
\small Monitoring and Controlling Resources &\small  -0.33 ($<10^{-9}$) &\small  -0.28 ($<10^{-6}$) &\small  0.01 ($<10^{0}$) \\ \hline 
\small Management of Personnel Resources &\small  -0.33 ($<10^{-9}$) &\small  -0.18 ($<10^{-2}$) &\small  -0.19 ($<10^{-2}$) \\ \hline 
\small Sociology and Anthropology &\small  -0.33 ($<10^{-9}$) &\small  -0.34 ($<10^{-10}$) &\small  -0.13 ($<10^{0}$) \\ \hline 
\small Equipment Selection &\small  -0.28 ($<10^{-6}$) &\small  0.30 ($<10^{-7}$) &\small  0.79 ($<10^{-80}$) \\ \hline 
\small Law and Government &\small  -0.27 ($<10^{-6}$) &\small  -0.27 ($<10^{-5}$) &\small  -0.08 ($<10^{0}$) \\ \hline 
\small Exposed to Radiation &\small  -0.27 ($<10^{-6}$) &\small  -0.09 ($<10^{0}$) &\small  0.41 ($<10^{-14}$) \\ \hline 
\small Troubleshooting &\small  -0.27 ($<10^{-6}$) &\small  0.32 ($<10^{-8}$) &\small  0.76 ($<10^{-72}$) \\ \hline 
\small Time Pressure &\small  -0.27 ($<10^{-6}$) &\small  0.07 ($<10^{0}$) &\small  0.60 ($<10^{-36}$) \\ \hline 
\small Work Schedules &\small  -0.26 ($<10^{-5}$) &\small  -0.01 ($<10^{0}$) &\small  0.71 ($<10^{-57}$) \\ \hline 
\small Operation Monitoring &\small  -0.25 ($<10^{-5}$) &\small  0.33 ($<10^{-9}$) &\small  0.75 ($<10^{-67}$) \\ \hline 
\small Mechanical &\small  -0.25 ($<10^{-5}$) &\small  0.38 ($<10^{-12}$) &\small  0.73 ($<10^{-62}$) \\ \hline 
\small Telephone &\small  -0.23 ($<10^{-4}$) &\small  -0.44 ($<10^{-17}$) &\small  -0.17 ($<10^{-2}$) \\ \hline 
\small Persuasion &\small  -0.23 ($<10^{-4}$) &\small  -0.32 ($<10^{-8}$) &\small  -0.27 ($<10^{-6}$) \\ \hline 
\small Management of Material Resources &\small  -0.22 ($<10^{-3}$) &\small  -0.07 ($<10^{0}$) &\small  -0.24 ($<10^{-4}$) \\ \hline 
\small Production and Processing &\small  -0.20 ($<10^{-3}$) &\small  0.35 ($<10^{-10}$) &\small  0.60 ($<10^{-36}$) \\ \hline 
\small Building and Construction &\small  -0.20 ($<10^{-3}$) &\small  0.32 ($<10^{-8}$) &\small  0.51 ($<10^{-24}$) \\ \hline 
\small Wear Specialized Protective or Safety Equipment such as Breathing Apparatus, Safety Harness, Full Protection Suits, or Radiation Protection &\small  -0.20 ($<10^{-3}$) &\small  0.32 ($<10^{-8}$) &\small  0.66 ($<10^{-47}$) \\ \hline 
\small Management of Financial Resources &\small  -0.19 ($<10^{-2}$) &\small  -0.18 ($<10^{-2}$) &\small  -0.26 ($<10^{-5}$) \\ \hline 
\small Depth Perception &\small  -0.18 ($<10^{-2}$) &\small  0.34 ($<10^{-10}$) &\small  0.73 ($<10^{-61}$) \\ \hline 
\small Establishing and Maintaining Interpersonal Relationships &\small  -0.18 ($<10^{-2}$) &\small  -0.41 ($<10^{-15}$) &\small  -0.35 ($<10^{-10}$) \\ \hline 
\small Exposed to Hazardous Conditions &\small  -0.17 ($<10^{-2}$) &\small  0.37 ($<10^{-12}$) &\small  0.69 ($<10^{-53}$) \\ \hline 
\small Hearing Sensitivity &\small  -0.17 ($<10^{-2}$) &\small  0.40 ($<10^{-14}$) &\small  0.60 ($<10^{-37}$) \\ \hline 
\small Coaching and Developing Others &\small  -0.16 ($<10^{-1}$) &\small  -0.13 ($<10^{-1}$) &\small  -0.39 ($<10^{-13}$) \\ \hline 
\small Finger Dexterity &\small  -0.16 ($<10^{-1}$) &\small  0.24 ($<10^{-4}$) &\small  0.71 ($<10^{-57}$) \\ \hline 
\small Installation &\small  -0.16 ($<10^{-1}$) &\small  0.37 ($<10^{-11}$) &\small  0.58 ($<10^{-34}$) \\ \hline 
\small Developing and Building Teams &\small  -0.15 ($<10^{-1}$) &\small  -0.11 ($<10^{0}$) &\small  -0.39 ($<10^{-13}$) \\ \hline 
\small Guiding, Directing, and Motivating Subordinates &\small  -0.15 ($<10^{-1}$) &\small  -0.08 ($<10^{0}$) &\small  -0.30 ($<10^{-7}$) \\ \hline 
\small Equipment Maintenance &\small  -0.15 ($<10^{-1}$) &\small  0.38 ($<10^{-12}$) &\small  0.69 ($<10^{-54}$) \\ \hline 
\small Repairing &\small  -0.14 ($<10^{-1}$) &\small  0.40 ($<10^{-14}$) &\small  0.68 ($<10^{-51}$) \\ \hline 
\small Medicine and Dentistry &\small  -0.13 ($<10^{-1}$) &\small  -0.27 ($<10^{-6}$) &\small  -0.11 ($<10^{0}$) \\ \hline 
\small Third Interest High-Point &\small  -0.13 ($<10^{0}$) &\small  -0.07 ($<10^{0}$) &\small  -0.18 ($<10^{-2}$) \\ \hline 
\small Negotiation &\small  -0.12 ($<10^{0}$) &\small  -0.27 ($<10^{-6}$) &\small  -0.42 ($<10^{-15}$) \\ \hline 
\small Inspecting Equipment, Structures, or Material &\small  -0.12 ($<10^{0}$) &\small  0.39 ($<10^{-13}$) &\small  0.64 ($<10^{-43}$) \\ \hline 
\small Operation and Control &\small  -0.12 ($<10^{0}$) &\small  0.36 ($<10^{-11}$) &\small  0.69 ($<10^{-52}$) \\ \hline 
\small Realistic &\small  -0.11 ($<10^{0}$) &\small  0.31 ($<10^{-8}$) &\small  0.70 ($<10^{-55}$) \\ \hline 
\small Coordination &\small  -0.11 ($<10^{0}$) &\small  -0.20 ($<10^{-3}$) &\small  -0.50 ($<10^{-23}$) \\ \hline 
\small Controlling Machines and Processes &\small  -0.10 ($<10^{0}$) &\small  0.38 ($<10^{-12}$) &\small  0.67 ($<10^{-48}$) \\ \hline 
\small Coordinating the Work and Activities of Others &\small  -0.10 ($<10^{0}$) &\small  -0.10 ($<10^{0}$) &\small  -0.45 ($<10^{-18}$) \\ \hline 
\small Social Perceptiveness &\small  -0.10 ($<10^{0}$) &\small  -0.30 ($<10^{-7}$) &\small  -0.46 ($<10^{-19}$) \\ \hline 
\small Auditory Attention &\small  -0.09 ($<10^{0}$) &\small  0.47 ($<10^{-20}$) &\small  0.58 ($<10^{-33}$) \\ \hline 
\small Speech Recognition &\small  -0.08 ($<10^{0}$) &\small  -0.38 ($<10^{-12}$) &\small  -0.50 ($<10^{-23}$) \\ \hline 
\small Repairing and Maintaining Mechanical Equipment &\small  -0.08 ($<10^{0}$) &\small  0.43 ($<10^{-17}$) &\small  0.63 ($<10^{-41}$) \\ \hline 
\small Foreign Language &\small  -0.07 ($<10^{0}$) &\small  -0.21 ($<10^{-3}$) &\small  -0.33 ($<10^{-9}$) \\ \hline 
\small Psychology &\small  -0.07 ($<10^{0}$) &\small  -0.27 ($<10^{-5}$) &\small  -0.42 ($<10^{-16}$) \\ \hline 
\small Degree of Automation &\small  -0.07 ($<10^{0}$) &\small  0.18 ($<10^{-2}$) &\small  0.36 ($<10^{-11}$) \\ \hline 
\small Administration and Management &\small  -0.06 ($<10^{0}$) &\small  -0.11 ($<10^{0}$) &\small  -0.46 ($<10^{-19}$) \\ \hline 
\small Staffing Organizational Units &\small  -0.06 ($<10^{0}$) &\small  -0.20 ($<10^{-3}$) &\small  -0.44 ($<10^{-18}$) \\ \hline 
\small Spend Time Using Your Hands to Handle, Control, or Feel Objects, Tools, or Controls &\small  -0.06 ($<10^{0}$) &\small  0.22 ($<10^{-3}$) &\small  0.65 ($<10^{-45}$) \\ \hline 
\small Sales and Marketing &\small  -0.03 ($<10^{0}$) &\small  -0.31 ($<10^{-8}$) &\small  -0.18 ($<10^{-2}$) \\ \hline 
\small Control Precision &\small  -0.03 ($<10^{0}$) &\small  0.34 ($<10^{-9}$) &\small  0.64 ($<10^{-43}$) \\ \hline 
\small Performing Administrative Activities &\small  -0.03 ($<10^{0}$) &\small  -0.29 ($<10^{-7}$) &\small  -0.51 ($<10^{-25}$) \\ \hline 
\small Exposed to Hazardous Equipment &\small  -0.03 ($<10^{0}$) &\small  0.49 ($<10^{-22}$) &\small  0.56 ($<10^{-30}$) \\ \hline 
\small Exposed to High Places &\small  -0.03 ($<10^{0}$) &\small  0.43 ($<10^{-16}$) &\small  0.47 ($<10^{-20}$) \\ \hline 
\small Rate Control &\small  -0.02 ($<10^{0}$) &\small  0.42 ($<10^{-15}$) &\small  0.61 ($<10^{-37}$) \\ \hline 
\small Indoors, Not Environmentally Controlled &\small  -0.01 ($<10^{0}$) &\small  0.52 ($<10^{-26}$) &\small  0.48 ($<10^{-21}$) \\ \hline 
\small Public Safety and Security &\small  -0.00 ($<10^{0}$) &\small  0.21 ($<10^{-3}$) &\small  -0.03 ($<10^{0}$) \\ \hline 
\small Reaction Time &\small  -0.00 ($<10^{0}$) &\small  0.44 ($<10^{-17}$) &\small  0.57 ($<10^{-32}$) \\ \hline 
\small Selling or Influencing Others &\small  0.00 ($<10^{0}$) &\small  -0.30 ($<10^{-7}$) &\small  -0.34 ($<10^{-9}$) \\ \hline 
\small Pace Determined by Speed of Equipment &\small  0.01 ($<10^{0}$) &\small  0.40 ($<10^{-14}$) &\small  0.53 ($<10^{-27}$) \\ \hline 
\small Clerical &\small  0.02 ($<10^{0}$) &\small  -0.24 ($<10^{-4}$) &\small  -0.56 ($<10^{-31}$) \\ \hline 
\small Required Level of Education &\small  0.02 ($<10^{0}$) &\small  -0.04 ($<10^{0}$) &\small  0.02 ($<10^{0}$) \\ \hline 
\small Glare Sensitivity &\small  0.03 ($<10^{0}$) &\small  0.43 ($<10^{-16}$) &\small  0.50 ($<10^{-23}$) \\ \hline 
\small In an Enclosed Vehicle or Equipment &\small  0.04 ($<10^{0}$) &\small  0.07 ($<10^{0}$) &\small  -0.00 ($<10^{0}$) \\ \hline 
\small Extremely Bright or Inadequate Lighting &\small  0.04 ($<10^{0}$) &\small  0.45 ($<10^{-18}$) &\small  0.46 ($<10^{-19}$) \\ \hline 
\small Therapy and Counseling &\small  0.04 ($<10^{0}$) &\small  -0.13 ($<10^{0}$) &\small  -0.49 ($<10^{-22}$) \\ \hline 
\small Frequency of Decision Making &\small  0.04 ($<10^{0}$) &\small  -0.09 ($<10^{0}$) &\small  -0.06 ($<10^{0}$) \\ \hline 
\small Outdoors, Under Cover &\small  0.05 ($<10^{0}$) &\small  0.25 ($<10^{-5}$) &\small  0.12 ($<10^{0}$) \\ \hline 
\small Personnel and Human Resources &\small  0.06 ($<10^{0}$) &\small  -0.10 ($<10^{0}$) &\small  -0.59 ($<10^{-35}$) \\ \hline 
\small Transportation &\small  0.06 ($<10^{0}$) &\small  0.29 ($<10^{-6}$) &\small  0.11 ($<10^{0}$) \\ \hline 
\small Wear Common Protective or Safety Equipment such as Safety Shoes, Glasses, Gloves, Hearing Protection, Hard Hats, or Life Jackets &\small  0.06 ($<10^{0}$) &\small  0.52 ($<10^{-25}$) &\small  0.45 ($<10^{-18}$) \\ \hline 
\small On-the-Job Training &\small  0.06 ($<10^{0}$) &\small  -0.02 ($<10^{0}$) &\small  -0.10 ($<10^{0}$) \\ \hline 
\small Responsibility for Outcomes and Results &\small  0.07 ($<10^{0}$) &\small  0.27 ($<10^{-5}$) &\small  -0.17 ($<10^{-2}$) \\ \hline 
\small Work With Work Group or Team &\small  0.07 ($<10^{0}$) &\small  -0.11 ($<10^{0}$) &\small  -0.56 ($<10^{-31}$) \\ \hline 
\small Arm-Hand Steadiness &\small  0.07 ($<10^{0}$) &\small  0.30 ($<10^{-7}$) &\small  0.51 ($<10^{-24}$) \\ \hline 
\small Coordinate or Lead Others &\small  0.08 ($<10^{0}$) &\small  -0.04 ($<10^{0}$) &\small  -0.63 ($<10^{-41}$) \\ \hline 
\small Exposed to Contaminants &\small  0.09 ($<10^{0}$) &\small  0.50 ($<10^{-23}$) &\small  0.44 ($<10^{-17}$) \\ \hline 
\small Related Work Experience &\small  0.09 ($<10^{0}$) &\small  0.16 ($<10^{-1}$) &\small  -0.34 ($<10^{-9}$) \\ \hline 
\small Spatial Orientation &\small  0.10 ($<10^{0}$) &\small  0.42 ($<10^{-16}$) &\small  0.34 ($<10^{-9}$) \\ \hline 
\small Sounds, Noise Levels Are Distracting or Uncomfortable &\small  0.10 ($<10^{0}$) &\small  0.56 ($<10^{-30}$) &\small  0.34 ($<10^{-9}$) \\ \hline 
\small Wrist-Finger Speed &\small  0.11 ($<10^{0}$) &\small  0.37 ($<10^{-12}$) &\small  0.44 ($<10^{-17}$) \\ \hline 
\small Response Orientation &\small  0.11 ($<10^{0}$) &\small  0.45 ($<10^{-18}$) &\small  0.40 ($<10^{-14}$) \\ \hline 
\small Night Vision &\small  0.12 ($<10^{0}$) &\small  0.42 ($<10^{-15}$) &\small  0.33 ($<10^{-9}$) \\ \hline 
\small Cramped Work Space, Awkward Positions &\small  0.12 ($<10^{0}$) &\small  0.46 ($<10^{-19}$) &\small  0.37 ($<10^{-12}$) \\ \hline 
\small Spend Time Climbing Ladders, Scaffolds, or Poles &\small  0.12 ($<10^{0}$) &\small  0.47 ($<10^{-20}$) &\small  0.26 ($<10^{-5}$) \\ \hline 
\small On-Site or In-Plant Training &\small  0.12 ($<10^{0}$) &\small  -0.05 ($<10^{0}$) &\small  -0.25 ($<10^{-5}$) \\ \hline 
\small Operating Vehicles, Mechanized Devices, or Equipment &\small  0.13 ($<10^{0}$) &\small  0.49 ($<10^{-22}$) &\small  0.33 ($<10^{-9}$) \\ \hline 
\small Economics and Accounting &\small  0.13 ($<10^{-1}$) &\small  -0.21 ($<10^{-3}$) &\small  -0.57 ($<10^{-31}$) \\ \hline 
\small Sound Localization &\small  0.14 ($<10^{-1}$) &\small  0.49 ($<10^{-23}$) &\small  0.31 ($<10^{-7}$) \\ \hline 
\small Manual Dexterity &\small  0.16 ($<10^{-1}$) &\small  0.33 ($<10^{-9}$) &\small  0.44 ($<10^{-17}$) \\ \hline 
\small Peripheral Vision &\small  0.17 ($<10^{-1}$) &\small  0.46 ($<10^{-19}$) &\small  0.26 ($<10^{-5}$) \\ \hline 
\small Importance of Repeating Same Tasks &\small  0.19 ($<10^{-2}$) &\small  0.04 ($<10^{0}$) &\small  -0.08 ($<10^{0}$) \\ \hline 
\small Outdoors, Exposed to Weather &\small  0.20 ($<10^{-3}$) &\small  0.27 ($<10^{-5}$) &\small  -0.06 ($<10^{0}$) \\ \hline 
\small Service Orientation &\small  0.21 ($<10^{-3}$) &\small  -0.28 ($<10^{-6}$) &\small  -0.75 ($<10^{-68}$) \\ \hline 
\small Resolving Conflicts and Negotiating with Others &\small  0.23 ($<10^{-4}$) &\small  -0.14 ($<10^{-1}$) &\small  -0.75 ($<10^{-68}$) \\ \hline 
\small Social &\small  0.24 ($<10^{-4}$) &\small  -0.21 ($<10^{-3}$) &\small  -0.73 ($<10^{-61}$) \\ \hline 
\small Spend Time Making Repetitive Motions &\small  0.25 ($<10^{-5}$) &\small  0.17 ($<10^{-2}$) &\small  0.18 ($<10^{-2}$) \\ \hline 
\small Multilimb Coordination &\small  0.26 ($<10^{-5}$) &\small  0.42 ($<10^{-15}$) &\small  0.28 ($<10^{-6}$) \\ \hline 
\small Deal With External Customers &\small  0.26 ($<10^{-5}$) &\small  -0.30 ($<10^{-7}$) &\small  -0.66 ($<10^{-47}$) \\ \hline 
\small Time Sharing &\small  0.27 ($<10^{-5}$) &\small  0.00 ($<10^{0}$) &\small  -0.68 ($<10^{-50}$) \\ \hline 
\small In an Open Vehicle or Equipment &\small  0.27 ($<10^{-6}$) &\small  0.56 ($<10^{-30}$) &\small  0.09 ($<10^{0}$) \\ \hline 
\small Exposed to Whole Body Vibration &\small  0.28 ($<10^{-6}$) &\small  0.46 ($<10^{-19}$) &\small  -0.06 ($<10^{0}$) \\ \hline 
\small Customer and Personal Service &\small  0.28 ($<10^{-6}$) &\small  -0.22 ($<10^{-3}$) &\small  -0.74 ($<10^{-64}$) \\ \hline 
\small Performing for or Working Directly with the Public &\small  0.29 ($<10^{-7}$) &\small  -0.32 ($<10^{-8}$) &\small  -0.62 ($<10^{-40}$) \\ \hline 
\small Exposed to Disease or Infections &\small  0.32 ($<10^{-9}$) &\small  -0.10 ($<10^{0}$) &\small  -0.69 ($<10^{-53}$) \\ \hline 
\small Contact With Others &\small  0.33 ($<10^{-9}$) &\small  -0.12 ($<10^{0}$) &\small  -0.73 ($<10^{-62}$) \\ \hline 
\small Exposed to Minor Burns, Cuts, Bites, or Stings &\small  0.38 ($<10^{-12}$) &\small  0.56 ($<10^{-30}$) &\small  0.03 ($<10^{0}$) \\ \hline 
\small Handling and Moving Objects &\small  0.38 ($<10^{-12}$) &\small  0.43 ($<10^{-16}$) &\small  0.10 ($<10^{0}$) \\ \hline 
\small Assisting and Caring for Others &\small  0.38 ($<10^{-12}$) &\small  -0.05 ($<10^{0}$) &\small  -0.78 ($<10^{-77}$) \\ \hline 
\small Gross Body Equilibrium &\small  0.38 ($<10^{-13}$) &\small  0.49 ($<10^{-22}$) &\small  -0.02 ($<10^{0}$) \\ \hline 
\small Frequency of Conflict Situations &\small  0.38 ($<10^{-13}$) &\small  0.01 ($<10^{0}$) &\small  -0.84 ($<10^{-98}$) \\ \hline 
\small Deal With Physically Aggressive People &\small  0.39 ($<10^{-13}$) &\small  0.04 ($<10^{0}$) &\small  -0.73 ($<10^{-61}$) \\ \hline 
\small Explosive Strength &\small  0.39 ($<10^{-13}$) &\small  0.20 ($<10^{-3}$) &\small  -0.44 ($<10^{-17}$) \\ \hline 
\small Very Hot or Cold Temperatures &\small  0.40 ($<10^{-13}$) &\small  0.54 ($<10^{-27}$) &\small  -0.05 ($<10^{0}$) \\ \hline 
\small Enterprising &\small  0.40 ($<10^{-14}$) &\small  -0.13 ($<10^{0}$) &\small  -0.86 ($<10^{-109}$) \\ \hline 
\small First Interest High-Point &\small  0.40 ($<10^{-14}$) &\small  -0.13 ($<10^{-1}$) &\small  -0.89 ($<10^{-127}$) \\ \hline 
\small Performing General Physical Activities &\small  0.41 ($<10^{-15}$) &\small  0.48 ($<10^{-21}$) &\small  0.01 ($<10^{0}$) \\ \hline 
\small Spend Time Keeping or Regaining Balance &\small  0.45 ($<10^{-18}$) &\small  0.47 ($<10^{-20}$) &\small  -0.11 ($<10^{0}$) \\ \hline 
\small Static Strength &\small  0.46 ($<10^{-19}$) &\small  0.43 ($<10^{-16}$) &\small  -0.07 ($<10^{0}$) \\ \hline 
\small Physical Proximity &\small  0.47 ($<10^{-20}$) &\small  -0.03 ($<10^{0}$) &\small  -0.61 ($<10^{-38}$) \\ \hline 
\small Food Production &\small  0.47 ($<10^{-20}$) &\small  0.04 ($<10^{0}$) &\small  -0.73 ($<10^{-63}$) \\ \hline 
\small Dynamic Strength &\small  0.48 ($<10^{-21}$) &\small  0.45 ($<10^{-18}$) &\small  -0.10 ($<10^{0}$) \\ \hline 
\small Responsible for Others' Health and Safety &\small  0.48 ($<10^{-22}$) &\small  0.50 ($<10^{-23}$) &\small  -0.49 ($<10^{-22}$) \\ \hline 
\small Trunk Strength &\small  0.53 ($<10^{-26}$) &\small  0.41 ($<10^{-14}$) &\small  -0.21 ($<10^{-3}$) \\ \hline 
\small Deal With Unpleasant or Angry People &\small  0.53 ($<10^{-27}$) &\small  0.03 ($<10^{0}$) &\small  -0.83 ($<10^{-98}$) \\ \hline 
\small Spend Time Kneeling, Crouching, Stooping, or Crawling &\small  0.53 ($<10^{-27}$) &\small  0.44 ($<10^{-17}$) &\small  -0.36 ($<10^{-11}$) \\ \hline 
\small Dynamic Flexibility &\small  0.53 ($<10^{-27}$) &\small  0.25 ($<10^{-4}$) &\small  -0.53 ($<10^{-26}$) \\ \hline 
\small Conventional &\small  0.54 ($<10^{-28}$) &\small  0.07 ($<10^{0}$) &\small  -0.88 ($<10^{-120}$) \\ \hline 
\small Extent Flexibility &\small  0.55 ($<10^{-29}$) &\small  0.45 ($<10^{-19}$) &\small  -0.21 ($<10^{-3}$) \\ \hline 
\small Spend Time Bending or Twisting the Body &\small  0.57 ($<10^{-31}$) &\small  0.47 ($<10^{-20}$) &\small  -0.26 ($<10^{-5}$) \\ \hline 
\small Spend Time Standing &\small  0.57 ($<10^{-32}$) &\small  0.37 ($<10^{-11}$) &\small  -0.42 ($<10^{-15}$) \\ \hline 
\small Gross Body Coordination &\small  0.58 ($<10^{-33}$) &\small  0.41 ($<10^{-15}$) &\small  -0.36 ($<10^{-11}$) \\ \hline 
\small Speed of Limb Movement &\small  0.60 ($<10^{-37}$) &\small  0.46 ($<10^{-20}$) &\small  -0.37 ($<10^{-12}$) \\ \hline 
\small Stamina &\small  0.60 ($<10^{-37}$) &\small  0.38 ($<10^{-13}$) &\small  -0.44 ($<10^{-17}$) \\ \hline 
\small Second Interest High-Point &\small  0.65 ($<10^{-44}$) &\small  0.20 ($<10^{-3}$) &\small  -0.71 ($<10^{-58}$) \\ \hline 
\small Spend Time Walking and Running &\small  0.67 ($<10^{-49}$) &\small  0.33 ($<10^{-9}$) &\small  -0.66 ($<10^{-46}$) \\ \hline 

\end{longtabu}
\subsection{Skill Types}
	\label{aggSkills}
\indent We provide example \onet skills from each of ten skill types.
These groups of skills are obtained from the co-occurrence of skills across jobs.
The left column provides a subjective labelling for each skill type based on the skills comprising that cluster of skills.
\begin{longtabu} to \textwidth {|c|X|}
	\hline
			\bf Skill Types & \bf O*NET Skills \\ \hline
		Computational and Analytical Skills &
		\small Active Learning, Analyzing Data or Information, Communications and Media, 
		Complex Problem Solving, Computers and Electronics, Developing Objectives 
		 and Strategies, Documenting/Recording Information, Fluency of Ideas, 
		 Instructing, Interacting With Computers, Interpreting the Meaning of
		 Information for Others, Judgement and Decision Making, Learning Strategies,
		 Making Decisions and Solving Problems, Mathematical Reasoning, Memorization, 
		 Number Facility, Originality, Processing Information, Provide Consultation and 
		 Advice to Others, Systems Analysis, Systems Evaluation, Updating and Using 
		 Relevant Knowledge \\ \hline
		
		Physical Planning and Construction &
		 \small Building and Construction, Chemistry, Design, Drafting, Engineering and 				 Technology, Estimating the Quantifiable Characteristics of Products, Events, 
		 or Information, Explosive Strength, Far Vision, Installation, Perceptual Speed,
		 Physics, Production and Processing, Public Safety and Security,Transportation, Visualization \\ \hline
		
		Harmful Workspace Management Skills &
		\small Cramped Work Space, Awkward Positions, Dynamic Flexibility, Exposed to 							 Contaminants, Exposed to Hazardous Conditions, Exposed to High Places, 
		 Exposed to Minor Burns, Cuts, Bites, or Stings, Exposed to Whole Body Vibration,
		 Extremely Bright or Inadequate Lighting, Finger Dexterity, Outdoors, Exposed to 			Weather, Pace Determined by Speed of Equipment, Responsible for Others' Health 
		 and Safety, Sounds, Noise Levels Are Distracting or Uncomfortable, Spend Time 
		 Bending or Twisting the Body, Spend Time Climbing Ladders, Scaffolds, or Poles,
 		 Spend Time Keeping or Regaining Balance, Spend Time Kneeling, Crouching, 
 		 Stooping, or Crawling, Spend Time Making Repetitive Motions, Spend Time 
 		 Standing, Spend Time Using Your Hands to Handle, Control, or Feel Objects, 
 		 Tools, or Controls, Spend Time Walking and Running, Very Hot or Cold  					Temperatures, Wear Specialized Protective or Safety Equipment such as Breathing
 		 Apparatus, Safety Harness, Full Protection Suits, or Radiation Protection \\ \hline	
 		
 		Relational/Social Skills &
 		 \small Customer and Personal Service, Education and Training, Fine Arts, Foreign 
 		 Language, History and Archeology, Identifying Objects, Actions, and Events,
 		 Law and Government, Performing for or Working Directly with the Public,
		 Philosophy and Theology, Psychology, Resolving Conflicts and Negotiating with
		 Others , Sales and Marketing, Selling or Influencing Others, Sociology and
		 Anthropology, Therapy and Counseling, Training and Teaching Others \\ \hline
		
		Control and Perceptual Skills &
		 \small Auditory Attention, Depth Perception, Equipment Maintenance, Equipment 
		 Selection, Glare Sensitivity, Hearing Sensitivity, Inspecting Equipment, 
		 Structures, or Material, Mechanical, Night Vision, Operating Vehicles, 
		 Mechanized Devices, or Equipment, Operation Monitoring, Operation and 
		 Control, Peripheral Vision, Quality Control Analysis, Rate Control, Repairing,
		 Repairing and Maintaining Electronic Equipment, Repairing and Maintaining 
		 Mechanical Equipment, Response Orientation, Sound Localization, Spatial 
		 Orientation, Troubleshooting, Visual Color Discrimination \\ \hline
		
		Emergency Response &
		\small Consequence of Error, Contact With Others ,Coordinate or Lead Others,
		 Deal With External Customers, Deal With Physically Aggressive People,
		 Deal With Unpleasant or Angry People, Degree of Automation, Exposed to 
		 Disease or Infections, Exposed to Radiation, Frequency of Conflict Situations,
		 Importance of Being Exact or Accurate, Importance of Repeating Same Tasks, 
		 Indoors, Environmentally Controlled, Near Vision, Responsibility for Outcomes 
		 and Results, Selective Attention \\ \hline
		
		Basic skills &
		\small  Artistic, Assisting and Caring for Others, Biology, Conventional, Duration of 
		 Typical, Work Week, Electronic Mail, Enterprising, Face-to-Face Discussions,
		 First Interest High-Point, Flexibility of Closure, Food Production, Freedom to 
		 Make Decisions, Frequency of Decision Making, Impact of Decisions on 
		 Co-workers or Company Results,  In an Enclosed Vehicle or Equipment, 
		 In an Open Vehicle or Equipment, Indoors, Not Environmentally Controlled,
		 Investigative, Letters and Memos, Level of Competition, Medicine and 
		 Dentistry, Monitor Processes, Materials, or Surroundings, On-Site or In-Plant
		 Training, On-the-Job Training, Outdoors, Under Cover, Physical Proximity, 
		 Public Speaking, Realistic, Related Work Experience, Required Level of 
		 Education, Second Interest High-Point, Social, Structured versus 
		 Unstructured Work, Technology Design, Telecommunications, Telephone, 
		 Third Interest High-Point, Time Pressure, Time Sharing,Work Schedules, 
		 Work With Work Group or Team \\ \hline
		
		Organization Skills &
		 \small Active Listening, Category Flexibility, Clerical, Communicating with 
		 Persons Outside, Organization, Communicating with Supervisors, 
		 Peers, or Subordinates, Coordination, Critical Thinking, Deductive
		 Reasoning, English Language, Establishing and Maintaining		
		 Interpersonal Relationships, Getting Information, Inductive 
		 Reasoning, Information, Ordering, Monitoring, Negotiation,
		 Oral Comprehension, Oral Expression, Organizing, Planning, and 
		 Prioritizing Work, Performing Administrative Activities, Persuasion, 
		 Problem, Sensitivity, Reading Comprehension, Service Orientation, 
		 Social Perceptiveness, Speaking, Speech Clarity, Speech 
		 Recognition, Spend Time Sitting, Time Management, Writing,
		 Written Comprehension, Written Expression \\ \hline
		
		Management Skills &
		 \small Administration and Management, Coaching and Developing Others,
		 Coordinating the Work and Activities of Others, Developing and 
		 Building Teams, Economics and Accounting, Evaluating Information 
		 to Determine Compliance with Standards, Geography, Guiding, 
		 Directing, and Motivating Subordinates, Judging the Qualities 
		 of Things, Services, or People, Management of Financial 
		 Resources, Management of Material Resources, Management of 
		 Personnel Resources, Mathematics, Monitoring and Controlling 
		 Resources, Operations Analysis, Personnel and Human Resources,
		 Programming, Scheduling Work and Activities, Science, Speed of 
		 Closure, Staffing Organizational Units, Thinking Creatively \\ \hline
		
		Physical Coordination &
		 \small Arm-Hand Steadiness, Control Precision, Controlling Machines 
		 and Processes, Dynamic Strength, Exposed to Hazardous 
		 Equipment, Extent Flexibility, Gross Body Coordination, 
		 Gross Body Equilibrium, Handling and Moving Objects,
		 Manual Dexterity, Multilimb Coordination, Performing 
		 General Physical Activities, Reaction Time, Speed of 
		 Limb Movement, Stamina, Static Strength, Trunk Strength, 
		 Wear Common Protective or Safety Equipment such as 
		 Safety Shoes, Glasses, Gloves, Hearing Protection, 
		 Hard Hats, or Life Jackets, Wrist-Finger Speed \\ \hline

\end{longtabu}
%%%%%%%%%%%%%%
\end{document}